\documentclass[12pt,a4paper]{book}

\makeatletter
\def\l@subsubsection#1#2{}
\def\l@subsubsubsection#1#2{}
\makeatother
\usepackage[margin=2.5cm,includeheadfoot]{geometry}
\usepackage[onehalfspacing]{setspace}
\usepackage{fancyhdr}

\usepackage{etoolbox}
\patchcmd{\chapter}{\thispagestyle{plain}}{\thispagestyle{fancy}}{}{}
\usepackage{subfigure}
\usepackage{ccaption}
\usepackage{emptypage}
\usepackage{comment}
\usepackage{etoolbox}
\usepackage{graphicx,amssymb,amsmath,amsthm,amsfonts,epsfig,epsf}
\usepackage[usenames]{color}
\usepackage{epstopdf}
\definecolor{darkred}{rgb}{0.5,0,0}
\usepackage{cmsd}
\usepackage{array}
\usepackage{afterpage}
\usepackage{bm}
\usepackage{dcolumn}
\usepackage[utf8]{inputenc}
\usepackage{latexsym}
\usepackage{rotating}
\usepackage{longtable}

\usepackage{enumerate}
\usepackage{tensor,multirow}
\usepackage{url}
\usepackage{placeins}
\usepackage[linktocpage]{hyperref}
\usepackage{float}
\usepackage{graphicx}
\usepackage[titletoc]{appendix}
\usepackage[numbers]{natbib}
\usepackage{mathrsfs,euscript}
\usepackage{acronym}

\def\blankpage{%
      \clearpage%
      \thispagestyle{empty}%
      \null%
      \clearpage}

\makeatletter
\renewcommand{\@chapapp}{}
\newenvironment{chapquote}[2][2em]
  {\setlength{\@tempdima}{#1}%
   \def\chapquote@author{#2}%
   \parshape 1 \@tempdima \dimexpr\textwidth-2\@tempdima\relax%
   \itshape}
  {\par\normalfont\hfill--\ \chapquote@author\hspace*{\@tempdima}\par\bigskip}
\makeatother

\begin{document}
\thispagestyle{empty}
\begin{figure}
\includegraphics[width=0.3\textwidth]{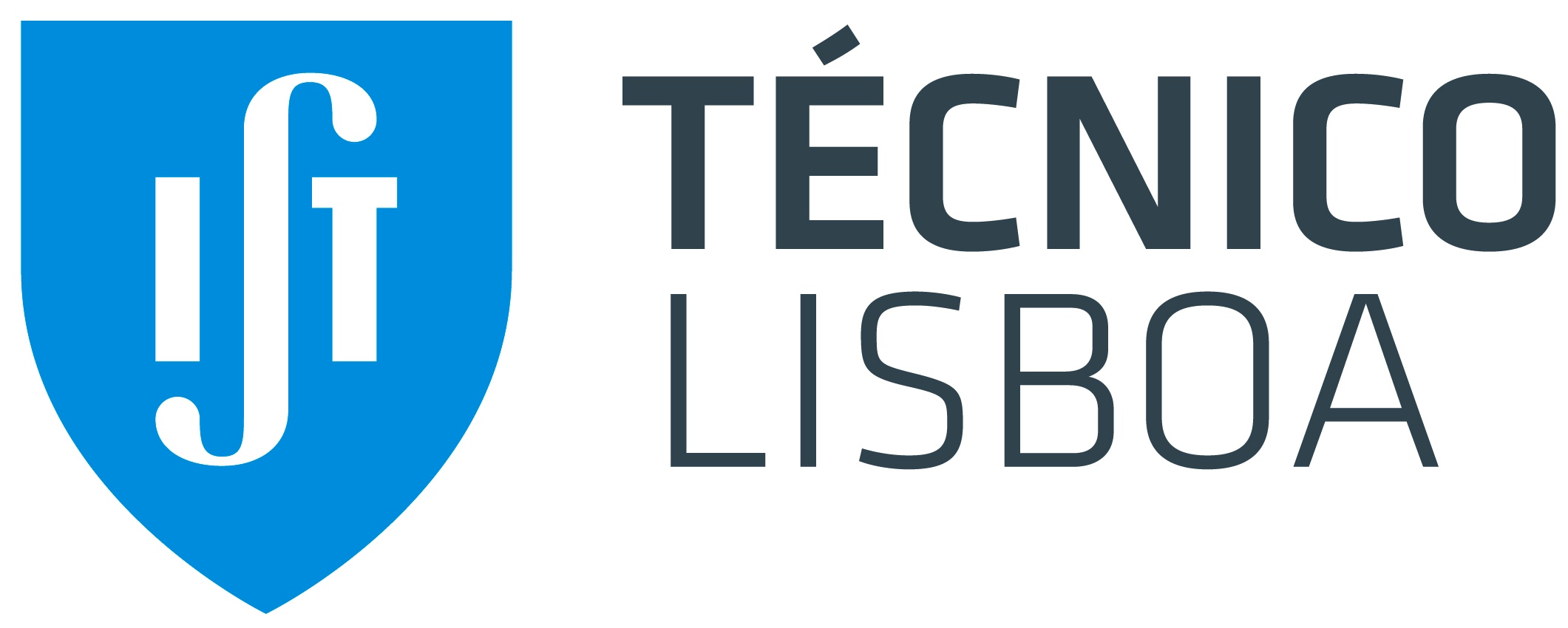}\\~\\
\end{figure}
\begin{center}
{\Large \bf UNIVERSIDADE DE LISBOA\vskip 1ex
INSTITUTO SUPERIOR T\'{E}CNICO}
\\~\\~\\~\\
{\bf \large Dynamical behavior of black-hole spacetimes}
\\~\\~\\~\\
{\bf \large Kyriakos Destounis}
\\~\\~\\~\\
{\bf Supervisor:} Doctor Vitor Manuel dos Santos Cardoso\\
{\bf Co-supervisor:} Doctor João Lopes Costa
\\~\\~\\~\\
{\bf Thesis approved in public session to obtain the PhD Degree in Physics}\\
{\bf Jury final classification: Pass with Distinction}
\\~\\~\\~\\~\\~\\~\\
{\bf 2019}
\\~\\~\\~\\
\end{center}
\thispagestyle{empty}
\begin{figure}
\vskip -7ex
\includegraphics[width=0.3\textwidth]{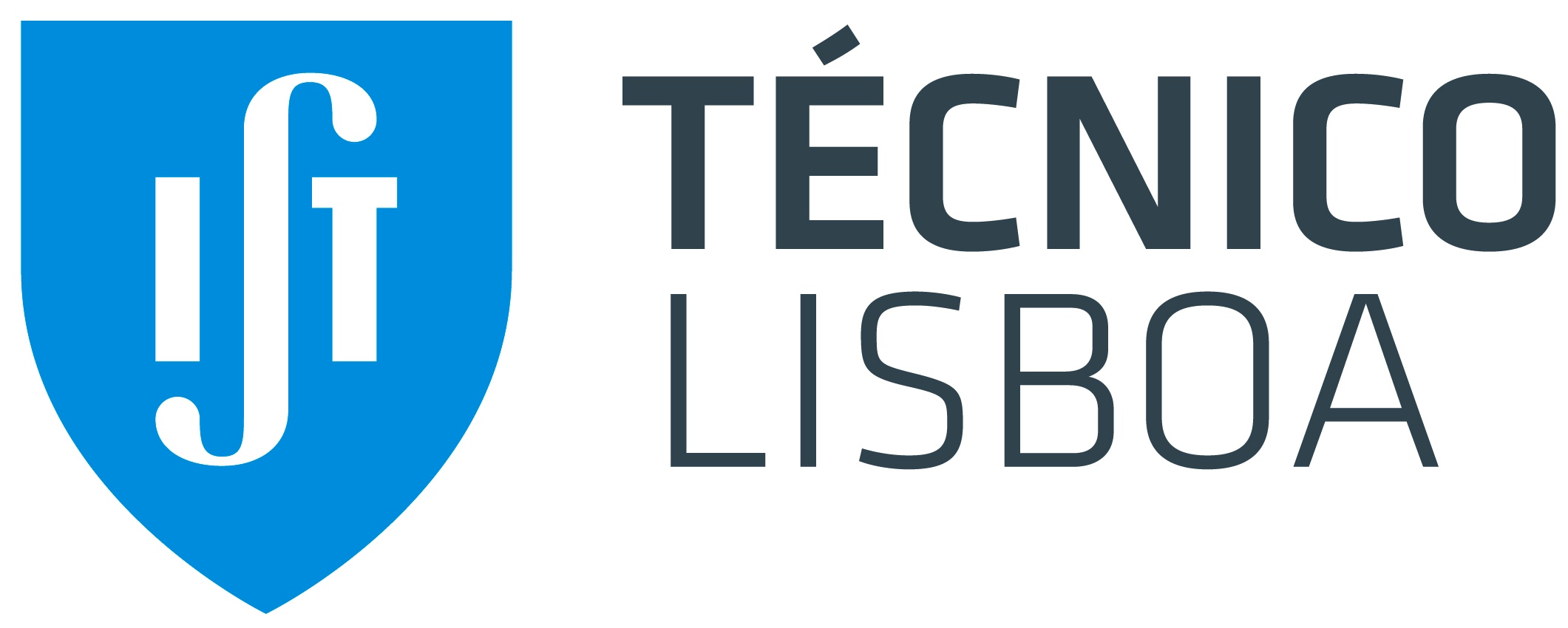}
\end{figure}
\begin{center}
\vskip -7ex{\Large \bf UNIVERSIDADE DE LISBOA\vskip 1ex
INSTITUTO SUPERIOR T\'{E}CNICO}
\\~\\
{\bf \large Dynamical behavior of black-hole spacetimes}
\\~\\
{\bf \large Kyriakos Destounis}
\\~\\
{\bf Supervisor:} Doctor Vitor Manuel dos Santos Cardoso\\
{\bf Co-supervisor:} Doctor João Lopes Costa
\\~\\
{\bf Thesis approved in public session to obtain the PhD Degree in Physics}\\
{\bf Jury final classification: Pass with Distinction}
\\~\\
{\bf Jury}\\
\begin{flushleft}
{\bf Chairperson:} \\Doctor Jos\'e Pizarro de Sande e Lemos, Instituto Superior T\'ecnico, Universidade de Lisboa\vskip 1ex 
{\bf Members of the Committee:}\\
Doctor Robertus Josephus Hendrikus Potting, Faculdade de Ci\^encias e Tecnologia, Universidade do Algarve \vskip 1ex
Doctor Vitor Manuel dos Santos Cardoso, Instituto Superior T\'ecnico, Universidade de Lisboa\vskip 1ex
Doctor Jos\'e Ant\'onio Maciel Nat\'ario, Instituto Superior T\'ecnico, Universidade de Lisboa\vskip 1ex
Doctor Caio Bezerra Macedo, Faculdade de F\'isica, Universidade Federal do Par\'a, Brasil\vskip 1ex
Doctor Aron Peter Jansen, Institut de Ci\`encies del Cosmos, Universitat de Barcelona, Espanha\\
\end{flushleft}
{\bf Funding Institution:}\\
H2020 ERC 
Consolidator Grant ``Matter and strong-field gravity: New frontiers in Einstein's theory'' grant 
agreement no. MaGRaTh--646597\\~\\
{\bf 2019}
\end{center}
\frontmatter
\chapter*{Resumo}
A teoria da Relatividade Geral prevê a existência de buracos negros como constituintes naturais do nosso Universo. Experiências recentes mostraram que os buracos negros não só existem, como interagem mutuamente, levando à emissão de energia gravitacional. Enquanto o binário de buracos negros relaxa para um estado final, o sistema oscila amortecidamente de tal forma que caracteriza completamente o objecto final. Estas oscilações são descritas por modos quasi-normais que, geralmente, estão presentes em sistemas dissipativos. A teoria de perturbações de buracos negros é um método crucial para investigar estes processos.

Os modos quasi-normais estão relacionados com a forma como um buraco negro responde a pequenas perturbações. A maioria dos buracos negros astrofísicos parece ser estável sobre pequenas perturbações, mas existem diversos casos em que um campo de perturbações pode crescer no tempo, em vez de decair. Estas instabilidades podem ter uma natureza superradiante, que se traduz numa amplificação da perturbação associada à diminuição da energia do buraco negro. A análise perturbativa da dinâmica de buracos negros é crucial em diversos contextos, desde a astrofísica até a aspectos mais fundamentais da Relatividade Geral.

Embora as equações de campo admitam um conhecimento completo da evolução futura do espaço-tempo, algumas soluções possuem uma fronteira, para lá da qual as equações de campo perdem o seu poder preditivo. Esta fronteira é chamada horizonte de Cauchy e define uma região na qual o espaço-tempo pode ser descrito de uma forma não-única. A existência de horizontes de Cauchy em soluções astrofísicamente relevantes das equações de campo pode ser uma ameaça à natureza preditiva das leis da física.

A censura cósmica forte serve, de alguma forma, para prevenir esses cenários ao afirmar que, genericamente, dados iniciais adequados são inextens\'iveis ao futuro para lá do horizonte de Cauchy como soluções das equações de campo. Estudos recentes indicam que o destino dos horizontes de Cauchy, como os encontrados no interior de buracos negros carregados ou em rotação, está intrinsecamente relacionado com o decaimento de perturbações no exterior do horizonte de eventos. Como tal, a validade da hipótese da censura cósmica forte depende de quão eficiente é a dissipação das fluctuações no exterior do horizonte de eventos.

Nesta tese vamos aplicar uma análise quantitativa à estabilidade dos horizontes de Cauchy em buracos negros electricamente carregados e imersos num Universo de de Sitter. Vamos mostrar que a censura cósmica forte é violada para buracos negros carregados quasi-extremos em de Siter, usando perturbações lineares. Para além disso, vamos investigar uma instabilidade superradiante para campos escalares carregados difundidos por buracos negros carregados em de Sitter a $d$-dimensões, e mostrar que o aumento do número de dimensões diminui a escala de tempo da instabilidade.
\\~\\
{\bf Palavras-chave:} Buracos negros, Modos quasi-normais, Horizontes de Cauchy, Censura cósmica forte, Instabilidades superradiantes, Dimensões mais altas
\chapter*{Abstract}
The theory of General Relativity predicts that black holes are a natural constituent of our Universe. Recent experiments revealed that black holes not only exist, but also interact with each other, leading to the emission of gravitational energy. Before the black hole binary relaxes to a final state, it undergoes damped oscillations which completely characterize the final object. These oscillations are described by quasinormal modes which are present in many dissipative systems. Black-hole perturbation theory is the key method to investigate such processes. 

Quasinormal modes are linked to the way a black hole spacetime responds to small fluctuations. Most astrophysical black holes appear to be stable under small perturbations but in various cases a perturbation field might not decay but rather grow in time. Such instabilities can have a superradiant nature which translates to the amplification of the perturbation in expense of the black hole's energy. A perturbative analysis of BH dynamics is crucial in several contexts, ranging from astrophysics to more fundamental aspects of General Relativity.
 
Although the field equations admit a well defined future evolution of spacetime, some solutions possess a boundary, beyond which the field equations lose their predictive power. This boundary is called a Cauchy horizon and it designates a region beyond which the spacetime can be described in a highly non-unique manner. The existence of Cauchy horizons in astrophysically relevant solutions of the field equations might pose a threat to the predictive nature of physical laws. 

Strong cosmic censorship states that, generically, suitable initial data are future inextendible beyond the Cauchy horizon, as a solution to the field equations. Recent studies indicate that the fate of Cauchy horizons, such as those found inside charged and rotating black holes, is intrinsically connected to the decay of perturbations exterior to the event horizon. As such, the validity of the strong cosmic censorship hypothesis is tied to how effectively the exterior damps fluctuations.

In this thesis we will perform a quantitative stability analysis of Cauchy horizons in electrically charged black holes immersed in a de Sitter Universe. We will provide strong evidence that strong cosmic censorship is violated for near-extremally charged de Sitter black holes under linear perturbations. Moreover, we shall investigate a superradiant instability of charged scalar fields scattering off a $d-$dimensional electrically charged-de Sitter black hole and show that the increment of dimensions decreases the timescale of the instability.
\\~\\
{\bf Key words:} Black holes, Quasinormal modes, Cauchy horizons, Strong cosmic censorship, Superradiant instabilities, Higher dimensions

\chapter*{Acknowledgements}
I am extremely grateful to my supervisor Vitor Cardoso and my co-supervisor Jo\~{a}o Costa for all the things they taught me the past 3 years. I am indebted for all the timeless hours they spend with me, explaining, collaborating, discussing and showing me how Physics is done. Most of our discussions led to successful collaborations and publications. Words cannot describe my gratitude, thanks Vitor and Jo\~{a}o!

I would also like to thank Aron Jansen, Peter Hintz, Lefteris Papantonopoulos, Grigoris Panotopoulos, George Pappas, Rodrigo Vicente and Mihalis Dafermos for very fruitful discussions and collaborations. I thank you all for brainstorming and collaborating with me in any way.

I am also grateful to all GRIT members, especially my office roommates, for the happy times we had together, the countless days we spent in the same room doing Physics and discussing about literally every single thing worth discussing (or not)!

I am more than grateful to my best friends during my stay in Lisbon, Lorenzo Annulli and Thanasis Giannakopoulos, for the amazing adventures we had together, the countless hours we spent exploring Lisbon and for the stories that will remain forever in my memory.

I want to warmly thank my family for encouraging me throughout my PhD years and giving me unconditional support. I know that without your physical and mental presence none of this would be possible. I love you all!

Fernanda, I don't know how to repay you for everything you have done for me throughout my PhD years. Thank you for being there all this time to listen to my nonsense and support me no matter what. It means the world to me! Half of this PhD is yours..

Finally, I would like to express my sincere gratitude to Ronnie James Dio, whose outstanding level of musical skills and godly voice was, is and will accompany my endeavors forever! May you rest in peace.
\tableofcontents
\chapter*{Preface}
\thispagestyle{empty}
The research included in this thesis has been carried out at the Center for Astrophysics and Gravitation (CENTRA) in the Physics department of Instituto Superior T\'{e}cnico. I declare that this thesis is not substantially the same as any that I have submitted for a degree or diploma or other qualification at any other university and that no part of it has already been or is being concurrently submitted for any such degree or diploma or any other qualification.

Chapters \ref{PRL}, \ref{PRD} are the outcome of collaborations with Prof. Vitor Cardoso, Prof. Jo\~{a}o Costa, Prof. Peter Hintz and Dr. Aron Jansen. Chapter \ref{JHEP} was done in collaboration with Prof. Lefteris Papantonopoulos, Prof. Bin Wang, Prof. Hongbao Zhang, Hang Liu and Ziyu Tang. All chapters are published on international scientific journals.

A list of the works included in this thesis are shown below:
\begin{itemize}
\item V. Cardoso, J. L. Costa, K. Destounis, P. Hintz, A. Jansen, ``Quasinormal modes and Strong Cosmic Censorship", {\it Phys. Rev. Lett. 120, 031103 (2018)}; \href{https://arxiv.org/abs/1711.10502}{1711.10502} (Chapter \ref{PRL})
\item V. Cardoso, J. L. Costa, K. Destounis, P. Hintz, A. Jansen, ``Strong Cosmic Censorship in charged black-hole spacetimes: still subtle", {\it Phys. Rev. D 98, 104007 (2018)}; \href{https://arxiv.org/abs/1808.03631}{1808.03631} (Chapter \ref{PRD})
\item K. Destounis, ``Charged Fermions and Strong Cosmic Censorship", {\it Physics Letters B 795 (2019)}; \href{https://arxiv.org/abs/1811.10629}{1811.10629} (Chapter \ref{PLB})
\item H. Liu, Z. Tang, K. Destounis, B. Wang, E. Papantonopoulos, H. Zhang, ``Strong Cosmic Censorship in higher-dimensional Reissner-Nordstr\"{o}m-de Sitter spacetime", {\it J. High Energ. Phys. (2019) 2019: 187}; \href{https://arxiv.org/abs/1902.01865}{1902.01865} (Chapter \ref{JHEP})
\item K. Destounis, ``Superradiant instability of charged scalar fields in higher-dimensional Reissner-Nordström-de Sitter black holes", {\it Phys.Rev. D100 (2019) no.4, 044054}; \href{https://arxiv.org/abs/1908.06117}{1908.06117} (Chapter \ref{higher instability})
\end{itemize}

Further publications by the author written during the development of this thesis but not discussed here are shown below:
\begin{itemize}
\item K. Destounis, G. Panotopoulos, \'{A}. Rincon, ``Stability under scalar perturbations and quasinormal modes of 4D Einstein-Born-Infeld dilaton spacetime: Exact spectrum", {\it Eur.Phys.J. C78 (2018) no.2, 139}; \href{https://arxiv.org/abs/1801.08955}{1801.08955}
\item K. Destounis, R. Fontana, F. Mena, E. Papantonopoulos, ``Strong Cosmic Censorship in Horndeski Theory", submitted to a scientific journal; \href{https://arxiv.org/abs/1908.09842}{1908.09842}
\end{itemize}
\thispagestyle{empty}
\blankpage
\vspace*{\fill} 
\thispagestyle{empty}
\begin{chapquote}{Robert John Arthur Halford}
\noindent``He said in the cosmos is a single sonic sound,\\
that is vibrating constantly.''
\end{chapquote}
\vspace*{\fill}
\blankpage
\thispagestyle{empty}
\noindent{\bf \Large Acronyms}\\
\begin{acronym}[MGHD]
\acro{GR}{General Relativity}
 \acro{BH}{Black hole}
 \acro{NS}{Neutron star}
 \acro{GW}{Gravitational wave}
 \acro{CH}{Cauchy horizon}
 \acro{LIGO}{Laser Interferometer Gravitational-Wave Observatory}
 \acro{QNM}{Quasinormal mode}
 \acro{SCC}{Strong Cosmic Censorship}
 \acro{WCC}{Weak Cosmic Censorship}
 \acro{dS}{de Sitter}
 \acro{AdS}{Anti-de Sitter}
 \acro{SdS}{Schwarzschild-de Sitter}
 \acro{RN}{Reissner-Nordstr\"{o}m}
 \acro{RNdS}{Reissner-Nordstr\"{o}m-de Sitter}
 \acro{KdS}{Kerr-de Sitter}
 \acro{WKB}{Wentzel-Kramers-Brillouin}
 \acro{MGHD}{Maximal globally hyperbolic development}
\end{acronym}
\mainmatter
\chapter{Introduction}\label{Chapter 1}

Einstein's theory of General Relativity (GR) is the most successful theory of gravitation \cite{Einstein:1913br,Einstein:1916vd}; it predicts and correctly describes black-hole (BH) spacetimes, gravitational-wave (GW) emission, cosmic expansion and many more phenomena \cite{Barack:2018yly}. It is undoubtedly a cornerstone of modern theoretical physics and astronomy. GR states that accelerating masses should produce GWs \cite{Einstein:1918btx,Eddington:1922ds}. GWs are ``ripples" in spacetime caused by some of the most violent and energetic processes in the Universe. Albert Einstein predicted the existence of GWs in 1916 and since then many tests have been conducted to understand if GR is the ideal theory which describes gravitation. Very recently, a considerable amount of effort has been dedicated to the detection of GWs of binary mergers. To be able to detect such events, though, the objects have to be extremely massive, and moving very quickly. 

For decades, scientists have hoped they could ``listen in" on violent astrophysical events by detecting their emission of GWs. The waves had never been observed directly until recently \cite{Abbott:2016blz}. In 2015, scientists reported that they detected such waves at the Laser Interferometer Gravitational-Wave Observatory (LIGO) in the United States, while the first GW signal at the Virgo interferometer, in Italy, was detected in 2017 \cite{Abbott:2017oio}. The waves were produced by two BHs that orbit each other and finally collide to merge into a single BH that ``shudders" a bit before settling down. GWs were emitted through all these stages. Several subsequent detections have occurred till date coming from BH \cite{Abbott:2016nmj,Abbott:2017gyy,Abbott:2017vtc} and neutron star (NS) binary mergers \cite{TheLIGOScientific:2017qsa}. With decades of hard work, theoretical predictions for these entire processes have been worked out in detail, following the fundamental rules of Einstein's theory. These predictions were used to translate the pattern of detected waves into an understanding of what produced them. The study that deals with the final stage of a binary merger is based on perturbation theory.

In GR, isolated BHs in equilibrium are simple objects; they are described by only a few parameters: their mass, angular momentum and charge \cite{hawking1972,Hawking:1973uf}. Such simplicity, however, is lacking in realistic scenarios due to the extraordinary complicated dynamics of gravitational collapse in BH formation. Additional complexities are introduced by active galactic nuclei, accretion disks, strong magnetic fields and other stars or planets around BHs.

Hence, a realistic dynamically evolving BH might never be fully described by only its basic parameters and can potentially be in a perturbed state indefinitely. Due to these technical and conceptual difficulties, one must almost always make approximations to generate physically interesting predictions. In many instances, one is interested in describing a BH spacetime as a superposition of the unperturbed background plus a very small perturbation. This approximation is based on perturbation theory. It applies in situations where the gravitational field is dominated by a known solution of the Einstein equations. Perturbation theory allows additional sources to be present in the problem as long as they are weak enough that they generate only a small correction to the overall geometry. 

Once a BH is perturbed, it responds by emitting GWs whose evolution in time is divided into three stages \cite{Gundlach1,Gundlach2}; an initial outburst of radiation, followed by a long period of damped proper oscillations dominated by quasinormal-mode (QNM) frequencies, and finally a late-time tails which suppress the oscillatory phase. Due to the dissipative nature of perturbed BH, their oscillatory phase cannot be analyzed with standard normal-mode theory, since the system is not time-symmetric. QNMs \cite{Kokkotas:1999bd,Berti:2009kk,Konoplya:2011qq}, thus, are solutions of a particular eigenvalue problem concerning perturbed BHs and describe the time-dependent proper oscillations that a BH undergoes. In general, quasinormal frequencies are complex, the real part being associated with the oscillatory frequency and the imaginary part with the decay timescale of the perturbation. The corresponding eigenfunctions are usually non-normalizable, and in general, they do not form a complete basis. Several real-world physical systems are dissipative, so one might reasonably expect QNMs to be omnipresent in physics.

A perturbative analysis of BH dynamics is crucial in several contexts, ranging from astrophysics to high-energy physics \cite{Pani:2013pma}. The stability analysis of BH spacetimes, BH ringdown after binary mergers, GW emission in astrophysical processes and even the gravity/gauge correspondences are just some noteworthy contexts in which BH perturbation theory is relevant. One of the most interesting phenomena resulting from the interaction between test fields and BHs are associated with BH superradiance \cite{Brito:2015oca}. This phenomenon takes place in scattering processes of waves off spinning or charged BHs. Under certain conditions, the test field can be superradiantly amplified at the expense of the BH rotational or electromagnetic energy \cite{Penrose:1971uk,Bekenstein:1973mi} which might lead to instabilities. Time evolutions of perturbations and QNM analyses can unravel such instabilities and predict the potential end-state of a nonlinearly developed spacetime including a back-reacting matter field. Besides instabilities of the external BH region, similar phenomena arise in BH interiors, such as those of charged and rotating BH spacetimes. In many cases, such instabilities are crucial on the examination of more fundamental aspects of GR.

The fate of an observer plunging into a BH is both an interesting and important problem within the framework of classical GR. If the observer falls into a neutral static BH his fate is clear: he will be crushed by the spacelike singularity lurking in the interior of the BH. The issue of infalling observers takes an unpredictable turn if the BH is charged or rotating; the observer's journey appears to continue unaffected through the interior of the BH and ``beyond", eventually emerging in ``another Universe", however, neither the geometry of this new region nor the fate of the observer can be determined uniquely by initial data. Such a journey would lead to a region where the observer will receive signals which do not emanate from initial data. The aforementioned scenario might be critical for determinism, which in the language of GR states that given suitable initial data on a spacelike hypersurface, the Einstein equations should predict, unambiguously, the evolution of spacetime. 

Therefore, we are all faced with a very fundamental problem: the loss of the deterministic nature of the field equations. It is well known that GR admits a well-posed initial value problem. For some situations though, the maximal development of suitable initial data can be extended in a highly non-unique way. 
Consequently, GR is sometimes unable to predict the global evolution of the spacetime. In such case, the boundary of the maximal development is called a Cauchy horizon (CH) and marks the division between the region where GR is able to forecast the evolution and the region where predictability of the field equations is lost.

BHs such as the (maximal analytic extension of) Reissner-Nordstr\"{o}m (RN) and Kerr solutions are known to have such horizons. For these BHs the inner horizon is a CH that hides the timelike singularity. In contrast with static BHs, the journey of the infalling observer beyond the event horizon of charged or rotating BHs seems, at first, eminently vague. In finite proper time the observer will reach and cross the CH unaffected. Since the maximal analytic extension of RN and Kerr is not the only possible extension, beyond this horizon, the observer's trajectory cannot be determined uniquely. 

Quite surprisingly, nature appears to abhor such a situation and conspires to prevent the passage beyond the CH. Since realistic BHs form through dynamical gravitational collapse, it is natural to assume that arbitrarily small time-dependent perturbations will propagate on the exterior and eventually plunge into the BH. These perturbations can be thought of as infalling radiation that scatters off the curvature potential of the BH and gets sucked inside the event horizon. The CH is expected to be unstable under such small perturbations, yielding a spacetime singularity which effectively seals off the tunnel to regions in which the field equations cease to make sense \cite{Simpson:1973ua}. The source of this instability is a blueshift effect which results from infinite proper time compression. Because in asymptotically flat spacetimes the causal past of the CH contains the entire Universe external to the BH, any observer approaching the CH sees an infinite amount of events in a finite proper time which leads to the blow-up of energy density \cite{Hartle}. Therefore, the instability turns the CH into a singular boundary and prevents the violation of predictability in asymptotically flat spacetimes. That this happens generically is the essence of strong cosmic censorship (SCC) \cite{Penrose69,1999math......1147C,Dafermos:2012np}, which states that solutions of the field equations that arise from proper initial data are future inextendible beyond the CH. Although asymptotically flat spacetimes seem to describe well the local structure of spacetime outside a collapsing star, if one wants to explore the mathematical aspects of GR in astrophysically appropriate scenarios it is mandatory to include a positive cosmological constant.

The most natural way of describing the accelerated expansion of the Universe is with a positive cosmological constant. According to the current standard model of cosmology \cite{Akrami:2018vks} the value of the cosmological constant $\Lambda$ is very small (of order $10^{-52}\, m^{-2}$) but positive. Hence, to appropriately consider our physical Universe an its constituents, we need to study astrophysical objects immersed in de Sitter (dS) space. Asymptotically dS BHs are part of a closed Universe where infinity lies beyond a cosmological horizon. Most intriguingly, the cosmological horizon is a redshift surface where perturbations decay exponentially fast \cite{Hintz:2016gwb,Hintz:2016jak}. This effect might compete and counter-balance the blueshift effect at the CH and, hence, perturbed charged and/or rotating BHs in dS spacetime might lead to possible counter-examples of SCC \cite{Costa:2014yha,Costa:2014zha,Costa:2014aia,Costa:2017tjc}.

Many studies have emerged so far concerning the classical stability of CHs in BH-dS spacetimes. Till this point, the most recent calculations indicated that the CH of Reissner-Nordstr\"{o}m-de Sitter (RNdS) BHs are unstable for the whole subextremal parameter space of the BH \cite{Brady:1998au}. The aforementioned studies are based on a faulty intuitive assumption that the exponential decay of perturbations on the exterior of dS BHs was governed by the minimum between the event and cosmological horizon's surface gravity \cite{Brady:1996za}. More recently, though, it was proven rigorously that the decay of perturbations in such spacetimes is, in fact, governed by the dominant QNM of the BH exterior \cite{Hintz:2016gwb,Hintz:2016jak}, where proper knowledge of these quantities was lacking. Such a fact has re-opened the issue of extendibility of solutions beyond the CH of asymptotically dS BHs and the validity of SCC.

Currently, the most modest way to tackle such an issue is with BH perturbation theory. The stability of CHs has proven to be intrinsically connected to how well the spacetime damps perturbations. In dS BHs with such horizons, a delicate competition between the damping rate and the blueshift amplification of small perturbations is inevitable \cite{Hintz:2015jkj}. Will the BH damp perturbations slowly enough leading to the blow-up of energy densities at the CH, or will it damp them so rapidly than the CH will remain stable enough as to allow the field equations to determine, in a highly non-unique manner, the evolution of gravitation and the SCC hypothesis to be violated? In this thesis, we will provide strong numerical evidence 

In this final introductory paragraph, the structure of the thesis is summarized. The thesis is divided in three parts. On the first part, we introduce the basic background needed to follow the upcoming research chapters. Topics such as the causal structure of BHs with CHs, perturbation theory, superradiance and SCC are covered in detail. In the second part of the thesis, we study the validity of the SCC conjecture in electrically charged-dS spacetimes at the linearized level. By using state-of-the-art numerical simulations, we probe a fixed RNdS background with neutral, charged and massive scalar fields, as well as charged fermionic fields. We demonstrate that the linearized analogue of SCC is violated for near-extremally charged RNdS BHs. We also extend the study of massless neutral scalar perturbations in higher-dimensional RNdS spacetimes and still find violation of SCC near extremality. In the third and final part of the thesis, a novel supperadiant instability of massive, charged scalar fields scattering off of a higher-dimensional RNdS BH is discovered and analyzed in detail. The aforementioned material is properly supported with appendices where we provide rigorous proofs of the CH stability condition for every configuration of perturbations discussed, we describe the master equations used to calculate the QNMs of such perturbations in spherically-symmetric spacetimes, and present a highly detailed analysis of the superradiant instability in RNdS spacetime. Throughout the thesis we will use geometrized units $c=G=1$.

\part{Black Holes, Quasinormal Modes and Strong Cosmic Censorship}
\chapter{Causal structure of black holes}\label{Chapter 2}
The description of the global geometry of the large scale distribution of matter in the Universe, as well as the local geometry of spacetime around a star, like our Sun, is a long-standing problem. Consider the Einstein-Hilbert action of a $4-$dimensional spacetime with a cosmological constant $\Lambda$,
\begin{equation}
\label{classical lagrangian1}
S=\int d^4x\sqrt{-g}\left(\frac{\mathcal{R}-2\Lambda}{16\pi}+\mathcal{L}_m\right),
\end{equation}
where $g=\det(g_{\mu\nu})$ is the determinant of the metric tensor $g_{\mu\nu}$, $\mathcal{R}$ is the Ricci scalar and $\mathcal{L}_m$ is the Lagrangian density representing the contribution of the matter fields minimally coupled to gravity. Varying \eqref{classical lagrangian1} with respect to the metric tensor yields the Einstein's field equations
\begin{equation}
\label{Einstein equations1}
G_{\mu\nu}+\Lambda g_{\mu\nu}=8\pi T_{\mu\nu},
\end{equation}
where $G_{\mu\nu}=\mathcal{R}_{\mu\nu}-\frac{1}{2}g_{\mu\nu}\mathcal{R}$ is the Einstein tensor, $\mathcal{R}_{\mu\nu}$ is the Ricci tensor and $T_{\mu\nu}$ is the energy-momentum tensor associated with the matter Lagrangian $\mathcal{L}_m$. Supplementing (\ref{Einstein equations1}) with the proper equations of motion for the matter fields leads to a complicated system of non-linear partial differential equations describing the evolution of all fields including the spacetime metric.
Some of the most astrophysically relevant solutions of \eqref{Einstein equations1} describe BHs with gravitational mass, electric charge and angular momentum. In this chapter, we review the causal structure of three exact solutions of Einstein equations; the Schwarzschild, RN and RNdS spacetimes. These metrics are extensively reported and utilized throughout this thesis. The following chapter is based on \cite{Hawking:1973uf,Wald:1984rg,Carroll:2004st}.

\section{The Schwarzschild solution}\label{section Schwarzschild}
The Schwarzschild solution \cite{Schwarzschild} represents, in view of Birkhoff's theorem, the most general spherically symmetric vacuum, with $\Lambda=0$. The metric can be given in the form
\begin{equation}
\label{Schwarzschild}
ds^2=-\left(1-\frac{2M}{r}\right)dt^2+\left(1-\frac{2M}{r}\right)^{-1}dr^2+r^2(d\theta^2+\sin^2\theta d\varphi^2),
\end{equation}
where $M$ is the Arnowitt-Deser-Misner (ADM) mass of the body as measured from infinity and $r>2M$. This spacetime is static and spherically symmetric. The solution is asymptotically flat as the metric has the form $g_{\mu\nu}=\eta_{\mu\nu}+\mathcal{O}(1/r)$ for large $r$, where $\eta_{\mu\nu}$ is the Minkowski metric tensor. It is natural to consider this metric for $r>R>2M$ as a solution outside a spherical body, with $R$ its radius, where the metric inside the body has a different form determined by the energy-momentum tensor of the matter in the body. However, as a BH solution, it is interesting to see what happens when the metric is regarded as an empty space solution for all $r$.

By inspection, \eqref{Schwarzschild} appears to be singular for $r=0$ and $r=2M$. A calculation shows that although this occurs at $r=2M$ in the Schwarzschild coordinates $(t,r,\theta,\varphi)$, no curvature invariants diverge there. This suggests that the ``singularity" at $r=2M$ is not a physical, but rather, a coordinate one. To confirm this, we define the tortoise coordinate
\begin{equation}
r_*=\int \frac{dr}{1-\frac{2M}{r}}=r+2M\log(r-2M).
\end{equation}
Then, $\upsilon\equiv t+r_*$ is the advanced null coordinate (or ingoing Eddington-Finkelstein coordinate) and $u\equiv t-r_*$ is the retarded null coordinate (or outgoing Eddington-Finkelstein coordinate). Using coordinates $(\upsilon,r,\theta,\varphi)$ the metric takes the ingoing Eddington-Finkelstein form
\begin{equation}
\label{ED1}
ds^2=-\left(1-\frac{2M}{r}\right)d\upsilon^2+2d\upsilon dr+r^2(d\theta^2+\sin^2\theta d\varphi^2).
\end{equation}
The metric \eqref{ED1} is non-singular at $r=2M$ and is analytic on a larger manifold for which $0<r<\infty$. Thus, by using different coordinates we can extend \eqref{Schwarzschild} so that it is no longer singular at $r=2M$. The same can be achieved by writing \eqref{Schwarzschild} in the coordinates $(u,r,\theta,\varphi)$, where
\begin{equation}
\label{ED2}
ds^2=-\left(1-\frac{2M}{r}\right)du^2-2du dr+r^2(d\theta^2+\sin^2\theta d\varphi^2).
\end{equation}
\begin{figure}[H]
\centering
\includegraphics[scale=0.7]{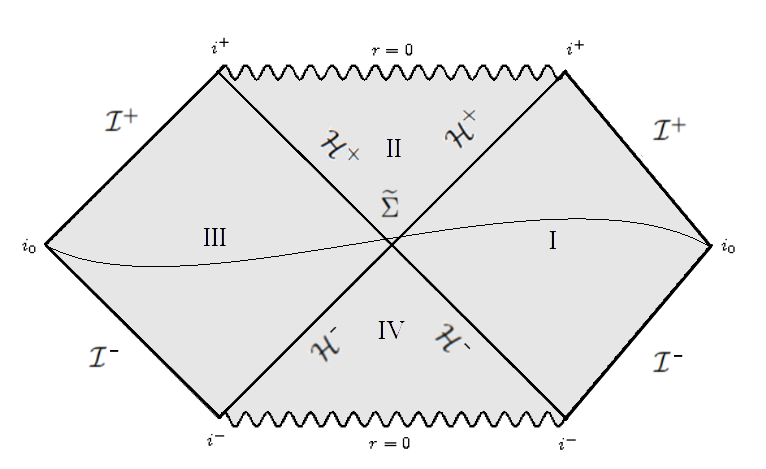}
\caption[1]{The Penrose-Carter diagram of the maximally extended analytic Schwarzschild solution. Null lines are at $\pm 45^\circ$. The diagram shows the future and past singularity $r=0$, the spacelike infinity $i_0$, the future and past timelike infinities $i^+$, $i^-$, the future and past null infinities $\mathcal{I}^+$, $\mathcal{I}^-$ and the future and past event horizon $\mathcal{H}^+$, $\mathcal{H}^-$, respectively. $\tilde{\Sigma}$ designates an initial Cauchy hypersurface.}
\label{Schwarzschild_diagram}
\end{figure}
The surface $r=2M$ is a null surface. As $r\rightarrow 0$ the scalar curvature diverges designating that $r=0$ is a curvature singularity. More importantly, beyond $r=0$ there are no continuous extensions of the metric and the field equations cease to make sense there \cite{Sbierski:2015nta}. This deserves to be called a terminal boundary. 

The Penrose-Carter diagram of the maximally extended Schwarzschild spacetime is shown in Fig. \ref{Schwarzschild_diagram}. There are no timelike or null curves which go from region I to the parallel region III. All future-directed timelike or null curves which cross the surface $r=2M$ $(\mathcal{H}^+)$ approach the future singularity at $r=0$, in region II. Similarly, past-directed timelike or null curves which cross $r=2M$ $(\mathcal{H}^-)$ approach the past singularity at $r=0$, in region IV.

If we consider the future light cone of any point outside of $r=2M$, the radial outwards curve reaches infinity but the inwards one reaches the future singularity; if the point lies in region II, both these curves hit the singularity and the entire future of the point ends at the singularity. Therefore, the singularity can be avoided only by observers outside $r=2M$. Thus, the surface $r=2M$ separates the spacetime in an external region where observers can travel to infinity and an internal region where observers are causally disconnected to infinity. This surface is called the event horizon and its radius is called the Schwarzschild radius. This spacetime is globally hyperbolic, meaning that it contains a Cauchy hypersurface $\tilde{\Sigma}$ such that every inextendible causal curve in the spacetime manifold intersects $\tilde{\Sigma}$ exactly once.

\section{The Reissner-Nordstr\"{o}m solution}\label{section RN}
The RN solution \cite{Reissner,Weyl,Nordstrom,Jeffery} represents the spacetime outside a spherically symmetric massive body, with $\Lambda=0$, carrying an electric charge. The energy momentum tensor is that of the electromagnetic field in the spacetime which results from the charge of the body. It is the only spherically symmetric asymptotically flat solution of the Einstein-Maxwell equations and is, locally, rather similar to the Schwarzschild solution. The metric in coordinates $(t,r,\theta,\varphi)$ has the form
\begin{equation}
\label{RN_metric}
ds^2=-\left(1-\frac{2M}{r}+\frac{Q^2}{r^2}\right)dt^2+\left(1-\frac{2M}{r}+\frac{Q^2}{r^2}\right)^{-1}dr^2+r^2(d\theta^2+\sin^2\theta d\varphi^2),
\end{equation}
where $M$ represents the ADM mass and $Q$ the electric charge of the object.
If $Q^2>M^2$ the metric is non-singular everywhere except for the naked singularity at $r=0$. We will not be interested in such parameters. If $Q^2< M^2$, the metric also appears to have ``singularities" at 
\begin{equation}
r_\pm=M\pm\sqrt{M^2-Q^2},
\end{equation}
where $r_+>r_-$, while if $Q^2=M^2$ then $r_+=r_-$ and we say that the BH is extremal. Following the logic of the previous section $r=r_+$ can be proven to be the event horizon of the BH and $r=r_-$ a horizon that lies inside the BH.  These coordinate singularities can be removed by introducing suitable coordinates and extending the manifold. Defining the tortoise coordinate
\begin{equation}
r_*=\int\frac{dr}{1-\frac{2M}{r}+\frac{Q^2}{r^2}},
\end{equation}
then for $r>r_+$
\begin{figure}[H]
\centering
\includegraphics[scale=0.8]{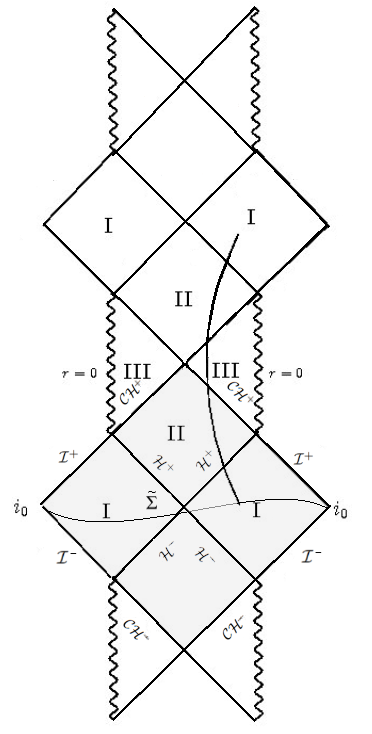}
\caption{The Penrose-Carter diagram of the maximally extended analytic Reissner-Nordstr\"{o}m solution for subextremal $Q^2<M^2$ parameters. Null lines are at $\pm 45^\circ$. The diagrams designate the future and past timelike singularities at $r=0$, the spacelike infinity $i^0$, the future and past lightlike infinities $\mathcal{I}^+$, $\mathcal{I}^-$, the future and past event horizon $r=r_+$ ($\mathcal{H}^+$, $\mathcal{H}^-$) and the future and past Cauchy horizon $r=r_-$ ($\mathcal{CH}^+$, $\mathcal{CH}^-$). $\tilde{\Sigma}$ designates an initial Cauchy hypersurface for the shaded region.}
\label{RN_diagram}
\end{figure}
\begin{align}
r_*&=r+\frac{r_+^2}{(r_+-r_-)}\log(r-r_+)-\frac{r_-^2}{r_+-r}\log(r-r_-), \,\,\,\,\,\,\,\quad\quad\quad\,\,\,\,\,\,\,\,\,\, \text{if}\,\,\,\,\,\, Q^2<M^2,\\
r_*&=r+M\log((r-M)^2)-\frac{2}{r-M},\,\,\,\,\,\,\,\quad\quad\quad\quad\quad\quad\quad\quad\quad\quad\quad\,\,\,\,\, \text{if}\,\,\,\,\,\, Q^2=M^2,\\
r_*&=r+M\log(r^2-2Mr+Q^2)+\frac{2}{Q^2-M^2}\arctan\left(\frac{r-M}{Q^2-M^2}\right),\,\, \text{if}\,\,\,\,\,\, Q^2>M^2.
\end{align}

By utilizing the advanced and retarded null coordinates $\upsilon, u,$ \eqref{RN_metric} can be written in $(\upsilon,r,\theta,\varphi)$ or $(u,r,\theta,\varphi)$ coordinates as in the Schwarzschild case.

The Penrose-Carter diagram of the maximally extended analytic RN metric is shown in Fig. \ref{RN_diagram}. Region I contains an infinite number of asymptotically flat regions where $r>r_+$. Region I is connected with regions II and III where $r_-<r<r_+$ and $0<r<r_-$, respectively. There is still an irremovable singularity at $r=0$ in region III, but unlike the Schwarzschild solution, it is timelike, thus it can be avoided by a timelike curve starting from region I, which crosses $r=r_+$ ($\mathcal{H}^+$) and $r=r_-$ ($\mathcal{CH^+}$). Such a curve can pass through regions II, III and II to emerge into another asymptotically flat region I. This raises the intriguing possibility that one might be able to travel to other Universes by passing through maximally extended analytic RN BHs.

The timelike character of the singularity also means that given any initial Cauchy hypersurface $\tilde{\Sigma}$, one can find past-directed causal curves in region III which do not intersect $\tilde{\Sigma}$. For example, the surface $\tilde{\Sigma}$ crosses two asymptotically flat regions I (see Fig. \ref{RN_diagram}). This is a Cauchy hypersurface for the shaded region. However, in the regions III, to the future, there are past-directed null causal curves which approach the singularity and do not cross the surface $r=r_-$, therefore, not intersecting with $\tilde{\Sigma}$ either. The surface with radius $r=r_-$ is, thus, said to be the future CH for $\tilde{\Sigma}$, that is, the boundary of the maximal globally hyperbolic development (MGHD) of initial data on $\tilde{\Sigma}$. The CH, therefore separates the spacetime in a region ($r_-<r<\infty$) where events emanate from initial data on $\tilde{\Sigma}$ and a region ($0<r<r_-$) where events do not emanate from initial data on $\tilde{\Sigma}$. The continuation of the solution beyond $r=r_-$ is, therefore, highly non-unique.

A maximal analytic extension of the RN solution is, of course, one of the many extensions one can obtain. Although it possesses timelike singularities and tunnels to other Universes, the Cauchy problem that GR poses states that by choosing appropriate initial data on $\tilde{\Sigma}$, the MGHD of the initial data 
is uniquely determined everywhere in the shaded region of Fig. \ref{RN_diagram}. The question that one should, therefore, ask is if the solutions are extendible beyond the CH. If an observer could smoothly cross the CH he/she would end up in a region where the spacetime is not uniquely determined by the initial data anymore. Hence, such a scenario will inevitably lead to the failure of the deterministic nature and the predictability of the field equations.

\begin{figure}[H]
\centering
\includegraphics[scale=0.55]{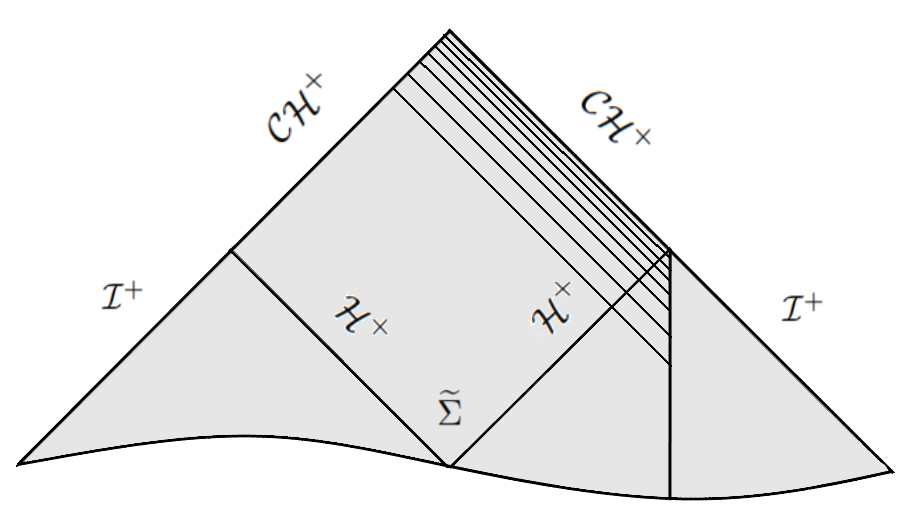}
\caption{An observer traveling to future timelike infinity is emitting an infinite amount of radiation into a RN BH. At the right $\mathcal{CH}^+$, an infinite proper time compression effect leads to the blow up of energy densities.}
\label{blueshift}
\end{figure}

Although the existence of CHs might lead to the loss of predictability, when observers cross them, a natural process first proposed by Penrose \cite{Simpson:1973ua}, may turn CHs unstable. One can consider an observer traveling to future timelike infinity emitting radiation with a constant frequency into the BH. Since the observer is living forever in the external region of the BH, he/she will send an infinite amount of radiation into the BH. On the other hand, a second observer plunging into the BH will reach the CH in finite proper time, while receiving the radiation from the external observer (see Fig. \ref{blueshift}). 
This will lead to an accumulation of energy as the observer reaches the CH and at the CH the energy density will eventually blow up. This phenomenon is called the blueshift effect and is based on infinite proper time compression.

Therefore, a hypothetical observer crossing the CH would see the entire history of one of the external Universes in finite proper time. Objects in this region would then appear to be infinitely blue-shifted. This suggests that the CH would be unstable against small perturbations in the initial data on a spacelike surface $\tilde{\Sigma}$, and that such perturbations would lead, in general, to singularities on $r=r_-$, where no extensions of the metric are physically possible \cite{Simpson:1973ua,Hartle}.

\section{The Reissner-Nordstr\"{o}m-de Sitter solution}\label{section RNdS}
\begin{figure}[H]
\centering
\includegraphics[scale=0.7]{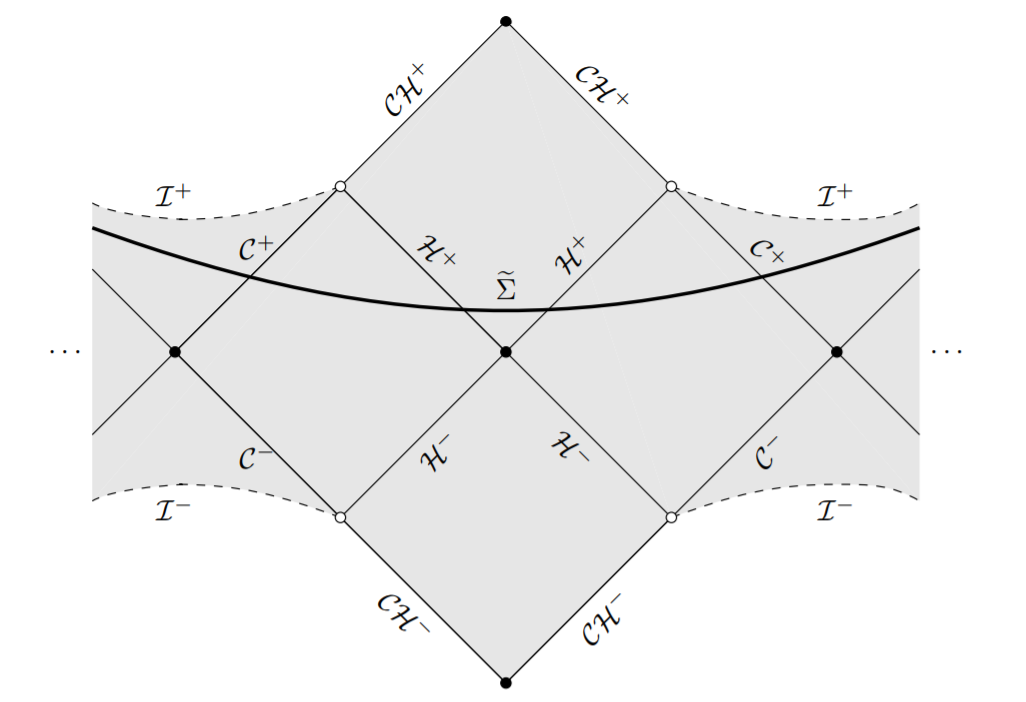}
\caption{The Penrose-Carter diagram of the extended subextremal Reissner-Nordstr\"{o}m-de Sitter solution. Null lines are at $\pm 45^\circ$. The diagram designates the future and past spacelike infinities $\mathcal{I}^+$, $\mathcal{I}^-$, the future and past event horizon $r=r_+$ ($\mathcal{H}^+$, $\mathcal{H}^-$), the future and past Cauchy horizon $r=r_-$ ($\mathcal{CH}^+$, $\mathcal{CH}^-$) and the future and past cosmological horizon $r=r_c$ ($\mathcal{C}^+$, $\mathcal{C}^+$). $\tilde{\Sigma}$ denotes an initial Cauchy hypersurface for the shaded region. The figure was taken from \cite{Dafermos:2018tha}.}
\label{RNdS_diagram}
\end{figure}

The generalization of the RN solution to include a cosmological constant was first provided by Carter \cite{Carter}. The RNdS spacetime is the solution to the Einstein-Maxwell field equations with $\Lambda>0$ and describes the asymptotically de Sitter spacetime outside an electrically charged massive body. In the coordinate system $(t,r,\theta,\varphi)$ it has the form
\begin{equation}
\label{RNdS_metric}
ds^2=-\left(1-\frac{2M}{r}+\frac{Q^2}{r^2}-\frac{\Lambda r^2}{3}\right)dt^2+\left(1-\frac{2M}{r}+\frac{Q^2}{r^2}-\frac{\Lambda r^2}{3}\right)^{-1}dr^2+r^2(d\theta^2+\sin^2\theta d\varphi^2),
\end{equation}
where $M$ is the ADM mass of the object, $Q$ the electric charge and $\Lambda$ the cosmological constant. The RNdS metric possesses three horizons which satisfy $r_-<r_+<r_c$. As in the previous cases, these are coordinate ``singularities" which can be removed by introducing the tortoise coordinate
\begin{equation}
r_*=\int\frac{dr}{1-\frac{2M}{r}+\frac{Q^2}{r^2}-\frac{\Lambda r^2}{3}}
\end{equation}
and utilize the standard advanced and retarded null coordinate $\upsilon$, $u$. Then, for $r>r_+$
\begin{equation}
r_*=\frac{1}{2\kappa_0}\log(r-r_0)-\frac{1}{2\kappa_-}\log(r-r_-)+\frac{1}{2\kappa_+}\log(r-r_+)-\frac{1}{2\kappa_c}\log(r_c-r).
\end{equation}
The constants $\kappa_i$ represent the surface gravity at the corresponding $r_i$ surface, where $r_0=-(r_-+r_++r_c)$. 
In the case of static, spherically symmetric solutions, the surface gravity is defines as \cite{Wald:1984rg}
\begin{equation}
\kappa_i=\frac{1}{2}\Big|\frac{df(r)}{dr}\Big|_{r=r_i},\,\,\,\,\,\, i\in\{0,-,+,c\},
\end{equation}
where $f(r)$ is the metric function $-g_{tt}$. The Penrose-Carter diagram of the standard RNdS metric is shown in Fig. \ref{RNdS_diagram}. We observe that the causal structure shares many similarities with the RN solution, that is, an event horizon at $r=r_+$ ($\mathcal{H}^\pm$), and a CH at $r=r_-$ ($\mathcal{CH}^\pm$). Interestingly, the existence of a positive cosmological constant gives rise to a cosmological horizon at $r=r_c$ ($\mathcal{C}^\pm$) and a cosmological region $r_c<r<\infty$. In this region, the spatial coordinates expand with faster-than-the-speed-of-light rates, therefore any radiation emerging from the cosmological region into the observable Universe is going to be infinitely red-shifted, hence it will be unobservable. Future infinity is now spacelike.

Although the external region to the BH is a truncation of the whole Universe, an observer can still ``live" forever by traveling to future timelike infinity. Therefore, the CH might still be an infinite blueshift surface. The difference between asymptotically flat and dS spacetimes, however, is that they damp perturbations in a different manner and dS spacetimes have a finite volume of radiation-emitting observers. As we will discuss in the next chapters, the ability of asymptotically dS BHs to damp perturbations exponentially fast might lead to stable enough CHs where the field equations may be satisfied weakly, and therefore, lead to a potential failure of SCC. 

\chapter{Linear perturbations of black holes}\label{Chapter 3}
QNMs appear naturally in the analysis of linear perturbations on fixed BH backgrounds. These fluctuations obey linear second-order differential equations, where the geometry of the background defines the properties of the equations. Usually, symmetries allow the separation of variables of the solutions to such equations, leading to linear ordinary differential equations. If the equations obey physically motivated boundary conditions then their eigenvalues will encode properties of the background. The boundary conditions are, usually, imposed at the event horizon and spatial infinity (or cosmological horizon for $\Lambda>0$), but the precise choice depends on the physical problem in study. Usually, we impose purely ingoing waves at the event horizon and purely outgoing waves at spatial infinity (or cosmological horizon for $\Lambda>0$).

The methods used to reduce equations to simple ordinary differential equations depend on the choice of the geometry and the symmetries it possesses. In this chapter, we review the basic method of BH perturbation theory which reduces the coupled non-linear equation describing the evolution of spacetime and matter into equations of motion of matter fields propagating in curved backgrounds. We discuss the choice of proper physically motivated boundary conditions and review all stages of the ringdown waveform. Finally, we discuss in depth the computational methods used throughout the thesis to extract QNMs from spherically-symmetric backgrounds. A more rigorous analysis of BH perturbation theory, QNMs and numerical methods can be found in  \cite{Kokkotas:1999bd,Berti:2009kk,Konoplya:2011qq,Pani:2013pma}.

\section{Perturbations of black-hole spacetimes}\label{section perturbations}
Consider the Einstein-Hilbert action \ref{classical lagrangian1}. Variations with respect to the metric tensor $g_{\mu\nu}$ yield the Einstein's field equation \eqref{Einstein equations1}. Supplementing (\ref{Einstein equations1}) with appropriate equations of motion for the remaining fields gives a system of non-linear partial differential equations describing the evolution of all fields including the spacetime metric. A particular solution of this system forms a set of background fields $g^\text{BG}_{\mu\nu}$, $A^\text{BG}_\mu$, $\Phi^\text{BG}$, where $A_\mu$ is the electromagnetic potential and $\Phi$ the matter field. If we introduce small perturbations in all fields then we can expand to first order
\begin{align}
\label{metric perturbations}
g_{\mu\nu}&=g^\text{BG}_{\mu\nu}+h_{\mu\nu},\\
\label{electromagnetic perturbations}
A_\mu&=A^\text{BG}_\mu+a_\mu,\\
\label{matter perturbations}
\Phi&=\Phi^\text{BG}+\Psi,
\end{align}
where the perturbation $h_{\mu\nu}$, $a_\mu$ and $\Psi$ are much smaller than the background solutions of the unperturbed fields. Therefore, by linearizing the full system of equations with respect to the perturbations, we obtain a set of linear differential equations satisfied by the perturbations. 

Typically one considers solutions of (\ref{Einstein equations1}) which are asymptotically flat, dS or AdS spacetimes. dS space is the maximally symmetric vacuum solution of Einstein's field equations with a positive cosmological constant ($\Lambda>0$) corresponding to a positive vacuum energy density and negative pressure. dS spacetimes are favorable in cosmology because they provide a simple model for the accelerated expansion of the Universe.  Hence, BHs in dS spacetime form a very interesting and physically motivated class of solutions which describe potential cosmological scenarios. 

Let us, first, demonstrate the decoupling of equations for scalar field perturbations. The action for a complex scalar field coupled with electromagnetism is given by 
\begin{equation}
S_m=\int d^4x\sqrt{-g}\mathcal{L}_m,\,\,\,\,\,\,\,\, \text{with}\,\,\,\,\,\, \mathcal{L}_m=-\frac{1}{2}g^{\mu\nu}\left(D_\mu\Phi\right)^\dagger D_\nu\Phi-\frac{1}{2}\mu^2\Phi^\dagger\Phi-\frac{1}{4}F^{\mu\nu}F_{\mu\nu},
\end{equation}
where $\mu$ is the mass of the scalar field, $F_{\mu\nu}=\nabla_\mu A_\nu -\nabla_\nu A_\mu$ is the electromagnetic tensor, $D_\mu\equiv \nabla_\mu-iqA_\mu$ is the covariant derivative associated with the electromagnetic potential $A_\mu$ and the electrostatic coupling constant $q$ and $\dagger$ is the complex conjugation operator. The equations of motion satisfied by the fields $g_{\mu\nu}$, $A_\mu$ and $\Phi$\footnote{A similar equation of motion is satisfied by $\Phi^\dagger$.} are
\begin{align}
\label{system}
G_{\mu\nu}+\Lambda g_{\mu\nu}&=8\pi T_{\mu\nu},\\
\nabla_\nu F^{\mu\nu}=q^2 \Phi^\dagger\Phi A^\nu&-i\frac{q}{2}(\Phi\nabla^\nu\Phi^\dagger-\Phi^\dagger\nabla^\nu\Phi),\\
\left(D^\mu D_\mu-\mu^2\right)\Phi&=0,
\end{align}
where the energy-momentum tensor is
\begin{align}
\nonumber
T_{\mu\nu}&=\frac{1}{4}\left(\left(D_\mu\Phi\right)^\dagger D_\nu\Phi+\left(D_\nu\Phi\right)^\dagger D_\mu\Phi\right)
-g_{\mu\nu}\left(\frac{1}{4}\left(D_\alpha\Phi\right)^\dagger D^\alpha\Phi+\frac{1}{4}\mu^2\Phi^\dagger\Phi\right)\\&+\frac{1}{2}F_{\mu{\alpha}}F_\nu^{\alpha}-\frac{1}{8}g_{\mu\nu}F^{\alpha\beta}F_{\alpha\beta},
\end{align}
Considering perturbations of the form (\ref{metric perturbations})-(\ref{matter perturbations}) and assuming that the scalar field introduces only a very small perturbation, we can set $\Phi^\text{BG}=0$. Consequently, we observe that the linearized equations of motion for $h_{\mu\nu}$, $a_\mu$ and $\Psi$ decouple, and thus the metric and electromagnetic fluctuations $h_{\mu\nu}$, $a_\mu$ can be consistently set to zero. The background satisfies
\begin{equation}
\label{linearized Einstein}
G_{\mu\nu}+\Lambda g_{\mu\nu}=8\pi\left(\frac{1}{2}F_{\mu{\alpha}}F_\nu^{\alpha}-\frac{1}{8}g_{\mu\nu}F^{\alpha\beta}F_{\alpha\beta}\right),\,\,\,\,\,\,\,\,\,\,\,\,\,\, \nabla_\nu F^{\mu\nu}=0,
\end{equation}
where $G_{\mu\nu}$, $g_{\mu\nu}$ and $F_{\mu\nu}$ depend on $g^\text{BG}_{\mu\nu}$, $A^\text{BG}_{\mu}$, and the scalar perturbation satisfies the Klein-Gordon equation
\begin{equation}
\label{KG}
\left(D^\mu D_\mu-\mu^2\right)\Psi=0,
\end{equation}
where the operator $D^\mu D_\mu$ depends on $g^\text{BG}_{\mu\nu}$, $A^\text{BG}_\mu$. 
Through similar procedures we can decouple the equations of motion of the spacetime metric and fermionic fields. 
The Lagrangian density of a fermionic field $\Phi$ coupled with electromagnetism is
\begin{equation}
\label{Dirac_lagrangian}
\mathcal{L}_m=\frac{i}{2}g^{\mu\nu}\left[\Phi^\dagger G^\mu \mathcal{D}_\nu\Phi-(\mathcal{D}_\nu\Phi)^\dagger G^\mu\Phi\right]-m_f\Phi^\dagger\Phi-\frac{1}{4}F^{\mu\nu}F_{\mu\nu},
\end{equation}
where $m_f$ is the mass parameter of the fermion, $G_\mu$ are the curved $\gamma-$matrices and $\mathcal{D}_\mu$ the covariant derivative
\begin{equation}
\mathcal{D}_\mu=D_\mu+\Gamma_\mu,
\end{equation}
with $\Gamma_\mu$ the spin connection coefficients (see Appendix \ref{appB}).
The energy momentum tensor is 
\begin{align}
\nonumber
T_{\mu\nu}&=\frac{i}{4}\left(\Psi^\dagger G^\mu (\mathcal{D}_\nu\Psi)+\Psi^\dagger G^\nu (\mathcal{D}_\mu\Psi)-(\mathcal{D}_\mu\Psi)^\dagger G^\nu\Psi-(\mathcal{D}_\nu\Psi)^\dagger G^\mu\Psi\right)\\&+\frac{1}{2}F_{\mu{\alpha}}F_\nu^{\alpha}-\frac{1}{8}g_{\mu\nu}F^{\alpha\beta}F_{\alpha\beta}.
\end{align}
Varying \eqref{Dirac_lagrangian} with respect to $\Phi$ leads to the Dirac equation in curved spacetime\footnote{A similar Dirac equation can be obtained by varying with respect to $\Phi^\dagger$.} \cite{Fock:1929vt,Chandra:1983}
\begin{equation}
\label{dirac equation_1}
(iG^\mu \mathcal{D}_\mu-m_f)\Phi=0.
\end{equation}
Using \eqref{metric perturbations}-\eqref{matter perturbations}, we can, again, decouple the system of coupled equations \eqref{system} to obtain
\begin{equation}
\label{dirac equation_2}
(iG^\mu \mathcal{D}_\mu-m_f)\Psi=0,
\end{equation}
where $\mathcal{D}_\mu$ and $G^\mu$ depend on $g_{\mu\nu}^\text{BG}$, $A^{\text{BG}}_\mu$.
A detailed analysis on Eqs. (\ref{KG}), (\ref{dirac equation_2}) and how to transform them into one-dimensional radial Sch\"{o}dinger-like equations can be found in Appendices \ref{appA} and \ref{appB}.

\section{Boundary conditions and quasinormal modes}\label{section QNMs}
To determine the oscillatory damped modes of a BH, which correspond to the eigenvalues of (\ref{KG}) or (\ref{dirac equation_2}), we first need to recast them into radial ordinary differential equations. Since we are going to be interested in spherically-symmetric spacetimes we can decompose $\Psi$ into a radial, an angular and a temporal part as follows:
\begin{equation}
\Psi=\sum_{l m}^{}{\psi(r)}\,Y_{l m}(\theta,\varphi)\,e^{-i\omega t},
\end{equation}
where $l,\,m$ are the angular the magnetic quantum numbers of the spherical harmonic $Y_{lm}$ and $\omega$ is a frequency.
Through proper manipulation (see Appendices \ref{appA} and \ref{appB}) we are usually able to recast many equations of motion (scalar or fermionic) into the following radial Schr\"{o}dinger-like form
\begin{equation}
\label{master}
\frac{d^2\psi(r)}{dr_*^2}+(\omega^2-V)\psi(r)=0,
\end{equation}
where $\psi(r)$ is the radial part of the perturbation $\Psi$, which generally depends on the coordinates of the spacetime, $V$ is the effective potential and $r_*$ is the tortoise coordinate. To determine and impose physically motivated boundary conditions we first examine the effective potential at the boundaries. For all studies ahead, we will be interested in asymptotically dS spacetimes which possess cosmological horizons. Therefore, the proper boundaries lie at the event ($r=r_+$) and cosmological horizon ($r=r_c$) of the spacetime in study. Thus, when
\begin{align}
r\rightarrow r_+,\,\,\,\,\,\, r_*\rightarrow-\infty,\,\,\,\,\,\,\,V\rightarrow 0,
\end{align}
hence the asymptotic solution to (\ref{master}) will have the general form $\Psi\sim e^{-i\omega(t\pm r_*)}$. Classically, anything that lies beyond the event horizon is causally disconnected with infinity, so the only accepted solution is purely ingoing waves there:
\begin{equation}
\label{bch}
\Psi\sim e^{-i\omega(t+ r_*)},\,\,\,\,\,\,r_*\rightarrow-\infty\,\,\,\,(r\rightarrow r_+).
\end{equation}
Similarly, when
\begin{align}
r\rightarrow r_c,\,\,\,\,\,\, r_*\rightarrow+\infty,\,\,\,\,\,\,\,V\rightarrow 0,
\end{align}
hence the asymptotic solution will have the general form $\Psi\sim e^{-i\omega(t\pm r_*)}$. Here, the only accepted wave solution is purely outgoing waves at the cosmological horizon; nothing can emerge from the cosmological region into the observable Universe , so:
\begin{equation}
\label{bcc}
\Psi\sim e^{-i\omega(t- r_*)},\,\,\,\,\,\,r_*\rightarrow+\infty\,\,\,\,(r\rightarrow r_c).
\end{equation}
By imposing the boundary conditions (\ref{bch}) and (\ref{bcc}) in (\ref{master}) we obtain a discrete set of eigenvalues $\omega$ which are called QNMs. The main difference between QNMs and other physical problems involving small fluctuations, such as a vibrating ideal string, is that the system in study is now dissipative, thus losing energy. This translates to waves escaping either at the event or cosmological horizon leading to damped time-dependent oscillations. The corresponding frequencies $\omega=\omega_R+i\omega_I$ usually have both real and imaginary parts which correspond to the frequency and damping rate of the mode, respectively. The modes are labeled with an integer $n$ called the overtone number. The fundamental mode $n=0$ corresponds to the lowest-lying dominant QNM and usually will dominate the ringdown waveform at late times. Astrophysical BHs are expected to be stable under linear perturbations, thus the modes must have a negative imaginary part for their amplitude to decay in time. If modes with positive imaginary parts appear as solutions to the eigenvalue problem then we say that the spacetime is modally unstable under linear perturbations.
 
\section{Quasinormal ringing and late-time tails}\label{section tails}
After years of analytical and numerical calculations, the physical picture of the behavior of radiative fields during gravitational collapse has become clear. The picture is similar for scalar, electromagnetic and gravitational radiation and it holds for non-rotating and rotating BHs. There are, generically, three distinct stages to the dynamical evolution of perturbations; initially, an outburst of radiation occurs which carries away energy through GW emission, later, the perturbation evolves in accordance with damped oscillations characteristic of the central BH. This behavior is associated with the QNM frequencies of the perturbing field, with the real part being linked to the frequency and the imaginary part to the damping rate of the emitted radiation. This stage has been extensively studied and is usually called quasinormal ringing. It has been proved that this stage's evolution through time does not depend on the initial configuration of the perturbation field. 
\begin{figure}[H]
\centering
\includegraphics[scale=0.6]{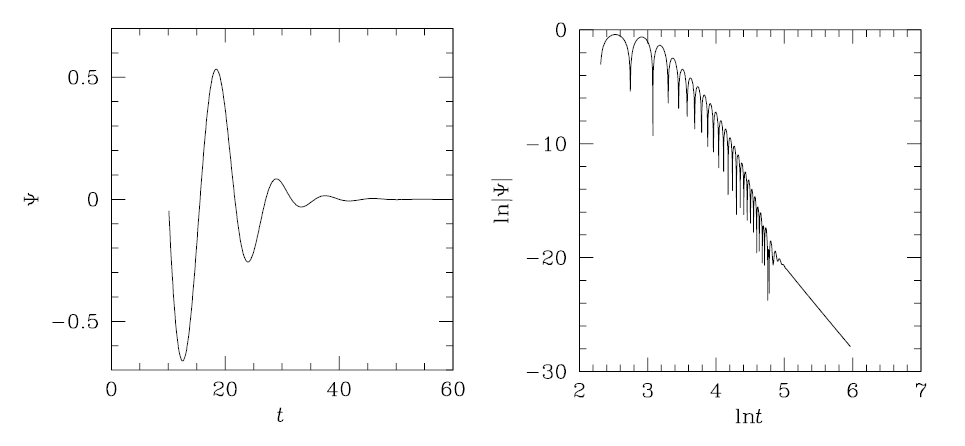}
\caption{Ringdown waveform of a gravitational Gaussian wave packet $\Psi$ propagating on a fixed Schwarzschild background. The figures were taken from \cite{Cardoso:2003pj}.}
\label{wave_packet}
\end{figure}

For asymptotically flat spacetimes, the quasinormal ringing gives way to the inverse power-law tail first described in \cite{Price1,Price2,Leaver:1986gd}. These investigations were extended to other BH solutions until the most complete picture to date was provided in a combination of papers \cite{Gundlach1,Gundlach2,Dafermos:2010hb,Dafermos:2014cua}. The behavior of radiation at fixed distances from the BH, at future null infinity and the event horizon was first described in \cite{Gundlach1,Gundlach2}. Moreover, numerical simulations of the collapse of a self-gravitating scalar field showed that the inverse power-law tails are a generic feature of radiative decay, even if a BH does not form. A definitive proof of boundedness and decay results for the wave equation in Kerr spacetime, without symmetry assumptions, for the general subextremal case was given in \cite{Dafermos:2010hb,Dafermos:2014cua}. Thus, if the radiative field $\Psi$ is observed at fixed radius $r$ and the field is static prior to collapse, then at $t\rightarrow\infty$
\begin{equation}
\label{flat_tail}
\Psi\sim t^{-(2l+2)}.
\end{equation}
In Fig. \ref{wave_packet}, the evolution of a gravitational perturbation on a fixed Schwarzschild background is presented.
From the left panel of Fig. \ref{wave_packet} it is clear that the gravitational radiation exhibits damped oscillations until it relaxes. On the right panel, the log-log plot reveals the distinct stages described above. The perturbations undergo a long stage of quasinormal ringing after the initial prompt response of radiation, where the QNMs dominate. After finite time, the ringing stage is suppressed, giving its turn to the inverse power-law tail which is depicted with straight lines in log-log scale. It has been argued that the nature of the tails in asymptotically flat spacetimes is primarily due to the power-law form of the effective potential as $r_*\rightarrow\infty$ \cite{Gundlach1,Gundlach2} and also due to backscattering off the background curvature \cite{Price1,Price2}. This is the universal picture of the dynamical evolution of perturbations in asymptotically flat spacetimes when physically motivated boundary conditions are imposed.

It is realized, from the above, that the asymptotic structure of spacetime dictates how the field behaves at late times. This observation begs the question of how the field's evolution will be affected if the conditions at infinity are altered. For example, what would happen when the BH is immersed in an expanding Universe? This question was first addressed in \cite{Brady:1996za,Mellor:1990,Brady:1999wd}. The dynamical evolution of a scalar field in Schwarzschild-de Sitter (SdS) and RNdS spacetime revealed that the field does not decay polynomially, but rather, exponential at late times. The numerical results in \cite{Brady:1996za} suggested, incorrectly, that a scalar field $\Psi$ at $t\rightarrow\infty$ will be compatible with the formula
\begin{equation}
\label{de Sitter tail}
\Psi\sim e^{-l\kappa_c t},
\end{equation}
where the decay rate $\kappa_c$ is the surface gravity of the cosmological horizon of the BH under consideration. The erroneous behavior described by \eqref{de Sitter tail} was derived by the assumption that the evolution of perturbations on a spacetime with a small cosmological constant is enough to infer for the whole subextremal parameter space of such spacetimes. 

In Chapters \ref{PRL}, \ref{PRD}, \ref{PLB} and \ref{JHEP} ( based on \cite{Cardoso:2017soq,PhysRevD.98.104007,Destounis:2018qnb,Liu:2019lon}) we will extensively shown that Eq. \eqref{de Sitter tail} is indeed incorrect. In fact, our findings indicate that there are three distinct families of QNMs in RNdS spacetimes which dominate the ringdown waveform, at late times, in different regions of the subextremal parameter space. The first one is directly linked to the photon sphere which dominates for sufficiently large cosmological constants and the second one's existence and timescale is related to the accelerated expansion of the Universe which dominates for small cosmological constants. The final family of modes appears as the event horizon approaches the CH and dominates the ringdown signal as the BH charge reaches extremality. 

The correct mathematical picture of the exponential decay of perturbations in subextremal asymptotically dS charged and rotating BHs was given recently in  \cite{Hintz:2016gwb,Hintz:2016jak}. There, it was rigorously proven that for some $\Psi_0\in \mathbb{C}$,
\begin{equation}
\label{exponential_tail}
|\Psi-\Psi_0|\leq C e^{-\alpha t},
\end{equation}
where $\alpha$ is the spectral gap, i.e. the QNM-free strip below the real axis. The spectral gap is the smallest  (in absolute value) non-zero imaginary part of all QNMs, that is
\begin{equation}
\alpha\equiv \text{inf}_{ln}\{-\text{Im}(\omega_{ln})\},\,\,\,\,\omega\neq 0.
\end{equation} 
Thus, \eqref{exponential_tail} describes, in general, the exponential late time behavior of the waveform for all subextremal parameters of asymptotically dS charged and rotating BH spacetimes. Eq. \eqref{exponential_tail} implies that QNMs always dominate the asymptotic behavior of perturbations in dS BHs and is completely compatible with our findings.

The presence, and slow decay, of tails at late times in asymptotically flat BHs is a key ingredient for the instability of CHs inside charged and rotating BHs. In contrast with the asymptotically flat case, the exponential decay at late times in asymptotically dS BHs is an indication that the CH might be stable for a finite volume of the parameter space of charged and rotating BHs immersed in a Universe with a positive cosmological constant. 

Therefore, the nature of the late time behavior of perturbations plays a major role on the study of CH stability. Furthermore, the novel results in \cite{Hintz:2016gwb,Hintz:2016jak} indicate that the late time behavior of perturbations in such spacetimes is governed by the QNMs of the BH and thus gives us the opportunity to conduct, for the first time, a quantitative analysis of CH stability at the linearized level by appropriately calculating $\alpha$ numerically.  

In Chapters \ref{PRL}, \ref{PRD}, \ref{PLB} and \ref{JHEP} we will present the outcome of quantitative analyses of CH stability in RNdS spacetime. These results put directly into question the validity of SCC.

\section{Computational methods of quasinormal modes}\label{setion methods}
BH perturbation theory is a useful tool to investigate various issues in astrophysics, high-energy theory and fundamental aspects of gravitation. Several modern applications require advanced numerical tools to study the linear dynamics of generic small perturbations propagating on stationary backgrounds. Such analyses are often complementary to nonlinear evolutions and serve as interpreters of dynamic numerical simulations of realistic scenarios. In this section, we give an overview of the numerical methods which are regularly utilized throughout this thesis to solve the eigenvalue problem in curved spacetime and extract the QNMs. For a more complete review of the numerical methods in perturbation theory see \cite{Berti:2009kk,Pani:2013pma}.

\subsection{The WKB approximation} One of the most widely used and tested method to calculate QNMs on fixed BH backgrounds is the Wentzel-Kramers-Brillouin (WKB) approximation. BH QNMs can be thought of as waves traveling around the BH, and more precisely, some families of QNMs have been connected with null particles trapped at the unstable circular null geodesic orbit (the photon sphere) \cite{Cardoso:2008bp}. When the spacetime is perturbed the unstable null particles leak out from the photon sphere, thus giving out their preferred oscillatory states. The instability timescale of null geodesics are connected to the decay timescale of QNMs, while the oscillation frequency is connected with the radius of the photon sphere \cite{Ferrari:1984zz,Cardoso:2008bp}.

The outstanding work of \cite{Schutz:1985zz} resulted in a derivation closely parallel to the Bohr-Sommerfeld quantization rule from quantum mechanics. The QNMs are calculated semi-analytically, using the WKB approximation. Although it is an approximation, this approach is powerful due to its ability to be carried out in higher orders, to improve the accuracy and estimate the errors explicitly.

The motivation for using the WKB approximation is the similarity between the equation of BH perturbation theory and the one-dimensional Schr\"{o}dinger equation which describes a particle encountering a potential barrier. In both cases, the master equation has the form (similar to \eqref{master})
\begin{equation}
\label{Schrodinger like}
\frac{d^2\psi}{dx^2}+Q(x,\omega)\psi=0.
\end{equation}

In the BH case, $\psi$ represents the radial part of the perturbation, assumed to have harmonic time dependence $e^{-i\omega t}$ and angular dependence $Y(\theta,\varphi)$ appropriate to the particular perturbation and the BH background under study. The coordinate $x$ is called the tortoise coordinate (also designated as $r_*$) which ranges from $-\infty$ (at the event horizon) to $+\infty$ (at spatial infinity or cosmological horizon). The function $-Q(x)$ is constant at the boundaries and possess a maximum at $x=x_0$. $Q(x)$ depends on the BH parameters, the angular harmonic indices and the frequency $\omega$.

Since $Q(x)$ tends to a constant for $|x|\rightarrow\infty$, $\psi\sim e^{\pm i\omega x}$, where $\text{Re}(\omega)>0$. As $x\rightarrow +\infty$, outgoing (ingoing) waves correspond to the negative (positive) sign, and as $x\rightarrow -\infty$, outgoing (ingoing) waves correspond to the positive (negative) sign. 
Here, ``outgoing" means moving away from the potential barrier, so in the BH case, ``outgoing at $x\rightarrow -\infty$" corresponds to waves falling into the BH event horizon.
The search for a solution of the normal mode problem involves matching two WKB solutions across both of the turning points $x_1,\,x_2$, simultaneously. 
\begin{figure}[H]
\centering
\includegraphics[scale=0.4]{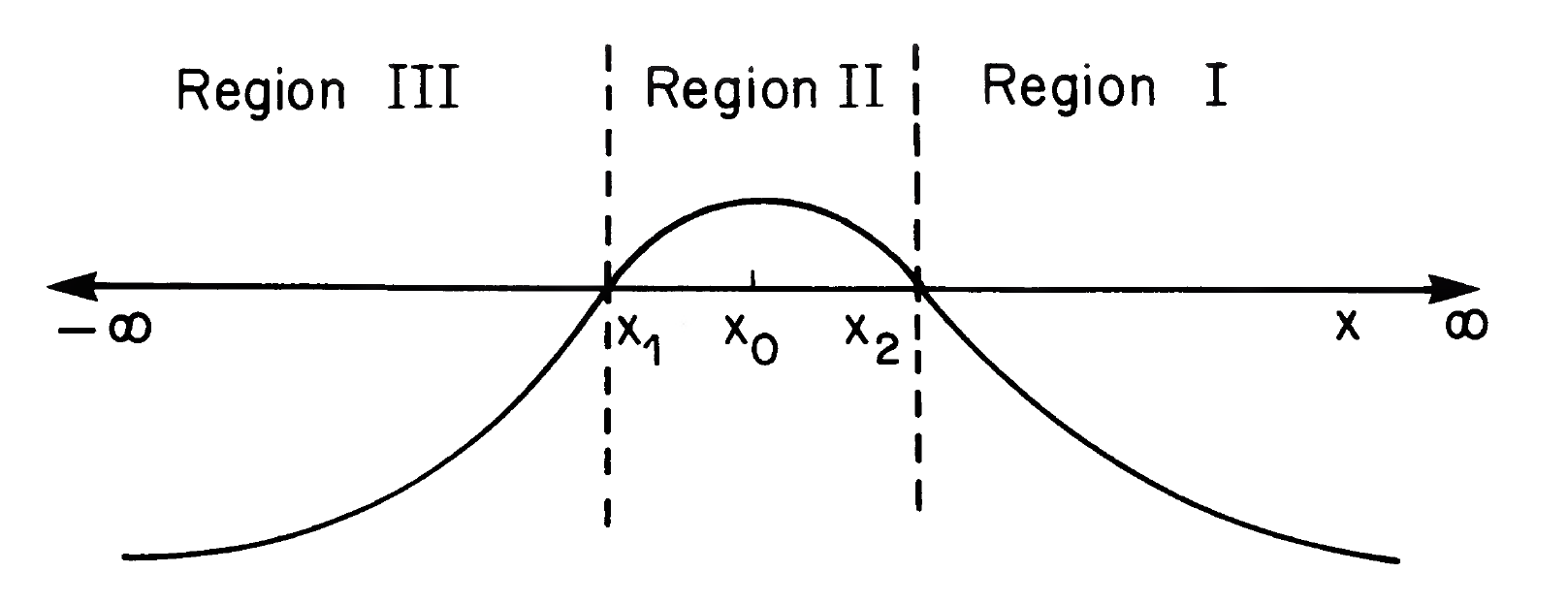}
\caption{Plot of the function -$Q(x)$. The figure was taken from \cite{Schutz:1985zz}}
\label{Q}
\end{figure}
Outside the turning points (regions I and III in Fig. \ref{Q}), the WKB solutions are given by \cite{Abramowitz,bender}
\begin{align}
\label{region I}
\psi_\text{I}(x)&\approx Q^{-1/4}e^{\pm i\int_{x_2}^{x}\sqrt{Q(t)}dt},\\
\label{region II}
\psi_\text{III}(x)&\approx Q^{-1/4}e^{\pm i\int_{x}^{x_1}\sqrt{Q(t)}dt}.
\end{align}
In region II we approximate $Q(x)$ by a parabola. This is justified provided that the turning points $x_1$, $x_2$ are closely spaced. Then $Q(x)$ has the form
\begin{equation}
\label{pot}
Q(x)=Q_0+\frac{1}{2}Q^{\prime\prime}_0(x-x_0)^2+\mathcal{O}(x-x_0)^3,
\end{equation}
where $Q_0=Q(x_0)<0$, and $Q^{\prime\prime}_0\equiv d^2Q/dx^2|_{x_0}>0$. Using
\begin{equation}
k\equiv \frac{1}{2}Q^{\prime\prime}_0,\,\,\,\,\, t\equiv (4k)^{1/4}e^{i\pi/4}(x-x_0),\,\,\,\,\, n+\frac{1}{2}\equiv-\frac{iQ_0}{\sqrt{2Q^{\prime\prime}_0}},
\end{equation}
brings \eqref{Schrodinger like} to the appropriate form
\begin{equation}
\frac{d^2\psi}{dt^2}+\left(n+\frac{1}{2}-\frac{1}{4}t^2 \right)\psi=0,
\end{equation}
whose solutions are parabolic cylinder functions $D_n(t)$ \cite{Abramowitz,bender}, with the general solution given by
\begin{equation}
\label{parabolic cylindrical}
\psi=A D_n(t)+B D_{-n-1}(it).
\end{equation}
For large $|t|$ the asymptotic forms of \eqref{parabolic cylindrical} yield \cite{Abramowitz,bender}
\begin{align}
\nonumber
\psi&\approx B e^{-3i\pi (n+1)/4}(4k)^{-(n+1)/4}(x-x_0)^{-(n+1)}e^{i\sqrt{k}(x-x_0)^2/2}\\
&+\left[A+\frac{B\sqrt{2\pi}e^{-in\pi /2}}{\Gamma(n+1)}\right]e^{i\pi n/4}(4k)^{n/4}(x-x_0)^n e^{-i\sqrt{k}(x-x_0)^2/2},
\end{align}
for $x\gg x_2$ and
\begin{align}
\nonumber
\psi&\approx A e^{-3i\pi n/4}(4k)^{n/4}(x_0-x)^{n}e^{-i\sqrt{k}(x-x_0)^2/2}\\
&+\left[B-i\frac{A\sqrt{2\pi}e^{-in\pi /2}}{\Gamma(-n)}\right]e^{i\pi (n+1)/4}(4k)^{-(n+1)/4}(x_0-x)^{-(n+1)} e^{i\sqrt{k}(x-x_0)^2/2},
\end{align}
for $x\ll x_1$, where $\Gamma(n)$ is the gamma function. It is straightforward to show that the $e^{-i\sqrt{k}(x-x_0)^2/2}$ factors of both solutions match the ``outgoing" waves of the WKB solutions \eqref{region I} and \eqref{region II}. By demanding only ``outgoing" waves, the coefficients of $e^{i\sqrt{k}(x-x_0)^2/2}$ factors must vanish. This is achieved if $B=0$ and $\Gamma(-n)=\infty$. The latter conditions implies that $n$ must be an integer. This leads to the simple condition for the QNMs
\begin{equation}
\label{WKB}
\frac{Q_0}{\sqrt{2Q^{\prime\prime}_0}}=i\left(n+\frac{1}{2}\right), \,\,\,\,\,\,\,\,\,\,\, n=0,1,2,\dots.
\end{equation}
Since $Q$ is frequency dependent, this condition leads to discrete complex values for the QN frequencies. This result is completely general and can be applied to any one-dimensional potential barrier problem.

Eq. \eqref{WKB} was first utilized in \cite{WKBI,WKBII,WKBIII,WKBIV} where a third-order WKB expansion was introduced and tested for Schwarzschild, RN and Kerr geometries. In \cite{Konoplya:2003ii}, the expansion was extended to sixth order. Although there is no proof of convergence, results improve for higher order WKB approximations. The WKB approximation works best for low overtone numbers $n$ and in the eikonal limit $l\rightarrow\infty$ (which corresponds to large $\omega_R/\omega_I$). The method assumes that the potential has a single extremum, which is the case for many BH effective potentials.
\\~\\~\\~\\
\subsection{A matrix method for eigenvalue problems}
There are plenty of numerical computation techniques to solve eigenvalue problems of differential equations. Some of the most powerful computational schemes arise by transforming the Schr\"{o}dinger-like equation into a matrix equation. The basis of these methods is the proper manipulation of the ordinary differential equation that governs perturbation propagation through proper decomposition of the spatial derivatives. The equation is discretized and transformed into a matrix equation which can be recast in terms of eigenvalues and eigenvectors. One of the methods utilized in this thesis is based on a non grid-based interpolation scheme, used in \cite{KaiLin1}. 

This method makes use of data points in a small region of a query point to estimate its derivatives by employing a Taylor expansion. The data points can be scattered, therefore they do not sit on a specified grid.  A key step of the method is to discretize the unknown eigenfunction in order to transform the differential equation and its boundary conditions into a homogeneous matrix equation. Based on the information about $N$ scattered data points, Taylor series are carried out for the unknown eigenfunction up to $N-$th order for each data point. The resulting homogeneous system of linear algebraic equations is solved for the eigenvalues. A huge advantage of this method is that the discretization of the wave function and its derivatives are made to be independent of any specific metric through well-motivated coordinate transformations.

The Taylor series for a univariate function $f(x)$ around a query point $x_0$ is expressed as:
\begin{equation}
\label{taylor}
f(x)=f(x_0)+(x-x_0)f^\prime(x_0)+\frac{1}{2!}(x-x_0)^2f^{\prime\prime}(x_0)+\frac{1}{3!}(x-x_0)^3 f^{\prime\prime\prime}(x_0)+\dots
\end{equation}
Since the goal is to carry out an interpolation based on the information of a set of $N$ scattered points distributed in a small vicinity, we apply (\ref{taylor}) $N$ times to each one of the data points around $x_0$. The result is written into a matrix form as
\begin{equation}
\mathcal{F}=\mathcal{M} D,
\end{equation}
where $\mathcal{F}$ and $D$ are $N\times 1$ column vectors and $\mathcal{M}$ is a $N\times N$ matrix. $\mathcal{F}$ contains the value of $f(x)$ at each of the $N$ data points, $D$ contains values of the function $f(x)$ and its derivatives at the query point $x_0$ and the $i-$th row of the matrix $\mathcal{M}$ consists of increasing power functions of the coordinate relative difference between the $i-$th data point and the query point $x_0$. For instance, for a univariate function $f(x)$ with data points the function values at coordinates $x_i$, where $i=1,2,\dots,N,\, \mathcal{F},\, D$ and $\mathcal{M}$ can be written as
\begin{equation}
\mathcal{F}=\left(f(x_1),f(x_2),f(x_3),f(x_4),\dots,f(x_N)\right)^T
\end{equation}
\begin{equation}
\mathcal{M}=\left(\begin{matrix}
1 & x_1-x_0 & \frac{(x_1-x_0)^2}{2} & \frac{(x_1-x_0)^3}{3!} & \frac{(x_1-x_0)^4}{4!} & \dots & \frac{(x_1-x_0)^N}{N!} \\
1 & x_2-x_0 & \frac{(x_2-x_0)^2}{2} & \frac{(x_2-x_0)^3}{3!} & \frac{(x_2-x_0)^4}{4!} & \dots & \frac{(x_2-x_0)^N}{N!}\\
1 & x_3-x_0 & \frac{(x_3-x_0)^2}{2} & \frac{(x_3-x_0)^3}{3!} & \frac{(x_3-x_0)^4}{4!} & \dots & \frac{(x_3-x_0)^N}{N!} \\
1 & x_4-x_0 & \frac{(x_4-x_0)^2}{2} & \frac{(x_4-x_0)^3}{3!} & \frac{(x_4-x_0)^4}{4!} & \dots & \frac{(x_4-x_0)^N}{N!} \\
\dots & \dots & \dots & \dots & \dots & \ddots & \vdots\\
1 & x_N-x_0 & \frac{(x_N-x_0)^2}{2} & \frac{(x_N-x_0)^3}{3!} & \frac{(x_N-x_0)^4}{4!} & \dots & \frac{(x_N-x_0)^N}{N!}
\end{matrix}\right)
\end{equation}
\begin{equation}
D=\left(f(x_0),f^\prime(x_0),f^{\prime\prime}(x_0),f^{\prime\prime\prime}(x_0),\dots,f^{(N)}(x_0)\right)^T
\end{equation}
This implies that the column vector $D$ can be expressed in terms of $\mathcal{F}$ and $\mathcal{M}$, once $\mathcal{M}$ has a nonzero determinant, $D=\mathcal{M}^{-1}\mathcal{F}$. In particular, if we are only interested in specific elements of $D$, we can use Cramer's rule to evaluate them as follows
\begin{equation}
\label{Cramer}
D_i=\frac{\text{det}(\mathcal{M}_i)}{\text{det}(\mathcal{M})},
\end{equation}
where $\mathcal{M}_i$ is the matrix formed by replacing the $i-$th column of $\mathcal{M}$ by the column vector $\mathcal{F}$. For instance, $f^{\prime\prime}(x_0)=\text{det}(\mathcal{M}_3)/\text{det}(\mathcal{M})$. To rewrite an $n-$th order derivative $f^{(n)}(x)$, we write the respective derivate as in (\ref{Cramer}) by using each data point of the small vicinity as query points. This way, we are able to rewrite all the derivatives at the above $N$ points as linear combinations of the function values $f(x_i)$. Substituting the derivatives into the eigenequation in study, we obtain $N$ equations with $f(x_1),\dots,f(x_N)$ as its variables. 

The corresponding BH radial master equation of the form \eqref{master} depends on the tortoise coordinate $r_*\in (-\infty,+\infty)$. To study the QNMs in this region we perform a proper coordinate transformation (depending on the BH spacetime) to recast the radial domain to $x\in[0,1]$, where $x$ depends on the coordinate transformation. After properly applying the boundary conditions by multiplying with the correct asymptotic behavior of the solutions of the master equation \eqref{master}, we end up with a homogeneous differential equation
\begin{equation}
\label{homogeneous}
a(x)\phi^{\prime\prime}(x)+b(x)\phi^\prime(x)+c(x)\phi(x)=0,
\end{equation}
where $a(x),b(x),c(x)$ are $\omega-$dependent functions and $\phi(x)$ a general radial function. By discretizing the interval $x\in[0,1]$ we introduce $N$ randomly distributed points with $x_1=0$ and $x_N=1$. By utilizing \eqref{Cramer} we discretize all derivatives and rewrite \eqref{homogeneous} in the matrix form
\begin{equation}
\mathscr{M}\mathcal{F}=0,
\end{equation}
where $\mathscr{M}$ is a square matrix that depends on the eigenvalues of the system in study and $\mathcal{F}=\left(f_1,f_2,\dots,f_i,\dots,f_N\right)^T$ with $f_i=\phi(x_i)$. As a homogeneous matrix equation, for it to have non-trivial solutions, the following equation must hold
\begin{equation}
\label{studied eq}
\text{det}(\mathscr{M})=0.
\end{equation}
Eq. (\ref{studied eq}) is the desired algebraic equation to calculate the eigenvalues of any particular eigenvalue system, which can be solved numerically using $\it{Mathematica}$ or $\it{Matlab}$.

The aforementioned method has been benchmarked in \cite{KaiLin1,Lin:2016sch,Lin:2017oag} and the results were compared with the sixth-order WKB approximation \cite{WKBI}, the continued fraction method \cite{Leaver:1985ax,Leaver2} and the Horowitz-Hubeny method \cite{Horowitz:1999jd} achieving great accuracy and precision.

\subsection{A pseudospectral collocation method}
To be able to justify that our numerical calculations are valid we need to utilize multiple methods. The {\it Mathematica} package {\it QNMSpectral} developed in \cite{Jansen:2017oag} is a spectral method utilized frequently in this thesis. The package is based on numerical methods first introduced in \cite{Dias:2010eu}. This method essentially discretizes the master equation using pseudospectral collocation methods and then directly solves the resulting generalized eigenvalue equation.

The choice of boundary conditions defines the asymptotics of our setup which we incorporate to the master equations by hand. This step normalizes the solutions so they behave smoothly at the boundaries. Having derived the master equation in a form with no divergences at the boundaries, the code discretizes the equation we input with pseudospectral methods (see \cite{2001cfsm.book.....B} for a complete review of these methods) by replacing continuous variables by a discrete set of collocation points on a grid. A function can then be represented as the values the function takes when evaluated on the gridpoints. A useful way of viewing this set of numbers representing a particular function is as coefficients of the Lagrange interpolation polynomials. The Lagrange interpolation polynomials corresponding to the grid $x_i$, where $i=0,\dots,N$ are polynomials $C_j(x)$ of degree $N$, with $j=0,\dots,N$ satisfying $C_j(x_i)=\delta_{ij}$. The choice of a grid uniquely determines the Lagrange polynomial as
\begin{equation}
C_j(x)=\prod_{j=0,j\neq i}^{N}\frac{x-x_j}{x_i-x_j}.
\end{equation}
A function $f(x)$ is then approximated by the Lagrange polynomials as
\begin{equation}
f(x)\approx\sum_{j=0}^{N}f(x_j)C_j(x_i).
\end{equation}
The expansion in terms of Lagrange interpolation polynomials allows one to construct the first derivative matrix $D^{(1)}_{ij}=C^\prime_i(x_j)$, and work similarly for higher order derivatives. Solving the resulting discretized master equation will lead to a function $f(x)$ which solves the master equation exactly at the collocation points. As the number of gridpoints is increased, it is expected that the function will also solve the equation at other points.

Numerical test indicate that the code is more efficient when one chooses the Chebyshev grid:
\begin{equation}
\label{grid}
x_i=\cos\left(\frac{i\pi}{N}\right),\,\,\,\,\,\,\,i=0,\dots,N.
\end{equation}
For this grid, it can be proven that any analytic function can be approximated with exponential convergence in $N$. These points, as given in (\ref{grid}), lie in $[-1,1]$ but with a proper transformation this can be rescaled to the domain of interest $[0,1]$. For this grid, the Lagrange polynomials are linear combination of the Chebyshev polynomials $T_n(x)$
\begin{equation}
C_j(x)=\frac{2}{N p_j}\sum_{m=0}^{N}\frac{1}{p_m}T_m(x_j)T_m(x),\,\,\,\,\,\,p_0=p_N=2,\,\,\,p_j=1.
\end{equation}
These functions are perfectly smooth and at the endpoints they are either $0$ or $1$.

Applying the above discretization technique, the code turns the problem of solving a linear ordinary differential equation, subject to specific boundary conditions, into solving a matrix equation, where the boundary conditions are already implicitly incorporated. The code then writes the matrix equation into a generalized eigenvalue problem. The simplest type of master equation will be of the form
\begin{equation}
c_0(x,\omega)\phi(x)+c_1(x,\omega)\phi^\prime(x)+c_2(x,\omega)\phi^{\prime\prime}(x)=0,
\end{equation}
where each of the $c_i$ are linear in $\omega$: $c_i(x,\omega)=c_{i,0}(x)+\omega c_{i,1}(x)$. Each of the coefficients $c_{i,j}(x)$ is turned into a vector by evaluating it on the gridpoints. These vectors are multiplied with the corresponding derivative matrices $D_{ij}^{(n)}$ and the resulting matrices are added, to bring the equation into the form
\begin{equation}
\label{generalized}
(M_0+\omega M_1)\phi=0,
\end{equation}
where the $M_i$ are now purely numerical matrices, independent of $\omega$. Explicitly, $(M_0)_{ij}=c_{0,0}(x_i)\delta_{ij}+c_{1,0}(x_i)D^{(1)}_{ij}+c_{2,0}(x_i)D^{(2)}_{ij}$ and similarly for $M_1$. Eq. \eqref{generalized} describes a generalized eigenvalue problem. The code implements the built-in function {\it Eigenvalues} or {\it Eigensystem} and returns the QNMs and its associated eigenfunctions upon request.

The aforementioned method has been benchmarked in \cite{Jansen:2017oag} achieving high accuracy and precision. Various tests with the sixth-order WKB approximation \cite{WKBI}, the continued fraction method \cite{Leaver:1985ax,Leaver2} and the matrix method \cite{KaiLin1} have been employed, reaching high-precision agreement.

\chapter{Superradiance: an overview}\label{superradiance_over}
Perturbing a BH with small fields could lead in two possible outcomes; the BH is stable under perturbations due to damping mechanisms that act on the BH exterior and will relax after the initial disruption or the BH is unstable under perturbations and will inevitably disappear or turn into another stable object. Although astrophysical BHs are expected to be stable under small fluctuations, a lot of studies have been performed to BH solutions that might be prone to instabilities due to new phenomena that might be possibly unveiled. Quite strikingly, one can extract energy from BHs \cite{Penrose:1971uk,Bekenstein:1973mi} by scattering test field off the BH horizon. This mechanism is called superradiance and it is mostly known for rotating and charged BHs. During this process, the test field grows on the expense of the BH electric or rotational energy, leading to instabilities and spontaneous symmetry breaking phenomena which give rise to hairy BH solutions, that is, BHs with non-vanishing order parameters which would be absent before the symmetry breaking occurred. In this chapter, we take a closer look to the mechanism that leads to superradiance in BH spacetimes. This chapter is based on \cite{Brito:2015oca}.

\section{Scalar field scattering off a potential barrier}
The first indication of superradiant amplification of scattered waves off electrostatic potentials originates from the early work of Klein who pioneered the study of the Dirac equation including a step potential \cite{Klein:1929zz}. Consequent studies performed by Hund \cite{Hund1941}, but now dealing with the wave equation of a charged scalar field, illustrated that the potential barrier can give rise to the production of charged particle pairs when the potential is sufficiently strong. In this section, we present a simplistic treatment of scalar field scattering to demonstrate the phenomenon of superradiance. 

Lets consider a massive scalar field $\Psi$ minimally coupled to an electromagnetic potential $A_\nu$ in $(1+1)$-dimensions. The equation of motion of the scalar field will be governed by
\begin{equation}
\label{wave equation}
\left(D^\nu D_\nu-\mu^2\right) \Psi=0,
\end{equation}
where $D_\nu\equiv \partial_\nu-iq A_\nu$ the covariant derivative, $q$ and $\mu$ the charge and mass of the scalar field, respectively, and $A^\nu=(A_0(x),0)$, with asymptotic behavior
\begin{equation}
\label{elpot}
A_0(x)\rightarrow\begin{cases}
0, & \text{for}\,\,\, x\rightarrow -\infty\\
V, & \text{for}\,\,\, x\rightarrow +\infty
\end{cases}
\end{equation}
If we decompose the scalar field as $\Psi=e^{-i\omega t}\psi(x)$, then \eqref{wave equation} becomes (following the procedure demonstrated in Appendix \ref{appA})
\begin{equation}
\label{scatter}
\frac{d^2 \psi(x)}{dx^2}+\left[\left(\omega-q A_0\right)^2-\mu^2\right]\psi(x)=0.
\end{equation}
If we consider an incident beam of bosons coming from $-\infty$ with amplitude $\mathcal{I}$, scattering off the potential with reflection and transmission coefficients $\mathcal{R}$ and $\mathcal{T}$, respectively, then the asymptotic behavior of the solution of \eqref{scatter} will be
\begin{equation}
\psi(x)\rightarrow\begin{cases}
\mathcal{I}e^{i k_{-\infty} x}+\mathcal{R}e^{-i k_{-\infty} x}, &\text{as}\,\,\, x\rightarrow -\infty\\
\mathcal{T}e^{ik_\infty x},&\text{as}\,\,\, x\rightarrow +\infty
\end{cases}
\end{equation}
where $k_{-\infty}=\sqrt{\omega^2-\mu^2}$, $k_\infty=\sqrt{(\omega-qV)^2-\mu^2}$  and $\omega>0$. The reflection and transmission coefficients depend on the shape of the potential $A_0$, however the Wronskian
\begin{equation}
W=\psi \partial_x \psi^\dagger-\psi^\dagger \partial_x \psi
\end{equation}
of two independent solutions of \eqref{scatter} is conserved, where $\psi^\dagger$ the complex conjugate spatial part of the solution of
\begin{equation}
\left((D^\nu D_\nu)^\dagger-\mu^2\right)\Psi^\dagger=0.
\end{equation}
Evaluating the Wronskian, or equivalently, the particle current density, for $\psi$ and $\psi^\dagger$ at the asymptotic regions we get
\begin{align}
W(x\rightarrow-\infty)&=2i\sqrt{\omega^2-\mu^2}(|\mathcal{I}|^2-|\mathcal{R}|^2),\\
W(x\rightarrow+\infty)&=2i\sqrt{(\omega-qV)^2-\mu^2}|\mathcal{T}|^2.
\end{align}
Since the Wronskian is conserved, we find
\begin{equation}
\label{rel}
|\mathcal{R}|^2=|\mathcal{I}|^2-{\frac{\sqrt{(\omega-qV)^2-\mu^2}}{\sqrt{\omega^2-\mu^2}}}|\mathcal{T}|^2.
\end{equation}
Considering a beam of massless particles, \eqref{rel} simplifies to
\begin{equation}
\label{rel0}
|\mathcal{R}|^2=|\mathcal{I}|^2-\frac{\omega-qV}{\omega}|\mathcal{T}|^2.
\end{equation}
Therefore, if
\begin{equation}
\label{sup}
0<\omega<qV
\end{equation}
then the beam of bosons will be superradiantly amplified, since $|\mathcal{R}|>|\mathcal{I}|$. It is noteworthy to point out that \eqref{sup} generalizes to massive case as
\begin{equation}
\label{supm}
\mu<\omega<qV-\mu,
\end{equation}
provided that $qV>2\mu$. 

Eq. \eqref{scatter} shares a striking resemblance with the master equation of scalar waves propagating on a fixed BH background, and therefore we expect that a similar phenomenon will occur in charged BH spacetimes when charged scalar waves scatter off the BH curvature potential.
\section{Black-hole scattering and superradiance}
Lets consider an asymptotically flat and spherically symmetric spacetime. As we have already seen, the propagation of various type of perturbations on fixed backgrounds obeys a Schr\"{o}dinger-like equation of the form \eqref{master}, with an effective potential which encodes the curvature of the BH background and the properties of the matter fields. Given the symmetries of the background, we consider a scattering experiment of a monochromatic wave with frequency $\omega$ with time dependence $e^{-i\omega t}$. The potential will possibly vanish at the boundaries (or is a constant). For this reason, we will assume the potentials constancy at the boundaries. Then, \eqref{master} will have the following asymptotic behavior
\begin{equation}
\psi\sim \left\{ \begin{matrix}
\mathcal{T}e^{-i k_+ r_*}+\mathcal{O}e^{i k_+ r_*},\,\,\,\,\,\, r\rightarrow r_+\\
\mathcal{R}e^{i k_{\infty} r_*}+\mathcal{I}e^{-i k_{\infty} r_*},\,\,\,\,\,\, r\rightarrow \infty ,
\end{matrix}\right.
\end{equation}
where $k^2_+=\omega^2-V(r\rightarrow r_+)$ and $k^2_\infty=\omega^2-V(r\rightarrow \infty)$. The previous boundary conditions correspond to an incident wave of amplitude $\mathcal{I}$ from spatial infinity giving rise to a reflected wave of amplitude $\mathcal{R}$ and a transmitted wave of amplitude $\mathcal{T}$ at the event horizon. Due to the presence of the event horizon $\mathcal{O}\equiv0$, meaning that nothing escapes from the BH.

If we assume that $V$ is real, which actually holds for scalar perturbations in rotating and charged BHs, then since the background is stationary, Einstein's equations are invariant under the $t\rightarrow-t$, $\omega\rightarrow-\omega$ transformations. Hence, another complex conjugate solutions of \eqref{master} exists, ${\psi^\dagger}$, which satisfies the complex conjugate boundary conditions. Solutions $\psi$, ${\psi^\dagger}$ are linearly independent and therefore, ordinary differential equation theory implies that the Wronskian $W$ should be $r_*-$independent. Thus, the Wronskians evaluated near the event horizon $W_+=-2ik_+(|\mathcal{T}|^2)$ and infinity $W_\infty=-2ik_\infty(|\mathcal{R}|^2-|\mathcal{I}|^2)$ must be equal, leading to
\begin{align}
\label{sup_relation}
|\mathcal{R}|^2=|\mathcal{I}|^2-\frac{k_+}{k_\infty}|\mathcal{T}|^2.
\end{align}
It is evident from \eqref{sup_relation} that when $k_+/k_\infty>0$, then $|\mathcal{R}|^2<|\mathcal{I}|^2$ which means that the reflected wave amplitude is smaller than the incident pulse's amplitude. This is expected from scattering experiments off perfect absorbers. However, if $k_+/k_\infty<0$, the reflected wave is superradiantly amplified, $|\mathcal{R}|^2>|\mathcal{I}|^2$. 

At this point, we state that dissipation is highly crucial for superradiance to occur. The purely absorbing ability of the event horizon is the driving force of the effect, since in absence of dissipation, energy conservation would imply that the outgoing flux would be equal to the transmitted one, and therefore $|\mathcal{R}|^2=|\mathcal{I}|^2$.

The phenomenon of superradiant scattering seems to imply that energy is being extracted from the BH. Considering back-reaction effects, energy is indeed extracted from the BH which leads to the decrease of mass and angular momentum from rotating BHs and the decrease of mass and charge from charged BHs \cite{Penrose:1971uk,Bekenstein:1973mi}. Such a decrease in BH energy leads to the increment of the energy of the test field, and possibly the rise of instabilities. The endpoints of such instabilities might lead to the evacuation of matter from the BH or to the formation of a novel constant observable configuration around the BH. This phenomenon is called spontaneous scalarization, if the test field is a scalar, and designates the violation of the no-hair theorem, which states that BHs can be completely characterized by three classical externally observable quantities; the mass, charge and angular momentum.

\section{Concluding remarks}
So far, an interesting study have suggested that higher-dimensional RNdS spacetimes are prone to instabilities under gravitational perturbations \cite{Konoplya:2008au}. Specifically, $d-$dimensional RNdS BHs with $d>6$ and large enough mass and charge, are gravitationally unstable. Such an instability has been further examined in \cite{Cardoso:2010rz,Konoplya:2013sba,Tanabe:2015isb}.  Why only $d=4,5$ and $6$-dimensional RNdS BHs are favorable to be stable, is still unknown. Due to the consideration of a BH charge in such spacetimes, though, one could argue that for them to form through gravitational collapse, charged matter should be present.

More recently, a new instability was found in $4-$dimensional RNdS BHs in \cite{Zhu:2014sya} and further analyzed in \cite{Konoplya:2014lha}. The $l=0$ charged scalar perturbation was proven to be unstable for various regions of the parameter space of RNdS BHs. The addition of an arbitrarily small amount of mass acts as a stabilization factor, as well as the increment of the scalar field charge beyond a critical value. It has been proven that this instability has a superradiant nature; the charged field is trapped in the potential well between the photon sphere and the cosmological horizon and is able to extract electric energy from the BH.

In Chapter \ref{higher instability} (based on \cite{Destounis:2019hca}), we extend the study of the dynamical instability emerging from a spherically symmetric charged scalar perturbation scattering off RNdS curvature potential in higher-dimensions. Since the spacetime dimensions directly affect the effective potential of the BH, it is interesting to examine if their increment prevents the instability to occur or enhances it. 

We will perform a thorough frequency-domain analysis of higher-dimensional RNdS BHs under charged scalar perturbations and conclude that the source of instability is directly linked with the existence of the cosmological horizon, as well as the QNMs of pure $d-$dimensional de Sitter space. Such modes satisfy the superradiant condition even when stable configurations occur. We show that the increment of dimensions, amplifies the instability leading to larger regions in the subextremal parameter space of RNdS where superradiant instabilities occurs. Interestingly, even though the introduction of mass stabilizes the system, there are still regions in the parameter space where the family remains unstable and superradiant. 

An open, and still interesting, problem is the nonlinear development of such a system to grasp the end-state of the evolving BH spacetime. A huge challenge in such nonlinear evolutions is the very large timescale of the instability which requires highly precise numerical developments. Since the increment of dimensions reduces the timescale of the instability, it would be more feasible for such an instability to be tested in higher-dimensional RNdS non-linearly and realize if it leads to a novel scalarized BH.

\chapter{Black-hole interiors and Strong Cosmic Censorship}\label{Chapter 4}
For many decades, SCC has been one of the most intriguing problem in GR. SCC addresses the issue of the nature of singularities as well as the predictability of Einstein's field equations. It states that given suitable initial data on a spacelike hypersurface, the laws of GR should determine, completely and uniquely, the future evolution of the spacetime. In this chapter, we review the history and the current state-of-the-art of the SCC conjecture and discuss its modern formulation. This chapter is based on \cite{Dafermos:2012np,Chambers:1997ef,Poisson:1997my}.

\section{Introduction}\label{section intro to SCC}

One of the most fundamental questions concerning the study of black holes and their internal structure is ``what is the anatomy of the interior of a BH?". A lot of progress has been made during the last decades on answering such an inquiry. Many spacetimes have been studied to grasp the nature of singularities lying deep inside charged and rotating BHs. Their structure is crucial in answering elementary questions about the global uniqueness of solutions in BH interiors, following the laws of GR, given suitable initial data placed in the external BH region. In this thesis, we will try to answer this question by following purely classical laws.

\begin{figure}[H]
\centering
\includegraphics[scale=0.5]{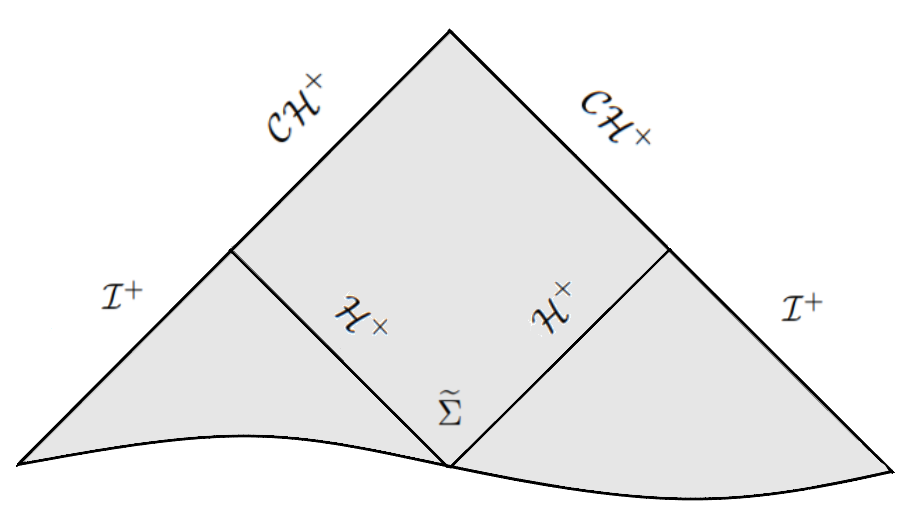}
\caption{The maximal globally hyperbolic development of initial data on a Cauchy hypersurface $\tilde{\Sigma}$. For simplicity, only the future development and Cauchy horizons $\mathcal{CH^+}$ are shown.}
\label{domain od dependence}
\end{figure}

The Cauchy initial value problem that GR poses guarantees the existence of a unique, up to isometries, MGHD, given suitable initial data on a spacelike Cauchy hypersurface. If we denote this surface as $\tilde{\Sigma}$ then its future MGHD coincides with the shaded region in Fig. \ref{domain od dependence}. Therefore, any complete past-directed causal curve in this region intersects $\tilde{\Sigma}$ exactly once. The inquiry we are trying to answer then is whether the shaded region is extendible, in a meaningful way, to a larger spacetime manifold $\mathcal{M}$. If so, the boundary of the MGHD is the CH, $\mathcal{CH}^+$. As we saw in Chapter \ref{Chapter 2}, CHs do exist in BH spacetimes and in fact might pose a threat to the predictability of physical laws. The question posed by SCC is then if stable CHs exist at all in BH spacetimes. Of course, equivalent arguments hold for the past MGHD and the past CH, $\mathcal{CH}^-$, which is not shown in Fig. \ref{domain od dependence}. 

\section{Cauchy horizons in asymptotically flat black holes}\label{section CH in flat BHs}

The first quantitative study of CH stability dates back to the early 70s when Penrose and Simpson \cite{Simpson:1973ua} first considered the possibility of observers crossing the CH that lies inside charged BHs. They investigated the effect of small asymmetries that emerge from dynamical gravitational collapse on the stability of CHs in RN BHs. By considering a test electromagnetic field on a fixed RN background, they found that instabilities on the probe field arise at the CH, though not at the event horizon. This led to the reasonable conclusion that in the full coupled Einstein-Maxwell theory, the inner horizon will not survive as a non-singular hypersurface when perturbations are present, but will instead become a curvature singularity. 

A decade later Chandrasekhar and Hartle \cite{Hartle} investigated the behavior on the CH of a flux of gravitational and electromagnetic radiation crossing the event horizon of a RN BH. They showed that the flux of null radiation received by an observer crossing the CH, along a radial timelike geodesic, diverges exponentially for some physically reasonable perturbations, thus leading to a blueshift effect. More precisely, physical quantities associated with the perturbations, such as the energy density measured by free-falling observers, diverge at the CH. Put differently, CHs of RN spacetimes are unstable under small time-dependent perturbations (see also \cite{McNamara1,Starobinsky1,Starobinsky2,Matzner}). It has been shown that the same is true for Kerr-Newman spacetimes \cite{McNamara2}. This is a clear indication that CHs of asymptotically flat spacetimes should not be smoothly accessible to free-falling observers.

Although linear studies have been proven very insightful, the concept of SCC requires a deeper understanding with models that are able to explore the possible back-reaction effect of the diverging perturbations at the CH. This question was tackled by Israel and Poisson \cite{Poisson:1990eh}. To simplify their analysis, they modeled the infinitely blue-shifted radiation by an ingoing spherically symmetric stream of massless particles, thus converting the RN geometry in study into the charged Vaidya solution \cite{Bonnor:1970zz}. In this solution, the CH is a weak curvature singularity, since the metric function approach a regular limit on the CH, for appropriately chosen coordinates. In addition, none of the curvature invariants were divergent there. 

However, they showed that the situation drastically changes if one considers, in addition, a flux of outgoing massless particles, that models a piece of the ingoing field that has been backscattered by the BHs curvature. Although there is no explicit geometry for such a case, they showed that the internal mass function (or Hawking mass) blows up at the CH. This mass divergence was exponential in outgoing null coordinates and therefore they called it mass inflation singularity. The divergence of the mass function guarantees that the Kretschmann scalar blows up at the CH. However, this is not what characterizes the strength of this singularity because curvature, per se, does not imply the breakdown of the field equations \cite{Klainerman:2012wt} nor the destruction of macroscopic observers \cite{Ori:1991zz}. What makes the mass-inflation phenomenon so relevant is that it is related to the Christodoulou criteria, that we will discuss later, which guarantees the breakdown of field equations \cite{Dafermos:2012np,Costa:2017tjc}.

Mass inflation was further studied in spherical symmetric null dust models \cite{Brady_Smith,Burko:1997tb}, while the evidence in rotating BHs, although less firm, is still quite convincing \cite{Ori_rotating,Brady_rotating,Droz}. The aforementioned studies share a common factor; they use null dust models to examine the infinite blueshift effect at the CH. Unfortunately, this forces one to input backscattering of the null streams by hand. Moreover, backscattering only occurs strictly after horizon formation and the rate of decay of radiation before this artificial backscattering occur is also put in by hand. It is clear that a more realistic scenario would be necessary. A natural candidate is the Einstein-Maxwell-scalar field model. Here, the field equations are coupled to the Maxwell and wave equations, but the scalar field is uncharged, thus the latter two interact only through the gravitational coupling. The above model admits true wave-like behavior and allows for the decay rate of radiation on the horizon to arise dynamically. It is a generalization of the spherically symmetric Einstein-scalar field model whose mathematical study was initiated by Christodoulou leading to his seminal proofs of both weak and strong cosmic censorship \cite{1999math......1147C}.

On a more mathematical point of view, SCC initially stated that generic initial data should be future inextendible as a Lorentzian manifold with continuous metric. The instability at the CH, due to the blueshift effect of scattered radiation, should then lead to a singularity when non-linearities kick in. By taking a closer look at solutions of the wave equation, instead of null dust models, on RN or Kerr we realize that they decay polynomially outside the BH and this might compete with the blueshift effect. However, Dafermos proved that the energy measured by a local observer at the CH, which is proportional to the gradient of the wave equation solution, is indeed infinite \cite{Dafermos:2003wr}. The blow-up, however, is in a sense weak because the amplitude of the solutions remain bounded at the CH \cite{Franzen:2014sqa}. In fact, the solutions of the wave equation in subextremal RN spacetime are globally bounded in the BH interior up to and including the CH, to which the solutions extend continuously but the energy is infinite. The same holds for subextremal Kerr BHs \cite{Dafermos:2014cua,LukOhStrongI,LukOhStrongII,Dafermos:2017dbw}. For results concerning extremal RN BHs see \cite{Aretakis:2012ei,Gajic:2015csa}. 

If one then naively extrapolates the linear behavior of the wave equation to the non-linear behavior of the field equations \cite{Dafermos:2012np,LukOhStrongI,LukOhStrongII,Dafermos:2003yw}, one can identify that the metric may extend continuously to the CH whereas the Christoffel symbols (gradients of the metric) blow up. Therefore, a more proper formulation of SCC, proposed by Christodoulou \cite{Christodoulou:2008nj}, is the following: \emph{the MGHD of generic initial data  should be future inextendible as a spacetime with square-integrable Christoffel symbols}. This statement is not at the level of blow up of the curvature, but rather at the non-square integrability of the Christoffel symbols and essentially forbids the Einstein equations from making sense at the CH even in a weak manner. This notion is perhaps less familiar than the traditional classifications of singularity \cite{Ellis:1974ug,TIPLER19778}, but more relevant to the partial differential equations properties governing the dynamics of the Einstein equations, for which the pointwise blow-up of curvature per se is of no particular importance \cite{Klainerman:2012wt}. 

Many theorems support the claim that the naive extrapolation of linear theory is indeed correct \cite{Dafermos:2003wr,LukOhStrongI,LukOhStrongII}; the blueshift instability is not sufficiently strong to destroy the spacetime earlier into a spacelike singularity, but does give rise to a null singularity at the CH across which the metric extends continuously but the mass function diverges. Therefore, although the initial SCC conjecture seems to fail, the modern formulation by Christodoulou may still be true in subextremal asymptotically flat spacetimes.

From the former, we conclude that there exists strong evidence that the existence of CHs in charged and rotating asymptotically flat BHs do not pose a threat in classical GR, since slight deviations in the initial conditions produce spacetimes with drastically different causal structures. Taking the deviations into account, they destroy the CH and replace it with a singularity, beyond which the field equations cease to make sense. Therefore, we may conclude that perturbed Kerr-Newman spacetimes do not represent a counter-example to the modern formulation of SCC.

\section{Cauchy horizons in asymptotically de Sitter black holes}\label{section CH in dS BHs}
Although it appears that SCC holds true for BHs in asymptotically flat spacetimes, the same cannot be stated for BHs in asymptotically dS spacetimes. Charged or rotating BHs in dS, like the RNdS or KdS space, also exhibit a CH. The major difference between the dS class of spacetimes and asymptotically flat ones is that earlier studies have indicated both stability and instability of the CH under small perturbations. Therefore, there is still no definite decision on the matter. A crucial ingredient for such studies is the existence of a cosmological horizon in the causal structure which forces the perturbations to decay exponentially in the exterior rather than polynomially. This effect can now compete with the exponential blueshift at the CH and possibly counterbalance it.

The first investigative study of CH stability in BH-dS spacetimes was performed by Mellor and Moss in 1990 \cite{Mellor:1990}. Confining their attention to spherically symmetric RNdS spacetime, they considered the effect of gravitational perturbations on the CH. They examined the flux of radiation due to these perturbations, as seen by an infalling observer that crosses the CH, and concluded that it remains finite. Unfortunately, a very strict assumption (vanishing flux of radiation at the cosmological horizon) in their analysis was recognized later which led Brady and Poisson to revisit the problem \cite{BradyPoisson}. Brady and Poisson's model mimics the perturbations propagating from the exterior to the interior of the BH as a spherical inflow of null dust. The requirement of a finite flux of energy at the cosmological horizon led to a significantly different stability condition, namely the CH is stable provided that the surface gravity at the CH is less than that of the cosmological horizon. This was the first indication that near-extremally charged RNdS BHs might violate SCC, at least in null dust models. However, if one includes backreaction, when the CH was stable, a divergent flux was still present there but the mass function would be bounded \cite{Brady:1992cz}. The generalization of the stability analysis by Mellor and Moss to the case of rotating BHs in dS was performed by Chambers and Moss \cite{Chambers_1994}. By studying linear perturbations of scalar, electromagnetic and gravitational fields on KdS spacetime they found that the CH is stable, provided that the surface gravity at the CH is larger than the one of the cosmological horizon. 

While we are primarily concerned with the classical stability of the CH in dS BHs, it is of interest to review the first quantum analysis. It is natural to ask whether a CH is quantum mechanically stable, when classical indications point to that direction. The intriguing partial answer to this question was provided by Markovic and Poisson in 1995 \cite{Markovic}. By examining the quantum fluxes measured by an observer approaching the CH of a RNdS BH they were able to conclude that the horizon is quantum-mechanically unstable, except for the case where the surface gravity at the CH equals the one of the cosmological horizon. This is a much more strict demand, meaning that when quantum backreaction is considered, the regularity requirements at the CH are higher as to assure a safe passage beyond it. It is important to note that since the calculation of the quantum energy-momentum tensor $<T_{\mu\nu}>$ in 4 dimensions is extremely difficult, they considered, instead, a simpler, but still instructive, approach to the problem, by quantizing a conformally invariant scalar field on a 2-dimensional version of the RNdS spacetime. This made the calculation much easier. 

Going back to classical results, Brady, Moss and Myers revisited SCC in RNdS spacetimes \cite{Brady:1998au}, but this time they took into account the backscattering of ingoing modes which led to an extra contribution to the influx along the CH arising from the condition that observers should measure a finite flux of radiation at the event horizon. This additional flux ensured that the CH is, in general, unstable. The initial data considered, though, were in fact rough, instead of smooth. In fact, the choice of smooth initial data would cancel out the backscattering contribution, thus, near extremality there would still be a region in the parameter space where the CH might remain classically stable. For a rigorous proof that the introduction of rough initial data can lead to classically unstable CHs in RNdS see \cite{Dafermos:2018tha}.

More recent studies by Costa, Gir\~ao, Nat\'ario and Silva \cite{Costa:2014yha,Costa:2014zha,Costa:2014aia,Costa:2017tjc} considered the structure of the BH interior starting with data along the event horizon satisfying an exponential Price law decay. Then, the solutions to the Einstein-Maxwell-scalar field system can be extended continuously across the CH with continuous metric and square-integrable Christoffel symbols, if appropriate conditions hold and the decay rate of the exponential power law is chosen to be fast enough. Therefore, these qualitative studies indicated that the modern formulation of SCC might be violated for near-extremally charged dS BHs. The proof of exponential decay in RNdS and KdS was provided in \cite{Dyatlov:2011jd}, and it was also shown that linear scalar waves are bounded and extend continuously up to and including the CH \cite{Hintz:2015jkj}. 

Subsequent work by Hintz and Vasy \cite{Hintz:2016gwb,Hintz:2016jak} showed that the metric and electromagnetic field decay, also, exponentially at the exterior of KdS and Kerr-Newman-dS spacetime. Their non-linear analyses proves global stability under small perturbations of the initial data, without symmetry assumptions. Besides global stability, it was shown that the exponentially decaying perturbations are governed by the spectral gap $\alpha$, that is, the imaginary part of the lowest-lying/dominant, non-zero QNM. 

Such results gives us the ability to appropriately test, for the first time, the modern formulation of SCC in asymptotically dS spacetimes, quantitatively, by explicitly calculating the spectral gap $\alpha$. The classical stability of CHs, therefore, narrows down to the calculation of the dominant QNM families and their delicate interplay with the blueshift effect. 

From the former, we realize that the stability of CHs in asymptotically dS spacetimes is an open field of study, where modern quantitative investigations are lacking. In the following chapters of the thesis, we will tackle the stability of CHs in RNdS spacetimes with state-of-the-art numerical simulations. The only ingredient missing to perform such an investigation is the condition for which CHs in asymptotically dS BHs will maintain enough regularity for the solutions to the wave equation and the metric to extend continuously with square-integrable derivatives beyond the CH. This condition is provided in the next section.

\section{Weak solutions of Einstein's equations}\label{section weak solutions}
Let us consider a matter field $\Phi$ back-reacting on a spacetime metric $g_{\mu\nu}$, which satisfies proper equations of motion. The Einstein field equations reads
\begin{equation}
\label{field equation}
G_{\mu\nu}+\Lambda g_{\mu\nu}=8\pi T_{\mu\nu}.
\end{equation}
Schematically, $G_{\mu\nu}\sim\Gamma^2+\partial\Gamma$, where $\Gamma$ represents the Christoffel symbols proportional to gradients of the spacetime metric, $\partial g_{\mu\nu}$ (we have dropped the indexes in $\Gamma^i_{jk}$ for simplicity).

If we assume that $\Phi$ and $g_{\mu\nu}$ are not necessarily twice continuously differentiable, then we can still make weak sense of (\ref{field equation}) by multiplying it with an arbitrary smooth, compact supported, function $\psi_0$ and integrating in a small neighborhood $\mathcal{V}$
\begin{equation}
\label{field_pert}
\int_\mathcal{V} d^4x\sqrt{-g}\,(\Gamma^2+\partial\Gamma)\psi_0+\int_\mathcal{V} d^4x\sqrt{-g}\,\Lambda g_{\mu\nu}\psi_0=8\pi\int_\mathcal{V} d^4x\sqrt{-g}\,T_{\mu\nu}\,\psi_0.
\end{equation}
The purpose of the test function $\psi_0$ is to carry away derivatives from fields without leaving non-zero terms when we perform integration by parts. The function $\psi_0$ is smooth, meaning that it is infinitely differentiable, and therefore no matter how many derivative we carry out to $\psi_0$ it will always be integrable.
If (\ref{field_pert}) is satisfied for any smooth function $\psi_0$, then we have a weak solution of (\ref{field equation}). 

For massless neutral scalar fields the energy momentum tensor is proportional to $T_{\mu\nu}\sim(\partial\Phi)^2$. Therefore, \eqref{field_pert} becomes
\begin{equation}
\label{terms}
\int_\mathcal{V} d^4x\sqrt{-g}\left(\Gamma^2\psi_0-\Gamma \partial \psi_0+\Lambda g_{\mu\nu} \psi_0-8\pi\psi_0 (\partial\Phi)^2\right)=0.
\end{equation}
The first term of \eqref{terms} is finite if $(\partial g_{\mu\nu})^2$ is integrable, the second term if $\partial g_{\mu\nu}$ is integrable while the third term requests that $g_{\mu\nu}$ is continuous at the CH. The final term is finite
\begin{equation}
\label{field final}
\int_\mathcal{V} d^4x\sqrt{-g}\, \psi_0 (\partial\Phi)^2<\infty,
\end{equation}
provided we request that $(\partial \Phi)^2$ is integrable. A requirement for \eqref{terms} to be satisfied weakly is, thus, integrability of $(\partial g_{\mu\nu})^2$ and $(\partial \Phi)^2$. Hence, the spacetime metric $g_{\mu\nu}$ and scalar field $\Phi$ must belong to the Sobolev space $H^1_\text{loc}$, that is the space of functions with square integrable first derivatives. In vacuum ($\Phi=0$) the former translates to $g_{\mu\nu}\in C^0$, $\Gamma\in L^2_\text{loc}$, where $C^0$ the space of continuous functions and $L^2$ the space of square integrable functions. This requirement corresponds to square integrability of the Christoffel symbols \cite{1999math......1147C}. 

To derive the condition for which $(\partial \Phi)^2$ is square integrable at the CH of RNdS BHs, and the SCC hypothesis is violated, we study the asymptotics of scalar waves there. Consider a neutral massless scalar field obeying the wave equation 
\begin{equation}
\label{wave equation 1}
\nabla^\mu\nabla_\mu\Phi=0,
\end{equation}
Due to the spherical symmetry of the RNdS backgrounds we are interested in, the scalar field can be expanded as
\begin{equation}
\Phi=\sum_{lm}^{}\psi(r)Y_{lm}(\theta,\varphi)e^{-i\omega t}.
\end{equation}
It is convenient to employ the outgoing Eddington-Finkelstein coordinates $u\equiv t-r_*$ which are regular at the CH.
The solutions of \eqref{wave equation 1} have the following asymptotic form at the CH:
\begin{align}
\label{first}
\Phi^{(1)}&\sim e^{-i\omega(t+r_*)}=e^{-i\omega(u+2r_*)}=e^{-i\omega u}e^{-2i\omega r_*},\\
\label{second}
\Phi^{(2)}&\sim e^{-i\omega(t-r_*)}=e^{-i\omega u}.
\end{align}
Near the CH, the tortoise coordinate acquires the simplified form
\begin{equation}
r_*=\int f^{-1}dr\sim \frac{\log|r-r_-|}{f^\prime(r_-)},
\end{equation}
where $f(r\rightarrow r_-)\sim|r-r_-|$ modulo irrelevant terms. Obviously, (\ref{second}) is regular at the CH since for $r= r_- \,\,(u=\text{const.})$ $\Phi^{(2)}$ is a smooth function. The potential non-smoothness comes from \eqref{first}
\begin{align}
\label{expo}
\Phi^{(1)}\sim e^{-i\omega u}e^{-2i\omega \log(r-r_-)/f^\prime(r_-)}=e^{-i\omega u} |r-r_-|^{-2i\omega/f^\prime(r_-)}=e^{-i\omega u}|r-r_-|^{i\omega/\kappa_-},
\end{align}
where $\kappa_-=|f^\prime(r_-)|/2$ the surface gravity at the CH (note that $f^\prime(r_-)<0$). 
If we consider modes of the form $\omega=\omega_R+i\omega_I$, with $\omega_I<0$, then 
\begin{equation}
\label{shit}
\Phi^{(1)}\sim e^{-i\omega u}|r-r_-|^{i\omega/\kappa_-}=e^{-i\omega u}|r-r_-|^{i\omega_R/\kappa_-}|r-r_-|^{\beta}
\end{equation}
where we defined $\beta\equiv \alpha/\kappa_-$, with $\alpha=\inf\{-\text{Im}(\omega)\}$ the spectral gap, or the imaginary part of the dominant, non-zero QNM $\omega$. In \eqref{shit}, the first factor is smooth at the CH and the second is purely oscillatory. Hence, the potential non-smoothness comes from the third factor of \eqref{shit}.
Since we want $(\partial\Phi)^2$ to be integrable at the CH for SCC to be violated, and since $r$ is the only relevant coordinate to potentially introduce non-smoothness, we require
\begin{equation}
\label{int12}
\int_\mathcal{V} (\partial_r \Phi^{(1)})^2 dr\sim\int_\mathcal{V}|r-r_-|^{2(\beta-1)}dr \sim\frac{|r-r_-|^{2\beta-1}}{2\beta-1}<\infty.
\end{equation}
For \eqref{int12} to be satisfied at $r\sim r_-$
\begin{equation}
\label{beta_condition}
2\beta-1>0\Leftrightarrow \beta>\frac{1}{2}.
\end{equation}
If \eqref{beta_condition} is satisfied, and since $\Phi$ and $g_{\mu\nu}$ share similar regularity requirements \cite{Hintz:2016gwb,Hintz:2016jak,Dafermos:2012np,Costa:2017tjc,LukOhStrongI,LukOhStrongII,Dafermos:2017dbw,Dafermos:2003yw}, then \eqref{terms} is bounded and we can obtain weak solutions at the CH which can be extended smoothly. In this case we say that SCC is not respected.

It has been proven that \eqref{beta_condition} holds for fermionic \cite{Destounis:2018qnb} and gravitoelectric perturbations \cite{Dias:2018etb} in RNdS, as well as scalar perturbations in higher-dimensional RNdS \cite{Liu:2019lon}, and for scalar and gravitational perturbations \cite{Dias:2018ynt} in KdS spacetime. A more rigorous proof of \eqref{beta_condition} for charged scalar and fermionic fields in $4-$dimensional RNdS spacetime, as well as bosonic fields in higher-dimensional RNdS spacetime, is provided in Appendices \ref{appC}, \ref{appD} and \ref{appE}. 

\section{Concluding remarks}
The seminal mathematical studies in \cite{Costa:2017tjc,Costa:2014yha,Costa:2014zha,Costa:2014aia} have proved that the inclusion of a positive cosmological constant and an exponential Price law decay of perturbations leads to solutions to the Einstein-Maxwell-scalar field system which can be extended continuously across the CH with continuous metric and square-integrable Christoffel symbols, if the exponential decay is assumed to be fast enough. This provides a strong indication that near-extremally charged RNdS BHs might be counter-examples of SCC. 

The subsequent work in \cite{Hintz:2015jkj,Hintz:2016gwb,Hintz:2016jak} demonstrated that the metric and electromagnetic field decay, indeed, exponentially at the exterior of KdS and Kerr-Newman-dS spacetimes. The proof demonstrates the non-linear stability of small perturbations of the initial data, without symmetry assumptions. Besides global stability, these studies revealed that the exponentially decaying perturbations are governed by the QNMs of the BH spacetime while the stability of CHs in such spacetimes is intrinsically connected to the dominant QNMs. Such result gives us the ability to test the modern formulation of SCC in asymptotically dS spacetimes, quantitatively, by calculating, for the first time and with high accuracy, the dominant QNMs of RNdS spacetimes. 

In Chapter \ref{PRL} (based on \cite{Cardoso:2017soq}) we will perform a complete quantitative analysis of the linear stability of CHs lying inside RNdS BHs against neutral scalar perturbations. We have found three distinct families of QNMs which play an important role in the late time behavior of the ringdown waveform. We will demonstrate that near-extremally charged RNdS BHs are potential counter-examples of the linearized analogue of SCC. Previous work \cite{Dafermos:2012np} suggests that one can extrapolate from the analogy $\Phi\rightarrow g_{\mu\nu}$, $\partial\Phi\rightarrow\Gamma$ and realize that SCC is not respected for the Einstein-Maxwell theory. Many subsequent studies emerged after \cite{Cardoso:2017soq}.

In \cite{Hod:2018dpx}, it was argued that a charged scalar field would reinforce SCC if the scalar field is highly massive and charged. Such an argument led to the work illustrated in Chapter \ref{PRD} (based on \cite{PhysRevD.98.104007}). We demonstrate that even a charged massive scalar field can violate SCC for a finite volume of the subextremal parameter space of RNdS space. Two subsequent studies \cite{Mo:2018nnu,Dias:2018ufh} also demonstrated that charged scalar fields in RNdS provide a counter-example to the linearized analogue of SCC. 

Although scalar perturbations can be considered a good probe of metric perturbations, such a study was lacking till recently. In \cite{Dias:2018etb} it is demonstrated that metric perturbations pose even a more severe threat to the validity of SCC. Such perturbations not only reveal the possibility of extensions as weak solutions to the field equations but also extensions as classical solutions. In fact, one can extend the spacetime in an arbitrarily smooth way past the CH of ``large" near-extremal RNdS BHs.

An interesting suggestion to restore SCC in charged BHs with a positive cosmological constant, was proposed in \cite{Dafermos:2018tha}, where it was shown that the pathologies identified in Chapter \ref{PRL} become non-generic if one considerably enlarges the allowed set of initial data by weakening their regularity. The considered data are also compatible with Christodoulou's formulation of SCC.

At that point, all linear studies were performed in spherical symmetry. The work in \cite{Dias:2018ynt} demonstrated that scalar and gravitational perturbations of KdS BHs do not violate the linearized analogue of SCC! This means that astrophysical BHs are not counter-examples of SCC. In any case, if we wish to consider SCC as a mathematical tool of testing GR and its limits then our results are not to be taken lightly.

The generalization to charged fermionic perturbations in the context of SCC is performed in Chapter \ref{PLB} (based on \cite{Destounis:2018qnb}) and \cite{Zhang2}. It is shown that even Dirac fields in near-extremally charged RNdS provide a counter-example to the linearized analogue of SCC. In Chapter \ref{PLB} we further extend the analysis of fermions to highly near-extremal RNdS spacetime and provide numerical evidence which demonstrate that even for arbitrarily high fermionic charges, SCC is still violated. A thorough analysis of the different families of QNMs will also be provided.

Up until that point, only linearized studies were performed to test the validity of SCC. The first numerical calculations with a self-gravitating scalar field \cite{Luna:2018jfk} demonstrated that backreaction does not reinforce SCC but rather, the Einstein-Maxwell-scalar field model is a serious counter-example to the SCC conjecture, in accordance with the predictions in Chapter \ref{PRL}.

In Chapter \ref{JHEP} (based on \cite{Liu:2019lon}) we demonstrate that the linearized analogue of SCC is still violated, even in higher-dimensional RNdS spacetimes. The three families of modes exhibit a very delicate interplay which leads to the conclusion that ``small" higher-dimensional BHs are preferred against ``large" $4-$dimensional ones to the preserve SCC in larger volumes of the subextremal parameter space. 

Many more studies have emerged so far concerning the linearized analogue of SCC against different kinds of matter fields in BH spacetimes with a positive cosmological constant and a CH. All of them, including the ones discussed, are motivated from our initial study in Chapter \ref{PRL} and the subsequent ones in Chapters \ref{PRD}, \ref{PLB} and \ref{JHEP}. For an incomplete list see \cite{Hod:2018lmi,Gwak1,Rahman:2018oso,Gim:2019rkl,Gwak:2019ttv,Etesi:2019arr,Rahman:2019uwf,Guo:2019tjy,Dias:2019ery,Gan:2019jac,Destounis:2019omd}.

\part{Strong Cosmic Censorship in charged black-hole spacetimes}
\chapter{Quasinormal modes and Strong Cosmic Censorship}\label{PRL}
The fate of CHs, such as those found inside charged BHs, is  intrinsically connected to the decay of small perturbations
exterior to the event horizon. As such, the validity of the SCC conjecture is tied to how effectively the exterior damps fluctuations.
Here, we study massless scalar fields in the exterior of RNdS BHs. Their decay rates are governed by QNMs of the BHs. We identify three families of modes in these spacetimes: one directly linked to the photon sphere, well described by standard WKB-type tools; another family whose existence and timescale is closely related to the dS horizon. Finally, a third family which dominates for near-extremally-charged BHs and which is also present in asymptotically flat spacetimes.
The last two families of modes seem to have gone unnoticed in the literature.
We give a detailed description of linear scalar perturbations of such BHs, and conjecture that SCC is violated in the near extremal regime. The following chapter is based on \cite{Cardoso:2017soq}.
\section{Introduction}
The study of the decay of small perturbations has a long history in GR.
An increasingly precise knowledge of the quantitative form of the decay of fluctuations is required to advance our understanding of gravitation, from the interpretation of gravitational wave data to the study of fundamental questions like the deterministic character of GR.

The well-known appearance of CHs in astrophysically relevant solutions of Einstein's equations signals a potential breakdown
of determinism within GR---the future history of any observer that crosses such a horizon cannot be determined using the Einstein field equations and the initial data!
Nonetheless, in the context of 
BH spacetimes, one expects that perturbations of the exterior region might be infinitely amplified by a blueshift mechanism,
turning a CH in the BH interior into a singularity/terminal boundary beyond which the field equations cease to make sense.
Penrose's SCC conjecture substantiates this expectation.

On the other hand, astrophysical BHs are expected to be stable due to perturbation damping mechanisms acting in the exterior region. Therefore, whether or not SCC holds true hinges to a large extent on a delicate competition between the decay of perturbations in the exterior region and their (blueshift) amplification in the BH interior.
For concreteness, let $\Psi$ be a linear scalar perturbation  (i.e., a solution of the wave equation) on a fixed
subextremal RN, asymptotically flat or dS BH, with cosmological constant $\Lambda\geq 0$.
Regardless of the sign of $\Lambda$, in standard coordinates, the blueshift effect leads to an exponential divergence governed by the surface gravity of the CH $\kappa_-$.

Now the decay of perturbations depends crucially on the sign of $\Lambda$.
For $\Lambda=0$, $\Psi$ satisfies an inverse power law decay~\cite{Price1,Price2,Dafermos:2014cua,Angelopoulos:2016wcv}
which is expected to be sufficient to stabilize the BH while weak enough to be outweighed by the blueshift amplification. Various results~\cite{Poisson:1990eh,Dafermos:2003wr,Dafermos:2012np,LukOhStrongI,LukOhStrongII} then suggest that, in this case, the CH will become, upon perturbation, a mass inflation singularity, strong enough
to impose the breakdown of the field equations.

For $\Lambda>0$, the situation changes dramatically.
In fact, it has been shown rigorously that, for some $\Psi_0\in\mathbb{C}$~\cite{SaBarretoZworski,BonyHaefner,Dyatlov:2011jd,Dyatlov:2013hba},
\begin{equation}
\label{waveProfile}
|\Psi-\Psi_0|\leq C e^{-\alpha t}\,,
\end{equation}
with $\alpha$  the spectral gap, i.e., the size of the QNM-free strip below the real axis.
Moreover, this result also holds for non-linear coupled gravitational and electromagnetic perturbations of Kerr--Newman--dS (with small angular momentum)~\cite{Hintz:2016gwb,Hintz:2016jak}.
This is alarming as the exponential decay of perturbations might now be enough to counterbalance the blueshift amplification.
As a consequence the fate of the CH now depends on the relation between $\alpha$ and $\kappa_-$.
Will it still, upon perturbation, become a ``strong enough'' singularity in order to uphold SCC?

A convenient way to measure the strength of such a (CH) singularity is in terms of the regularity of the spacetime metric extensions it allows~\cite{Costa:2014aia,Ori:2000fi,Earman}. For instance, mass inflation is related to inextendibility in (the Sobolev space) $H^1$ which turns out to be enough to guarantee the non-existence of extensions as (weak) solutions of the Einstein equations~\cite{Christodoulou:2008nj}, i.e., the complete breakdown of the field equations. As a proxy for extendibility of the metric itself, we will focus on the extendibility of a linear scalar perturbation.
On a fixed RNdS, the results in~\cite{Hintz:2015jkj} (compare with~\cite{CostaFranzen}) show that $\Psi$ extends to the CH with regularity at least
\begin{equation}
\label{hintzRegEstimate}
H^{1/2+\beta} \,, \quad\quad\quad \beta \equiv \alpha/\kappa_-\;.
\end{equation}
Now the non-linear analysis of~\cite{Hintz:2016gwb,Hintz:2016jak,Costa:2017tjc,Dafermos:2017dbw} suggests that the metric will have similar
extendibility properties as the scalar field. It is then tempting to conjecture, as was done before in 
Refs.~\cite{Maeda:1999sv,Dafermos:2012np,Costa:2014yha}:
\emph{if there exists a parameter range for which $\beta>1/2$
then the corresponding (cosmological) BH spacetimes should be extendible beyond the CH with metric in $H^1$.}\footnote{see also Section \ref{section weak solutions} and Appendix \ref{appC} for a rigorous proof of the requirement $\beta>1/2$.} Even more strikingly, one may be able to realize some of the previous extensions as weak solutions of the Einstein equations. This would correspond to a severe failure of SCC, in the presence of a positive cosmological constant! 

The construction of bounded Hawking mass solutions of the Einstein-Maxwell-scalar field system with a cosmological constant allowing for $H^1$ extensions beyond the CH was carried out in~\cite{Costa:2017tjc}.

It is also important to note that if $\beta$ is allowed to exceed unity then (by Sobolev embedding)  the scalar field extends in $C^1$; the coupling to gravity should then lead to the existence of solutions with bounded Ricci curvature. Moreover, for spherically symmetric self gravitating scalar fields, the control of both the Hawking mass and the gradient of the field is enough to control the Kretschmann scalar~\cite{Costa:2014aia}.
We will henceforth relate $\beta<1$ to the blow up of curvature components.

At this moment, to understand the severeness of the consequences of the previous discussion,  what we are most lacking is an understanding of how the decay rate of perturbations $\alpha$ is related to $\kappa_-$. Since $\alpha$ is the spectral gap, this can be achieved by the computation of the QNMs of RNdS BHs.
The purpose of this work is to perform a comprehensive study of such modes and to discuss possible implications for SCC by determining $\beta$ throughout the parameter space of RNdS spacetimes.\\~\\~\\
\section{Setting}
We focus on charged BHs in dS spacetimes, the RNdS solutions.
In Schwarzschild-like coordinates, the metric reads
\begin{equation}
\label{RNdS_space}
ds^2=-f(r)dt^2+\frac{dr^2}{f(r)}+r^2(d\theta^2+\sin^2\theta d\phi^2)\,,
\end{equation}
where $f(r)=1-{2M}{r^{-1}}+{Q^2}{r^{-2}}-\Lambda r^2/3$.
 $M,\,Q$ are the BH mass and charge and $\Lambda$ is the cosmological constant.
The surface gravity of each horizon is then
\begin{equation}
\label{surfGrav}
\kappa_i= \frac{1}{2}|f'(r_i)|\;\;,\; i\in\{-,+,c\}\;,
\end{equation}
where $r_-<r_+< r_c$ are the CH, event horizon and cosmological horizon radius. A minimally coupled scalar field $\Psi$ on a RNdS background with harmonic time dependence can be expanded in terms of spherical harmonics,
\begin{equation}
\Psi\sim\sum_{l m}\frac{\psi_{l m}(r)}{r}Y_{lm}(\theta,\phi)e^{-i\omega t}\,.
\end{equation}
Dropping the subscripts on the radial functions, they satisfy the equation
\begin{equation}
\label{master_eq_RNdS2}
\frac{d^2 \psi}{d r_*^2}+\left(\omega^2-V_l(r)\right)\psi=0\,,
\end{equation}
where we introduced the tortoise coordinate $dr_*={dr}/{f(r)}$.
The effective potential for scalar perturbations is
\begin{equation}
\label{RNdS_general potential2}
V_l(r)=f(r)\left(\frac{l(l+1)}{r^2}+\frac{f^\prime(r)}{r}\right),
\end{equation}
where $l$ is an angular number, corresponding to the eigenvalue of the spherical harmonics. We will be mostly interested in the characteristic frequencies of this spacetime, obtained by imposing the boundary conditions
\begin{equation}
 \psi(r\to r_+)\sim e^{- i\omega r_*}\;\;,\;\;\psi(r \to r_c)\sim e^{ i\omega r_*}\,,\label{bcs}
\end{equation}
which select a discrete set of frequencies $\omega_{l n}$, called the QN frequencies~\cite{Berti:2009kk}.
They are characterized, for each $l$, by an integer $n\geq 0$
labeling the mode number. The fundamental mode $n=0$ corresponds, by definition, to the longest-lived mode, i.e., to the frequency with the smallest (in absolute value) imaginary part.

To determine the spectral gap $\alpha$, and hence the decay rate of perturbations, we will focus on the set of all modes $\omega_{l n}$~\footnote{For $l=0$ there is a zero mode, corresponding to $\Psi_0$ in Eq.~\eqref{waveProfile}, which we ignore here.}
and set
\begin{equation}
\alpha\equiv {\rm inf}_{l n}\left\{-\operatorname{Im}(\omega_{l n})\right\},\, \qquad \beta\equiv \alpha/\kappa_-  \,.\label{betaldef}
\end{equation}
We will henceforth drop the ``$l n$'' subscripts to avoid cluttering.
In previous works, we have used a variety of methods to compute the QNMs~\cite{Berti:2009kk,GRIT}.
The results shown here were obtained mostly with the Mathematica package of~\cite{Jansen:2017oag} (based on methods developed in~\cite{Dias:2010eu}), and checked in various cases with
a variety of other methods~\cite{Berti:2009kk,GRIT,KaiLin1,Iyer:1986np}.
%
\section{QNMs of RNdS BHs: the three families}
Our results are summarized in Figs.~\ref{Qdependence}-\ref{nearExtremalModes} where one can distinguish three families of modes:

\subsection{Photon sphere modes}
\begin{figure}[H]
\includegraphics[scale=0.55]{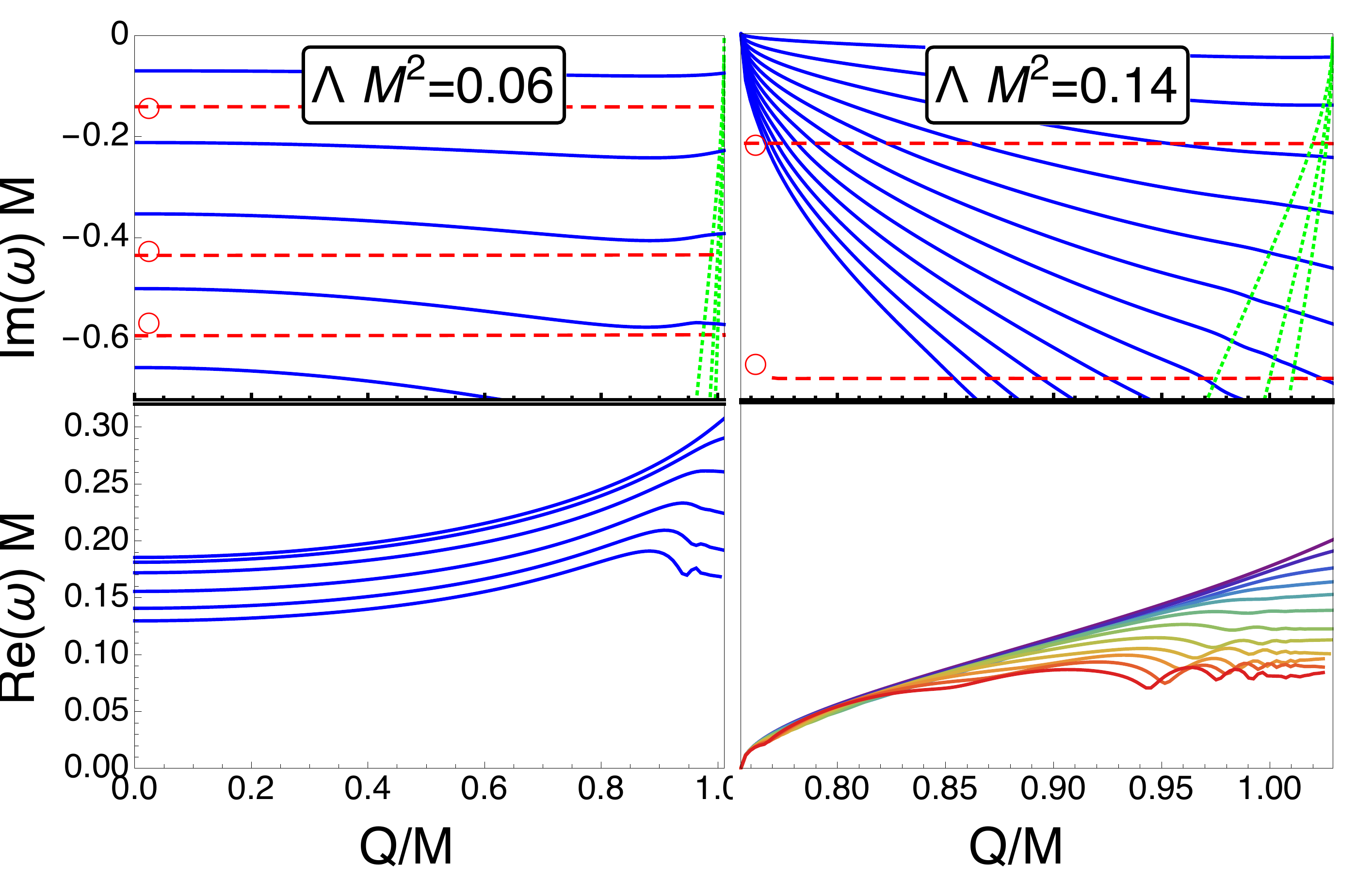}
\caption{
Lowest lying quasinormal modes for $l=1$ and $\Lambda M^2 = 0.06$ (left) and $0.14$ (right), as a function of $Q/M$.
The top plots show the imaginary part, with dashed red lines corresponding to purely imaginary modes, and solid blue to complex, ``PS'' modes, whose real part is shown in the lower plots.
The red circles in the top plots indicate the modes of empty de Sitter at the same $\Lambda$, which closely matches the first imaginary mode here, but lie increasingly less close to the higher modes.
Near the extremal limit of maximal charge, another set of purely imaginary modes (dotted green lines) comes in from $-\infty$ and approaches $0$ in the limit.
Only a finite number of modes are shown, even though we expect infinitely many complex and extremal modes in the range shown.
\label{Qdependence}}
\end{figure}

BHs and other sufficiently compact objects have trapping regions. Here, null particles can be trapped on circular unstable trajectories, defining the photon sphere.
This region has a strong pull in the control of the decay of fluctuations and the spacetime's QNMs which have large frequency (i.e., large $|{\rm Re}\,\omega|$)~\cite{Cardoso:2017njb,Cardoso:2008bp,saBarreto,Dyatlov:2011jd}. For instance, the decay timescale
is related to the instability timescale of null geodesics near the photon sphere.
For BHs in dS space, we do find a family of modes which can be traced back to the photon sphere.
We refer to them as ``photon sphere modes,'' or in short ``PS'' modes.
These modes are depicted in blue (solid line) in Figs.~\ref{Qdependence}-\ref{nearExtremalModes}. Different lines correspond to different overtones $n$; the fundamental mode
is determined by the large $l$ limit (and $n=0$); we find that $l=10$ or $l=100$ provide good approximations of the imaginary parts of the dominating mode; note however that the real parts do not converge when $l\rightarrow\infty$. These modes are well-described by a WKB approximation, and for very small cosmological constant they asymptote to the Schwarzschild BH QNMs~\cite{GRIT}.

For small values of the cosmological constant, PS modes are only weakly dependent on the BH charge. This is apparent from Fig.~\ref{Qdependence}.
For $\Lambda M^2 > 1/9$ there is now a nonzero minimal charge, at which $r_+ = r_c$. This limit is the charged Nariai BH and is shown as the blue dashed line in Fig.~\ref{ContourPlot}.
The corresponding QNMs are also qualitatively different, as seen in Fig.~\ref{Qdependence}. They in fact vanish in this limit, a result
that can be established by solving the wave equation analytically to obtain (see Ref.~\cite{Cardoso:2003sw} for the neutral case, we have generalized it to charged BHs, see Section \ref{supl})
\begin{equation}
\frac{\text{Im}(\omega)}{\kappa_+} = - i \left(n + \frac{1}{2}\right) \, . \label{nariai}
\end{equation}
Note that the results presented here are enough to disprove a conjecture~\cite{Brady:1998au} that suggested that $\alpha$ should be equal to $\min\{\kappa_+,\kappa_c\}$.
Such possibility is inconsistent with~\eqref{nariai} and
it is also straightforward to find other non-extremal parameters for which the WKB prediction yields smaller $\alpha$'s  (e.g. for $\Lambda M^2=0.1$ and $Q=0$ we have $\kappa_+=0.06759$, $\kappa_c=0.05249$, and $\alpha=0.03043$).

\subsection{dS modes}
Note that solutions with purely imaginary $\omega$ exist in pure dS spacetime~\cite{Du:2004jt,LopezOrtega:2006my,VasydS}
\begin{eqnarray}
\omega_{0, \rm pure \,\rm dS}/\kappa_c^{\rm dS} &=&-i l\,,\\
\omega_{n\neq0, \rm pure \,\rm dS}/\kappa_c^{\rm dS} &=&-i(l+n+1)\,.\label{pure_dS_scalar01}
\end{eqnarray}
Our second family of modes, the (BH) dS modes, are deformations of the pure dS modes~\eqref{pure_dS_scalar01}; the dominant mode ($l=1, n=0$) is almost identical and higher modes have increasingly larger deformations.

These modes are intriguing, in that they have a surprisingly weak dependence on the BH charge and seem to be described by the surface gravity $\kappa_c^{\rm dS}=\sqrt{{\Lambda}/{3}}$ of the cosmological horizon of pure dS space, as opposed to that of the cosmological horizon in the RNdS BH under consideration. This can, in principle, be explained by the fact that the accelerated expansion of the RNdS spacetimes is also governed by $\kappa_c^{\rm dS}$~\cite{Brill:1993tw,Rendall:2003ks}.

This family has been seen in time-evolutions~\cite{Brady:1996za,Brady:1999wd} but, to the best of our knowledge, was only recently identified in the QNM calculation of neutral BH spacetimes~\cite{Jansen:2017oag}. Furthermore, our results indicate that as the BH ``disappears'' ($\Lambda M^2\to 0$), these modes converge to the exact dS modes (both the eigenvalue and the eigenfunction itself).

\subsection{Near-extremal modes}
Finally, in the limit that the Cauchy and event horizon radius approach each other, a third ``near extremal'' family, labeled as $\omega_{\rm NE}$, dominates the dynamics.
In the extremal limit this family approaches
\begin{equation}
\label{NEanalytic}
\omega_{\rm NE} = - i (l + n + 1) \kappa_- = - i (l + n + 1) \kappa_+ \, ,
\end{equation}
independently of $\Lambda$, as shown by our numerics. As indicated by~\eqref{NEanalytic}, the dominant mode in this family is that for $l = 0$, this remains true away from extremality.

In the asymptotically flat case, such modes seem to have been described analytically in Refs.~\cite{Kim:2012mh,Hod:2017gvn}. Here we have shown numerically that such modes exist, and that they are in fact the limit of a new family of modes.
It is unclear if the NE family is a charged version of the Zero-Damping-Modes discussed recently in the context of rotating Kerr BHs~\cite{Richartz:2017qep}. It is also unclear if there is any relation between such long lived modes and the instability of exactly extremal geometries~\cite{Aretakis:2012ei,Casals:2016mel}.

\section{Maximizing $\beta$}
%
\begin{figure}[H]
\includegraphics[scale=0.75]{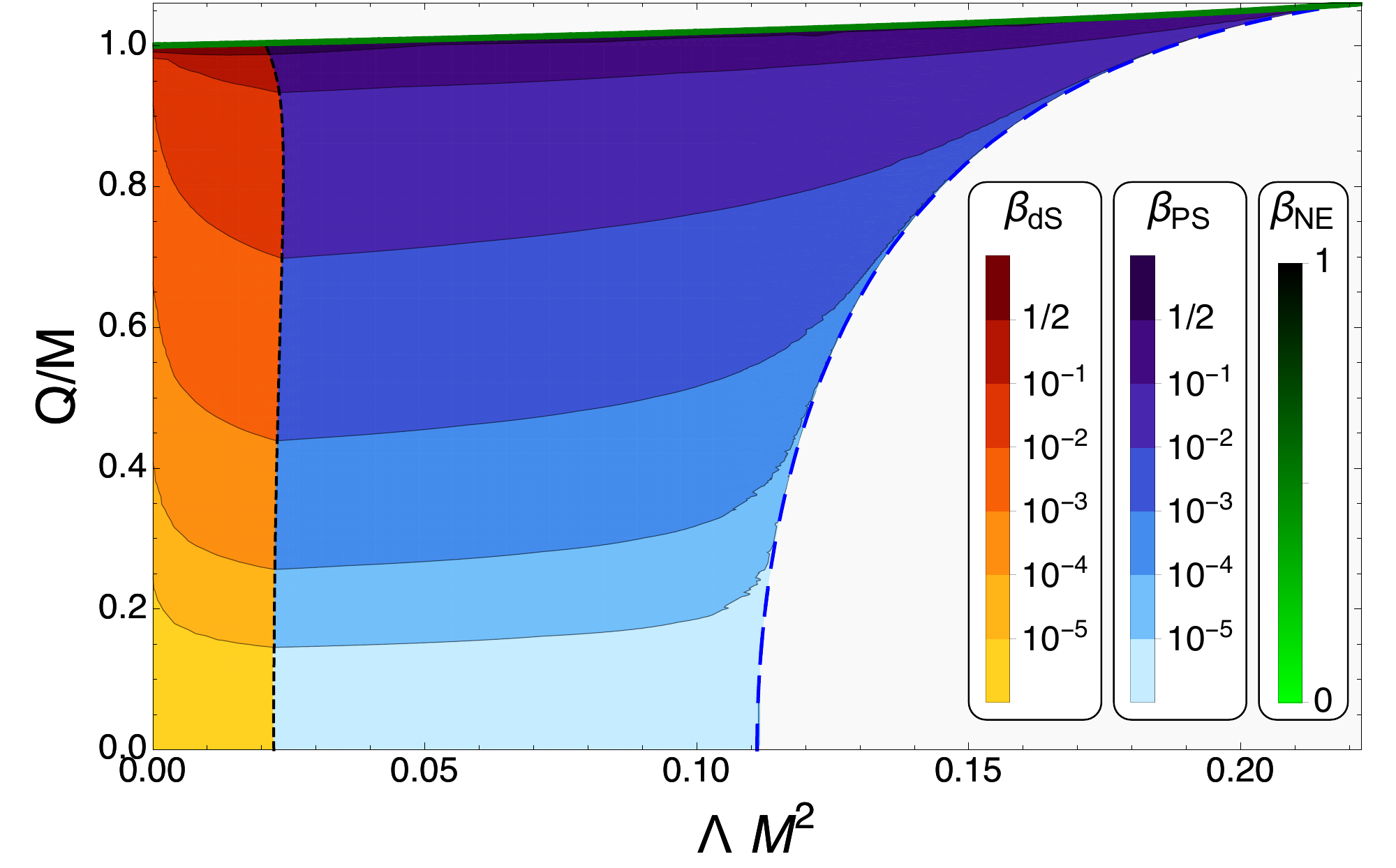}
\caption{Parameter space of the RNdS solutions, bounded by a line of extremal solutions of maximal charge where $r_- = r_+$ on top, and for $\Lambda M^2 > 1/9$ a line of extremal solutions where $r_c = r_+$.
In the physical region the value of $\beta$ is shown. For small $\Lambda M^2$ the dominant mode is the $l=1$ de Sitter mode, shown in shades of red.
For larger $\Lambda M^2$ the dominant mode is the large $l$ complex, PS mode, here showing the $l=100$ WKB approximated mode in shades of blue.
For very large $Q/M$ the $l=0$ extremal mode dominates. The green sliver on top where the NE mode dominates is merely indicative, the true numerical region is too small to be noticeable on these scales.
}
\label{ContourPlot}
\end{figure}
The dominating modes of the previous three families determine $\beta$, shown in Fig.~\ref{ContourPlot}.
Each family has a region in the parameter space where it dominates over the other families.
The dS family is dominant for ``small'' BHs (when $\Lambda M^2\ \lesssim 0.02$).
In the opposite regime the PS modes are dominant.
Notice that in the limit of minimal charge $\beta = 0$, since $\kappa_-$ remains finite while the imaginary parts of QNMs in the PS family approach 0 according to~\eqref{nariai} (since $\kappa_+\rightarrow 0$). More interesting is the other extremal limit, of maximal charge. In Fig.~\ref{ContourPlot}, the uppermost contours of the dS and PS families show a region where $\beta > 1/2$. 
Within this region as the charge is increased even further, the NE family becomes dominant.
In Fig.~\ref{ContourPlot} this is shown merely schematically, as the region is too small to plot on this scale, but it can already be seen in Fig.~\ref{Qdependence}.

\begin{figure}[H]
\includegraphics[scale=0.55]{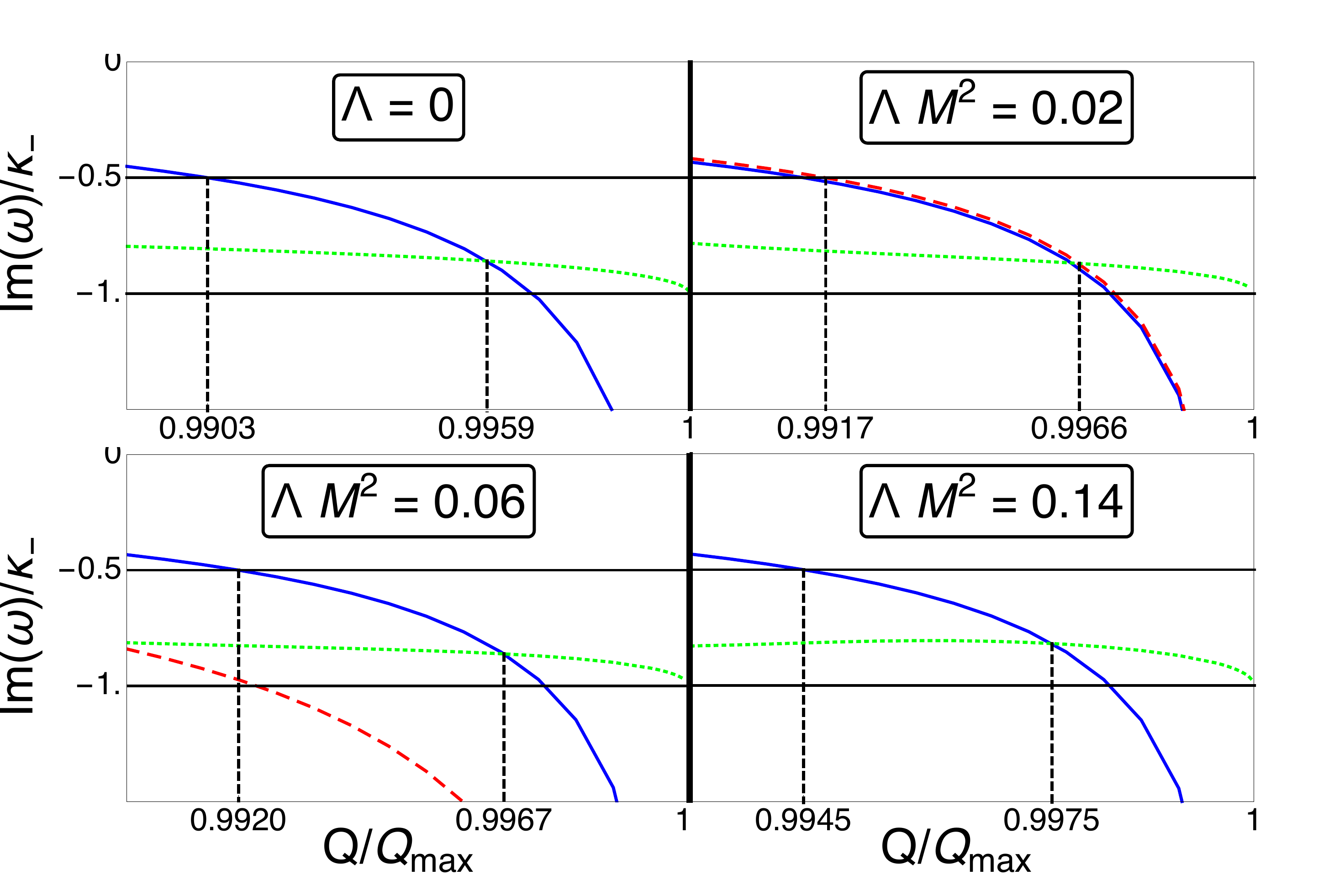}
\caption{
Dominant modes of different types, showing the (nearly) dominant complex PS mode (blue, solid) at $l= 10$, the dominant de Sitter mode (red, dotted) at $l=1$ and the dominant NE mode (green, dashed) at $l=0$.
The two dashed vertical lines indicate the points where $\beta \equiv -\operatorname{Im}(\omega)/\kappa_- =1/2$ and where the NE becomes dominant. (Note that the value of $\beta$ is only relevant for $\Lambda>0$.)
}
\label{nearExtremalModes}
\end{figure}
To see more clearly how $\beta$ behaves in the extremal limit we show 4 more constant $\Lambda M^2$ slices in Fig.~\ref{nearExtremalModes}.
Here one sees clearly how above some value of the charge $\beta>1/2$, as dictated by either the dS or the
PS family. Increasing the charge further, $\beta$ would actually diverge if it were up to these two families ($\omega M$ approaches a constant for both families, so $\omega/\kappa_-$ diverges).
However, the NE family takes over to prevent $\beta$ from becoming larger than 1. Further details on these modes and on the maximum $\beta$ are shown in Section \ref{supl}.

\section{Conclusions}
The results in~\cite{Hintz:2016gwb,Hintz:2016jak} show that the decay of small perturbations of dS BHs is dictated by the spectral gap $\alpha$. At the same time, the linear analysis in~\cite{Hintz:2015jkj} and the non-linear analysis in~\cite{Costa:2017tjc} indicate that the size of $\beta\equiv\alpha/\kappa_-$ controls the stability of CHs and consequently the fate of the SCC conjecture.  Recall that for the dynamics of the Einstein equations, and also for the destiny of observers, the blow up of curvature (related to $\beta<1$) per se is of little significance: it implies neither the breakdown of the field equations~\cite{Klainerman:2012wt} nor the destruction of macroscopic observers~\cite{Ori:1991zz}.

In fact, a formulation of SCC in those terms is condemned to overlook relevant physical phenomena like impulsive gravitational waves or the formation of shocks in relativistic fluids. For those and other reasons, the modern formulation of SCC, which we privilege here, makes the stronger request $\beta<\frac{1}{2}$ in order to guarantee the breakdown of the field equation at the CH.

Here, by studying (linear) massless scalar fields and searching through the entire parameter space of subextremal and extremal RNdS spacetimes, we find ranges for which $\beta$ exceeds $1/2$ but, remarkably, it doesn't seem to be allowed, by the appearance of a new class of ``near-extremal'' modes, to exceed unity! This opens the perspective of having CHs which, upon perturbation, can be seen as singular, by the divergence of curvature invariants, but nonetheless maintain enough regularity as to allow the field equations to determine (classically), in a highly non-unique way, the evolution of gravitation. 

This corresponds to a severe failure of determinism in GR that cannot be taken lightly in view of the importance that a cosmological constant has in cosmology and the fact that the pathologic behavior is observed in parameter ranges which are in loose agreement with what one expects from the parameters of some astrophysical BHs. Of course, astrophysical BHs are expected to be neutral and here we are dealing with charged BHs. This is justified by the standard charge/angular momentum analogy, where near-extremal charge corresponds to fast rotating BHs~\cite{2011ApJ...736..103B,Middleton:2015osa,FastSpin}.
\section{Supplementary material}\label{supl}
\subsection{The eigenfunctions}
\begin{figure}[H]
\includegraphics[scale=0.4]{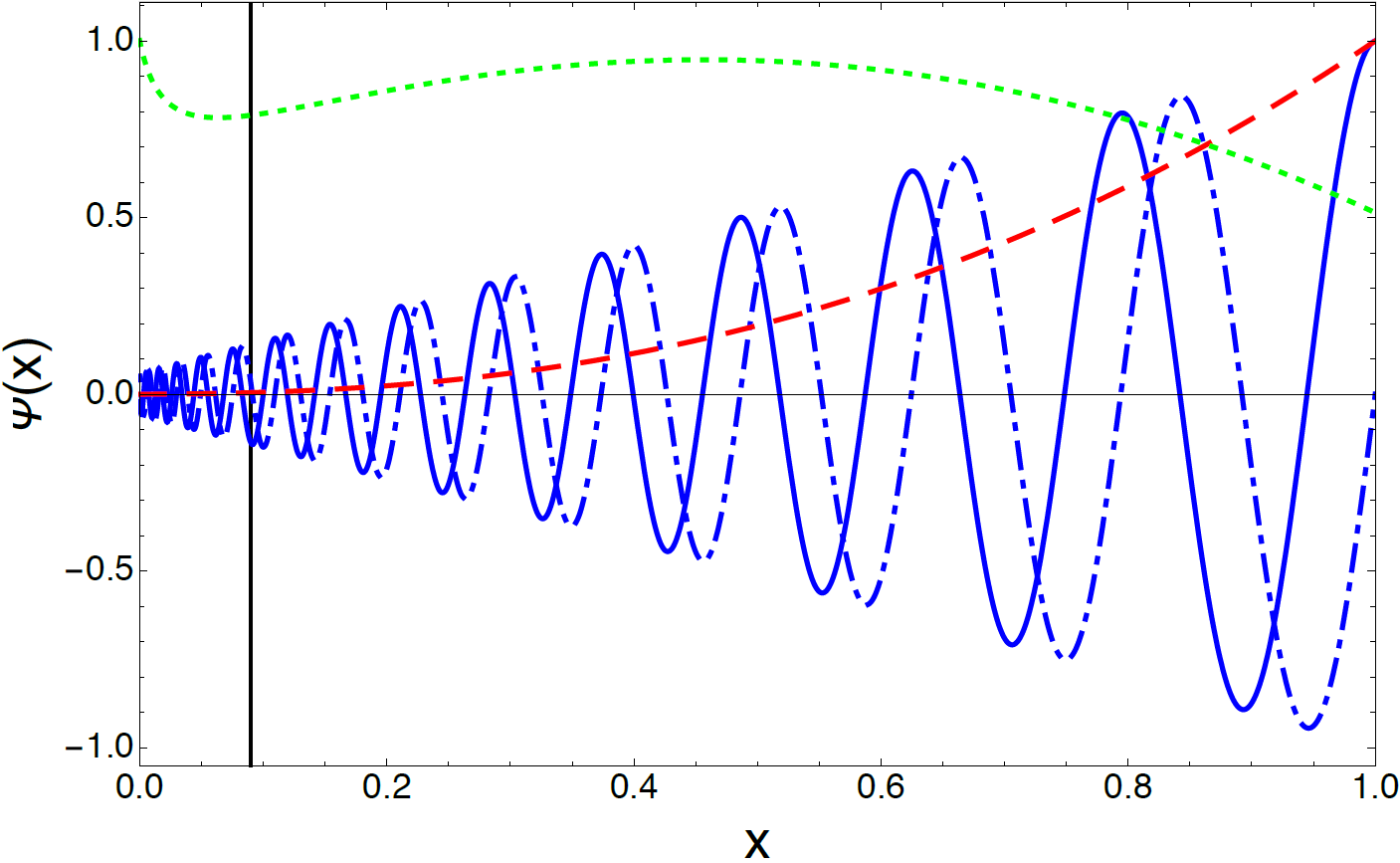}
\caption{
Scalar wavefunctions $\psi(x)$ (defined in Eq.~(\ref{phidef}), with $x = (r - r_+)/(r_c - r_+)$) of the dominant mode of each of the three families, at $\Lambda M^2 = 0.02$ and $Q/Q_{\text{max}} \approx 0.9966$, as indicated in the top right of Fig. \ref{nearExtremalModes}.
Shown are the near extremal mode with $\omega_{\text{NE}}/\kappa_- = -0.87005i$ (for $l=0$, green dotted line), 
the dS mode with $\omega_{\text{dS}}/\kappa_- = -0.87043i$ (for $l=1$, red dashed line), 
and the complex mode with $\omega_{\text{C}}/\kappa_- = 26.448 -0.89101i$ (for $l=10$, blue solid and dash-dotted line for real and imaginary parts). The solid vertical black line indicates the light ring.
}
\label{eigenfunctionPlot}
\end{figure}

The difference between PS, dS and NE modes is also apparent from the eigenfunction itself. It is useful to define a re-scaled function
$\psi(r)$ as,
\begin{equation}
\label{phidef}
\Psi(r) = (r - r_+)^{- i \omega / (2 \kappa_+) }  \psi(r)  (r_c - r)^{- i \omega / (2 \kappa_c) } \,.
\end{equation}
The conditions on $\psi(r)$ are that it approaches a constant as $r \rightarrow r_+, r_c$.
Figure~\ref{eigenfunctionPlot} shows the behavior of $\psi$ for different modes, for a specific set of RNdS parameters.
Although not apparent, there is structure close to the photon sphere for the PS eigenfunction. 

\subsection{Analytic solutions for $r_c = r_+$}

In the limit $r_c = r_+$, the limit of minimal charge for $\Lambda M^2 \geq 1/9$, the QNMs can be found analytically.
In this limit, Eq. \eqref{master_eq_RNdS2} with potential \eqref{RNdS_general potential2}, written in the coordinate $x=(r-r_+)/(r_c-r_+)$, becomes
\begin{equation}
\left( \frac{4  r_+ x (1-x) }{3-2 r_+} l (l+1) + \lambda^2 \right) \psi(x) + 4 x (1-x) (1-2 x) \psi^\prime(x) + 4 x^2 (1-x)^2  \psi^{\prime\prime}(x) = 0 \,
\end{equation}
where we defined $\lambda \equiv \omega / \kappa_+$ and we have set $M=1$, which can be restored in the end by dimensional analysis.
This equation has the solutions
\begin{equation}
\psi(x) = c_1 P_{\alpha }^{i \lambda }(2 x-1)+c_2 Q_{\alpha }^{i \lambda }(2 x-1) \, ,
\end{equation}
where $P, Q$ are the Legendre $P$ and $Q$ functions, and
\begin{equation}
\alpha = \frac{1}{2} \left( -1 + \sqrt{1 - \frac{2 r_+}{r_+ - 3/2} l (l+1)} \right) \, .
\end{equation}

Now, analysis of the asymptotic behavior near $x=0$ and $x=1$ shows that we can only satisfy the ``outgoing'' boundary conditions~\cite{Berti:2009kk} 
when $c_2 = 0$ and $\lambda$ is either $\lambda = - i (\alpha + n)$, or $\lambda = -i (n + 1 - \alpha)$.
These combine to give
\begin{equation}
\frac{\omega}{\kappa_+} = \pm \frac{1}{2} \sqrt{-1 +  l(l+1) \frac{2 r_+}{r_+ - 3/2}} - i \left( n + \frac{1}{2}\right) \, ,
\end{equation}
Restoring units we obtain,
\begin{eqnarray}
\label{modeMinQ}
\frac{\omega}{\kappa_+} &=& \pm \frac{1}{2} \sqrt{-1 + 2 l (l+1) \Upsilon}  - i \left(n + \frac{1}{2}\right) \, , \label{nariai_1} \\
\Upsilon&=&\frac{\gamma^2 + 2^{1/3} \Lambda M^2}{\gamma^2 +( 2^{1/3} - 3 \times 2^{-1/3} \gamma) \Lambda M^2}\,, \nonumber \\
\gamma & =& \left( - 3 (\Lambda M^2)^2 + (\Lambda M^2)^{3/2} \sqrt{(9 \Lambda M^2 - 2)} \right)^{1/3} \,. \nonumber
\end{eqnarray}
The argument of the square root in \eqref{nariai_1} is positive.
The imaginary part of these frequencies is exactly the same as that of the neutral Nariai BH~\cite{Cardoso:2003sw}.
The real part is different and now depends also on $\Lambda$, but asymptotes to the previous result for $Q=0$ (i.e., $\Lambda = 1/9$), as it should. It is an interesting feature that when $\Lambda M^2 \rightarrow 2/9$, $\Upsilon$, and therefore the real part of $\omega/\kappa_+$, diverges.

\subsection{Searching for long-lived modes}
Here we address the question whether there might be a more slowly decaying mode that we have missed and could save SCC.
If such a mode exists, it would be highly unlikely to be part of the three families we found, since we can follow their continuous change as the BH parameters are varied, as shown in the figures.
Furthermore, the known modes in the limiting cases are all accounted for, and we never observed any mode crossings for given family and angular momentum $l$.
\begin{table}[H]
\begin{center}
\begin{tabular}{|c||c|}
\hline
 $l$ & $\omega _ 0/\kappa _-$ \\ \hline\hline
 0 & \underline{\bf-0.8539013779 i(61)} \\ \hline
 1 & 3.2426164126 - 0.7958326323 i(67) \\ 
     & ( \underline{\bf -1.5003853731 i(64)}) \\ \hline
 2 & 5.4796067815 - 0.7754179185 i(69) \\ \hline
 3 & 7.7016152057 - 0.7699348317 i(70) \\ \hline
 4 & 9.9181960834 - 0.7676996545 i(73) \\ \hline
 5 & 12.1322536641 - 0.7665731858 i(75) \\ \hline
 10 & \underline{\bf23.1897597770 - 0.7649242108 i(80)} \\ \hline
 100 & 222.0602900249 - 0.7643094446 i(82) \\ \hline
\end{tabular}
\end{center}
\caption{
The dominant QNMs for a range of angular momenta $\ell$, in units of the surface gravity of the CH, for the BH with $\Lambda M^2 = 0.06$ and $Q/Q_\text{max} = 0.996$.
The bold, underlined modes at $l=0, 1$ and $10$ are the dominant modes for the NE, dS and PS modes respectively, as seen also in the bottom left of Fig. \ref{nearExtremalModes}.
Note that the dS mode is subdominant even for fixed $l=1$, and the PS mode at $l=10$ is only dominant to very good approximation, the true dominant mode being that with $l \rightarrow \infty$.
Numbers in brackets indicate the number of agreed digits in the computations with grid size and precision $(N,p) = (400,200)$ and $(450,225)$.
}
\label{extracheck}
\end{table}

It is theoretically possible, though also very unlikely, that there is a fourth family (with anywhere between a single and infinitely many members) that we have missed.

Typically, smaller eigenvalues are found more easily than larger eigenvalues, making it more unlikely to miss the dominant mode.
It could be however that the corresponding eigenfunction is either very sharply peaked or highly oscillatory, in which cases it would require a large number of grid points to be resolved accurately enough.
This again decreases the possibility that we have missed something. We will rule out this last scenario, as best as we can numerically, as follows. We pick a representative BH for which $\beta > 1/2$, indicating violation of SCC, namely $\Lambda M^2 = 0.06$ and $Q/Q_\text{max} = 0.996$. For these BH parameters we compute the QNMs for various angular momenta $l$ as shown in Table~\ref{extracheck}.
There is no new QNM that is more dominant than those found before, except as expected the $l= 100$ photon sphere mode, but not significantly, note the extremely rapid convergence with increasing $l$.

The main method we use essentially discretizes the equation and rearranges it into a generalized eigenvalue equation, whose eigenvalues are the QNMs (see~\cite{Jansen:2017oag} for more details). 
This has two technical parameters, the number of grid points $N$ and the precision $p$ (number of digits) used in the computation. 
To be sure that the obtained results are not numerical artefacts one has to repeat the computation at different $(N,p)$ and test for covergence, which we have done for all results shown.

The computation here was done at even higher accuracy than in the main results, with $(N,p) = (400,200)$ and $(450,225)$.
The most we used previously was  $(300,150)$ and $(350,175)$ (near extremality, away from extremality a much lower accuracy usually suffices). The number in brackets behind each mode in Table~\ref{extracheck} is the number of digits that agrees between the computations at these two accuracies. 

We checked that even before testing for convergence, there are no modes with imaginary part smaller (in absolute sense) than shown in Table~\ref{extracheck}.
This confirms our results with as much certainty as one can reasonably expect from a numerical result.

\chapter{Strong Cosmic Censorship in charged black-hole spacetimes: still subtle}\label{PRD}
In Chapter \ref{PRL} it was shown that SCC may be violated in highly charged BH spacetimes living in a universe with a positive cosmological constant.
Several follow-up works have since suggested that such result, while conceptually interesting, cannot be upheld in practice.
We focus here on the claim that the presence of charged massive scalars suffices to save SCC.
To the contrary, we show that there still exists a finite region in parameter space where SCC is expected to be violated. The following chapter is based on \cite{PhysRevD.98.104007}.
\section{Introduction}
In Chapter \ref{PRL} we presented an indication that SCC might be violated for charged, near-extremal RN BHs in a dS spacetime. More precisely, for linear massless and neutral scalar perturbations of RNdS BHs, the basic quantity controlling the stability of the CH, and therefore the fate of SCC, is given by~\cite{Hintz:2015jkj}
\begin{equation}
\label{betaOld1}
\beta \equiv -\text{Im}(\omega_0)/\kappa_-\;,
\end{equation}
where $\omega_0$ is the longest-lived non-zero QNM and $\kappa_-$ is the surface gravity of the CH.
Moreover, the results in~\cite{Hintz:2016gwb,Hintz:2016jak,Costa:2017tjc} suggest that $\beta$ remains the essential quantity in the non-linear setting: the higher $\beta$, the more stable the CH. Concretely, the modern formulation of SCC\footnote{inextendibility of metric coefficients in $H^1_{\rm loc}$ and of Christoffel symbols in $L^2_{\rm loc}$ across the CH.} demands that
\begin{equation}\label{theoremHintz1}
\text{SCC} \leftrightarrow  \beta < 1/2
\end{equation}
in order to guarantee the breakdown of the field equation at the CH. One should also recall that $\beta<1$ is related to the blow up of curvature invariants.
In Chapter \ref{PRL}, a thorough numerical study of $\beta$ for the full range of subextremal RNdS spacetimes revealed, quite surprisingly, that $\beta>1/2$ in the near-extremal regime. However, it turned out that $\beta\leq 1$ always, with equality at extremal charge.
This provides evidence for the existence of CHs which, upon perturbation, are rather singular due to the divergence of curvature invariants, but where the gravitational field can still be described by the field equations; the evolution of gravitation beyond the CH however is highly non-unique. This corresponds to a severe failure of determinism in GR.

There are different ways to interpret the results of Chapter \ref{PRL}. One could take the SCC conjecture in its conceptual version, where SCC is purely a mathematical question about GR and its limits. Then the results of Chapter \ref{PRL} either signify a failure of SCC, or are superseded by nonlinear effects. 
Here, we have nothing else to add on this purely mathematical question.

Alternatively, one can interpret the SCC conjecture in an anthropic-astrophysical sense, where restrictions arising from experimental or observational data (including gravitational waves, BH and cosmological observations, or information arising from particle physics) need to be taken into account. In other words, in such a viewpoint GR would need to be supplemented with all the fields of the Standard Model and perhaps even with quantum-gravity effects. In this context, the following are commonly accepted facts: 

\noindent {\bf i.} first, BHs in our universe are nearly neutral.
Electromagnetic charge is quickly neutralized by either environmental plasma, Schwinger pair-creation or Hawking evaporation~\cite{Cardoso:2016olt}. In light of this, one can question the relevance
of SCC violations in highly charged, non-spinning BH spacetimes;

\noindent {\bf ii.} to form a charged BH, charged matter is necessary. Thus, SCC violation with only neutral fields is unrealistic, and needs to be generalized to charged fields as well.

The results of Chapter \ref{PRL} were followed by various attempts to save the conjecture, supported by observation i.\ or ii.\ above. 
Firstly, it was shown that the CHs of rapidly rotating BHs in cosmological backgrounds behave differently from those of highly charged BHs~\cite{Dias:2018ynt}. 
According to Ref.~\cite{Dias:2018ynt}, in KdS Eq.~\eqref{theoremHintz1} remains valid, but now $\beta$ seems to be bounded exactly by $1/2$, with the bound being saturated at extremality. 
Such a result might suffice to save SCC in the context of astrophysical BHs. However, the behavior of rapidly spinning, but weakly charged BHs is unknown, and these may well exist in our Universe.

Here, we will discuss another work~\cite{Hod:2018dpx} providing evidence that when point ii.\ above is taken into account and charged scalars are considered, then $\beta<1/2$ in an appropriate region of parameter space, and consequently SCC is upheld. 
This last implication requires, first of all, the validity of~\eqref{theoremHintz1} for charged scalars, which does require a justification~\footnote{We stress the fact that the relation~\eqref{theoremHintz1} is not universal. In fact, for BHs with vanishing cosmological constant the value of $\beta$ seems to be irrelevant in the context of SCC \cite{LukOhStrongI,LukOhStrongII}.}.
In Appendix~\ref{appC} we show that Eq.~\eqref{theoremHintz1} does generalize in the expected way, with the critical value being, once again, $\beta = 1/2$.
In addition, the methods in Ref.~\cite{Hod:2018dpx} require working in the large-coupling regime $q Q\gg {\rm max}(\mu\,r_+,l+1)$, with $Q$ the BH charge, $q$ the field charge, $\mu$ the scalar field mass, and $r_+$ the radius of the event horizon.

We finish this section by acknowledging yet another interesting recent suggestion to remedy SCC, in the presence of a positive cosmological constant: in Ref.~\cite{Dafermos:2018tha}, it was shown that the pathologies identified in Chapter \ref{PRL} become non-generic if one considerably enlarges the allowed set of initial data by weakening their regularity. Although the considered data are compatible with the modern formulation of SCC, we believe that SCC is, in essence, a formation of singularities problem\footnote{In contrast, Weak Cosmic Censorship is concerned with the avoidance of naked singularities.} which is mainly of interest for regular initial data; the mechanism of SCC becomes obscured if one considers initial data which are too ``rough'' (compare with the problem of the formation of shocks in fluid mechanics~\cite{ChristodoulouShocks}).

\section{Charged Scalar Perturbations of RNdS}
The purpose of our work is to explore the decay of charged scalar fields in the full range of charge coupling $q Q$ and various choices of scalar masses $(\mu M)^2$ on those RNdS BH backgrounds which were identified as pathological in Chapter \ref{PRL}. We will then be addressing concern ii.\ above. Note that a deeper understanding of concern ii.\ would also require the study of fermions. Such study is provided in Chapter \ref{PLB}. We will show that it is not necessary to impose lower bounds on the scalar field mass to obtain $\beta<1/2$. On the other hand, we will demonstrate that for small charge coupling one can still find regions in parameter space where SCC is violated ($\beta>1/2$). 

The background spacetime is a charged RNdS,
\begin{equation}
\label{RNdS_space}
ds^2=-f(r)dt^2+\frac{dr^2}{f(r)}+r^2(d\theta^2+\sin^2\theta d\phi^2)\,,
\end{equation}
where $f(r)=1-{2M}{r^{-1}}+{Q^2}{r^{-2}}-\Lambda r^2/3$. Here, $M,\,Q$ are the BH mass and charge, respectively, and $\Lambda>0$ is the cosmological constant. The surface gravity of each horizon is then
\begin{equation}
\label{surfGrav}
\kappa_i= \frac{1}{2}|f'(r_i)|\;\;,\; i\in\{-,+,c\}\;,
\end{equation}
where $r_-<r_+<r_c$ are the CH, event horizon and cosmological horizon radius. A minimally coupled charged massive scalar field $\Psi$ on a RNdS background with harmonic time dependence can be expanded in terms of spherical harmonics,
\begin{equation}
\Psi\sim\sum_{lm}\frac{\psi_{l m}(r)}{r}Y_{lm}(\theta,\phi)e^{-i\omega t}\,.
\end{equation}
Dropping the subscripts on the radial functions, they satisfy the equation
\begin{equation}
\label{master_eq_RNdS}
\frac{d^2 \psi}{d r_*^2}+\left[\left(\omega-\Phi(r)\right)^2-V_l(r)\right]\psi=0\,,
\end{equation}
where $\Phi(r)=q Q/r$ is the electrostatic potential, $q$ the charge of the scalar field and $dr_*={dr}/{f(r)}$ the tortoise coordinate.
The effective potential for scalar perturbations is
\begin{equation}
\label{RNdS_general potential}
V_l(r)=f(r)\left(\mu^2+\frac{l(l+1)}{r^2}+\frac{f^\prime(r)}{r}\right),
\end{equation}
where $l$ is an angular number, corresponding to the eigenvalue of the spherical harmonics, and $\mu$ the mass of the scalar field.
We are interested in the characteristic quasinormal (QN) frequencies $\omega_{ln}$ of this spacetime, obtained by imposing the boundary conditions~\cite{Berti:2009kk}
\begin{equation}
 \psi(r\to r_+)\sim e^{- i\left(\omega-\Phi(r_+)\right) r_*}\,\,,\,\,
 \psi(r \to r_c)\sim e^{ i\left(\omega-\Phi(r_c)\right) r_*}\nonumber \label{bcs}\,.
\end{equation}
The QN frequencies are characterized, for each $l$, by an integer $n\geq 0$ labeling the mode number. The fundamental mode $n=0$ corresponds, by definition, to the \emph{non-vanishing} frequency with the largest imaginary part and will be denoted by $\omega_0\neq 0$. 

As shown in Appendix~\ref{appC}, for $q Q\neq0$ the stability of the CH continues to be determined by~\eqref{betaOld1}. We note that the only vanishing mode we find (see results below) corresponds to the trivial mode at $l=0$ and $q=\mu=0$. 
In fact, massless, neutral scalars can always be changed by an additive constant without changing the physics. 
Thus, the zero mode is irrelevant for the question of stability of the CH and consequently must be discarded in the definition of $\beta$.

The results shown in the following sections were obtained mostly with the Mathematica package of~\cite{Jansen:2017oag} (based on methods developed in~\cite{Dias:2010eu}), and checked in various cases with a WKB approximation~\cite{Iyer:1986np}.

\section{QNMs of massless, neutral scalar fields}
In~Chapter \ref{PRL},
we found 3 qualitatively different families of QNMs: the PS family, the dS family and the NE family.
The first two connect smoothly to the modes of asymptotically flat Schwarzschild and of empty dS, respectively, while the last family cannot be found in either of these spacetimes.
Finally, apart from the previous 3 families  (for $q=\mu=0$) there is also  a single orphan mode---the trivial zero mode at $l=0$.

\section{Charged Massless Scalars}
Since the main point of the current work is to investigate if the inclusion of charged matter saves SCC, we will restrict ourselves to choices of near extremal BH parameters identified as problematic in~Chapter \ref{PRL} from the point of view of SCC. Since the dependence on the cosmological constant was found to be minimal (provided that it is positive), we will restrict  to $\Lambda M^2 = 0.06$ throughout this chapter. We expect our results to be qualitatively independent of this choice.

The BH charges we consider are:
\begin{equation}
\label{EqBHC}
1 - Q/Q_\text{max}  = 10^{-3},\ 10^{-4},\ 10^{-5}\,.
\end{equation}

\begin{figure}[H]
\centering
\includegraphics[width=0.9\textwidth]{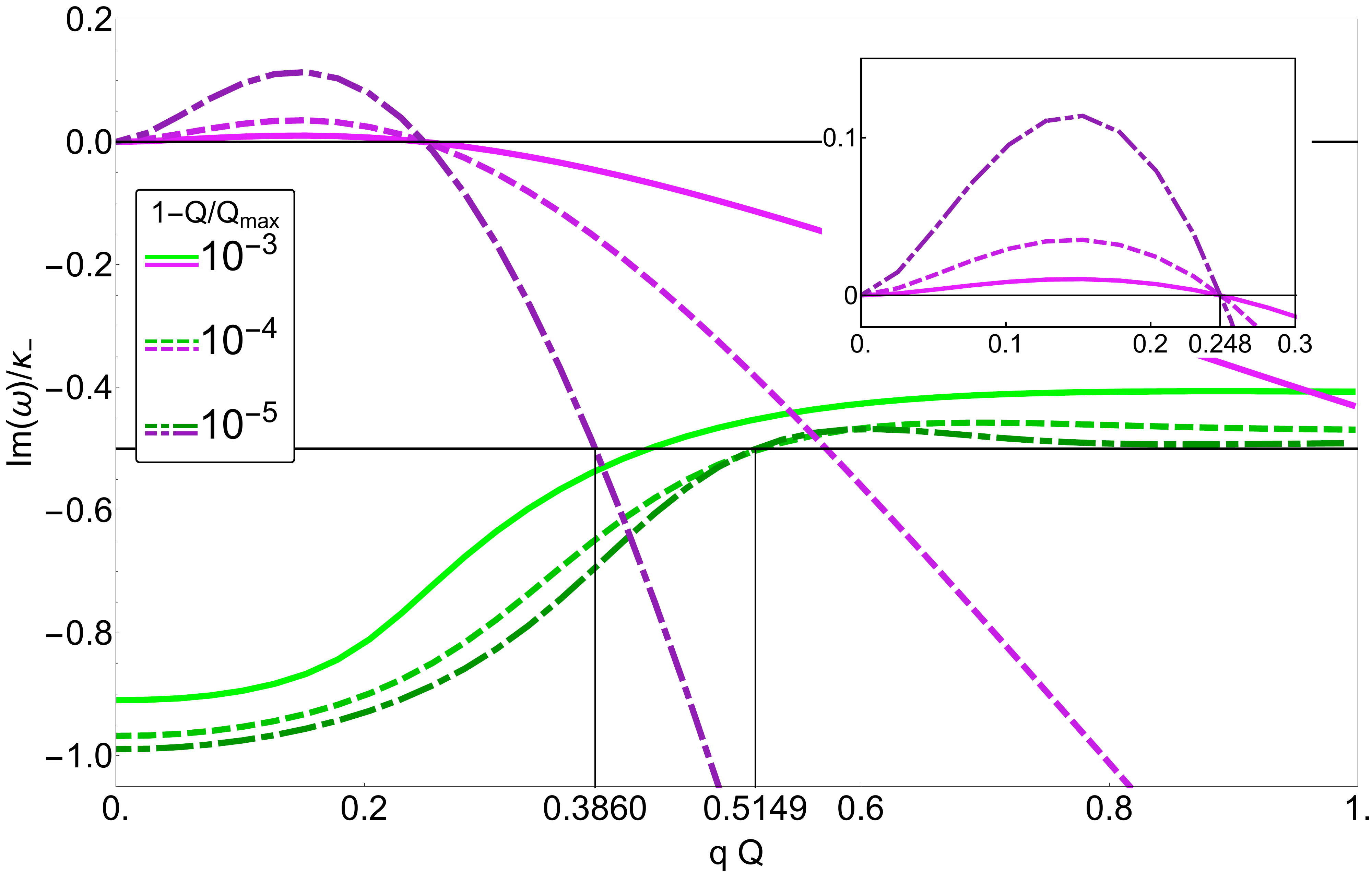}
\caption{The two lowest QNMs of a charged, massless scalar perturbation of a RNdS BH with $\Lambda M^2 = 0.06$ and  $1 - Q/Q_\text{max}  = 10^{-3}, 10^{-4}$, $10^{-5}$, as a function of the scalar charge $q$.
The purple modes originating from $(\omega,q) = (0,0)$ are called superradiant modes, while the green ones are the NE modes.}
\label{chargedScalar}
\end{figure}

\begin{figure}[H]
\centering
\includegraphics[width=0.9\textwidth]{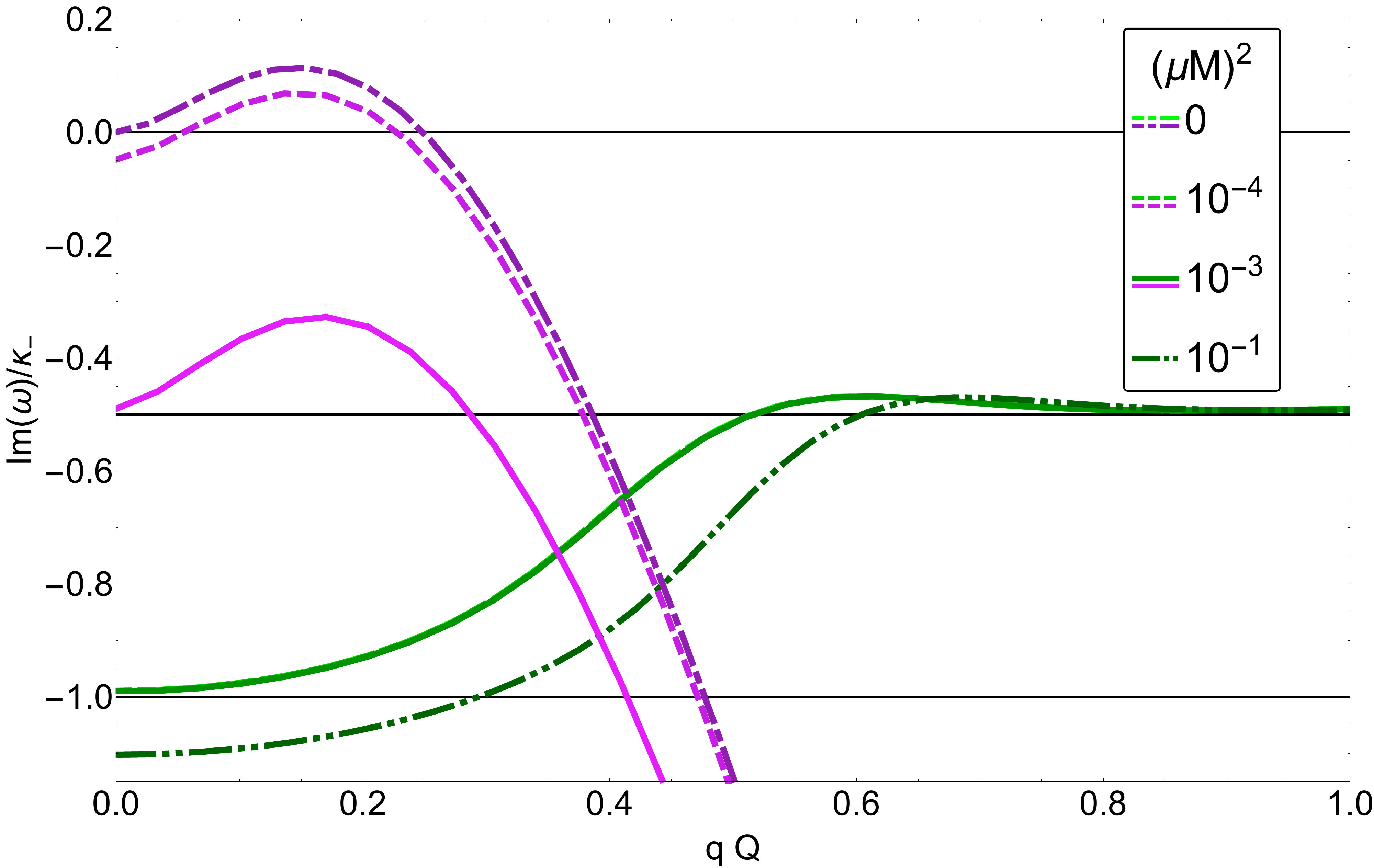}
\caption{Lowest QNMs of a charged scalar perturbation of a RNdS BH with $\Lambda M^2 = 0.06$ and  $1 - Q/Q_\text{max}  =10^{-5}$, as a function of the scalar charge $q$, for various scalar masses $(\mu M)^2 = 0, 10^{-4}, 10^{-3}$ and $10^{-1}$.
The top 3 purple lines are SR modes, with the very top one corresponding to the one with the same style in Fig.~\ref{chargedScalar}, and for the largest mass $(\mu M)^2 = 10^{-1}$ the SR mode lies outside the plotted range. 
Green lines are NE modes, of which the top three overlap.}
\label{chargedScalarSmallMu}
\end{figure}

According to our results (see Fig.~\ref{chargedScalar}, Fig.~\ref{chargedScalarSmallMu} and Section~\ref{app:largel}), in this parameter range, the dominant mode is a spherically symmetric $l=0$ mode. 
Note that this was already the case for the massless, neutral scalars studied in~Chapter \ref{PRL}. Note also that, in view of our focus on near extremal charges, the PS and dS families are considerably subdominant and, consequently, are not seen here.  

Consider first charged, massless scalars. In Fig.~\ref{chargedScalar}, the two most dominant QNMs are shown, for each of the BH charges in~\eqref{EqBHC}.
The green lines correspond to NE modes, $\omega_{\rm NE}$.
At $q=0$, we find $1/2<\beta<1$, in agreement with our previous results in Chapter \ref{PRL}. For $q>0$, the NE modes are initially subdominant but eventually, for sufficiently large $q Q$, become dominant and such that $\text{Im}(\omega_{\rm NE})/\kappa_->-1/2$, so $\beta<1/2$ for such $q Q$.
This corroborates the arguments of~\cite{Hod:2018dpx} and extends them to the massless scalar setting. In particular, we conclude that the large scalar mass condition of~\cite{Hod:2018dpx} is not necessary to guarantee that $\beta<1/2$: large $q Q$ suffices. 

The purple lines complicate the story. We call the corresponding QNMs superradiant (SR) modes, as they are associated, for small $q Q$, with a superradiant instability~\cite{Brito:2015oca} 
(for larger $q Q$ they are decaying modes). These modes were seen for the first time in Ref.~\cite{Zhu:2014sya} and further analyzed in Refs.~\cite{Konoplya:2014lha,Destounis:2019hca}. 
They originate from the trivial mode of the massless, neutral scalar, at $l=0$, which corresponds to nothing more than a constant shift in the scalar field. 
When we add charge or mass, the corresponding wave equation no longer admits constant solutions, the trivial mode disappears and gives rise to the dynamical mode seen in Fig.~\ref{chargedScalar}. 
For small coupling $q Q$, the SR modes are unstable, $\text{Im}(\omega)>0$, with the maximum imaginary part increasing with the size of the BH charge. 
This linear instability suggests that under evolution by the full Einstein equations, coupled with the fields under consideration here, even the exterior of our RNdS BH will be severely unstable; thus, we cannot infer anything about SCC in this case. A full spectral analysis of such unstable modes of RNdS BHs is provided in Appendix \ref{superradiance}, while in Chapter \ref{higher instability} these unstable modes are discussed in higher-dimensional RNdS spacetimes.
However, it is also apparent from Fig.~\ref{chargedScalar} that the SR modes cross the $\text{Im}(\omega) = 0$ at $q Q \approx 0.248$ (which to a good approximation is independent of
the particular NE BH charge). The modes then become stable and, eventually, subdominant---the dominant mode becomes the one arising from the NE family.
Very interestingly however, by inspection of Fig.~\ref{chargedScalar}, we find that there are choices of parameters for which $\beta>1/2$: this happens for instance when $1 - Q/Q_\text{max}  = 10^{-5}$ and $0.386 < q Q < 0.515$. 
We remark that for massless scalars, we once again find that $\beta$ is bounded, never reaching unity.    

\section{Charged Massive Scalars}
%
We now focus on a BH charge satisfying $1 - Q/Q_\text{max}  = 10^{-5}$, and study the effect of adding a scalar mass to the QNM landscape. 
We present part of our results in Fig.~\ref{chargedScalarSmallMu} by showing the two most dominant QNMs for a selection of scalar masses.
The effect of the mass is to decrease the imaginary part of both modes.
This means that massive scalars decay faster; consequently, the larger the scalar mass, the harder it becomes for its fluctuations to restore SCC. 

The strongest effect here is in the SR modes. These are highly sensitive to the mass, which moves the imaginary part downwards by an approximately constant shift
(in the uncharged case $q Q=0$, this was justified rigorously for small $\mu>0$ in \cite{HintzVasySemilinear}).
As a consequence, for non-zero mass, the SR instability is no longer present for sufficiently small charges $q$, but resurfaces when $q$ exceeds some positive lower bound. 
We note that the existence of unstable SR modes with non-vanishing scalar mass (which follows from their existence in the massless case by the continuous dependence of the QNM spectrum on all parameters) is, to the best of our knowledge, being numerically detected here for the first time. These modes are easy to miss in view of their large sensitivity to changes in the mass. In fact, for larger, but still very small, masses, the SR instability is no longer present. 

The NE mode is much less sensitive to the mass (the first 3 green lines in Fig.~\ref{chargedScalarSmallMu} lie on top of each other).
While it also moves down, it continues to be the case that for large enough $q Q$, of order 1, and the mass range considered, $\beta<1/2$.
The limiting value of $\beta$ at large $q Q$ seems to be independent of the mass.

Finally, notice that for the largest mass presented, $(\mu M)^2 = 10^{-1}$, the SR mode is outside of the plotted range, and the NE mode is below $\text{Im}(\omega/\kappa_-) = -1$, indicating a $\beta > 1$, found here for the first time.
Note that, although it seems that the NE mode might continue increasing its negative imaginary part with increasing scalar mass, we do expect $\beta$ to remain bounded, since at some point the PS mode will become dominant, 
and this is independent of the mass at large $l$. 
Furthermore, in empty dS, $\beta$ remains bounded when $\Lambda \mu^2\to\infty$, as follows from~\cite{VasyWaveOndS}. 

\section{Conclusions}
In Chapter \ref{PRL} we presented evidence for the failure of SCC for highly charged RNdS BHs under neutral scalar perturbations.  This was achieved by relying on Eq.~\eqref{theoremHintz1} and performing a thorough numerical computation of $\beta$. Here, following a suggestion in Ref.~\cite{Hod:2018dpx}, we extend our analysis to charged (massless and massive) scalars.  

To obtain a quantitative formulation of (a linearized version of)  SCC we started by showing that, by suitably extending the definition of $\beta$, Eq.~\eqref{theoremHintz1} remains valid for charged and massive scalar perturbations. 
We then performed a detailed numerical computation of the dominant QNMs in RNdS, for choices of BH parameters identified as problematic in Chapter \ref{PRL}, while taking into account the entire range of coupling constants $q Q\geq 0$ and several choices of scalar masses. 
From this we can then compute $\beta$ and infer about SCC, at least in the cases where we have mode stability $\text{Im}(\omega_0)<0$.    

Our main results are plotted in Figs.~\ref{chargedScalar} and~\ref{chargedScalarSmallMu}, and our conclusions can be summarized as follows: 
\begin{enumerate}
\item For all choices of scalar mass and large enough charge coupling we get $\beta<1/2$. Consequently, our linear analysis suggests that SCC should be valid, in the corresponding parameter ranges. This is in line with Ref.~\cite{Hod:2018dpx}, but here we show that the result also holds for small masses.

Superimposing neutral and charged scalar perturbations, the smaller of the two values of $\beta$ for the two types of perturbations (namely $\beta_{|q=0}>1/2$ and $\beta_{|q\gg 1}<1/2$) is the one relevant for SCC. Consequently the expected failure of SCC for uncharged scalars is put into question as soon as we add a charged scalar field with sufficiently strong coupling.

\item Nonetheless, for all choices of scalar masses we always find an interval of coupling charges for which $\beta>1/2$, which predicts a potential failure of SCC in this setting. Moreover, even if we add neutral perturbations we will get $\beta_{|q=0}>1/2$ and $\beta_{|q \neq0}>1/2$ and the situation remains alarming for SCC.  

\item Finally, for large scalar masses and small charges we get, for the first time, $\beta>1$. Recall that this is related to bounded curvature and therefore opens the possibility to the existence of solutions to the Einstein--Maxwell--Klein--Gordon system  with a scalar field satisfying Price's law and bounded curvature across the CH.~\footnote{Spherically symmetric solutions of the Einstein-Maxwell scalar system with bounded curvature were constructed in Ref.~\cite{Costa:2014aia}, but these have a compactly supported scalar field along the event horizon.} 

Nonetheless, this should be a non-generic feature: if we once again superimpose a neutral scalar perturbation---as a proxy for a linearized gravitational perturbation---we will get $\beta_{|q=0}<1$, which should be enough to guarantee the blow-up of curvature.

\end{enumerate}
We end with some final comments.

First, the charged matter could just as well be fermionic instead of scalar. Note that fermions do not have a SR instability, so the entire range of fermion charge parameters is open for the study of SCC at this linear level. In particular, it is interesting to see if charged fermions also have the potential to restore SCC at large charge. We pursue this study in Chapter \ref{PLB}. 

Second, during the last stages of this work we were informed by the authors of Ref.~\cite{Mo:2018nnu} that the SR mode, unlike the near extremal, is sensitive to the cosmological constant: 
for large enough cosmological constant it appears to be absent. 
A quick check indicates that our chosen value of $\Lambda M^2 = 0.06$ is close to the value where the SR mode is the most unstable (see Appendix \ref{superradiance}).
\footnote{For $1 - Q/Q_\text{max} = q Q = 10^{-2}$ the massless superradiant mode has $\text{Im}(\omega)/\kappa_- = 2.75 \times 10^{-5}$ when $\Lambda M^2 = 0.06$, and obtains its maximum of $\text{Im}(\omega)/\kappa_- = 3.07 \times 10^{-5}$ when $\Lambda M^2 = 0.0426$.}
Hence, for different $\Lambda M^2$ we expect the role of this mode to be either similar or smaller (for $\Lambda = 0$ the instability is absent~\cite{Zhu:2014sya,Konoplya:2014lha}), and thus for it to be equally hard or easier to find a regime where $\beta > 1/2$.

Third, one might also argue, as is done in \cite{Hod:2018dpx}, that for physical black holes made from charged matter coming from the standard model we must have $q Q \gg 1$; hence SCC is, according to the presented results, expected to be satisfied. We stress that even with this input we remain in the realm  of the conceptual version of SCC (as described in the introduction of this chapter) since the input is only relevant when the conjecture is in danger and this only happens for highly charged BHs.  

Finally, in the final stage of preparation \cite{Dias:2018etb} appeared, where the gravitational and electromagnetic perturbations are analyzed and found not to save SCC.
In fact for these perturbations it is found that $\beta$ can exceed not just $1/2$ but even 1 and 2, making the charged matter studied here the dominant mode.
 
Regardless of the approach to SCC that the reader subscribes to, the results presented here indicate at least a growing level of sophistication required for the Cosmic Censor, and the situation regarding SCC, in the presence of a positive cosmological constant, remains subtle!

%
\begin{figure}[H]
\centering
\includegraphics[width=0.9\textwidth]{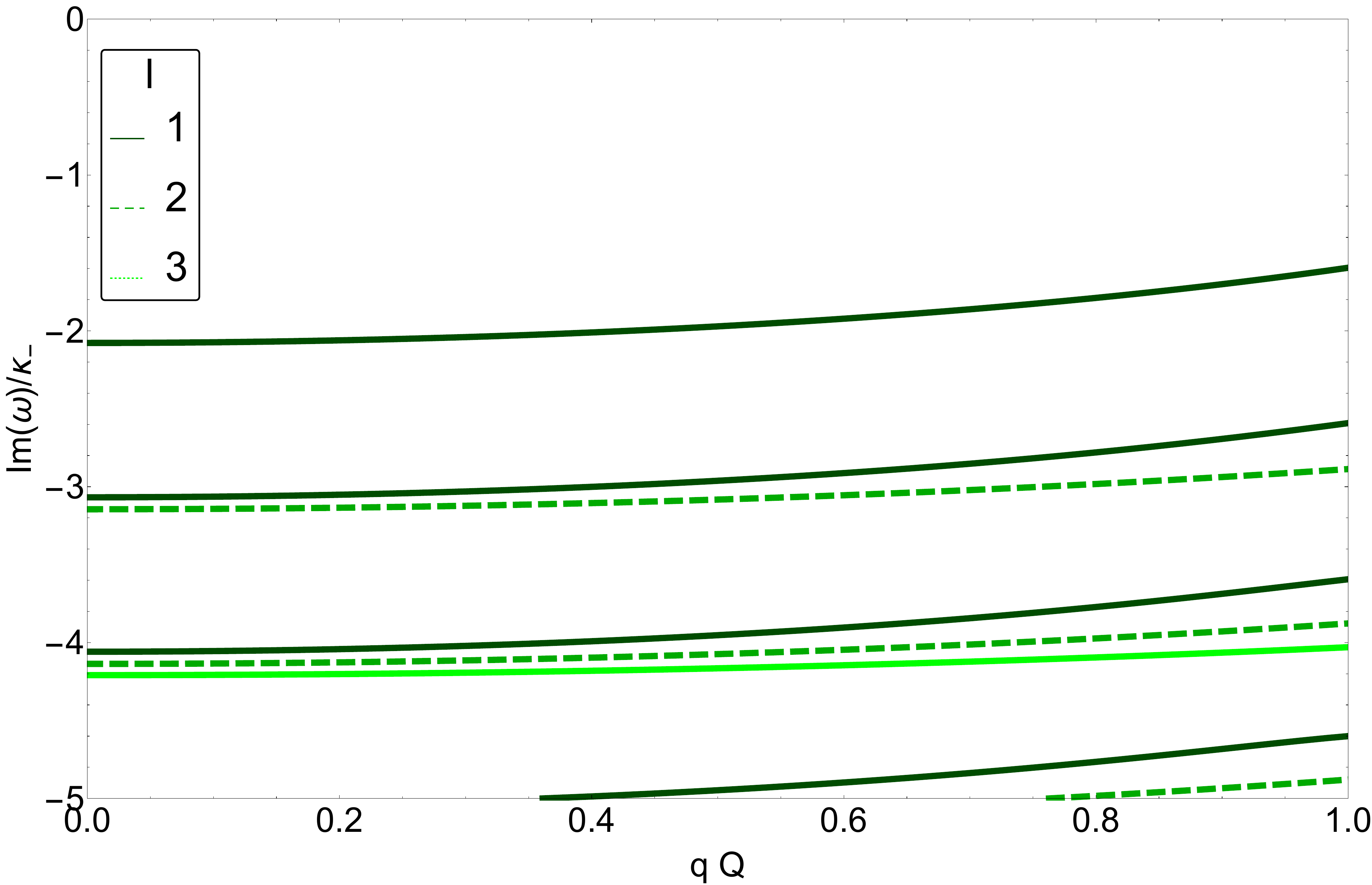}
\caption{Higher $l$ QNMs of a charged, massless scalar perturbation of a RNdS BH with $\Lambda M^2 = 0.06$ and  $1 - Q/Q_\text{max}  = 10^{-5}$, as a function of the scalar charge $q$.}
\label{higherlmassless}
\end{figure}
\section{Higher $l$ modes}\label{app:largel}
In this section we verify the expectation that the higher $l$ QNMs do not affect SCC. 
In Fig.~\ref{higherlmassless} we show the $l=1,2,3$ modes. 
These are all modes of the NE family, since they are present already at $q=0$ and they follow their pattern. No modes of the other families are present at this range.
The dependence on $q$ is rather mild, and in particular they do not come near the dominant $l=0$ mode.

While Fig.~\ref{higherlmassless} shows the modes for massless scalars, we have done the same check for the massive. Of the masses considered, the modes for $(\mu M)^2 = 10^{-4}$ and $10^{-3}$ are visually indistinguishable from the massless ones, and for $(\mu M)^2 = 10^{-1}$ they lie just below the ones presented.

Finally, one might worry if for even larger $l$ the PS modes will become dominant.
To address this we have computed by WKB approximation, which is expected to become very accurate in the large $l$ limit, the modes at $l=100$. 
The dominant mode that we find for the parameters of Fig.~\ref{higherlmassless} and $ q Q = 0.45$ is $\omega / \kappa_- = 5472 - 18.35 i$, which we confirmed using \cite{Jansen:2017oag}.
Furthermore, we have checked all other values of scalar charge and mass considered and found this value to be largely independent of those parameters.

This is very far from the $l=0$ mode, so we are convinced that throughout the parameter space considered, $l=0$ indeed gives the dominant mode.

\chapter{Charged Fermions and Strong Cosmic Censorship}\label{PLB}
In Chapters \ref{PRL} and \ref{PRD} we demonstrated that the SCC conjecture might be violated for near-extremally-charged BHs in dS space. Here, we extend our study to charged fermionic fields in the exterior of RNdS spacetime. We, again, identify three families of modes; one related to the PS, a second related to the dS horizon and a third which dominates near extremality. We show that for near-extremally-charged BHs there is a critical fermionic charge below which SCC may potentially be violated. Surprisingly enough, as one approaches extremality even more, violation of SCC may occur even beyond the critical fermionic charge. The following chapter is based on \cite{Destounis:2018qnb}.
\section{Introduction}
We recently demonstrated the implications of massless neutral scalar perturbations on SCC in RNdS BH spacetimes. Three different families of modes were identified in such spacetime; one directly related to the PS, well described by standard  WKB tools, another family whose existence and timescale is closely related to the dS horizon and a third family which dominates for near-extremally-charged BHs. Surprisingly enough, our results show that NE RNdS BHs might violate SCC at the linearized level, leading to a possible failure of determinism in GR. The key quantity controlling the stability of the CH, and therefore the fate of SCC, is given by~\cite{Hintz:2015jkj}
\begin{equation}
\label{betaOld2}
\beta \equiv -\text{Im}(\omega_0)/\kappa_-\;,
\end{equation}
where $\omega_0$ is the longest-lived/dominant non-zero QNM~\cite{Kokkotas:1999bd,Berti:2009kk,Konoplya:2011qq}) and $\kappa_-$ is the surface gravity of the CH.
The results in~\cite{Hintz:2016gwb,Hintz:2016jak,Costa:2017tjc} suggest that $\beta$ remains the key quantity in the non-linear setting: the higher $\beta$, the more stable the CH. Concretely, the modern formulation of SCC requires that
\begin{equation}\label{theoremHintz2}
\beta < 1/2
\end{equation}
in order to guarantee the breakdown of field equations at the CH \cite{Christodoulou:2008nj}.
In Chapter \ref{PRL}, a thorough linear numerical study of $\beta$ throughout the whole parameter space of subextremal RNdS spacetimes revealed that $1/2<\beta<1$ in the near-extremal regime, leading to a CH with enough regularity for extensions to be possible beyond it. 
This provides evidence for the existence of CHs which, upon perturbation, are rather singular due to the divergence of curvature invariants, but where the gravitational field can still be described by the field equations; the evolution of gravitation beyond the CH, however, is highly non-unique. Recent studies \cite{Liu:2019lon,Rahman:2018oso} have generalized the linear massless scalar field study in higher-dimensional RNdS BHs and RNdS BHs on the brane finding violation in the near-extremal regime.

There are different ways to interpret the previous results. If one takes the SCC conjecture as a purely mathematical question about GR then this either signifies a failure of SCC, or such phenomena are superseded by nonlinear effects. In fact, the results of \cite{Luna:2018jfk} proved that even nonlinear effects could not save the conjecture from failing for near-extremally-charged BHs. 

A subsequent study of metric fluctuations in RNdS BHs showed that such perturbations possibly exhibit a much worse violation of SCC. In \cite{Dias:2018etb} it was shown that for a sufficiently large NE RNdS BH, perturbations arising from smooth initial data can be extended past the CH in an arbitrarily smooth way.
Nevertheless, astrophysical BHs are expected to be nearly neutral~\cite{Cardoso:2016olt,Barausse:2014tra}. Taking this into consideration, one can question the relevance of SCC violations in highly charged, non-spinning BHs. In fact, a recent study suggests that rapidly rotating BHs in cosmological backgrounds do not violate SCC. According to \cite{Dias:2018ynt}, in KdS spacetime (\ref{betaOld2}) remains unchanged, but now $\beta$ seems to be bounded exactly by $1/2$, at extremal rotation. Similar results were obtained in \cite{Rahman:2018oso} for higher-dimensional KdS BHs.

Considering the formation of a charged BH, one would argue that charged matter has to be present. In \cite{Hod:2018dpx,Hod:2018lmi}, it was claimed that charged scalar fields would lead to the restoration of SCC in an appropriate region of the parameter space of RNdS and Kerr-Newmann-dS BHs. This implication requires working in the large-coupling regime for which $\beta<1/2$. Taking into account the whole parameter space, subsequent studies \cite{PhysRevD.98.104007,Mo:2018nnu,Dias:2018ufh} presented numerical evidence that SCC may still be violated in the setting of charged scalar perturbations in RNdS. 

In Chapter \ref{PRD}, we also take into account massive charged scalar fields which lead to strong evidence that $\beta>1$. Recall that this is related to bounded curvature and therefore opens the possibility to the existence of solutions to the Einstein-Maxwell-Klein-Gordon system with a scalar field satisfying Price's law and bounded curvature across the CH. Nonetheless, if the neutral scalar perturbations where superimposed to the charged massive ones, then the smaller of the two types of perturbations is the one relevant for SCC, thus getting $\beta<1$, which should be enough to guarantee the blow-up of curvature components. 

In \cite{Dias:2018ufh} it was shown that even for large scalar field charge there are NE BHs for which $\beta>1/2$. A key ingredient of the aforementioned studies was the existence of a superradiantly unstable mode. This unstable mode was the dominant one for small scalar charges rendering the question of the validity of SCC irrelevant for a significant region of the parameter space.

Is it natural to question then, if the charged matter could just as well be fermionic instead of scalar. Fermions do not superradiate, leaving the entire range of fermionic charge open for the study of SCC at the linearized level. 
The results of \cite{Zhang2} provide evidence that fermionic perturbations of RNdS BHs might violate SCC for sufficiently large BH charge. As a matter of fact, the family that seems to dominate the dynamics near extremality is, mostly, the PS family with a very small participation of a family which is purely imaginary for zero fermionic charge $q$ and quickly becomes subdominant as $q$ increases.

Unfortunately, there is no information about the classification of the latter family and if it will eventually dominate the dynamics for even higher BH charges. Moreover, since a dS horizon is present, the dS family of modes might be present as well and even dominate the dynamics for small cosmological constants in analogy with what was found in Chapter \ref{PRL}. The tool used to extract the modes in \cite{Zhang2} is time domain analysis. Although it is a very powerful tool for such calculations, there is a slight chance that long-lived modes may be missed either because of their timescale being larger than the evolution time of the system or because of improper choice of initial data which might not trigger the long-lived modes.

In this chapter, we study the propagation of massless charged fermions on a fixed RNdS background and extract the QNMs with a spectral method developed in \cite{Jansen:2017oag} which is based on numerical methods introduced in \cite{Dias:2010eu} (for a topical review see \cite{Dias:2015nua}). After characterizing the families of modes that are present, we will examine the implications on SCC for NE RNdS BHs.

\section{Charged Fermions in Reissner-Nordstr\"{o}m-de Sitter spacetime}
We focus on RNdS BHs, described by the metric
\begin{equation}
\label{RNdS_space1}
ds^2=-f(r)dt^2+\frac{dr^2}{f(r)}+r^2(d\theta^2+\sin^2\theta d\varphi^2)\,,
\end{equation}
where $f(r)=1-{2M}{r^{-1}}+{Q^2}{r^{-2}}-\Lambda r^2/3$. Here, $M,\,Q$ are the BH mass and charge, respectively, and $\Lambda>0$ is the cosmological constant. The surface gravity of each horizon is then
\begin{equation}
\label{surfGrav1}
\kappa_i= \frac{1}{2}|f'(r_i)|\;\;,\; i\in\{-,+,c\}\;,
\end{equation}
where $r_-<r_+<r_c$ are the Cauchy, event and cosmological horizon radius. Since fermions are described by spinors, we use the tetrad formalism to accommodate them in curved space. The tetrads by definition satisfy the relations
\begin{align*}
e^{(a)}_\mu \,e^{\nu}_{(a)}&=\delta^\nu_\mu,\\
e^{(a)}_\mu \,e^{\mu}_{(b)}&=\delta^{(a)}_{(b)},
\end{align*}
The choice of the tetrad field determines the metric through
\begin{align*}
\label{tetradgmn1}
g_{\mu\nu}&=e^{(a)}_\mu \,e^{(b)}_\nu\,\eta_{(a)(b)},\\
\eta_{(a)(b)}&=e^\mu_{(a)}\,e^\nu_{(b)}\,g_{\mu\nu},
\end{align*}
where $\eta_{(a)(b)}$ and $g_{\mu\nu}$ are the Minkowski and RNdS metric, respectively. In order to write the Dirac equation, we also introduce the spacetime-dependent gamma matrices $G^\mu$ which are related to the special relativity matrices, $\gamma^{(a)}$, by
\begin{equation*}
\label{gtet1}
G^\mu=e^\mu_{(a)}\gamma^{(a)},
\end{equation*}
and are chosen in a proper way to satisfy the anti-commutation relations
\begin{align*}
\{\gamma^{(a)},\gamma^{(b)}\}&=- 2\eta^{(a)(b)},\\
\{G^\mu, G^\nu\}&=- 2 g^{\mu\nu}.
\end{align*}
Consequently, we define $G^\mu$ with respect to a fixed tetrad
\begin{align*}
G^t&=e^t_{(a)}\gamma^{(a)}=\frac{\gamma^t}{\sqrt{f(r)}},\,\,\,\,\,\,\,\,G^r=e^r_{(a)}\gamma^{(a)}=\sqrt{f(r)}\gamma^r,\\
G^\theta&=e^\theta_{(a)}\gamma^{(a)}=\gamma^\theta,\,\,\,\,\,\,\,\,\,\,\,\,\,\,\,\,\,\,\,G^\varphi=e^\varphi_{(a)}\gamma^{(a)}=\gamma^\varphi,
\end{align*}
where $\gamma^t,\,\gamma^r,\,\gamma^\theta$ and $\gamma^\varphi$ are the $\gamma-$matrices in ``polar coordinates" \cite{Finster:1998ak}
\begin{align*}
\gamma^t&=\gamma^{(0)},\\
\gamma^r&=\sin\theta\cos\varphi\,\gamma^{(1)}+\sin\theta\sin\varphi\,\gamma^{(2)}+\cos\theta\,\gamma^{(3)}\\
\gamma^\theta&=\frac{1}{r}\left(\cos\theta\cos\varphi\,\gamma^{(1)}+\cos\theta\sin\varphi\,\gamma^{(2)}-\sin\theta\,\gamma^{(3)}\right),\\
\gamma^\varphi&=\frac{1}{r\,\sin\theta}\left(-\sin\varphi\,\gamma^{(1)}+\cos\varphi\,\gamma^{(2)}\right)
\end{align*}
and 
\begin{equation*}
\label{dirac_gamma1}
\gamma^{(0)}=\begin{pmatrix}
\mathbf{1}&0\\
0&-\mathbf{1}
\end{pmatrix},\,\,\,\,\,\,\,\,\,
\gamma^{(k)}=\begin{pmatrix}
0&\sigma^{k}\\
-\sigma^{k}&0
\end{pmatrix}
\end{equation*}
the standard Dirac $\gamma$-matrices, where $\sigma^k,\,k=1,2,3$ the Pauli matrices. The propagation of a spin $1/2$ particle of mass $m_f$ on a fixed RNdS background is then described by the Dirac equation in curved spacetime \cite{Fock:1929vt}
\begin{equation}
\label{dirac equation1}
(iG^\mu D_\mu-m_f)\Psi=0,
\end{equation}
with the covariant derivative 
\begin{equation*}
D_\mu=\partial_\mu-iqA_\mu+\Gamma_\mu,
\end{equation*}
where $q$ the charge of the Dirac particle, $A=-(Q/r)dt$ the electrostatic potential and $\Gamma_\mu$ the spin connection coefficients defined as
\begin{equation*}
\label{spin1}
\Gamma_\mu=-\frac{1}{8}\omega_{(a)(b)\mu}\left[\gamma^{(a)},\gamma^{(b)}\right].
\end{equation*}
The spin connection $\omega_{(a)(b)\mu}$ is defined as
\begin{align*}
\omega_{(a)(b)\mu}=\eta_{(a)(c)}\left(e^{(c)}_\nu e^\lambda_{(b)}\Gamma^\nu_{\mu\lambda}-e^\lambda_{(b)}\partial_\mu e^{(c)}_\lambda\right),
\end{align*}
with $\Gamma^\nu_{\mu\lambda}$ the Christoffel symbols. By choosing the ansatz $\Psi=f(r)^{-1/4}r^{-1}\psi$, (\ref{dirac equation1}) can be written as
\begin{align}
\label{dirac_21}
\left[\frac{i\gamma^t}{\sqrt{f(r)}}\frac{\partial}{\partial t}+i\sqrt{f(r)}\gamma^r\frac{\partial}{\partial r}-\frac{i\gamma^r}{r}+i\left(\gamma^\theta\frac{\partial}{\partial\theta}+\gamma^\varphi\frac{\partial}{\partial\varphi}\right)-\gamma^{t}\frac{q Q}{r\sqrt{f(r)}}-m_f\right]\psi=0.
\end{align}
Since the external fields are spherically symmetric and time-independent, we can separate out the angular and time dependence of the wave functions via spherical harmonics and plane waves, respectively. For the Dirac wavefunctions, we choose the ansatzes
\begin{align}
\label{spinor11}
\psi^+_{jk\omega}&=e^{-i\omega t}\begin{pmatrix}
\phi^k_{j-1/2}F^+(r)\\
i\phi^k_{j+1/2}G^+(r)
\end{pmatrix},\\
\label{spinor21}
\psi^-_{jk\omega}&=e^{-i\omega t}\begin{pmatrix}
\phi^k_{j+1/2}F^-(r)\\
i\phi^k_{j-1/2}G^-(r)
\end{pmatrix},
\end{align}
where we introduced the spinor spherical harmonics \cite{Finster:1998ak} 
\begin{align*}
\phi^k_{j-1/2}&=\begin{pmatrix}
\sqrt{\frac{j+k}{2j}}Y^{k-1/2}_{j-1/2}(\theta,\varphi)\\
\sqrt{\frac{j-k}{2j}}Y^{k+1/2}_{j-1/2}(\theta,\varphi)
\end{pmatrix},\,\,\,\,\,\,\,\,\,\,\,\,\,\,\,\,\text{for}\,\,\,\,j=l+\frac{1}{2},\\
\phi^k_{j+1/2}&=\begin{pmatrix}
\sqrt{\frac{j+1-k}{2j+2}}Y^{k-1/2}_{j+1/2}(\theta,\varphi)\\
-\sqrt{\frac{j+1+k}{2j+2}}Y^{k+1/2}_{j+1/2}(\theta,\varphi)
\end{pmatrix},\,\,\,\,\,\,\text{for}\,\,\,\,j=l-\frac{1}{2},
\end{align*}
with $j=1/2,3/2,\dots$, $k=-j,-j+1,\dots,j$ and $Y_l^m$ the ordinary spherical harmonics. By substituting (\ref{spinor11}) and (\ref{spinor21}) into (\ref{dirac_21}) and utilizing the identities
\begin{align*}
K=\vec{\sigma}\vec{L}+\mathbf{1}&=-r\sigma^r\left(\sigma^\theta\partial_\theta+\sigma^\varphi\partial_\varphi\right)+\mathbf{1},\\
K \phi^k_{j\mp 1/2}&=\pm(j+\frac{1}{2})\phi^k_{j\mp 1/2},\\
\sigma^r\phi^k_{j\mp 1/2}&=\phi^k_{j\pm 1/2},
\end{align*}
with $\vec{\sigma},\,\vec{L}$ the Pauli and angular momentum vectors, respectively, and $\sigma^r,\,\sigma^\theta,\,\sigma^\varphi$ the Pauli matrices in ``polar coordinates" \cite{Finster:1998ak}, we end up with the coupled Dirac equations
\begin{align}
\label{final11}
\frac{\partial F}{\partial r_*}-\frac{\xi\sqrt{f(r)}}{r}F+\left(\omega -\frac{qQ}{r}\right)G+m_f\sqrt{f(r)}G&=0,\\
\label{final21}
\frac{\partial G}{\partial r_*}+\frac{\xi\sqrt{f(r)}}{r}G-\left(\omega -\frac{qQ}{r}\right)F+m_f\sqrt{f(r)}F&=0,
\end{align}
where $\xi=\pm(j+1/2)=\pm 1,\,\pm 2,\dots$ and $dr_*=f/dr$. Since the charge-to-mass ratio of the electron is of order $10^{11} \text{C}/\text{kg}$, it is reasonable to explore massless fermions. By setting $m_f=0$ we can decouple (\ref{final11}), (\ref{final21}) by introducing a new coordinate
\begin{equation*}
\label{new_tort1}
d\bar{r}_*=\frac{\left(1-\frac{qQ}{r\omega}\right)}{f}dr,
\end{equation*}
to get
\begin{align}
\label{almdec11}
\frac{dF}{d\bar{r}_*}-WF+\omega G&=0,\\
\label{almdec21}
\frac{dG}{d\bar{r}_*}+WG-\omega F&=0,
\end{align}
and subsequently
\begin{align}
\frac{d^2F}{d\bar{r}_*^2}+\left(\omega^2-V_+\right)F&=0,\\
\frac{d^2G}{d\bar{r}_*^2}+\left(\omega^2-V_-\right)G&=0,
\end{align}
with 
\begin{equation*}
V_\pm=\pm\frac{dW}{d\bar{r}_*}+W^2,
\end{equation*}
where
\begin{equation}
\label{W1}
W=\frac{\xi\sqrt{f}}{r\left(1-\frac{qQ}{r\omega}\right)}.
\end{equation}
\begin{figure}[H]
\subfigure{\includegraphics[scale=0.225]{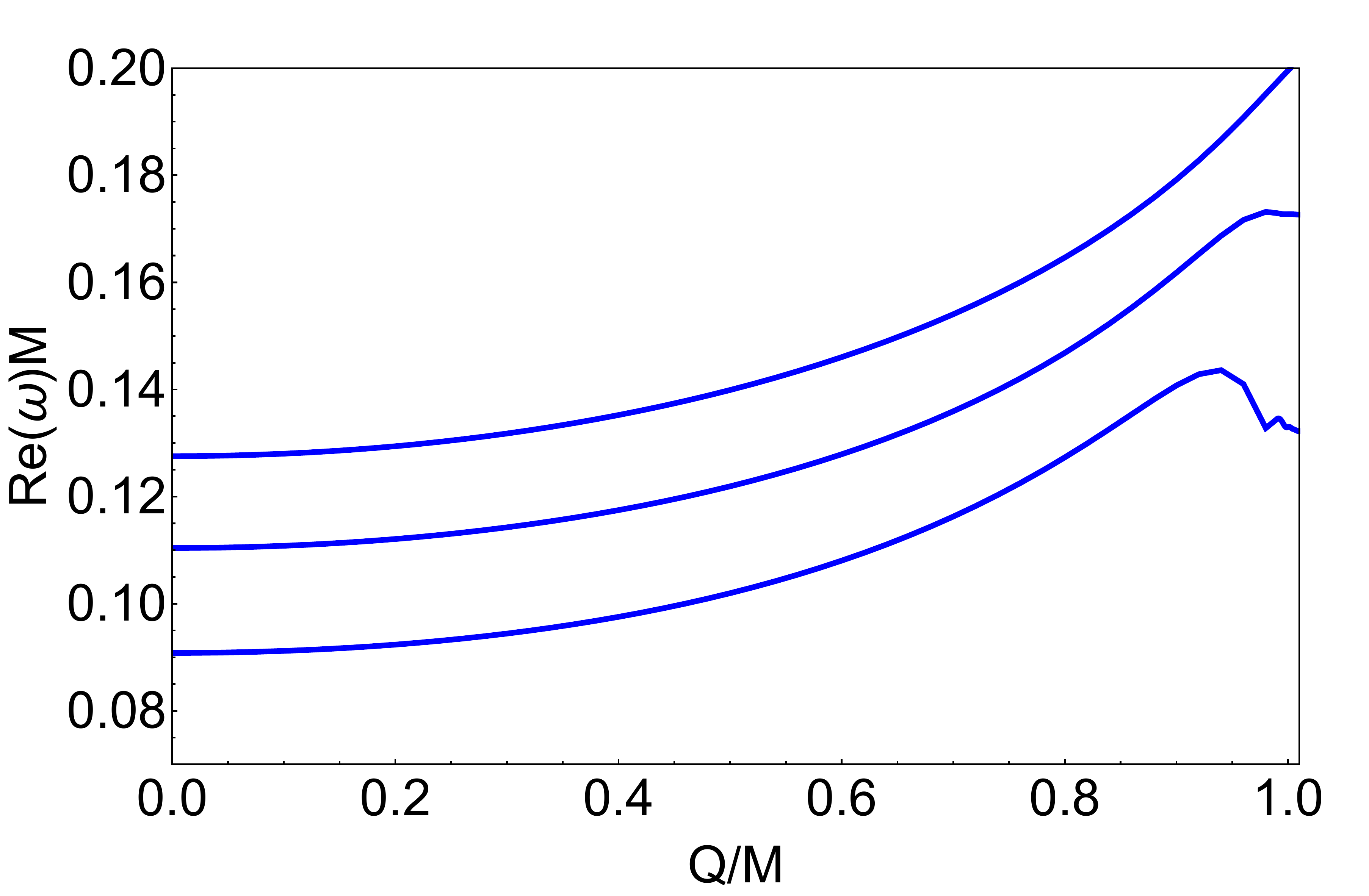}}
\hskip -2ex
\subfigure{\includegraphics[scale=0.225]{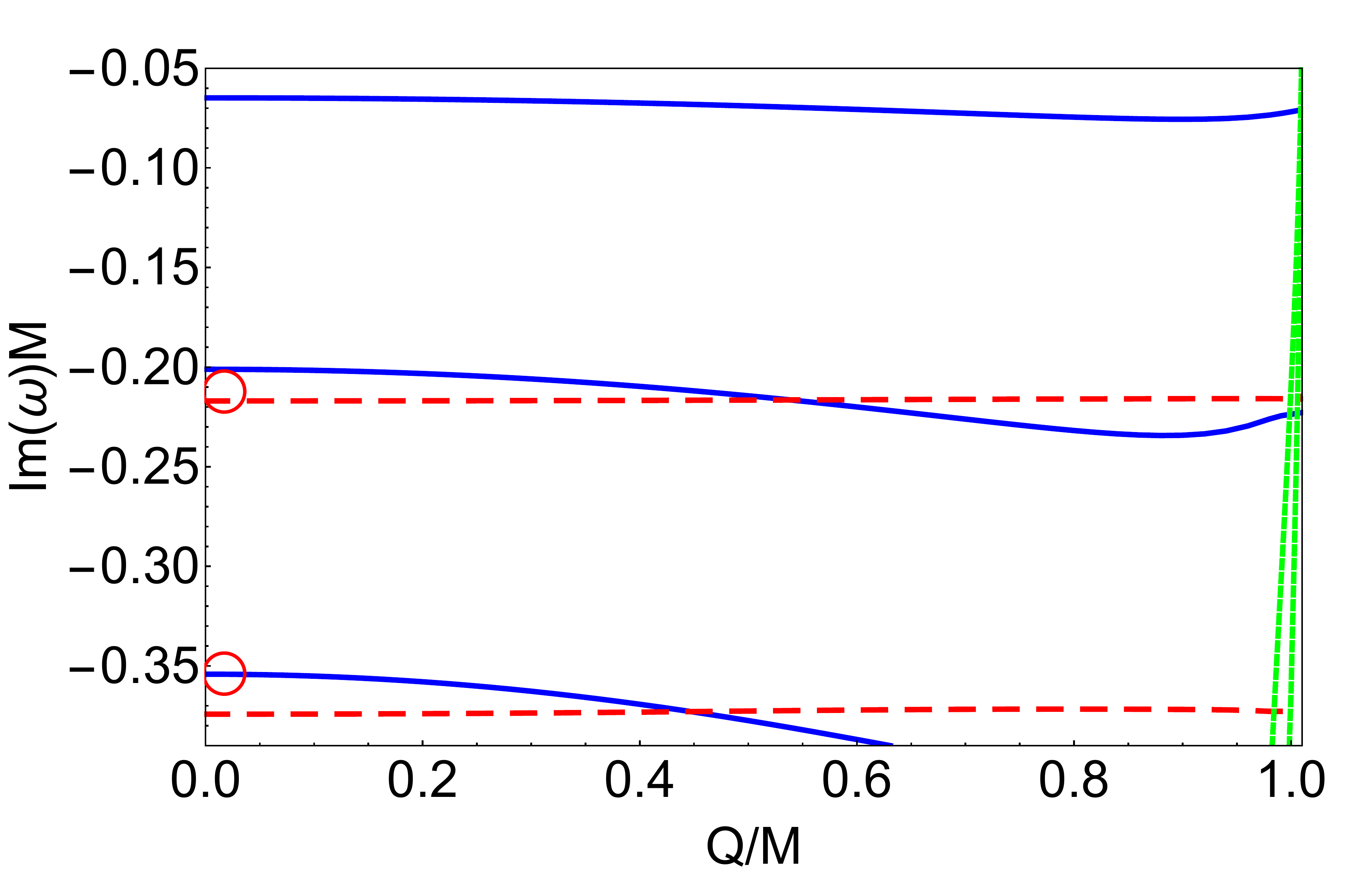}}
\caption{Lowest lying fermionic QNMs for $\xi=1$, $q=0$ and $\Lambda M^2=0.06$ as a function of $Q/M$. The right plot shows the imaginary part, with dashed red lines corresponding to purely imaginary modes, and solid blue lines to complex PS modes, whose real part is shown in the left plot. The red circles in the right plot designate the $\xi=1$ Dirac modes of empty dS space at the same $\Lambda$, which closely match the first imaginary mode shown here, but lie less close to the higher overtone. Near extremality, another set of purely imaginary modes (dotted green lines) come in from $-\infty$ and approach $0$ in this limit. Only a finite number of modes are shown, even though we expect infinitely many near-extremal modes in the range shown.}
\label{QNMs}
\end{figure}
It can be shown by utilizing \eqref{almdec11}, \eqref{almdec21} that potentials related in this manner and subjected to Sommerfeld conditions are isospectral, thus, allowing us to work only with the field $F$~\cite{Berti:2009kk,Anderson:1991kx}. Since we are interested in the characteristic frequencies of this spacetime, we impose the boundary conditions
\begin{equation*}
F(r\rightarrow r_+)\sim e^{-i\omega \bar{r}_*},\,\,\,\,\,\,\,\,\,\,\,\,\,\,\,F(r\rightarrow r_c)\sim e^{i\omega \bar{r}_*}
\end{equation*}
which select a discrete set of frequencies $\omega$ called the QNMs. The QN frequencies are characterized, for each $\xi$, by an integer $n\geq 0$ labeling the mode number. The fundamental mode $n=0$ corresponds, by definition, to the non-vanishing frequency with the smallest (in absolute value) imaginary part and will be denoted by $\omega\neq 0$. It is apparent from (\ref{almdec11}), (\ref{almdec21}) and (\ref{W1}) that the symmetry $\text{Re}(\omega)\rightarrow-\text{Re}(\omega)$, $q\rightarrow-q$, $\xi\rightarrow-\xi$ holds, enabling us to only study positive $\xi$.

As shown in Appendix \ref{appD}, for $q\neq 0$ the stability of the CH continues to be determined by~\eqref{betaOld2}.
The results shown in the following sections were obtained with the Mathematica package of~\cite{Jansen:2017oag}, and checked in various cases with a WKB approximation~\cite{Iyer:1986np} and with a code developed based on the matrix method \cite{KaiLin1}.

\section{QNMs of massless, charged fermionic fields: the three families}
In the previous chapters,
we found three qualitatively different families of QNMs: the PS, dS and NE family.
The first two connect smoothly to the modes of asymptotically flat Schwarzschild and of empty dS, respectively, while the last family cannot be found in either of these spacetimes.
Here we, again, distinguish three families of modes.

\subsection{Photon sphere modes}
The photon sphere is a spherical trapping region of space where gravity is strong enough that photons are forced to travel in unstable circular orbits around a BH. This region has a strong pull in the control of decay of perturbations and the QNMs with large frequencies. For asymptotically dS BHs, we find a family that can be traced back to the photon sphere and refer to them as PS modes. These modes are shown with blue colors in Figs. \ref{QNMs}-\ref{smallL} and \ref{higher_k}. They satisfy the symmetry $\text{Re}(\omega)\rightarrow-\text{Re}(\omega)$ for $q=0$ and the symmetry breaks as the fermionic charge is turned on, according to (\ref{W1}). For very small $\Lambda$, $q$ and $Q$, $\xi\rightarrow\infty$ defines the dominant mode which can be very well approximated by a WKB approximation and asymptote to the Schwarzschild BH Dirac QNMs \cite{Cho:2003qe}. The lowest lying PS modes are weekly dependent on the BH charge as it is apparent for the case presented in Fig. \ref{QNMs}. For sufficiently large $\Lambda$ the former does not hold. For large BH charges the $\xi=1$ PS modes dominate the family (see Section \ref{higher}). As the BH ``disappears" ($M\rightarrow 0$) we observe that the PS family has increasingly large frequencies and timescales until they abruptly vanish (see Section \ref{dS}).

It is important to note that at the eikonal limit the fermionic PS QNMs coincide with the scalar ones. This occurs since at the this limit the effective potentials for fermionic and scalar perturbations are dominated by the angular numbers $\xi$, $l$ thus gaining a similar form. A basic difference between scalar and fermionic perturbations
is that the eikonal scalar QNMs are the dominant ones for all BH parameters, in contrast with the eikonal fermionic QNMs which are dominant for a very small region of the BH parameter space.

\subsection{de Sitter modes}
In pure dS space solutions of the Dirac equation with purely imaginary $\omega$ exist \cite{LopezOrtega:2006ig}
\begin{equation}
\label{pure_dS}
\omega_\text{pure dS}/\kappa^\text{dS}_c=-i\left(\xi+n+\frac{1}{2}\right)
\end{equation}
where $\xi=1,2,\dots$. The second family of modes we find are the Dirac BH dS QNMs, which are deformations of pure dS QNMs (\ref{pure_dS}). The dominant BH dS mode ($\xi=1$, $n=0$) has almost identical imaginary part with (\ref{pure_dS}) and higher overtones have increasingly larger deformations.

These modes have weak dependence on the BH charge and are described by the surface gravity $\kappa_c^\text{dS}=\sqrt{\Lambda/3}$ of the cosmological horizon of pure dS space, as opposed to that of the cosmological horizon in the RNdS BH in study. This could be explained by the fact that the accelerated expansion of RNdS spacetimes is also governed by $\kappa_c^\text{dS}$ \cite{Brill:1993tw,Rendall:2003ks}.

To the best of our knowledge, this family of Dirac BH dS modes has been identified here for the first time. The scalar equivalent of these modes has been identified for the first time in the QNM calculations of \cite{Cardoso:2017soq,Jansen:2017oag}. Moreover, as the black hole vanishes ($M \rightarrow 0$), these modes converge smoothly to the exact pure dS modes (\ref{pure_dS}) (see Section \ref{dS}). 

A key similarity of fermionic dS QNMs and scalar dS QNMs is the fact that they are both
proportional to the surface gravity of the cosmological horizon of pure dS space. On
the other hand, the fermionic dS QNMs do not admit an $\omega=0$ mode, while scalars do.
Such mode has been seen in time evolutions \cite{Zhu:2014sya,Konoplya:2014lha} and rises from the fact that the effective
potential forms a potential well right outside the photon sphere, serving as a trapping region.
This region is connected to a superradiant instability in RNdS against charged scalar fluctuations
which effectively puts the validity of SCC out of discussion since the internal and external 
regions of the BH in study are effectively unstable. The effective potential of fermionic 
perturbations does not contain any potential wells which is a key indication of the inexistence of superradiance (together with the Pauli exclusion principle).

\subsection{Near-extremal modes}
In the limit where the Cauchy and event horizon approach each other, a third NE family dominates the dynamics. In the extremal limit and for sufficiently small fermionic charges this family approaches
\begin{equation}
\label{near extremal}
\omega_\text{NE}\approx\frac{q Q}{r_-}-i\kappa_-\left(\xi+n+\frac{1}{2}\right)\approx\frac{q Q}{r_+}-i\kappa_+\left(\xi+n+\frac{1}{2}\right),
\end{equation}
where $\xi=1,2,\dots$, with weak dependence on $\Lambda$ as shown by our numerics. As indicated by (\ref{near extremal}), the dominant mode of this family is the $\xi=1$, $n=0$. In the asymptotically flat case, such modes have been identified in \cite{Richartz2014}. Here, we show that these modes exist in RNdS BHs, and that they are the limit of a new family of modes.

Comparing the NE family of modes \eqref{near extremal} to the one discussed in Chapter \ref{PRL}, but initially found in \cite{Hod:2017gvn} for RN BHs, we realize that their real parts coincide, since they only depend on the choice
of BH parameters, but their imaginary parts differ slightly. In any case, as the extremal 
charge is approached, both families share the same fate; a vanishing imaginary part ($\kappa_-=
\kappa_+=0$ at extremality).
\section{Dominant modes and Strong Cosmic Censorship}
\begin{figure}[H]
\includegraphics[scale=0.225]{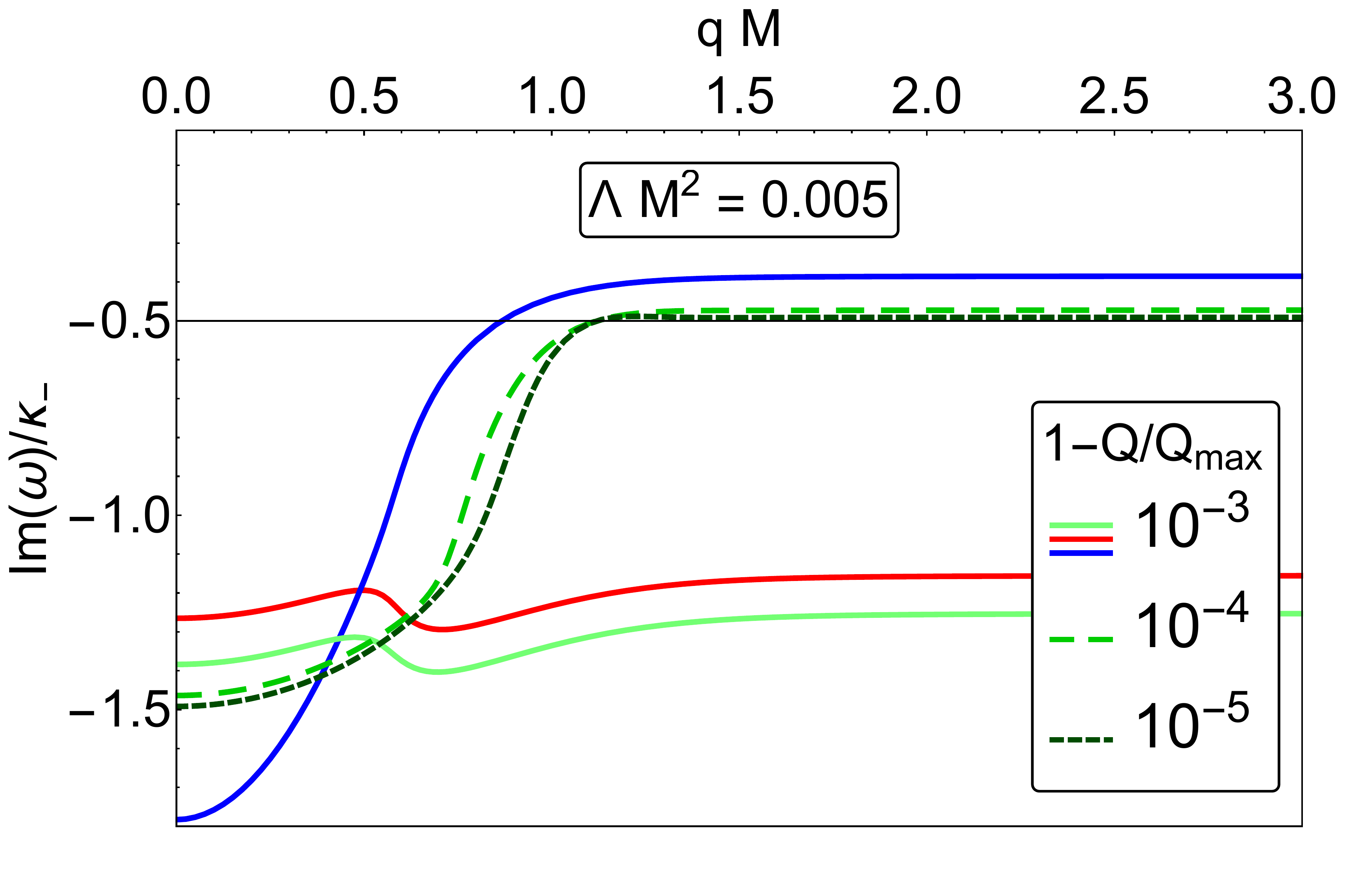}\hskip -1.5ex
\includegraphics[scale=0.225]{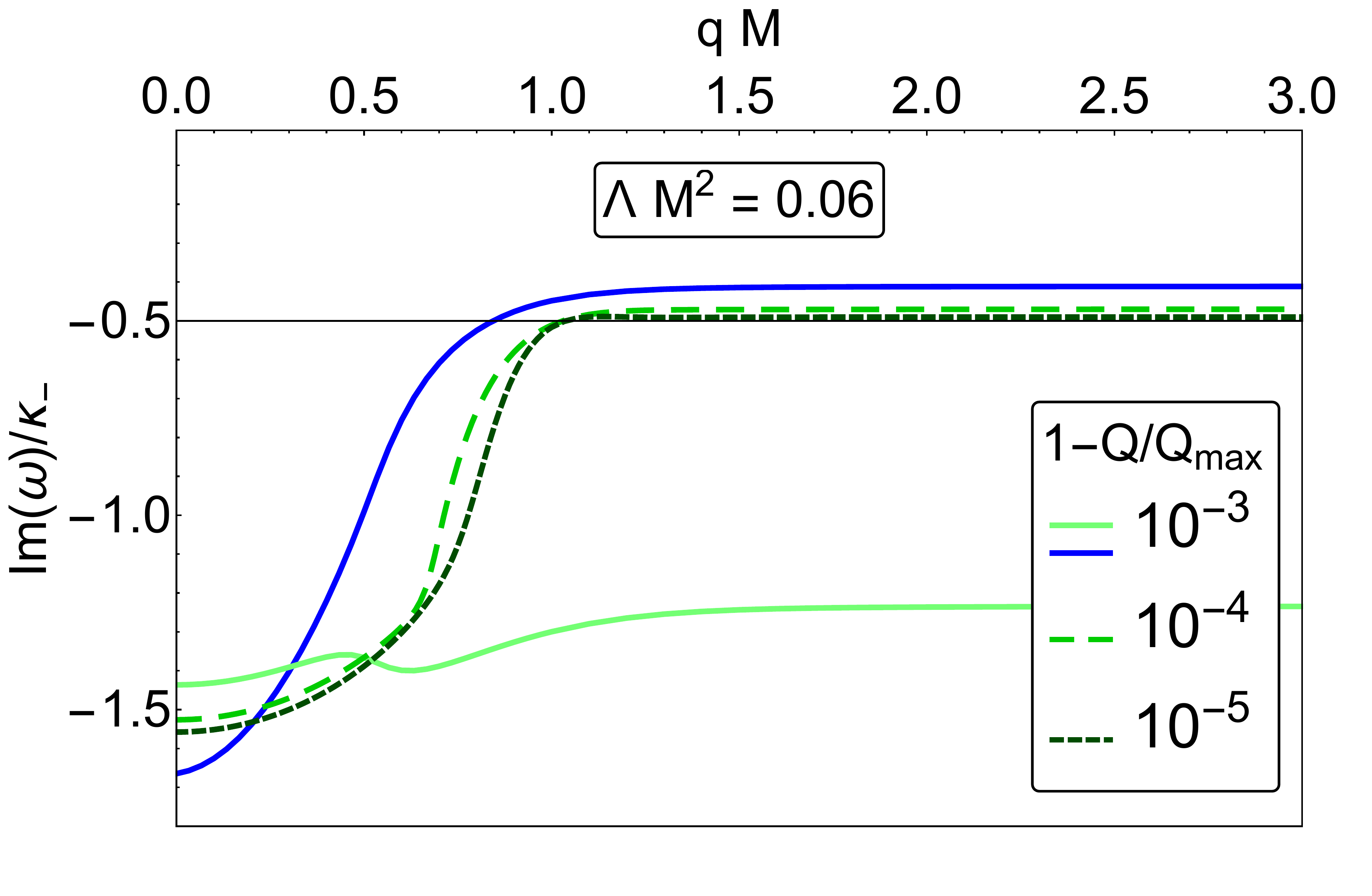}
\caption{Lowest lying QNMs of a charged, massless fermionic perturbation of a RNdS BH with $\xi=1$, $\Lambda M^2=0.005,\,0.06$ and $1-Q/Q_\text{max}=10^{-3},\,10^{-4},\,10^{-5}$ as a function of the fermionic charge $q M$. The modes denoted with blue, red and green colors belong to the PS, dS and NE family, respectively.}
\label{smallL}
\end{figure}
Since our purpose is to investigate the implications of charged fermions in SCC, we will restrict ourselves to choices of NE RNdS BH parameters which are problematic, since in this region $\kappa_-$ becomes comparable to the $\text{Im}(\omega)$ of the dominant QNM. For the region of interest, $\xi=1$ modes dominate all three families (see Section \ref{higher}).

In \cite{Zhang2} it was shown that for the choice of $\Lambda M^2=0.06$ and $Q/Q_\text{max}=0.996$ only the $\xi=1$ PS mode is relevant for SCC and
there is a region in the parameter space where $\beta>1/2$ (for $q M\lessapprox 0.53$) implying the potential violation of SCC. Interesting enough, there was no participation of the NE modes to the determination of $\beta$ for these parameters. On the other hand, for $\Lambda M^2=0.06$ and $Q/Q_\text{max}=0.999$ a family that originates from purely imaginary modes (for $q=0$) comes into play to dominate for very small fermionic charges and quickly becomes subdominant to give its turn to the $\xi=1$ PS mode. Again, $\beta>1/2$ (for $q M\lessapprox 0.85$) so SCC may be violated.

Our numerics completely agree with this picture. Here, we will be mostly interested in the case of even higher BH charges and the classification of the families originating from purely imaginary modes. To do so, we will study various choices of $\Lambda$. The BH charges we consider are:
\begin{equation}
1-Q/Q_\text{max}=10^{-3},\,\,10^{-4},\,\,10^{-5}.
\end{equation}
According to our results (see Fig. \ref{smallL}) for small BHs ($\Lambda M^2=0.005$) with $1-Q/Q_\text{max}=10^{-3}$ we see that $\beta$ is defined by the dS mode up to $q M\approx 0.5$; for larger $q$ the PS mode becomes dominant. Interestingly enough, for $q M<0.5$, the NE mode lies very close to the dS one being the first subdominant mode in this range. For larger BHs ($\Lambda M^2=0.06$) with $1-Q/Q_\text{max}=10^{-3}$ the dS mode moves rapidly to the subdominant side, giving its place to the NE mode to dominate up until $qM\approx 0.35$; for larger $q$ the PS mode dominates again. For BHs with $1-Q/Q_\text{max}\geq 10^{-4}$ the NE mode always dominate the dynamics, while the rest of the families lie out of the range of interest. 

\begin{figure}[H]
\begin{center}
\includegraphics[scale=0.22]{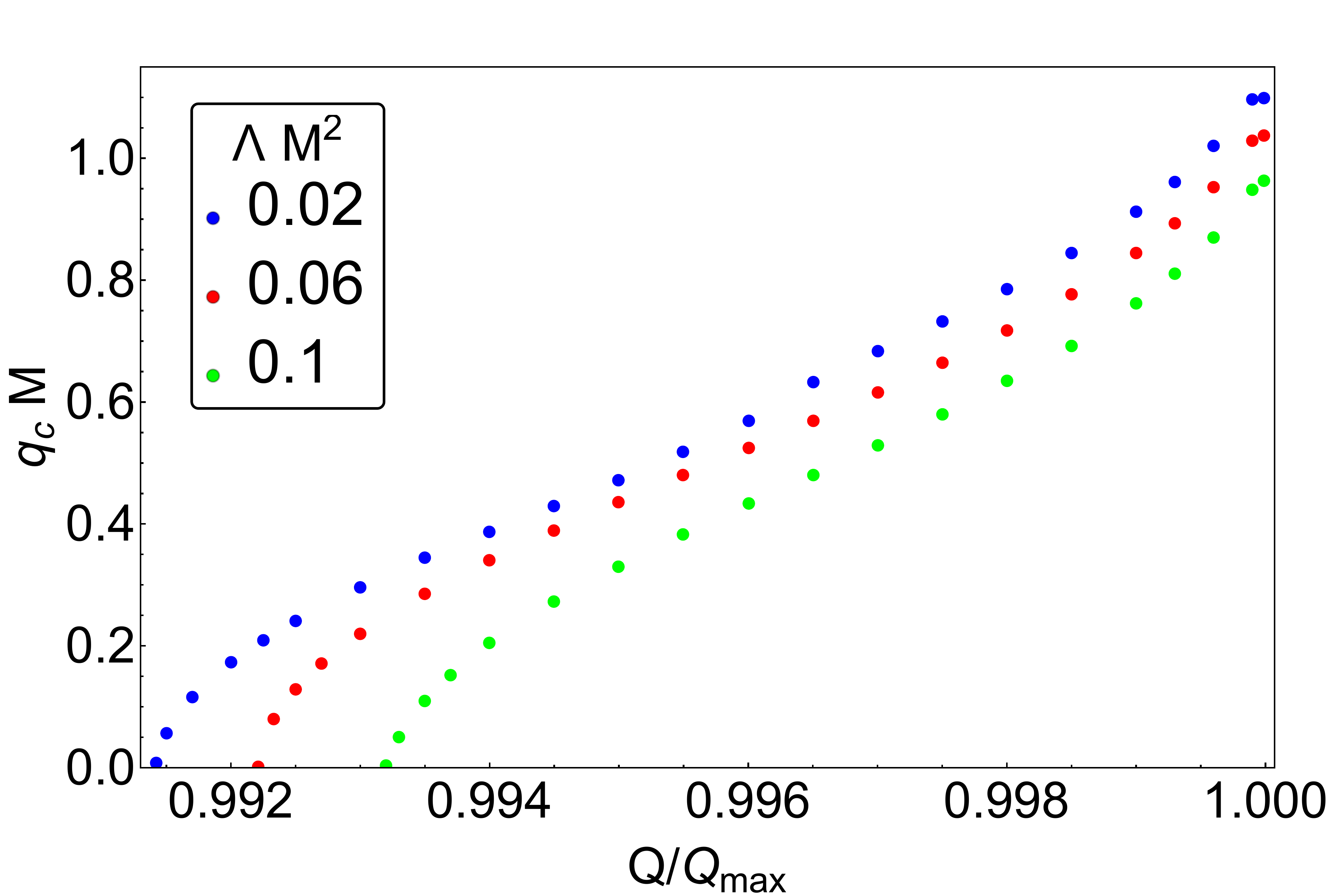}\hskip -1ex
\includegraphics[scale=0.22]{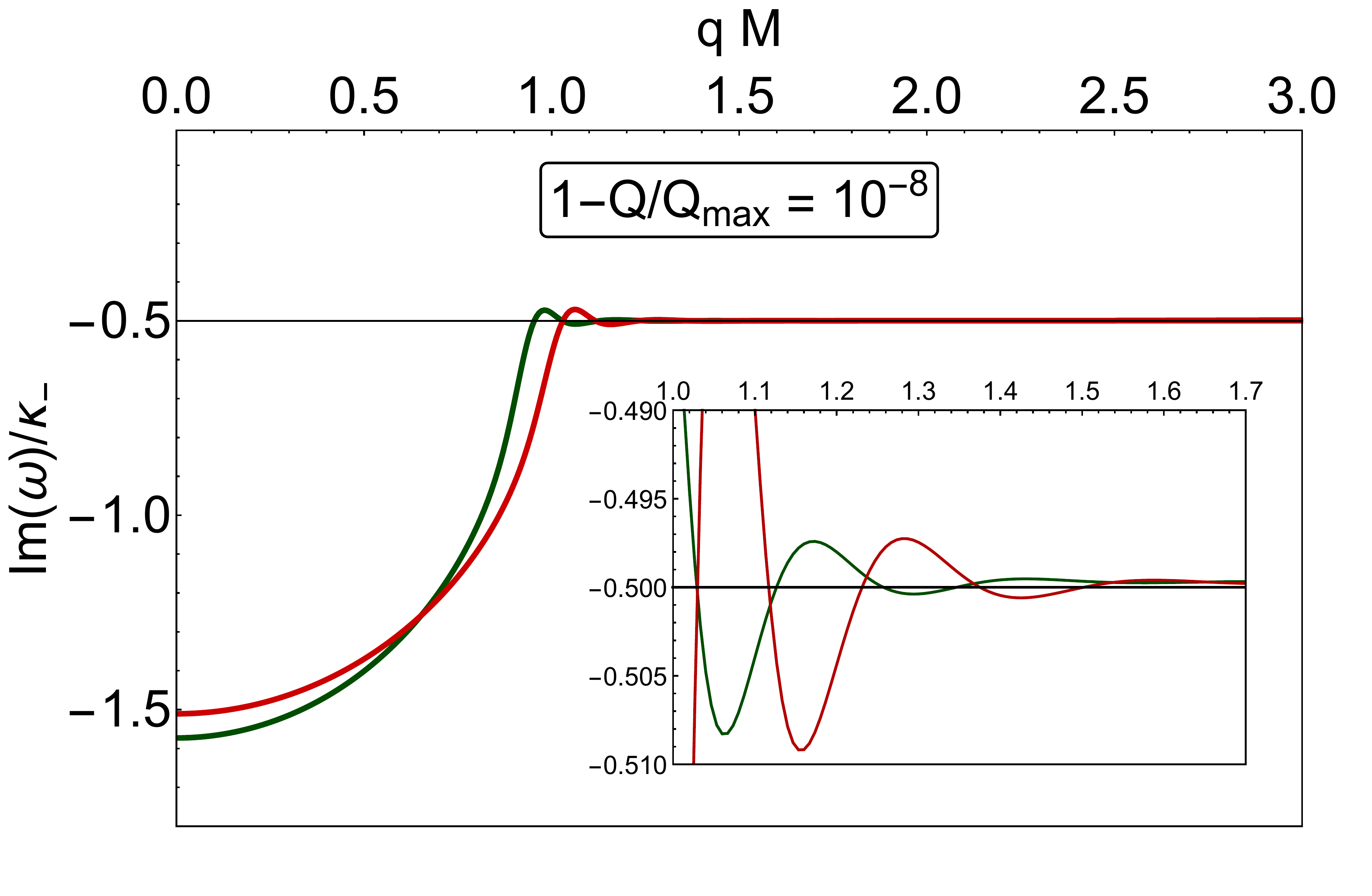}
\end{center}
\caption{{\bf Left:} Dependence of the critical fermionic charge $q_c$ on the BH charge $Q/Q_\text{max}$ and the cosmological constant $\Lambda M^2$. Here, different colors denote different choices of cosmological constants. {\bf Right:} Dominant $\xi=1$ NE QNMs of a charged, massless fermionic perturbation of a RNdS BH with $\Lambda M^2=0.005$ (red line) and $\Lambda M^2=0.06$ (green line) for $1-Q/Q_\text{max}=10^{-8}$ as a function of the fermionic charge $q M$.}
\label{qc_wiggles}
\end{figure}

For all cases presented, there is always a critical fermionic charge $q_c$ above which $\beta<1/2$ and SCC is preserved. In Fig. \ref{qc_wiggles} (left panel) we display the dependence of $q_c M$ on the $\Lambda M^2$ and $Q/Q_{\text{max}}$. We observe that as the BH becomes extremal a larger violation gap occurs in the parameter space. A larger $q_c M$ is also obtained for smaller cosmological constants. Similar results were obtained in \cite{PhysRevD.98.104007,Mo:2018nnu} for the case of charged scalar perturbations, although the absence of superradiance effect in fermionic fields leads to even larger regimes in the parameter space where violation of SCC may occur. 

By observing the cases with $1-Q/Q_\text{max}=10^{-5}$ we see that above $q_c$, $\beta$ lies very close to $1/2$. To examine if non-perturbative effects are present we plot $\beta$ for $\Lambda M^2=0.005,\,0.06$ and $1-Q/Q_\text{max}=10^{-8}$ versus the fermionic charge. In Fig. \ref{qc_wiggles} (right panel) we observe the existence of 
arbitrarily small oscillations of the imaginary part of the fundamental NE mode in 
highly near-extremal RNdS BHs. Such a phenomenon was previously observed for charged 
scalar perturbations in RNdS \cite{Dias:2018ufh} and gravitational perturbations in Kerr BHs \cite{Yang:2013uba}. 
These oscillations have been named ``wiggles" recently, and have very small amplitude. They are suppressed exponentially fast with increasing $q$ and are precisely the non-perturbative effect that an asymptotic series, such as the WKB approximation, can easily miss, since they are highly subdominant. We believe that these wiggles were missed from the analysis of \cite{Zhang2} because they did not consider highly NE BHs.  

The ramifications of the existence of wiggles are fierce for SCC. Our results indicate that, even for fermionic fields with $q>q_c$, there are still NE BHs for which $\beta>1/2$ and SCC may be violated, regardless of the cosmological constant, in contrast with the results in \cite{Zhang2}. Finally, notice that for all cases presented, all dominant modes originate below $\text{Im}(\omega/\kappa_-)=-1$, indicating that $\beta>1$, corresponding to a potential scenario of bounded curvature as explained in Chapter \ref{PRL}.
\section{Conclusions}

We have presented evidence in the previous chapters for the potential failure of SCC in NE RNdS BHs under neutral and charged scalar perturbations. By utilizing (\ref{betaOld2}) we performed thorough numerical analyses of $\beta$ through the calculation of QNMs of the system. Here, we extend our analysis to charged fermionic fields.

First, we provide justification that \eqref{betaOld2} remains valid for charged fermionic perturbations. Then, we perform a detailed numerical computation of the dominant modes of RNdS BHs and distinguish three families of QNMs. The first family is closely related to the PS of the BH while the second is related to the existence and timescale of the dS horizon of pure dS space. The final family dominates the dynamics when NE BH charges are considered. According to our study, the only relevant region for SCC is the NE, where the surface gravity of the CH, $\kappa_-$, becomes comparable with the decay rates of the dominant QNMs. We show that all families admit their dominant modes for $\xi=1$ in this region and search for potential violation, while taking into account the entire range of $qM\geq 0$. Finally, by computing $\beta$ we consider the implications on SCC.
\begin{figure}[H]
\begin{center}
\includegraphics[scale=0.4]{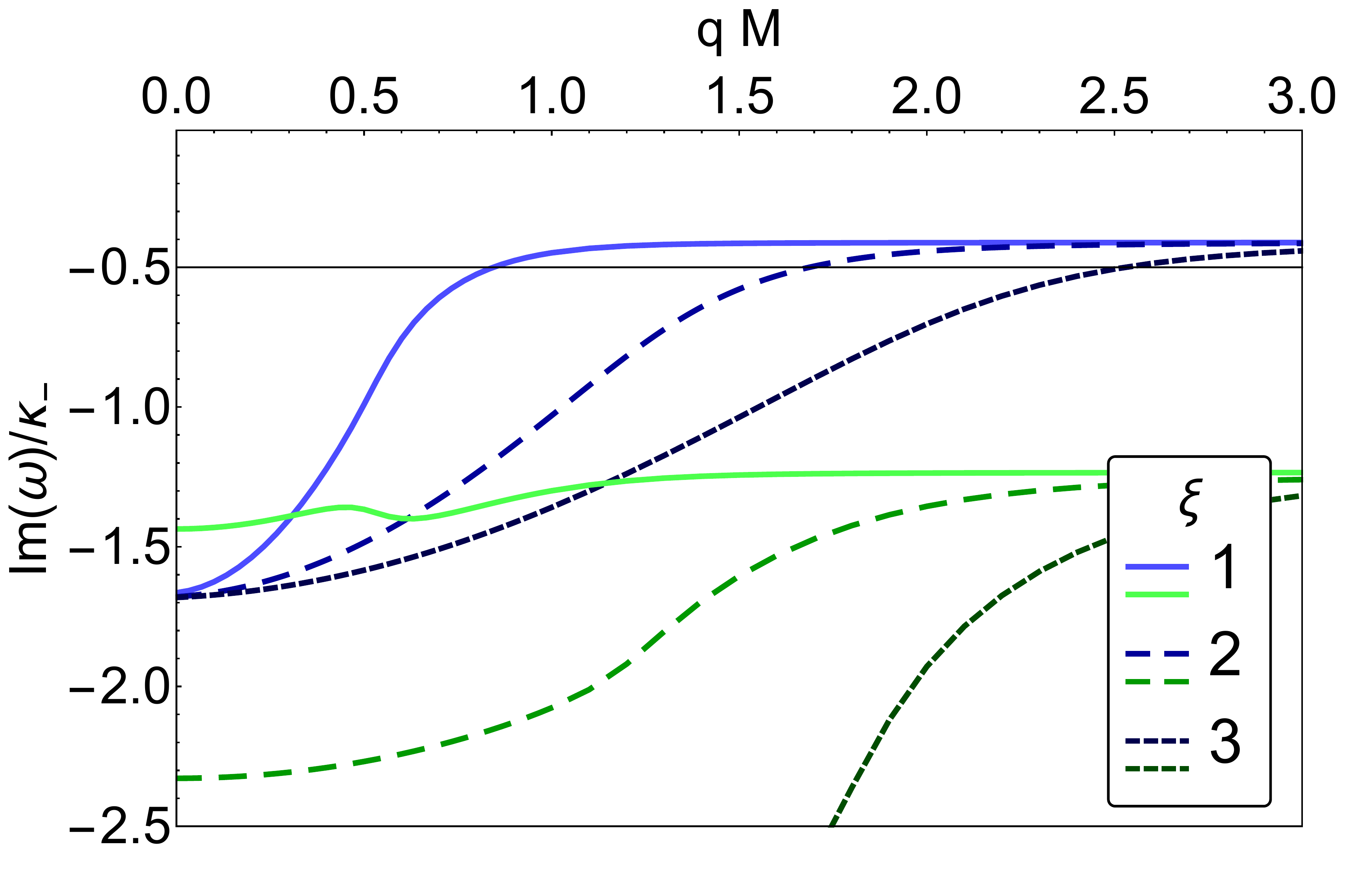}
\end{center}
\caption{Lowest lying QNMs of a charged, massless fermionic perturbation of a RNdS BH for various $\xi$ with $\Lambda M^2=0.06$ and $1-Q/Q_\text{max}=10^{-3}$ as a function of the fermionic charge $qM$.  The modes denoted with blue and green colors belong to the PS and NE family, respectively. The dS family is absent in this range.}
\label{higher_k}
\end{figure}

Our main results are shown in Figs. \ref{QNMs} - \ref{qc_wiggles} and our conclusions are summarized here. For all choices of $\Lambda M^2$ we always find a region of fermionic charges for which $\beta>1/2$ which predicts a potential failure of SCC, since the CH can be seen as singular due to the blow-up of curvature but maintain enough regularity for metric extensions to be possible beyond it. For sufficiently large fermionic charges the conjecture seems to be initially restored for highly charged RNdS BHs. After examining BHs even closer to extremality, we realize that even beyond the critical fermionic charge, violation can still occur due to the existence of wiggles.

We point out that for all cases presented, all dominant modes from the dS, PS or NE family admit $\beta>1$ for a small but significant regime of fermionic charges. This result is even more alarming for SCC since it is related to bounded curvature and therefore opens the possibility to the existence of solutions with even higher regularity across the CH. Nevertheless, if we superimpose all perturbations, then the smallest of all types of perturbations is the one relevant for SCC. Thus, the neutral scalar modes admit $1/2<\beta<1$, which is enough to guarantee the blow-up of curvature invariants at the CH.

\begin{figure}[H]
\includegraphics[scale=0.225]{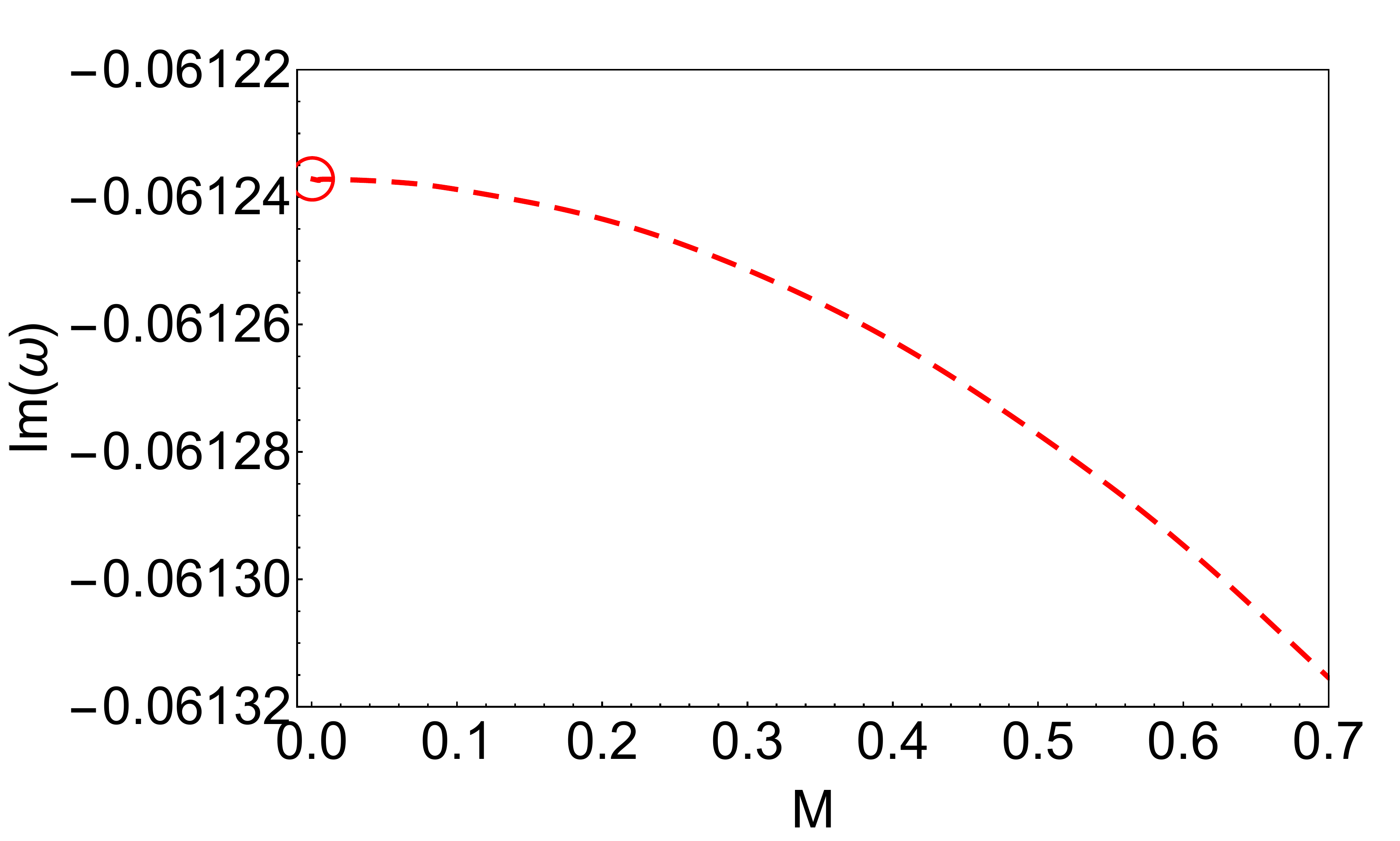}\hskip -1ex
\includegraphics[scale=0.22]{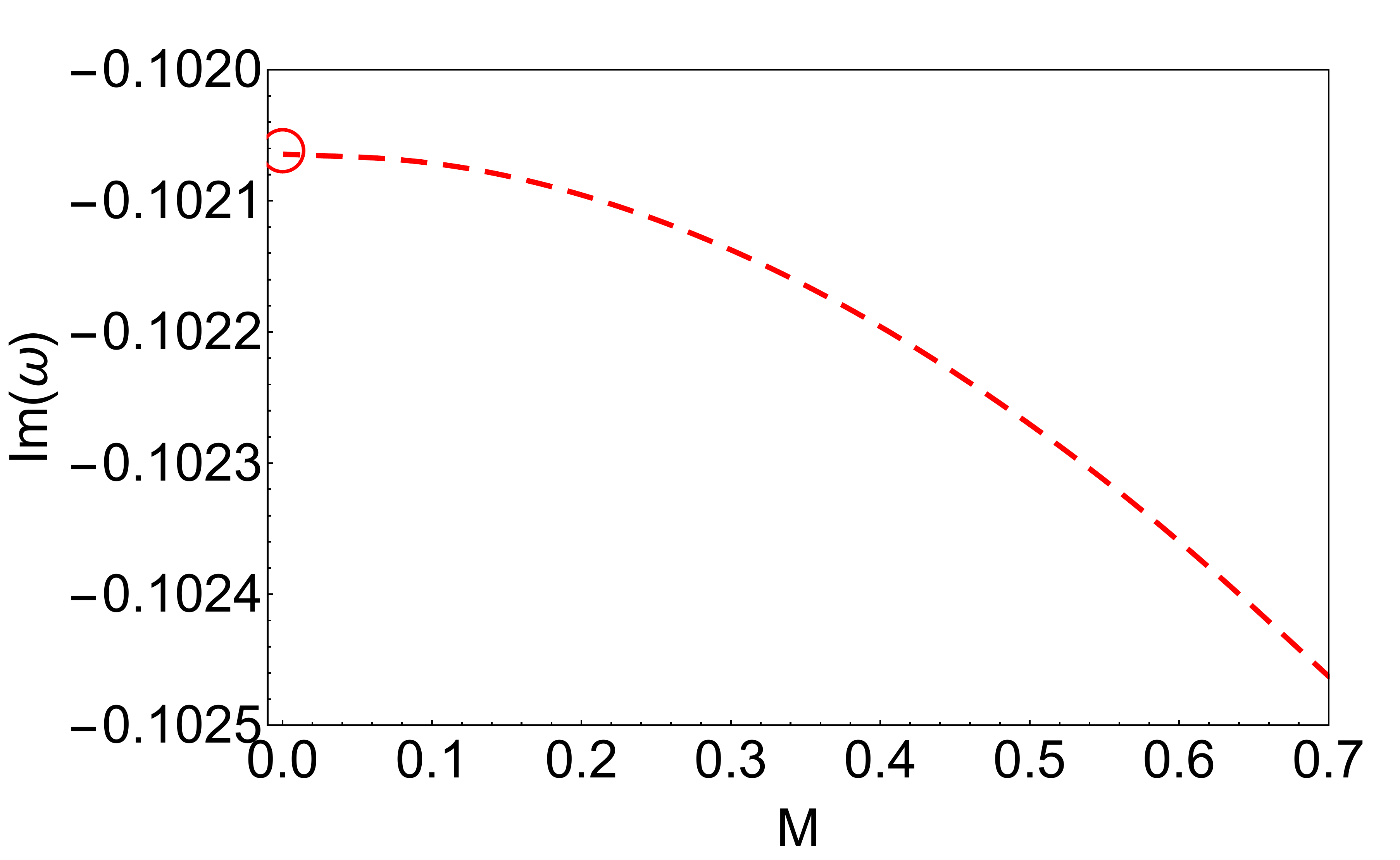}
\caption{Fundamental (left, dashed) and first overtone (right, dashed) $\xi=1$ BH dS QNM of a neutral, massless fermionic perturbation propagating on a fixed RNdS BH with $\Lambda=0.005$ and $Q=10^{-4}$ as a function of the BH mass $M$. The red circles in each plot designate the respective pure dS QNMs.}
\label{dSconv}
\end{figure}

\section{Higher $\xi$ modes}{\label{higher}}
In this section we verify the expectation that the higher $\xi$ QNMs do not affect SCC. In Fig. \ref{higher_k} we show the $\xi=1,2,3$ modes. The ones depicted with blue colors belong to the PS family, since they originate from complex modes for $q=0$ and follow their pattern. The ones depicted with green colors belong to the NE family, since they originate from purely imaginary modes for $q=0$ and follow their pattern. The dS modes are not present in the range of interest since they are too subdominant for the chosen cosmological constant. We clearly see that the modes defining $\beta$ according to (\ref{betaOld2}) will be the $\xi=1$ QNMs. The same holds for other choices of $\Lambda$.

For completeness, in Table \ref{table} we show various modes from different families with $\xi=1,10$ for various choices of $Q$, $q$ and $\Lambda$. We compare the neutral $\xi=10$ PS modes with a WKB approximation for arbitrarily large $\xi$ and verify that indeed the imaginary parts lie very close. It is apparent that for NE charges $\xi=1$ modes always dominate. It is also apparent that the only way for $\xi\rightarrow\infty$ modes to be the dominant ones of the PS family is for very small cosmological constants. Specifically, for $\Lambda M^2=0.001$, $\xi\rightarrow\infty$ modes are dominant up to a critical BH charge $Q_c/M\approx 0.866$. Above $Q_c/M$, $\xi=1$ modes dominate the PS family. E.g. for $Q/M=0.865$ and $q=0$ the dominant ($\xi\rightarrow\infty$) PS mode admits $\text{Im}(\omega_\text{PS})/\kappa_-=-0.0479$, while the dominant ($\xi=1$) dS mode admits $\text{Im}(\omega_\text{dS})/\kappa_-=-0.0135$; the NE family is too subdominant for this BH charge. None of those modes can potentially violate SCC so it becomes a necessity to search closer to extremality, where we are aware that $\kappa_-$ becomes comparable to $\text{Im}(\omega)$.\footnote{For the case presented, the surface gravity is $\sim 100$ larger than the decay rate of the dominant mode.} Since $\beta$ is maximal at $q=0$, any $qM>0$ will make $\text{Im}(\omega)/\kappa_-$ even smaller. Finally, for larger $\Lambda$, $Q_c$ decreases, moving even further away from extremality.

Considering the above, we are convinced that throughout the parameter space in study, $\xi=1$ indeed gives the dominant modes for all families.

\begin{table}[H]
\centering
\scalebox{0.645}{
\begin{tabular}{||c| c | c ||} 
\hline
  \multicolumn{3}{||c||}{$Q/M=10^{-1}$} \\
   \hline
\hline
  \multicolumn{3}{||c||}{$\Lambda M^2=0.005$} \\
   \hline
    $\xi$ & $qM=0$ & $qM=0.1$ \\ [0.5ex] 
   \hline
   1 &  $\omega_\text{PS}$= 0.1795 - 0.0947 i & $\omega_\text{PS}$= -0.1760 - 0.0941 i   \\  
   & $\omega_\text{dS}$= -0.0614 i & $\omega_\text{dS}$= -0.00003 - 0.0614 i \\
   \hline
   10 & $\omega_\text{PS}$= 1.8831 - 0.0941 i& $\omega_\text{PS}$= -1.8797 - 0.0940 i \\ 
   \hline
   WKB &$\,\,\,\,\,\,\,\,\,\,\,\,\,\,\,\,\,\,\,\,\,\,\,\,\,\,\,\,\,\,\,\,\,\,\,\,$- 0.0941 i & - \\
   \hline\hline
  \multicolumn{3}{||c||}{$\Lambda M^2=0.06$} \\
   \hline
    $\xi$ & $qM=0$ & $qM=0.1$ \\ [0.5ex] 
   \hline
      1 & $\omega_\text{PS}$= 0.1280 - 0.0650 i & $\omega_\text{PS}$= -0.1247 - 0.0647 i    \\  
      & $\omega_\text{dS}$= -0.2170 i & $\omega_\text{dS}$= -0.0003 - 0.2170 i \\
      \hline
      10 &$\omega_\text{PS}$= 1.3097 - 0.0654 i & $\omega_\text{PS}$= -1.3064 - 0.0654 i \\ 
      \hline
      WKB &$\,\,\,\,\,\,\,\,\,\,\,\,\,\,\,\,\,\,\,\,\,\,\,\,\,\,\,\,\,\,\,\,\,\,\,$ - 0.0654 i & - \\
      \hline
\end{tabular}
}
\scalebox{0.6}{
\begin{tabular}{||c| c | c ||} 
\hline
  \multicolumn{3}{||c||}{$Q/Q_\text{max}=1-10^{-3}$} \\
   \hline
\hline
  \multicolumn{3}{||c||}{$\Lambda M^2=0.005$} \\
   \hline
    $\xi$ & $qM=0$ & $qM=0.1$ \\ [0.5ex] 
   \hline
    & $\omega_\text{PS}$=0.2353 - 0.0865 i &   $\omega_\text{PS}$=-0.1875 - 0.0852 i  \\  
   1& $\omega_\text{dS}$=-0.0613 i & $\omega_\text{dS}$=-0.0003 - 0.0612 i \\
   & $\omega_\text{NE}$=-0.0671 i & $\omega_\text{NE}$=0.1004 - 0.0669 i\\
   \hline
   10 & $\omega_\text{PS}$=2.4650 - 0.0872 i & $\omega_\text{PS}$=-2.4152 - 0.0872 i \\ 
   \hline
   WKB &$\,\,\,\,\,\,\,\,\,\,\,\,\,\,\,\,\,\,\,\,\,\,\,\,\,\,\,\,\,\,\,\,$ - 0.0870 i& - \\
   \hline\hline
  \multicolumn{3}{||c||}{$\Lambda M^2=0.06$} \\
   \hline
    $\xi$ & $qM=0$ & $qM=0.1$ \\ [0.5ex] 
   \hline
      1 & $\omega_\text{PS}$=0.2016 - 0.0708 i & $\omega_\text{PS}$=0.2548 - 0.0692 i    \\  
      & $\omega_\text{NE}$=-0.0611 i & $\omega_\text{NE}$=0.0974 - 0.0609 i \\
      \hline
      10 &$\omega_\text{PS}$=2.0918 - 0.0716 i & $\omega_\text{PS}$=2.1436 - 0.0715 i \\ 
      \hline
      WKB & $\,\,\,\,\,\,\,\,\,\,\,\,\,\,\,\,\,\,\,\,\,\,\,\,\,\,\,\,\,\,\,\,\,\,$- 0.0718 i& - \\
      \hline
\end{tabular}
}
\caption{Lowest lying fermionic QNMs of RNdS BH for various $Q$, $q$, $\Lambda$ and $\xi$.}
\label{table}
\end{table}
\section{Convergence of the families}\label{dS}
In this section we demonstrate the convergence of the BH dS modes to the pure dS QNMs, as well as the behavior of the PS modes, for vanishing $M$. In Fig. \ref{dSconv} we plot with dashed lines the fundamental ($n=0$) and first overtone ($n=1$) BH dS modes and illustrate that as the BH disappears ($M\rightarrow 0$) the QNMs tend smoothly to the exact pure dS fermionic QNMs which are denoted with red circles. Here, we only demonstrate this for a specific choice of parameters but our numerics reveal that the smooth convergence happens for all BH parameters. We realize that the increment of the BH mass affects weakly the dS family. It is important to note that equivalent results for the BH dS family were obtained in RNdS under scalar perturbations.

\begin{figure}[H]
\includegraphics[scale=0.22]{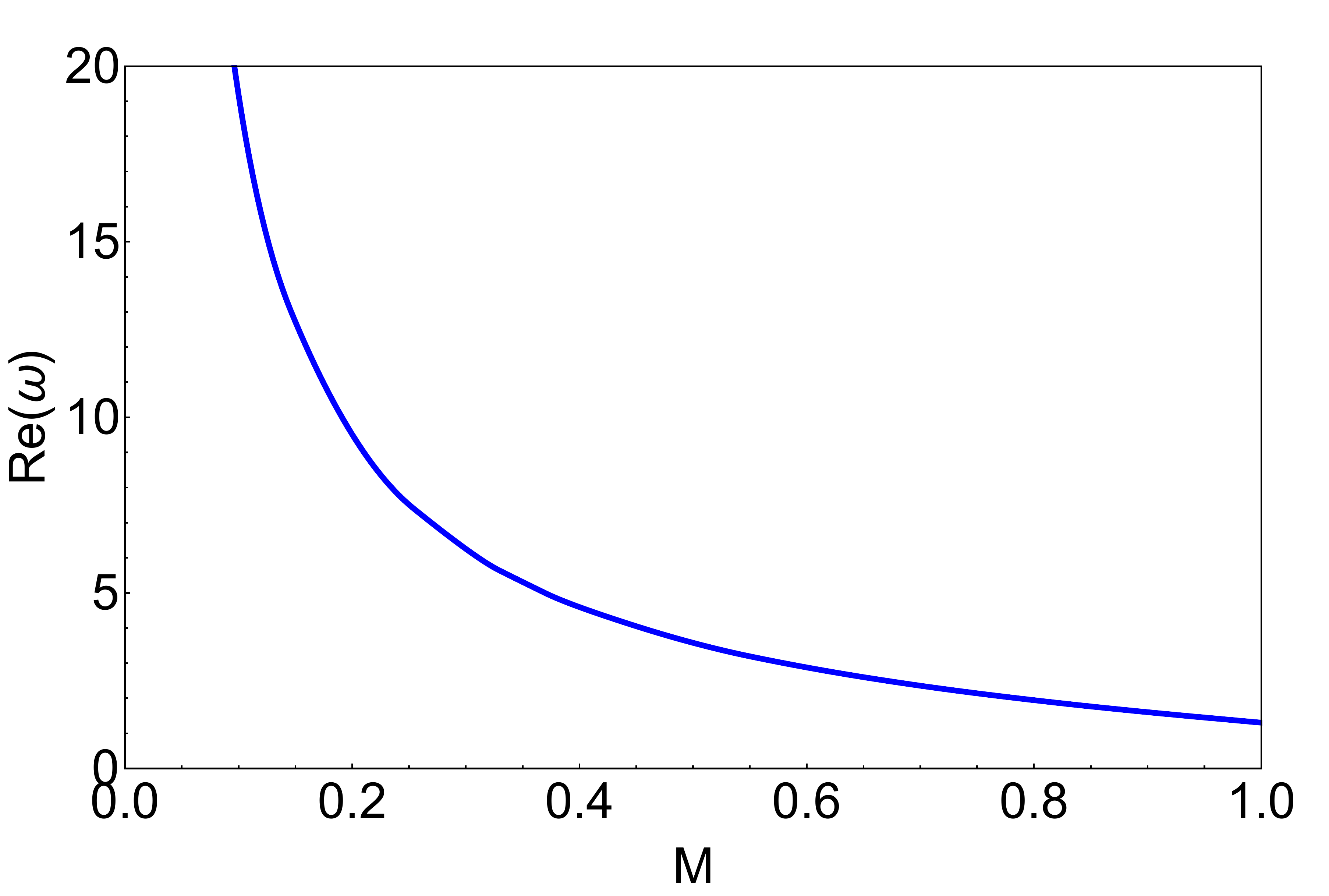}\hskip -1ex
\includegraphics[scale=0.225]{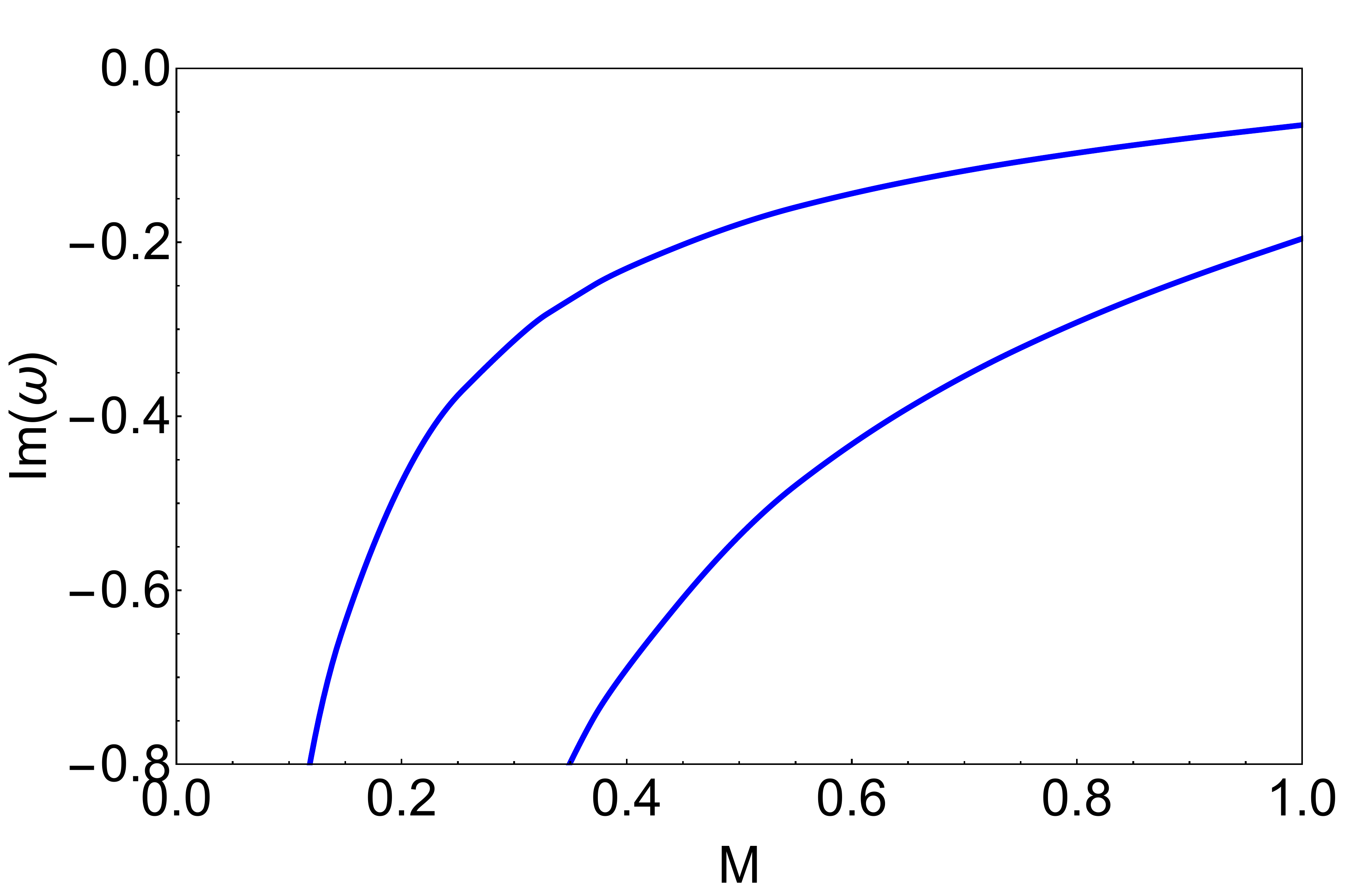}
\caption{Real (left) and imaginary (right) part of the $\xi=10$ fundamental and first overtone PS QNM of a neutral, massless fermionic perturbation propagating on a fixed RNdS BH with $\Lambda=0.06$ and $Q=10^{-4}$ as a function of the BH mass $M$. The real parts coincide in the range shown.}
\label{PSconv}
\end{figure}

The story is different for the PS family of modes (see Fig. \ref{PSconv}). As $M\rightarrow 0$ the real and imaginary parts diverge. This occurs due to the shrinking of the photon sphere. If we consider, for example, a perturbed string with a specific length that vibrates, then by continuously decreasing its length we will observe that the vibrations will have increasingly higher frequency and smaller timescales until the point where the string vanishes and oscillations cease to exist. The same occurs for a the BH in study as it ``disappears". 

\chapter{Strong Cosmic Censorship in higher-dimensional Reissner-Nordstr\"{o}m-de Sitter spacetime}\label{JHEP}
In the previous chapters we have demonstrated that SCC may be violated for near-extremally-charged black holes in 4-dimensional de Sitter space under scalar and fermionic perturbations. Here, we extend the study of neutral massless scalar perturbations in higher dimensions and discuss the dimensional influence on the validity of SCC hypothesis. By giving an elaborate description of neutral massless scalar perturbations of RNdS BHs in $d=4,5$ and $6$ dimensions we conclude that SCC is violated near extremality. The following chapter is based on \cite{Liu:2019lon}.
\section{Introduction}
It is well known that the would-be CH in asymptotically flat BHs is a singular boundary \cite{Simpson:1973ua,Hartle,Poisson:1990eh}. The remnant fields of gravitational collapse exhibit an inverse power-law decay behavior in the exterior
of asymptotically flat BHs \cite{Price1,Price2}, and will be amplified when propagating along the CH due to the
exponential blueshift effect occurring there. The gravitational collapse of matter fields cannot lead to stable enough CHs, leading to the preservation of the deterministic power of physical laws and the SCC hypothesis, proposed by Penrose \cite{Penrose69}.

However, for dS BH spacetimes, due to the change of the boundary conditions, remnant perturbation fields outside dS BHs decay exponentially instead of polynomially \cite{Hintz:2016gwb,Hintz:2016jak}.  The extendibility of the metric at the CH depends delicately on the competition
between the exponential decay outside the BH and the exponential blueshift
amplification along the CH. In such a scenario, the decay of perturbations could be fast enough such that the CH singularity can be so weak that the spacetime metric will be extendible beyond it as a weak solution to the Einstein field equations \cite{Christodoulou:2008nj}, meaning that SCC may be violated!  Mathematically, it was proven \cite{Hintz:2015jkj}  that SCC will not be respected under massless neutral scalar perturbations if the following condition is satisfied
\begin{equation}
\beta\equiv-\frac{\mathrm{Im}(\omega)}{\kappa_{-}}>\frac{1}{2},\label{eq15}
\end{equation}
where $\kappa_{-}$ is the surface gravity of the CH and $\mathrm{Im}(\omega)$ is the imaginary part of the lowest-lying/dominant QNM of the perturbation in the external region of the BH.

In particular, for NE RNdS BHs, it has been shown in Chapter \ref{PRL} that neutral massless scalar perturbations are extendible past the CH, since the blueshift amplification along the CH is dwarfed by the exponential decay outside of the dS BH. Such a violation of SCC
becomes even more severe in the case of the coupled electromagnetic and gravitational perturbations \cite{Dias:2018etb}. 

Later on, it was shown that the violation of SCC can be alleviated by considering a sufficiently charged scalar field on the exterior of RNdS BHs \cite{PhysRevD.98.104007,Hod:2018dpx,Mo:2018nnu,Dias:2018ufh}, although there was still a region in the parameter space where SCC may be violated (see Chapter \ref{PRD}). Similar results have been obtained for Dirac field perturbations \cite{Zhang2} (see Chapter \ref{PLB}). On the other hand, the nonlinear evolution of massive neutral scalar fields in RNdS space revealed that SCC might not be saved by such nonlinear effects \cite{Luna:2018jfk}. In addition, by investigating SCC in lukewarm RNdS and Martnez-Troncoso-Zanelli (MTZ) BH spacetimes, under non-minimally coupled massive scalar perturbations, it was demonstrated that the validity of the hypothesis depends on the characteristics of the scalar field \cite{Gwak1}.
Besides charged BHs, SCC has been examined in KdS BH backgrounds and interestingly enough no violation was found for linear scalar and gravitational perturbations \cite{Dias:2018ynt}.

All available studies of SCC in RNdS BH backgrounds are limited in 4 dimensions even though it has been found that in higher dimensions, physics becomes richer \cite{Emparan:2008eg}. In contrast to the $4-$dimensional results,
various instabilities have been found in higher-dimensional spacetimes. In a wide class of $d\geq 4$ configurations, such as black
strings and black branes, the Gregory-Laflamme instability against linear perturbations was discussed in
\cite{Gregory1, Gregory2}. For higher-dimensional BHs in the Einstein-Gauss-Bonnet theory, it was found
that instabilities occur for large angular quantum numbers $l$, while the lowest lying QNMs of the first few angular quantum numbers were found stable \cite{Konoplya2,Beroiz1}. In particular,  numerical investigations have uncovered
the surprising result that RNdS BH backgrounds are unstable for $d\geq 7$, if the values of black hole mass and charge are large enough \cite{Konoplya:2008au}, followed by the analytic confirmation of \cite{Cardoso:2010rz}. Moreover,  it was argued that the Weak Cosmic Censorship hypothesis could be restored easier in higher dimensions \cite{Mken} by examining the gravitational collapse of spherically symmetric generalized Vaidya spacetimes.
It is, thus, of great interest to generalize the discussion of SCC to higher-dimensional RNdS BHs and explore whether and how they affect the validity of the conjecture.

\section{Scalar fields in higher-dimensional RNdS spacetime}
The $d-$dimensional RNdS spacetime is described by the metric 
\begin{equation}
\label{dspace}
ds^2=-f(r)dt^2+\frac{1}{f(r)} dr^2+r^2 d\Omega_{d-2}^2,
\end{equation}
where
\begin{align}\label{metric1}
f(r)=1-\frac{m}{r^{d-3}}+\frac{q^2}{r^{2(d-3)}}-\frac{2\Lambda}{(d-2)(d-1)}r^2,
\end{align}
and
\begin{equation}
\Lambda=\frac{(d-2)(d-1)}{2L^2}, \quad d\Omega^2_{d-2}=d\chi_2^2+\sum_{i=2}^{d-2}\left(\prod_{j=2}^{i}\sin^2\chi_j\right)\, d\chi_{i+1}^2,
\end{equation}
in which $q$ and $m$ are related to the electric charge $Q$ and the ADM mass $M$ of the BH, and $L$ is the cosmological radius. $M$ and $Q$
are expressed as
\begin{equation}
M=\frac{d-2}{16\pi}\omega_{d-2}m,\quad Q=\frac{\sqrt{2(d-2)(d-3)}}{8\pi}\omega_{d-2}q,\quad \omega_{d}=\frac{2\pi^{\frac{d+1}{2}}}{\Gamma(\frac{d+1}{2})},
\end{equation}
with $\omega_{d}$ being the volume of the unit $d$-sphere. The causal structure of a subextremal $d-$dimensional BH described by (\ref{metric1}) admits three distinct horizons, where $r_-<r_+< r_c$ are the Cauchy, event and cosmological horizon radius, respectively. We denote the extremal electric charge of the BH as $Q_{\text{max}}$ at which the CH and event horizon coincide. The maximal cosmological constant is denoted as $\Lambda_{\text{max}}$ for each dimension,\footnote{e.g. for $d=4$, $\Lambda_\text{max}=2/9$, for $d=5$, $\Lambda_\text{max}=3\pi/4$ and for $d=6$, $\Lambda_{\text{max}}=\left(648\pi^2/25\right)^{\frac{1}{3}}$ provided that the black hole mass is set to $M=1$.} meaning that if $\Lambda>\Lambda_{\text{max}}$ holds then the spacetime would admit at most one horizon with positive radius, thus rendering our discussion irrelevant. To ensure the existence of  three distinct horizons, the cosmological constant must be restricted to $\Lambda<\Lambda_{\text{max}}$.
The surface gravity of each horizon is then
\begin{equation}
\label{surfGrav}
\kappa_i= \frac{1}{2}|f'(r_i)|\;\;,\; i\in\{-,+,c\}.
\end{equation}

The propagation of a neutral massless scalar field $\Psi$ on a fixed $d-$dimensional RNdS background is described by the Klein-Gordon equation. By expanding our field in terms of spherical harmonics
\begin{equation}
\Psi\sim\sum_{lm}e^{-i\omega t}\frac{\psi(r)}{r^{\frac{d-2}{2}}}Y_{lm}(\chi)\label{eq3},
\end{equation}
we end up with the master equation
\begin{equation}
\frac{d^2\psi}{dr_{\ast}^{2}}+(\omega^2-V)\psi=0,\label{eq7}
\end{equation}
where
\begin{equation}
V=f(r)\left(\frac{l(d+l-3)}{r^2}+\frac{(d-2)f'(r)}{2r}+\frac{f(r)(d-4)(d-2)}{4r^2}\right),\label{eq9}
\end{equation}
and $dr_{\ast}={dr}/{f(r)}$ the tortoise coordinate. By imposing the boundary conditions
\begin{equation}\psi(r\rightarrow r_+)\sim e^{-i\omega  r_*},\,\,\,\,\,\,\,\,\,\,\,\,\,\,\,\psi(r\rightarrow r_c)\sim e^{i\omega {r}_*},\label{eq4}
\end{equation}
we select a discrete set of QN frequencies called QNMs. Due to the similarity of characteristics of (\ref{eq9}) and the effective potential for odd (Regge-Wheeler \cite{Regge}) and even (Zerilli \cite{Zerilli,Zerilli2}) gravitational perturbations, the study of massless neutral scalar fields propagating on spherically symmetric backgrounds is a good proxy for more physically relevant gravitational field perturbations. 

As shown in Appendix \ref{appE}, for $d\geq 4$ the stability of the CH continues to be determined by (\ref{eq15}). The results shown in the following sections were obtained with the Mathematica package of \cite{Jansen:2017oag}, the asymptotic iteration method (AIM) \cite{AIM,Cho1}, and checked in various cases with a WKB approximation \cite{Iyer:1986np} and with a code developed based on the matrix method \cite{KaiLin1}.
\section{Dominant families of modes in higher-dimensional RNdS spacetime}
According to Chapter \ref{PRL}, the region of interest in $4-$dimensional RNdS, where violation of SCC may occur, lies close to extremality. There, the decay rate of perturbations in the exterior becomes comparable with the surface gravity of the CH $\kappa_-$ leading to $\beta>1/2$. Motivated by the aforementioned study, we scan the parameter space of higher-dimensional RNdS spacetimes for near-extremal parameters. By applying our numerics in the region of interest we discover three distinct families of modes.

The photon sphere is a spherical trapping region of space where gravity is strong enough that photons are forced to travel in unstable circular orbits around a BH. This region has a strong pull in the control of decay of perturbations and the QNMs with large frequencies. For instance, the decay timescale is related to the instability timescale of null geodesics near the photon sphere. For asymptotically dS BHs, we find a family that can be traced back to the photon sphere and refer to them as PS modes. The dominant modes of this family are approached in the eikonal limit, where $l\rightarrow\infty$, and can be very well approximated with the WKB method (see Section \ref{app_WKB}). For vanishing $\Lambda,\,Q$ they asymptote to the Schwarzschild BH QNMs in $d\geq 4$ dimensions. We find that $l=10$ provides a good approximation of the imaginary parts of the dominant modes which we depict in our plots with solid blue lines.

Our second family of modes, the BH dS family, corresponds to purely imaginary modes which can be very well approximated by the pure $d$-dimensional scalar dS QNMs \cite{Du:2004jt,LopezOrtega:2006my,VasydS}:
\begin{align}
\label{dS1}
\omega_{\text{pure dS}}/\kappa_c^\text{dS}&=-i (l+2n),\\
\label{dS2}
\omega_{\text{pure dS}}/\kappa_c^\text{dS}&=-i (l+2n+d-1).
\end{align}
The dominant mode of this family ($n=0,\,l=1$) is almost identical to (\ref{dS1}) which we denote in our figures with red dashed lines. These modes are intriguing, in the sense that they have a surprisingly weak dependence on the BH charge and seem to be described by the surface gravity of $d$-dimensional dS $\kappa_c^{\text{dS}}=\sqrt{2\Lambda/(d-2)(d-1)}$ of
the cosmological horizon of pure $d$-dimensional dS space, as opposed
to that of the cosmological horizon in the RNdS BH under consideration.

Finally, as the CH approaches the event horizon, a new family of modes appears to dominate the dynamics. In the extremal limit of a $d$-dimensional RNdS BH the dominant ($n=l=0$) mode of this family approaches (see Section \ref{app_NE})
\begin{equation}
\label{NE}
\omega_\text{NE}=-i\kappa_{-}=-i\kappa_{+},
\end{equation}
where $\kappa_-,\,\kappa_+$ the surface gravity of the Cauchy and event horizon in $d$-dimensional RNdS spacetime. We call this family the NE family of modes. Higher angular numbers $l$ admit larger (in absolute value) imaginary parts, thus rendered subdominant. In the asymptotically flat case, these modes seem to have been described analytically in the eikonal limit \cite{Zhang:2018jgj}.

\section{Strong Cosmic Censorship in higher-dimensional RNdS spacetime}

In Fig. \ref{beta} we depict the dominant modes of each of the previous families versus $\kappa_-$. We have chosen $d=4,\,5,$ and $6$-dimensional near-extremal RNdS BHs with various $\Lambda/\Lambda_\text{max}$. It is evident that for sufficiently ``small" BHs\footnote{Usually we use $r_+/r_c$ to measure the size of ``small/large" BHs, but in our discussion we compare BHs in different dimensions by fixing $\Lambda/\Lambda_{\text{max}}$ which has some connection with the size of BHs. It turns out that the value of $r_+/r_c$ would be notably influenced by $Q/Q_{\text{max}}$. On the other hand, if $Q/Q_{\text{max}}$ is fixed, one can find that the difference between $r_c$ and $r_+$ would increase with the decrease of $\Lambda/\Lambda_{\text{max}}$. For these reasons, the ``small/large" BHs in this chapter are only referred to BHs with small/large $\Lambda/\Lambda_{\text{max}}$.} (very small $\Lambda/\Lambda_{\text{max}}$), the increment of dimensions fortifies SCC for a larger region of the parameter space of $Q/Q_\text{max}$. On the other hand, for sufficiently ``large" BHs (large $\Lambda/\Lambda_{\text{max}}$), the increment of dimensions work against the validity of SCC admitting violations for smaller $Q/Q_\text{max}$. To deepen into the understanding of this complex situation we denote the degree of difficulty of SCC violation with $d_4$ for $d=4$, $d_5$ for $d=5$ and $d_6$ for $d=6$. For example, for the case of $\Lambda/\Lambda_\text{max}=0.05$, the degree of difficulty of SCC violation follows $d_6>d_5>d_4$, meaning that 6-dimensional RNdS BHs require the highest BH charge to be violated. The second hardest BH to be violated is the 5-dimensional and, finally, the easiest to be violated is the 4-dimensional.

In the ``intermediate" region, where $\Lambda/\Lambda_\text{max}$ is neither too small nor too large, the picture becomes obscured by the delicate interplay of the QNMs of the dominant PS and dS family. To that end, we have depicted two interesting cases. In the first case, for $\Lambda/\Lambda_\text{max}=0.15$, the degree of difficulty to violate SCC follows $d_6>d_4>d_5$, while in the second case for $\Lambda/\Lambda_\text{max}=0.25$ we have $d_4>d_6>d_5$. This perplex picture appears due to the opposite behavior that the dominant PS and dS family possess. As shown in Fig. \ref{beta}, higher dimensions oblige $\beta_\text{PS}$, as it gets determined by the dominant modes of the PS family ($l=10$), to move upwards in the plots, thus becoming subdominant, while $\beta_\text{dS}$, as it gets determined by the dominant modes of the dS family ($l=1$), moves downwards. On the other hand, the increment of $\Lambda/\Lambda_{\text{max}}$ has the opposite effect on these families as expected by the results demonstrated in Chapter \ref{PRL}. It is easy to realize (see the pattern in Fig. \ref{beta}) that the inclusion of even higher than 6 dimensions will make the picture of ``intermediate" and ``large" BHs even richer and much more perplex\footnote{E.g. for $\Lambda/\Lambda_{\text{max}}=0.4$ we can see that if $d=7$ or 8 where to be included, then the dS family would eventually dominate for such dimensions, thus changing the picture into a richer version of $\Lambda/\Lambda_{\text{max}}=0.15$ or $0.25$.}. The only solid case is the one for ``small" BHs. There, $\beta$ is essentially (for the largest part of the parameter space) determined by the dominant modes of the dS family which will become even more dominant for increasing dimensions if no new families or instabilities occur in $d>6$ dimensions.\footnote{It is natural to question whether more slowly decaying modes might appear in the dimensions considered. For this purpose, we have used calculations of very high accuracy to rule out the possibility of lost dominant modes. This means that, if more families do exist, they should be subdominant thus are irrelevant for SCC.}
\begin{table}[H]
\centering
\begin{tabular}{||c|c||}
\hline
parameter regions  & degree of difficulty of SCC violation\\
\hline
\text{\bf I.\,\,}$\Lambda/\Lambda_{\text{max}}\gtrsim 0.279$ & $d_4>d_5>d_6$ \\
\hline
\text{\bf II.\,\,}$0.179\lesssim\Lambda/\Lambda_{\text{max}}\lesssim 0.279$ & $d_4>d_6>d_5$ \\
\hline
\text{\bf III.\,\,}$0.135\lesssim\Lambda/\Lambda_{\text{max}}\lesssim 0.179$ & $d_6>d_4>d_5$\\
\hline
\text{\bf IV.\,\,\,}$\Lambda/\Lambda_{\text{max}}\lesssim 0.135$ & $d_6>d_5>d_4$ \\
\hline
\end{tabular}
\caption{Comparison of $\Lambda/\Lambda_{\text{max}}$ with respect to the degree of difficulty of SCC violation in $d=4,\,5$ and $6-$dimensional RNdS BHs.
\label{table21}}
\end{table}
To distinguish between ``small", ``intermediate" and ``large" BHs, we scan thoroughly the parameter space of $d=4,\,5$ and $6-$dimensional RNdS BHs to find critical values of $\Lambda/\Lambda_{\text{max}}$ where different violation configurations are introduced. We find 3 critical values which divide the range of $0<\Lambda/\Lambda_{\text{max}}\leq 1$ into 4 regions. In Table \ref{table21}, we summarize the division of our parameter space into the regions of interest and display the degree of difficulty of SCC violation at each region. We realize that region I corresponds to ``large" BHs, regions II and III correspond to ``intermediate" BHs and, finally, region IV corresponds to ``small" BHs. These regions can be directly seen in Fig. \ref{beta} and arise due to the existence and competition between the dominant PS and dS family, as discussed above.

In any case, we clearly see that $\beta>1/2$ above some value of the BH charge, no matter the choice of the cosmological constant. This leads to CHs which upon scalar perturbations maintain enough regularity for the scalar field (and thus the metric) to be extendible past it, resulting to a potential violation of SCC. Moreover, if it was up to the PS and dS family, $\beta$ would always diverge at extremality. However, the dominant modes of the NE family ($l=n=0$) will always take over to keep $\beta$ below 1.

\section{Conclusions and Discussions}
The study of \cite{Hintz:2015jkj} indicate that the stability of the CH in asymptotically dS spacetimes is governed by $\beta$ defined in (\ref{eq15}). Subsequently, the results of Chapter \ref{PRL} indicate a potential failure of determinism in GR when near-extremal $4-$dimensional RNdS BHs are considered. Under massless neutral scalar perturbations, the CH might seem singular, due to the blow-up of curvature components, but maintain enough regularity as to allow the field equations to be extended beyond a region where the evolution of gravitation is classically determined in a highly non-unique manner.

\begin{figure}[H]
\centering
\includegraphics[height=1.66in,width=2.26in]{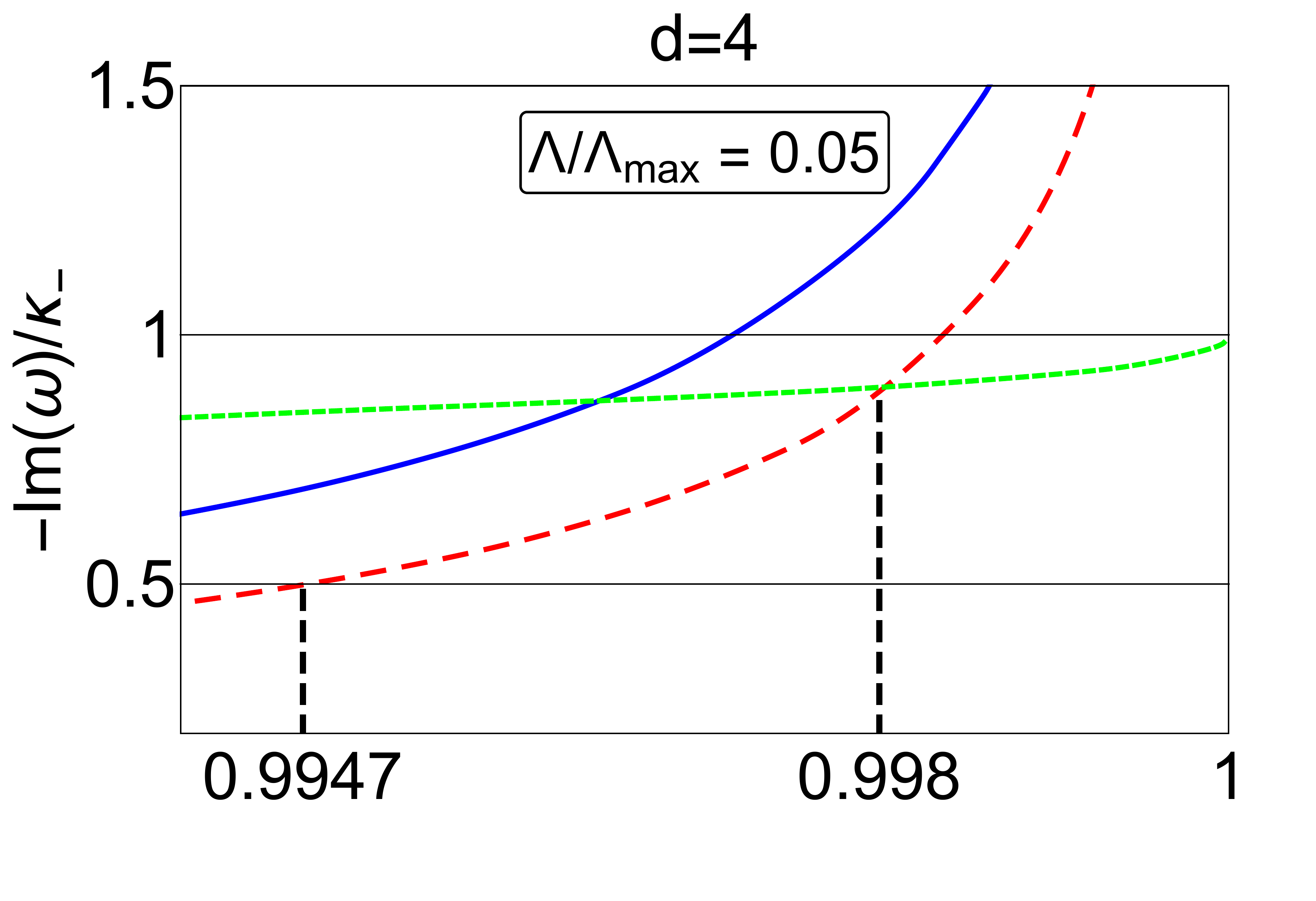}
\hskip -4ex
\includegraphics[height=1.65in,width=2.25in]{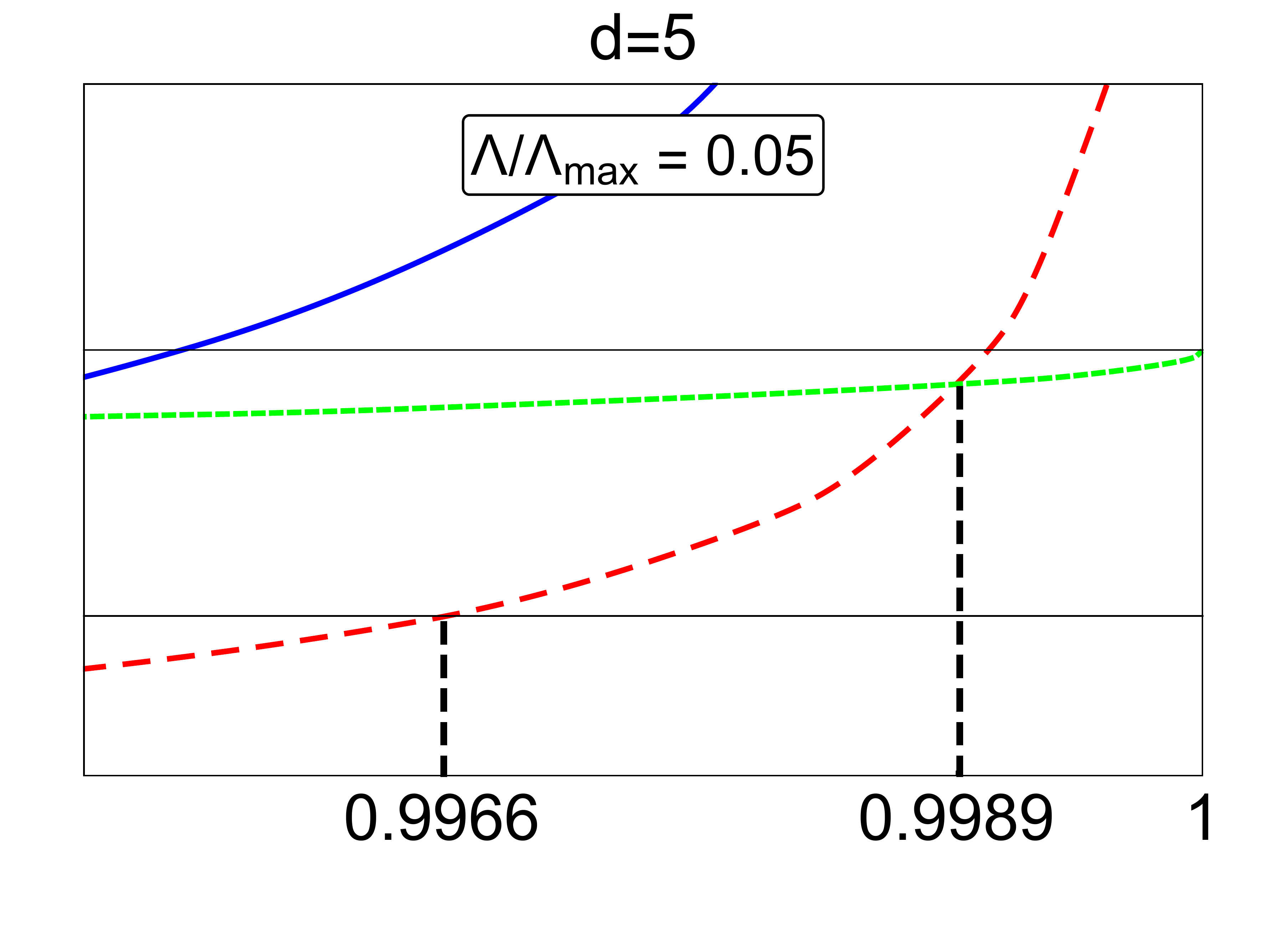}
\hskip -4ex
\includegraphics[height=1.65in,width=2.25in]{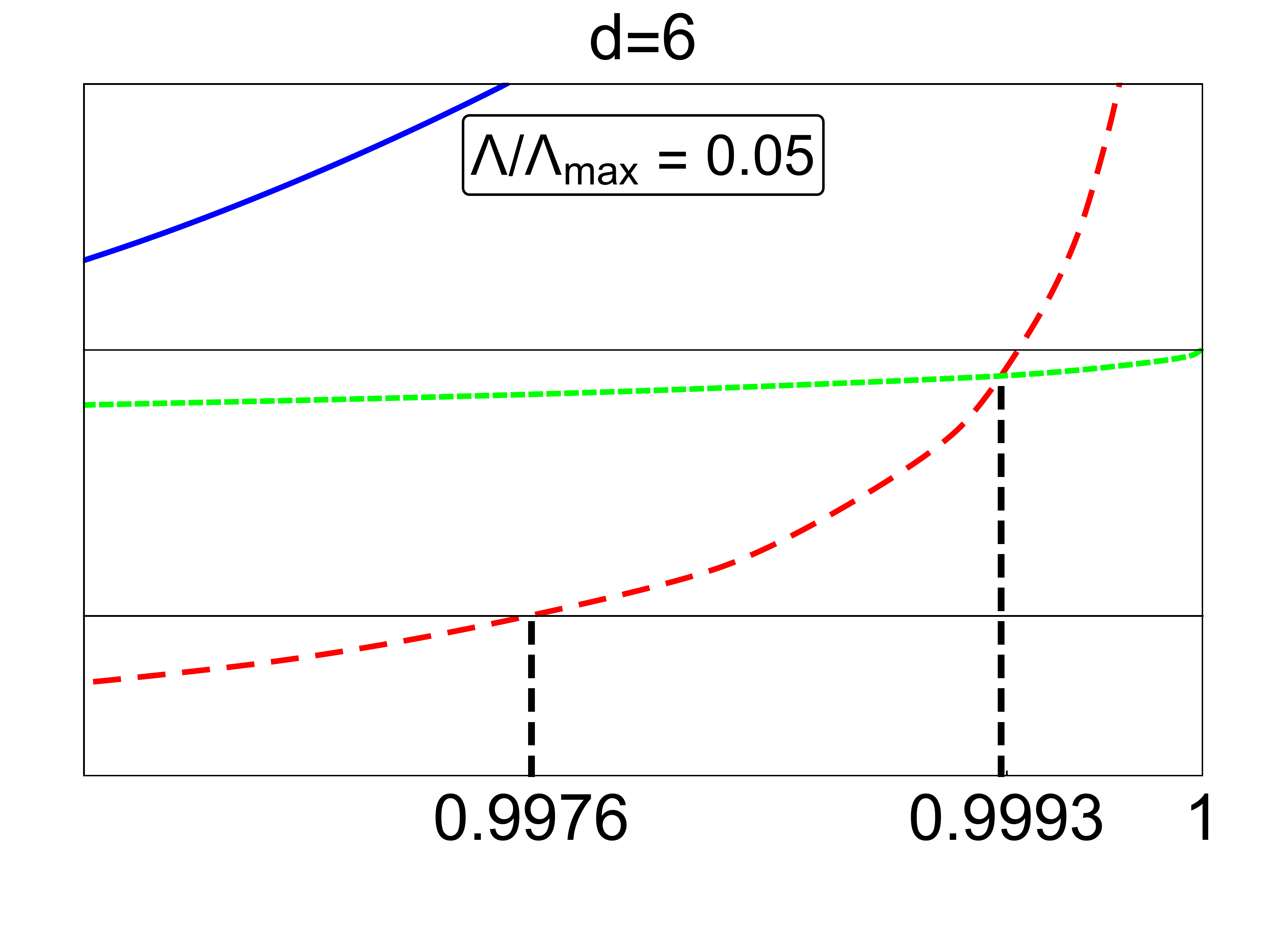}
\vskip -3.5ex
\includegraphics[height=1.66in,width=2.26in]{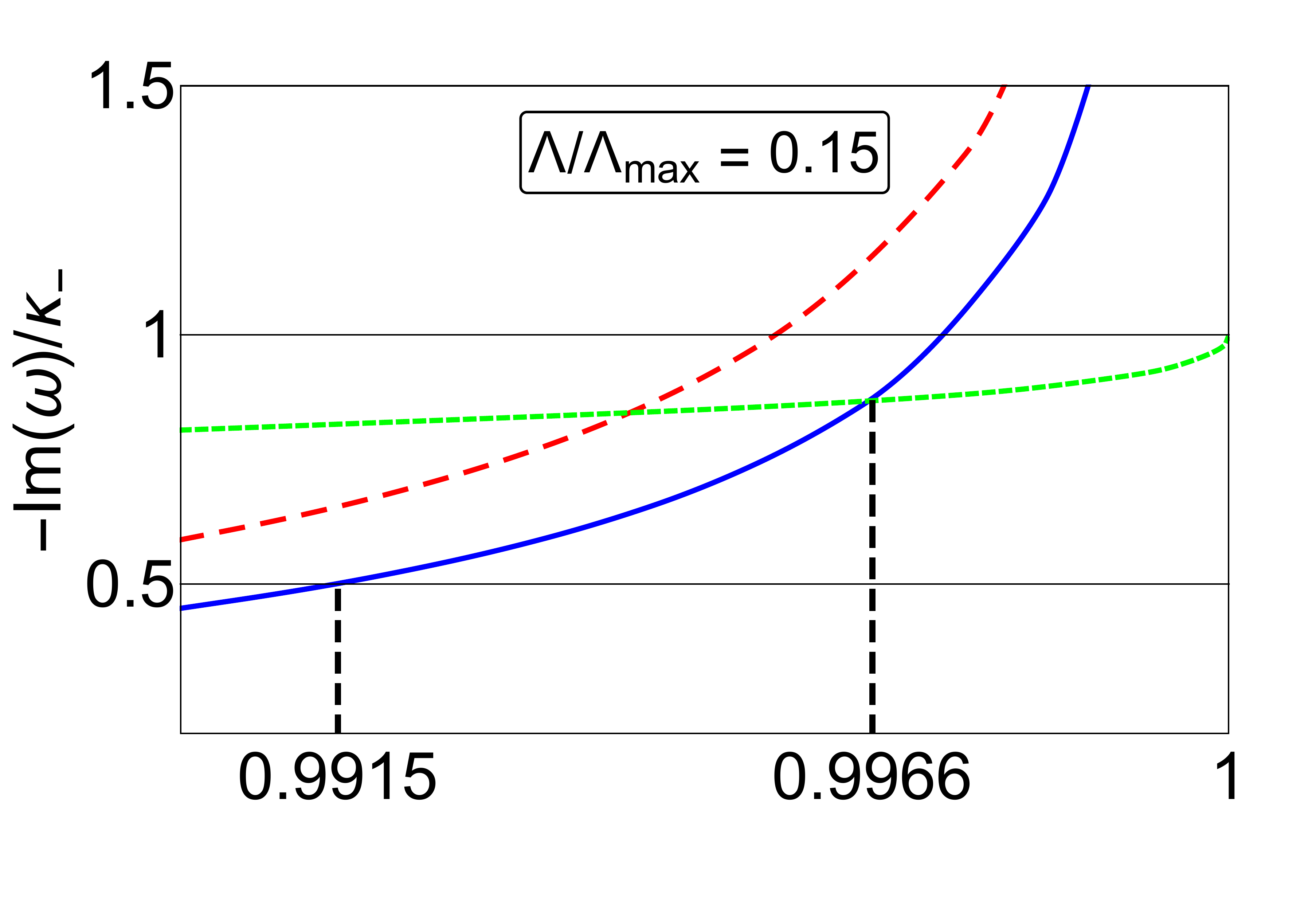}
\hskip -4ex
\includegraphics[height=1.65in,width=2.25in]{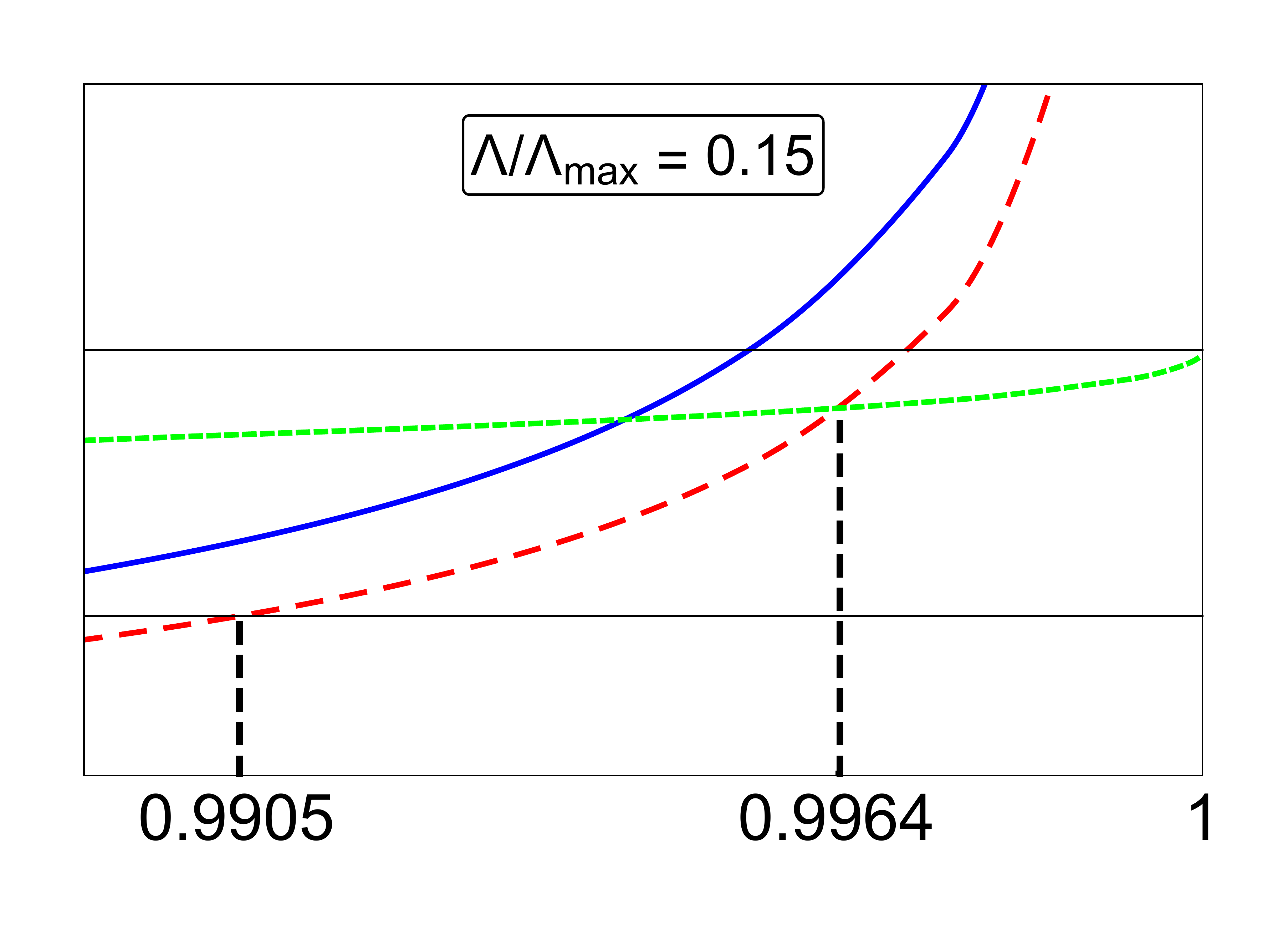}
\hskip -4ex
\includegraphics[height=1.65in,width=2.25in]{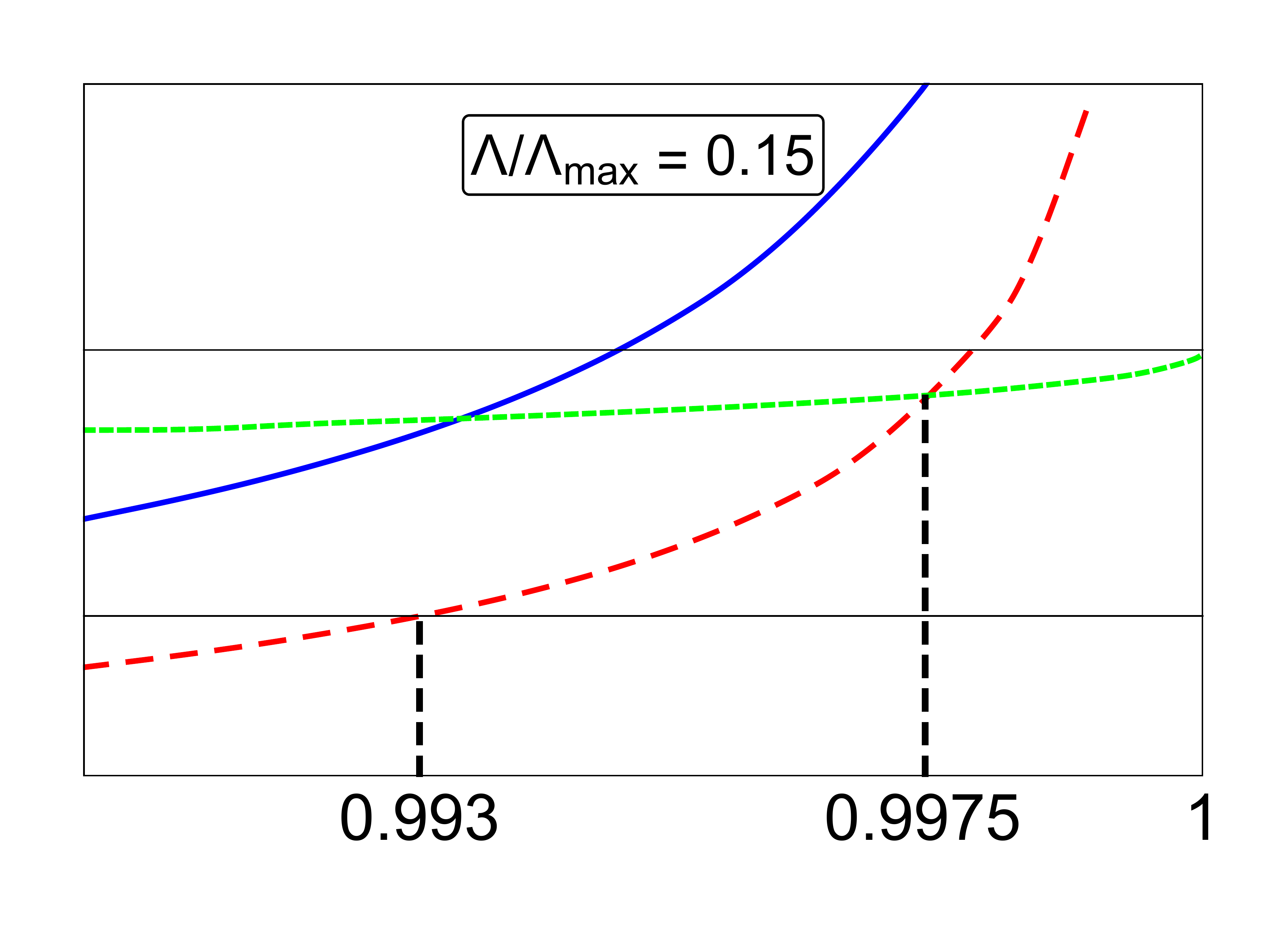}
\vskip -3.5ex
\includegraphics[height=1.66in,width=2.26in]{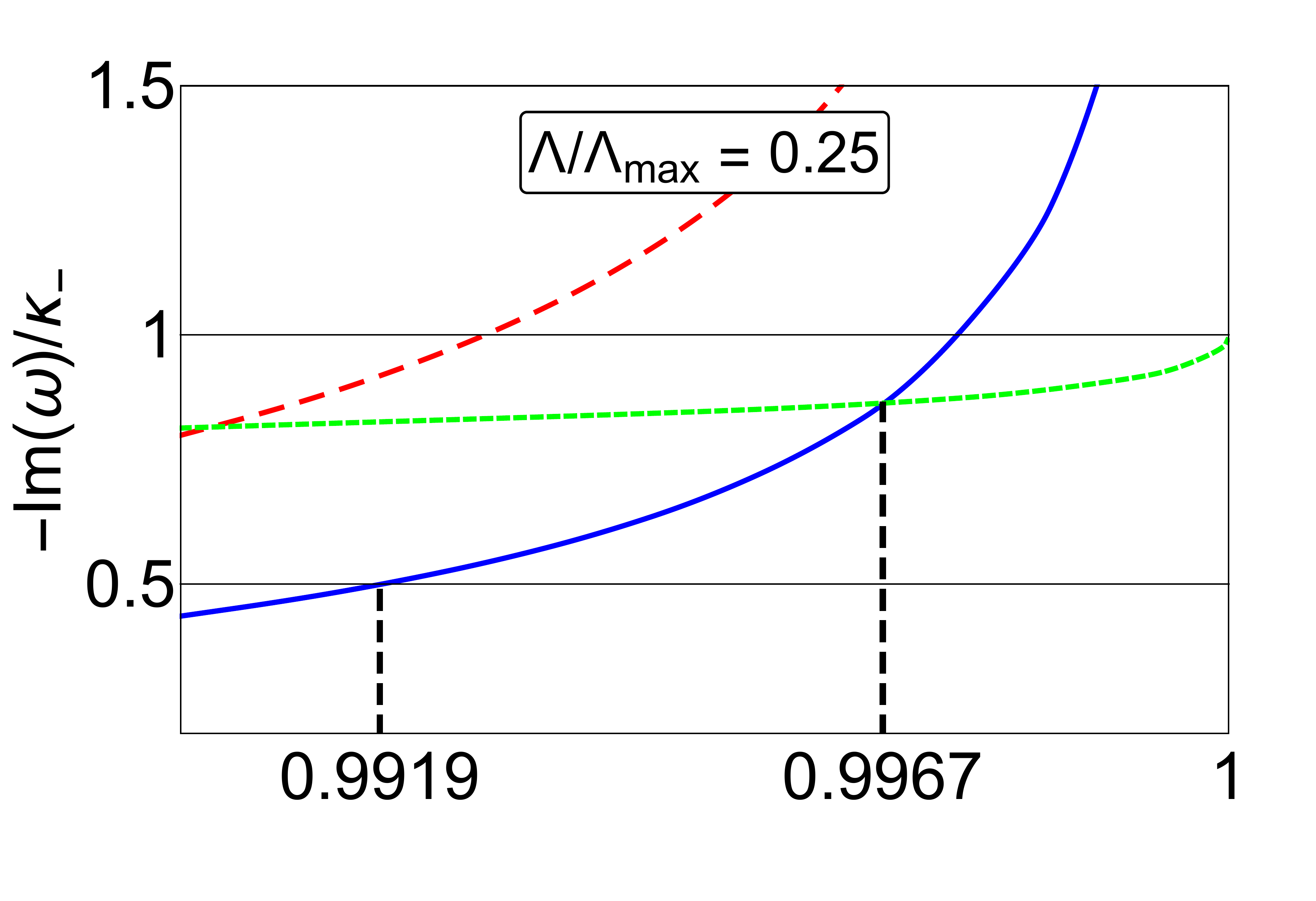}
\hskip -4ex
\includegraphics[height=1.65in,width=2.25in]{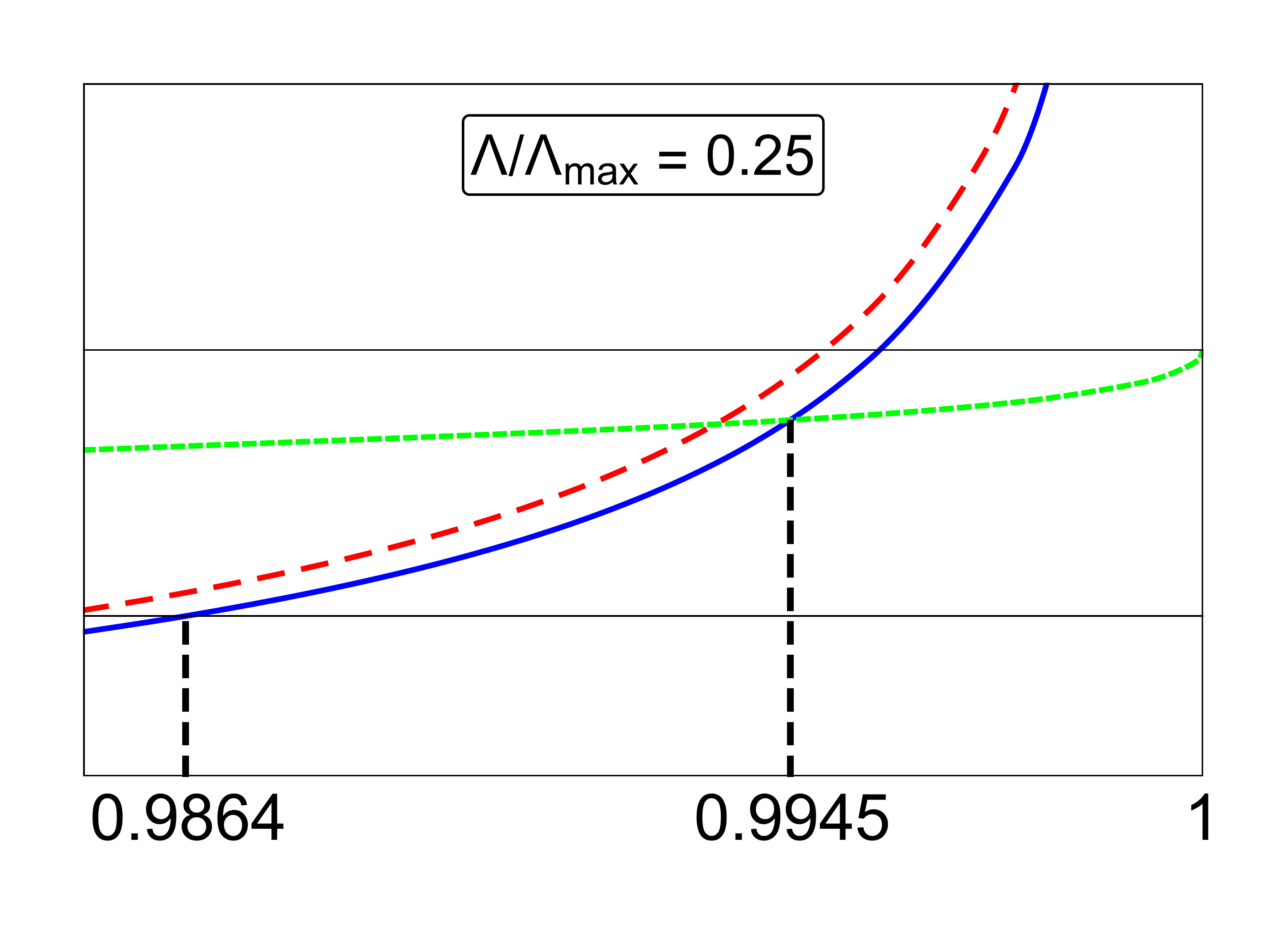}
\hskip -4ex
\includegraphics[height=1.65in,width=2.25in]{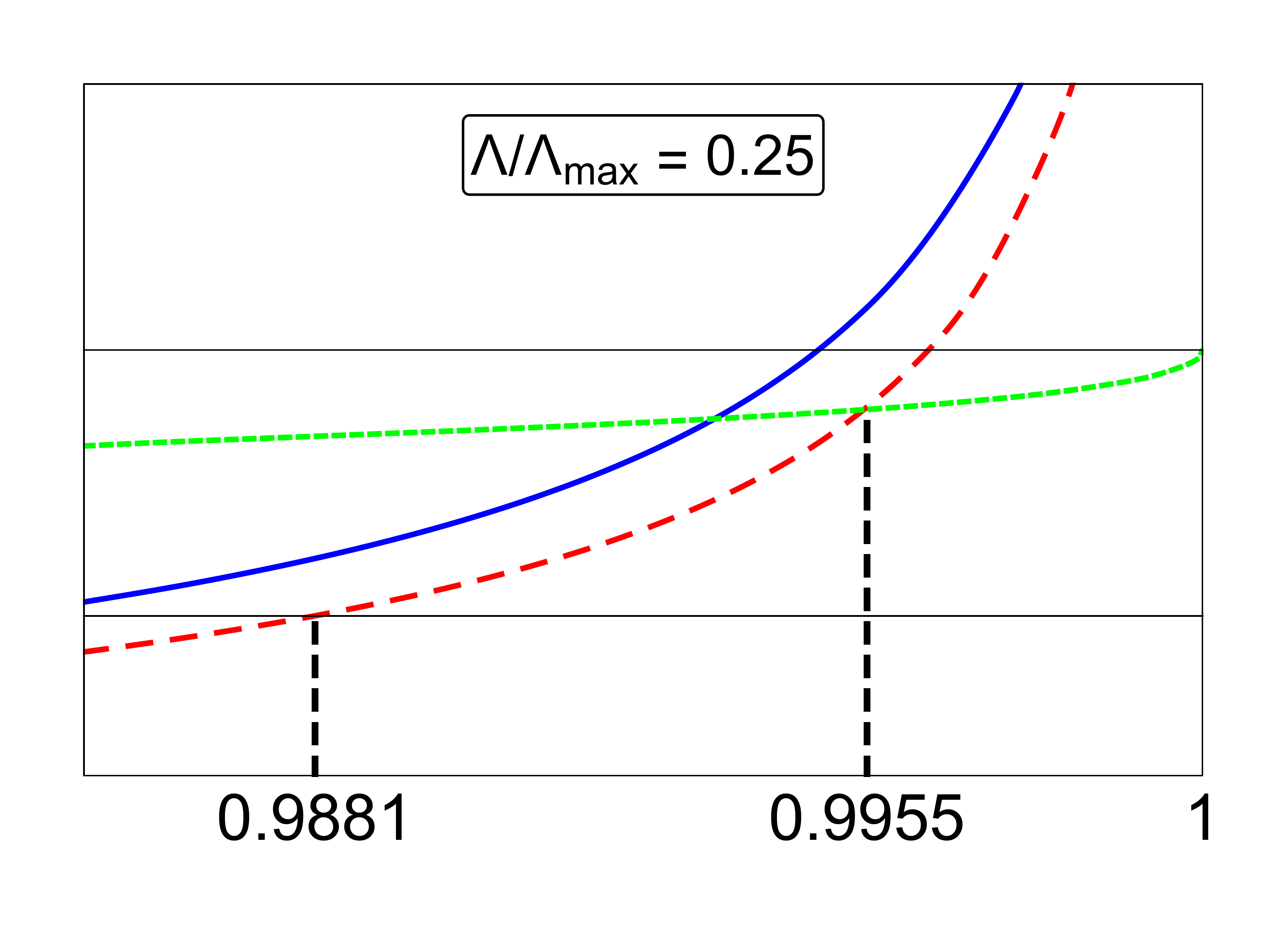}
\vskip -3.5ex
\includegraphics[height=1.66in,width=2.26in]{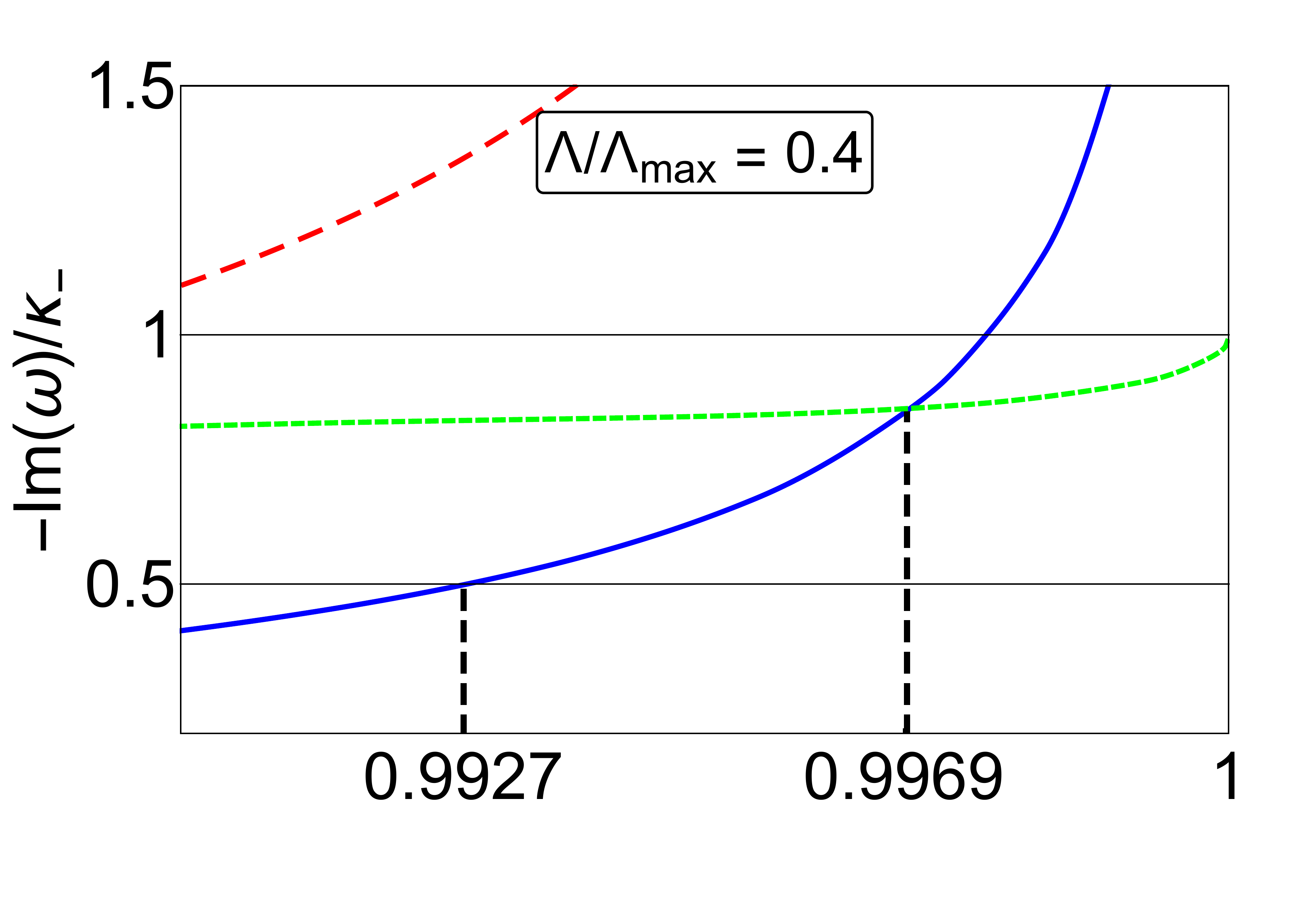}
\hskip -4ex
\includegraphics[height=1.65in,width=2.25in]{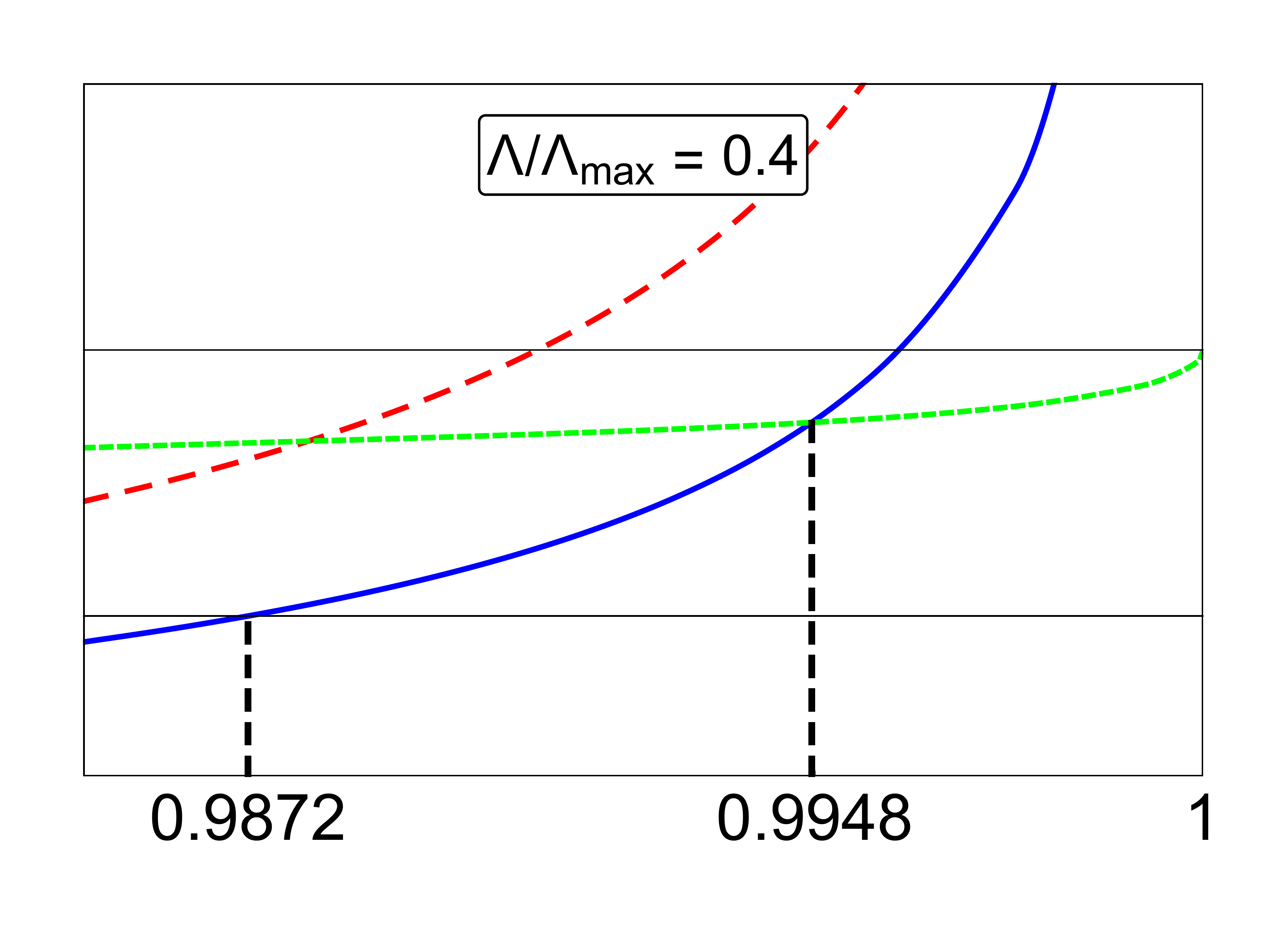}
\hskip -4ex
\includegraphics[height=1.65in,width=2.25in]{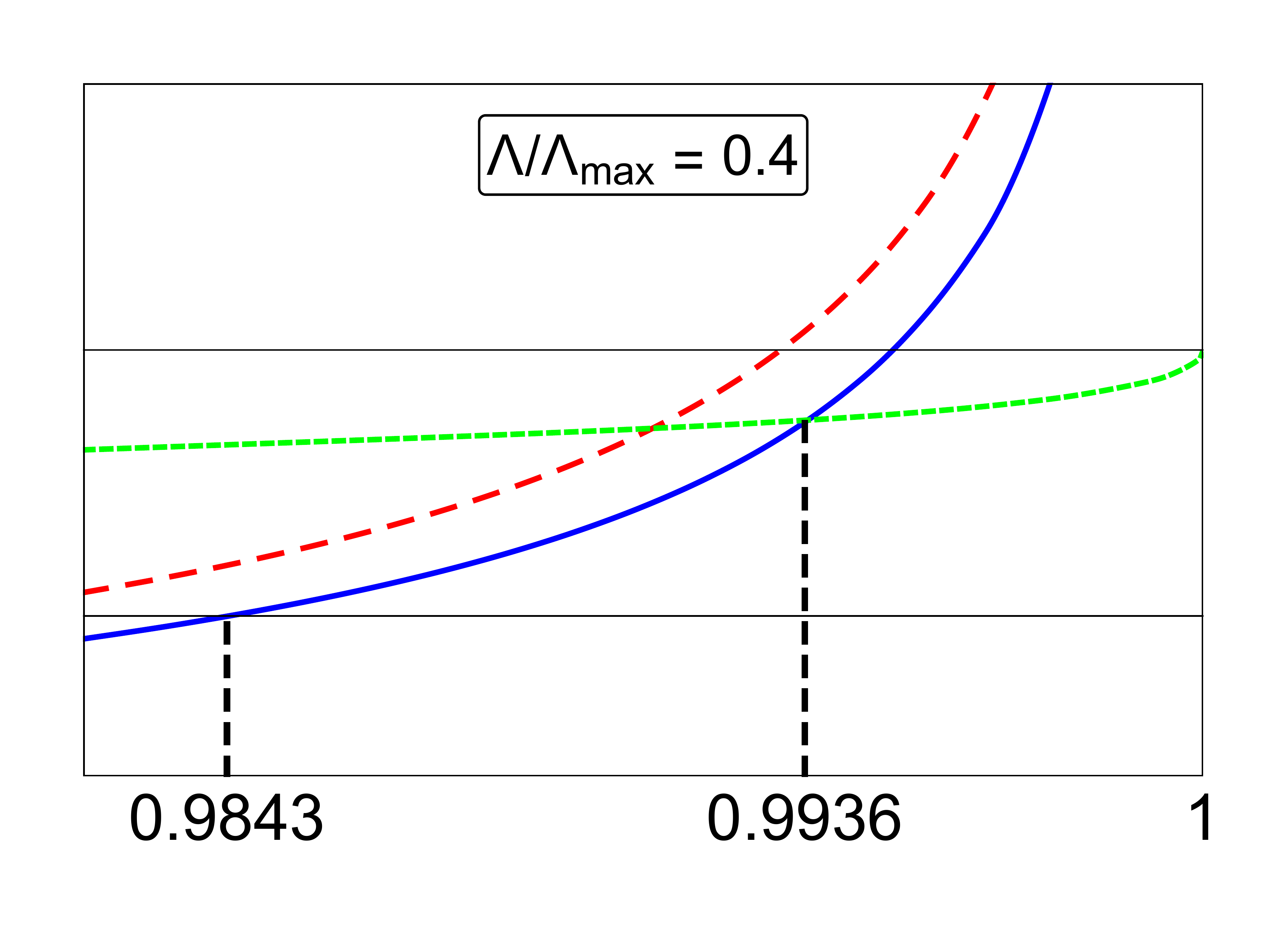}
\vskip -3.5ex
\includegraphics[height=1.66in,width=2.26in]{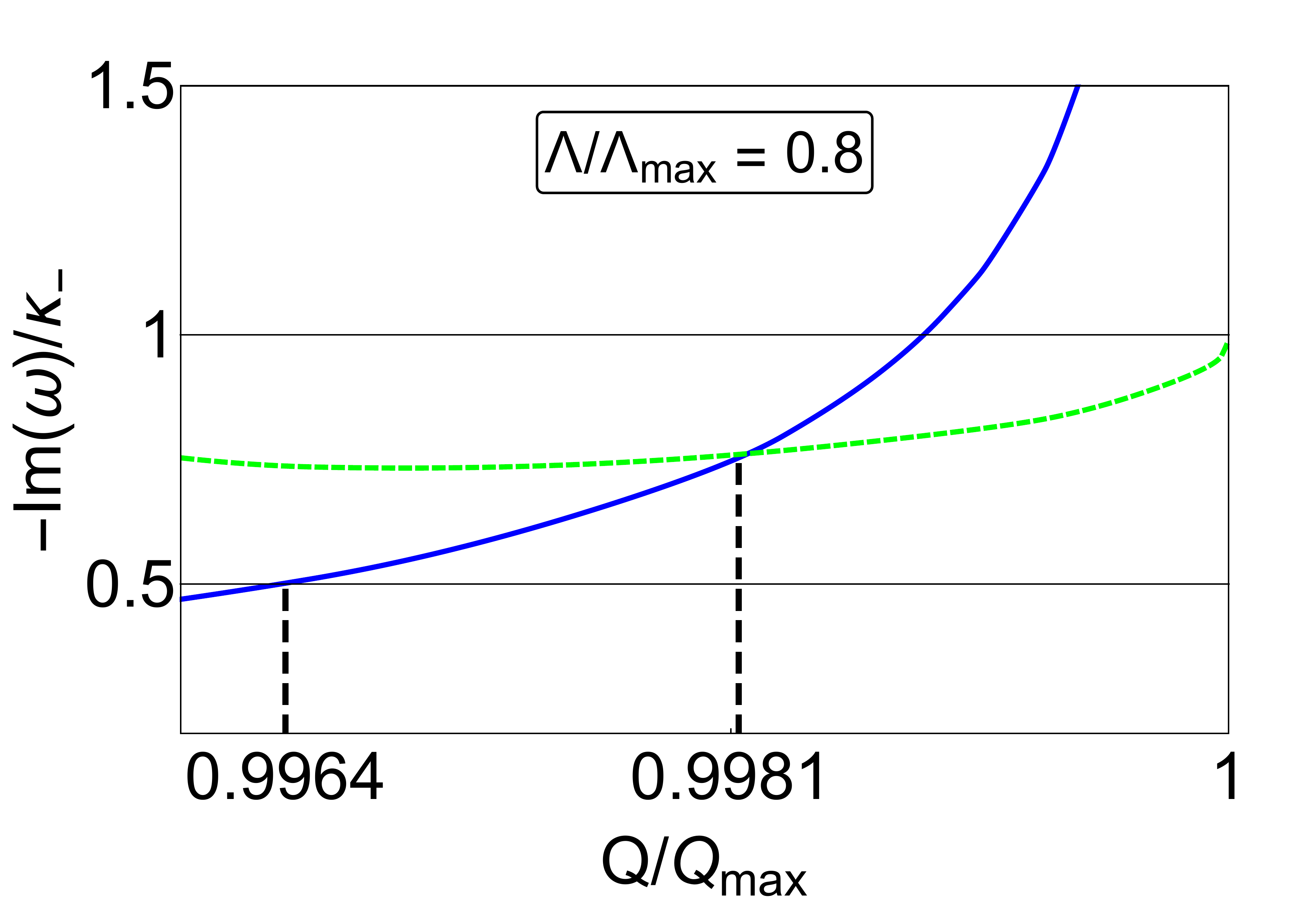}
\hskip -4ex
\includegraphics[height=1.65in,width=2.25in]{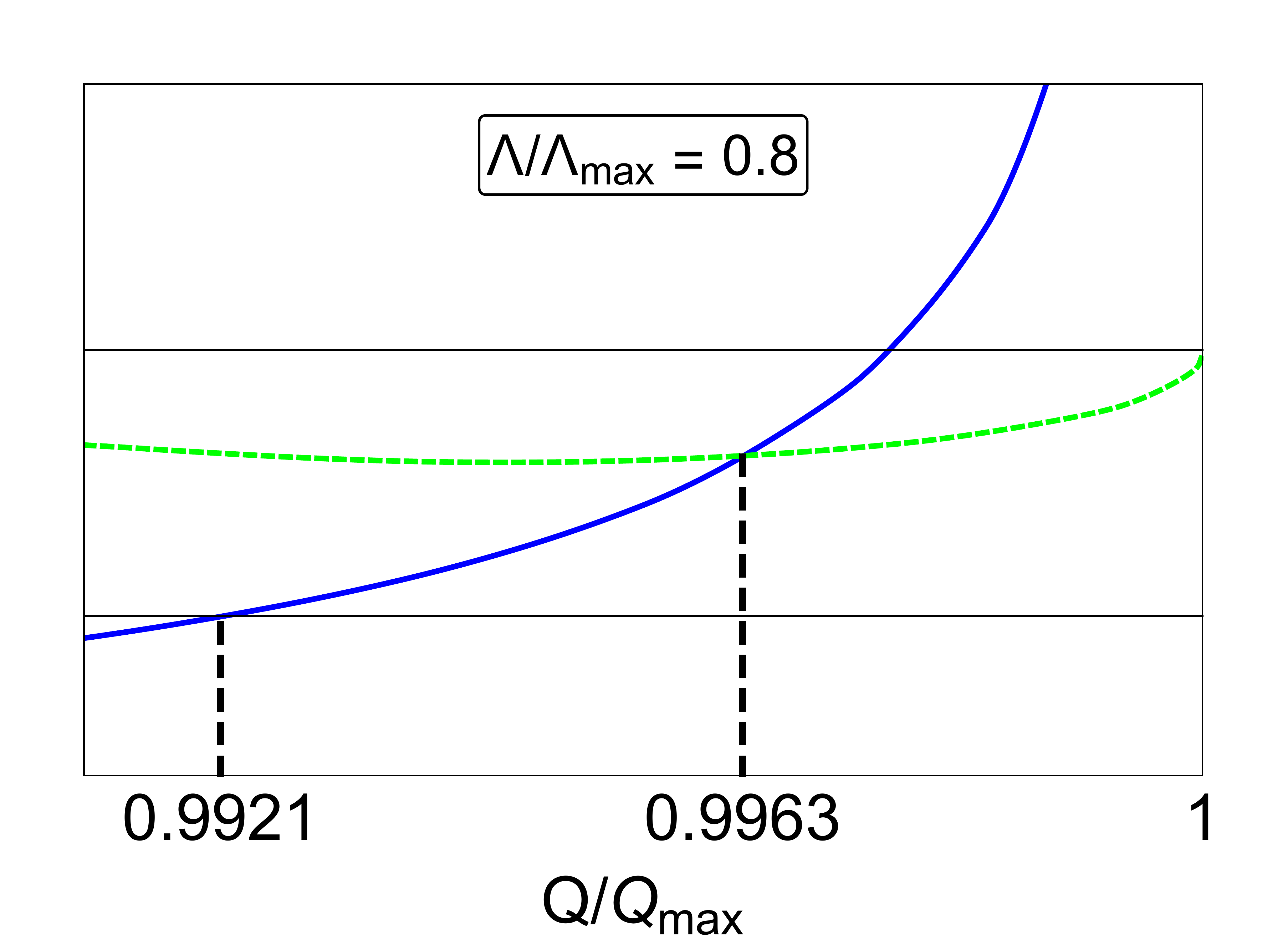}
\hskip -4ex
\includegraphics[height=1.65in,width=2.25in]{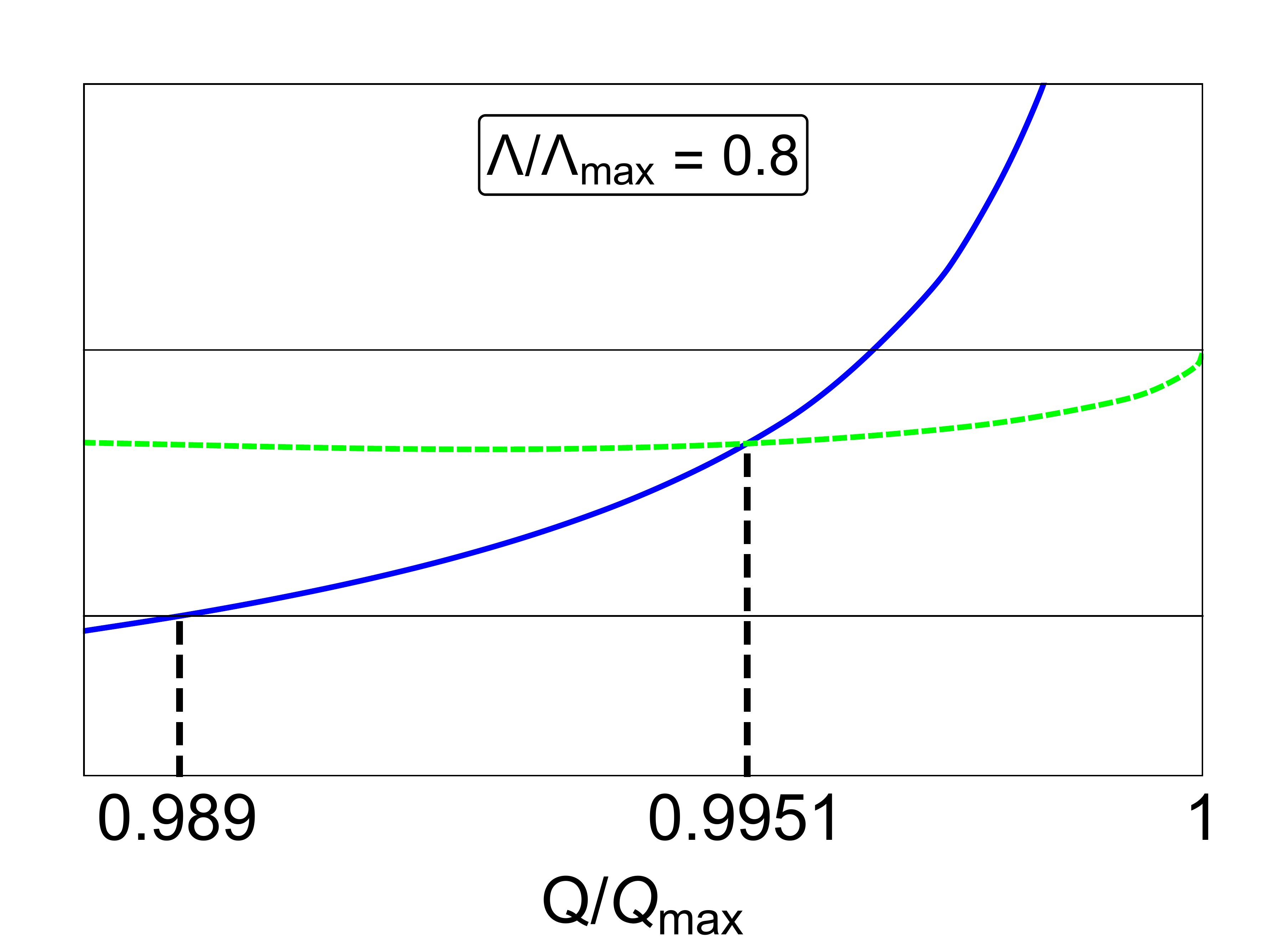}
\caption{Dominant modes of different families, showing the (nearly) dominant complex PS mode (blue, solid) at $l=10$, the dominant BH dS mode (red, dashed) at $l=1$ and the dominant NE mode (green, dashed) at $l=0$ for $d=4,\,5$ and $6$-dimensional RNdS spacetime with $M=1$. The two dashed vertical lines designate the points where $\beta=-\text{Im}(\omega)/\kappa_-=1/2$ and where the NE mode becomes dominant. The dS family in the final row of plots is too subdominant, thus lying outside of the region of interest.}
\label{beta}
\end{figure}

Here, we extend the study to higher-dimensional RNdS BHs and find that the same picture occurs when scalar fields are considered. We have proven that (\ref{eq15}) remains unchanged for $d-$dimensions. By inferring to $d=4,\,5$ and $6-$dimensional RNdS BHs we realize that the introduction of higher dimensions will fortify SCC for sufficiently ``small" BHs ($\Lambda/\Lambda_{\text{max}}\lesssim 0.135$), by the introduction of higher BH charges beyond which $\beta>1/2$. Moreover, we observe that ``intermediate" and ``large" BHs ($\Lambda/\Lambda_{\text{max}}\gtrsim 0.135$) possess a much more complex picture with some dimensions being preferred over others to fortify SCC. This perplexity arises due to the delicate competition of the PS and dS family of modes. Even though for ``large" BHs we see that the preferred dimension to fortify SCC, with higher $Q/Q_\text{max}$ beyond which $\beta>1/2$, is $d=4$, we understand that the introduction of even higher than 6 dimensions will eventually change the picture due to the behavior of the dS family demonstrated in Fig. \ref{beta}, if no instabilities occur in our region of interest \cite{Konoplya:2008au}.

In any case, we can always find a region in the parameter space of the higher-dimensional RNdS BHs in study for which $\beta$ exceeds $1/2$, but still not exceeding unity\footnote{$\beta>1$ would correspond to extensions of the scalar field in $C^1$ at the CH, thus the coupling to gravity should lead to the existence of solutions with bounded curvature.}. 
This still leaves as with CHs which upon perturbations might seem singular, due to the blow-up of curvature components, but that doesn't imply the breakdown of Einstein's field equations \cite{Klainerman:2012wt} nor the destruction of macroscopic observers \cite{Ori:1991zz} at the CH.

It is important to mention that SCC in higher-dimensional RNdS spacetime was also discussed in \cite{Rahman:2018oso}, with a wishful premise that the large $l$ mode always dominates, i.e., the value of $\beta$ always decreases monotonously with the increase of angular number $l$. However, this is not the case, as we have seen in Fig. \ref{beta}, due to the existence of three different families of modes. In fact, the existence of more families highly affects $\beta$ according to the choice of our cosmological constant. This led to an updated version of \cite{Rahman:2018oso} which was published recently. In the new version, their improved results are in agreement with ours.

\section{WKB approximation of the dominant photon-sphere modes}\label{app_WKB}
The WKB method can provide accurate approximation of QNMs in the eikonal limit. The QNMs of BHs in the eikonal limit under massless scalar perturbations are related to the Lyapunov exponent $\lambda$ of the null unstable geodesic, which is inversely-proportional to the instability timescale associated with the geodesic motion of null particles near the photon sphere. For $d-$dimensions we have \cite{Cardoso:2008bp}
\begin{equation}
\begin{split}
\omega_\text{WKB}&=l\sqrt{\frac{f(r_s)}{r_s^2}}-i\left(n+\frac{1}{2}\right)\sqrt{-\frac{1}{2}\frac{r_s^2}{f(r_s)}\left(\frac{d^2}{dr_\ast^2}\frac{f(r)}{r^2}\right)_{r_s}}\\
&=\Omega_c l-i\left(n+\frac{1}{2}\right)\left|\lambda\right|,
\end{split}
\end{equation}
where $r_s$ is the radius of the null circular geodesic, and $\Omega_c$ the coordinate angular velocity of the geodesic. By focusing on the modes with overtone number $n=0$, we have $\beta=\left|\lambda\right|/{2\kappa_-}$ for the dominant PS modes at the eikonal limit.

In Table \ref{table11}, we compare the value of $\beta$ obtained by AIM and the spectral method \cite{Jansen:2017oag} at $l=10$ and the value evaluated by the WKB method at large $l$ for the same BH parameters. We observe that the difference between $\beta_\text{WKB}$, $\beta_\text{spectral}$ and $\beta_\text{AIM}$ is very small, meaning that the choice of $l=10$ in our numerics can be regarded as a good approximation of $\beta_\text{WKB}$ of the dominant PS modes at the large $l$ limit.

\begin{table}[H]
\centering
\begin{tabular}{||c|c|c|c||}
\hline
$\beta_\text{PS}$& $d=4$& $d=5$&$d=6$\\
\hline
$\beta_\text{WKB}\,\,\,\,\,\,(l\rightarrow\infty)$&0.328192 & 0.687518 & 0.775677\\
\hline
$\beta_\text{AIM}\,\,\,\,\,\,\,\,(l=10)$&0.328304& 0.689089 & 0.778164\\
\hline
$\beta_\text{spectral} \,(l=10)$& 0.328304&0.689089&0.778164\\
\hline
\end{tabular}
\caption{Comparison of $\beta_\text{PS}$ obtained with WKB, AIM and a spectral method for a RNdS BH with $M=1/(3\sqrt{2})$, $\Lambda=3$ and $Q/Q_\text{max}=0.992$.}\label{table11}
\end{table}

\section{Approximation of the dominant near-extremal modes}\label{app_NE}
In \cite{Hod:2017gvn} it has been proven that long-lived modes (or quasi-bound states) can be supported by a $4-$dimensional NE RN BH. In Chapter \ref{PRL} it was realized that this family of modes exist in NE RNdS BHs and is weakly dependent on the choice of $\Lambda$. Particularly, for neutral massless scalar fields, these modes can be very well approximated as
\begin{equation}
\omega_\text{d=4,NE}=-i(l+n+1)\kappa_-=-i(l+n+1)\kappa_+
\end{equation}
when $r_-=r_+$. Motivated by this result, we realize that for any dimension the dominant NE modes should be approximated by Eq. (\ref{NE}). For the sake of proving the validity of our approximate results, in Table \ref{table3} we show various dominant NE modes extracted from our spectral code versus the approximate Eq. (\ref{NE}). Higher overtones and angular numbers are not approximated by (\ref{NE}) anymore, but in any case, they are subdominant, thus they do not play any role for SCC.

\begin{table}[H]
\centering
\begin{tabular}{||c|c|c|c||}
\hline
$\beta_\text{NE}$& $d=4$& $d=5$& $d=6$\\
\hline
$\beta_\text{approx}$&1 & 1 & 1\\
\hline
$\beta_\text{spectral}\,\,(l=n=0)$& 0.996&0.997&0.999\\
\hline
\end{tabular}
\caption{Comparison of $\beta_\text{approx}$ derived from (\ref{NE}) versus $\beta_\text{spectral}$ obtained with a spectral method for a $d$-dimensional RNdS BH with $M=1$, $\Lambda=0.1$ and $Q/Q_\text{max}=0.999999$.}\label{table3}
\end{table}

\part{Superradiant instabilities in charged black-hole spacetimes}

\chapter{Instability of higher-dimensional de Sitter black holes}\label{higher instability}
BHs possess trapping regions which lead to intriguing dynamical effects. By properly scattering test fields off a BH, one can extract energy, leading to a dynamical instability called superradiance. Here, we study the superradiance effect of an electrically-charged BH immersed in a $d-$dimensional dS Universe under charged scalar fluctuations. By performing a thorough spectral analysis of charged scalar perturbations on $d-$dimensional RNdS BHs we compute the unstable quasinormal resonances and link their nature with a novel family of QNMs associated with the existence and timescale of the cosmological horizon of pure dS space. Our results indicate that the instability has a superradiant nature, is enhanced in higher dimensions and occurs for a larger region of the parameter space, for both massless and massive, charged scalar perturbations. This chapter is based on \cite{Destounis:2019hca}.
\section{Introduction}
The study of linear perturbations has a long  history in GR.  A perturbative analysis of BH spacetimes was pioneered by Regge and Wheeler \cite{Regge}, and has proven crucial  in  several  contexts, ranging  from  astro-physics to high-energy physics \cite{Barack:2018yly,Kokkotas:1999bd,Berti:2009kk,Pani:2013pma}.  The stability analysis of BH spacetimes, an understanding of ringdown signals in the post-merger phase of a binary coalescence and their use in tests of GR, or even the analysis of fundamental light fields in the vicinities of BHs are some noteworthy examples where BH perturbation theory plays an important role \cite{Kokkotas:1999bd,Berti:2009kk,Pani:2013pma}.

Perturbing a BH with small fluctuations could lead to two possible outcomes; the BH is stable under perturbations, due to damping mechanisms that act on the BH exterior, and will relax after the initial disruption or the BH is unstable under perturbations and will inevitably disappear or evolve to another stable object. Although astrophysical BHs are expected to be stable under small fluctuations, a lot of concern has been given to BH solutions that might be prone to instabilities due to new phenomena that might be possibly unveiled. Quite strikingly, one can extract energy from BHs through scattering techniques \cite{Penrose:1971uk,Bekenstein:1973mi} by properly probing them with test fields, and under certain circumstances the test fields can grow in time. This effect is called superradiance \cite{Brito:2015oca} and was explored in the context of rotating and charged BHs \cite{Detweiler,Vitor1,Vitor2,Vitor3,Vitor4,Vitor5,Kerr1,Kerr2,Kerr3,Kerr4,Kerr5,RN1,RN2,RN3,RN4,RN5,RN6,KN1,KN2}, stars \cite{Vicente:2018mxl,stars1,stars2} and other compact objects \cite{Maggio:2018ivz}.

Perturbation theory has revealed that BHs vibrate in a well described manner, exhibiting a discrete spectrum of preferable oscillatory modes, called QNMs \cite{Chandrasekhar:1975zza,PhysRevLett.52.1361,Kokkotas:1999bd,Berti:2009kk,Konoplya:2011qq}. Linear perturbation theory has been an active field of study for many decades which uncovered various methods of testing the modal stability of compact objects, at the linearized level, both analytically and numerically. Various studies have brought to light spacetimes which upon perturbations become unstable(for an incomplete list see \cite{Furuhashi:2004jk,Konoplya:2008au,Dolan:2012yt,Destounis:2018utr,Zhu:2014sya,Konoplya:2014lha,Dias:2010eu,Cardoso:2010rz,Konoplya:2013sba,Herdeiro:2013pia,Degollado:2013bha,Sanchis-Gual:2015lje,Li:2012nd,Li:2012rx,Li:2014fna,Li:2014gfg,Li:2014xxa,Li:2015mqa}).

An interesting study \cite{Konoplya:2008au} suggests that higher-dimensional RNdS spacetimes are prone to instabilities under gravitational perturbations. Such a gravitational instability has been further examined in \cite{Cardoso:2010rz,Konoplya:2013sba,Tanabe:2015isb}. Specifically, $d-$dimensional RNdS BHs with $d>6$ and large enough mass and charge, are unstable under gravitational perturbations. Why only $d=4,5$ and $6-$dimensional RNdS BHs are favorable as to the BH stability, is still unknown. 

More recently, a new instability was found in $4-$dimensional RNdS BHs \cite{Zhu:2014sya,Konoplya:2014lha}. The $l=0$ charged scalar perturbation was proven to be unstable for various regions of the parameter space of RNdS BHs. The addition of an arbitrarily small amount of mass acts as a stabilization factor, as well as the increment of the scalar field charge beyond a critical value. Furthermore, the modes were proven to be superradiant. 

In this chapter, we investigate such an instability in higher-dimensions by employing a frequency-domain analysis. We analyze both massless and massive charged scalar perturbations in subextremal $d-$dimensional RNdS spacetime and show that the instability still persists. For simplicity, we narrow down our study in $d=4,5$ and $6$ dimensions. 

Intriguingly, we will show that the instability originates from a new family of QNMs that arise only in asymptotically dS BHs and can be very well approximated by the scalar QNMs of pure $d-$dimensional dS space. This novel family was very recently identified in RNdS for both scalar (see Chapter \ref{PRL}) and fermionic perturbations (see Chapter \ref{PLB}). Finally, we will demonstrate that as the spacetime dimensions increase, the instability is amplified, occurs for a larger region of the subextremal parameter space and still satisfies the superradiant condition. 

\section{Charged scalar fields in higher-dimensional RNdS}
The purpose of our work is to explore massless and massive charged scalar fields in the full range of their charge and mass on $d-$dimensional RNdS backgrounds, specifically the modes that are prone to instabilities. The $d-$dimensional RNdS spacetime is described by the metric 
\begin{equation}
ds^2=-f(r)dt^2+\frac{1}{f(r)}dr^2+r^2d\Omega^2_{d-2},
\end{equation}
with 
\begin{equation*}
d\Omega^2_{d-2}=d\chi_2^2+\sum_{i=2}^{d-2}\left(\prod_{j=2}^{i}\sin^2\chi_j\right)\, d\chi_{i+1}^2.
\end{equation*}
The metric function reads
\begin{align}
 f(r)&=1-\frac{m}{r^{d-3}}+\frac{q_0}{r^{2(d-3)}}-\frac{2\Lambda r^2}{(d-2)(d-1)},
\end{align}
where $\Lambda$ is the cosmological constant and $m$, $q_0$ are functions related to the ADM mass $M$ and electric charge $Q$ of the BH, respectively,
\begin{equation}
M=\frac{d-2}{16\pi}w_{d-2}m,\,\,Q=\frac{\sqrt{2(d-2)(d-3)}}{8\pi}w_{d-2}q_0,
\end{equation}
with $w_d=2\pi^{\frac{d+1}{2}}/\Gamma(\frac{d+1}{2}),$ the volume of the unit $d-$sphere.
The causal structure of a subextremal higher-dimensional RNdS BH possesses three distinct horizons, namely the Cauchy $r=r_-$, event $r=r_+$ and cosmological horizon $r=r_c$, where $r_-<r_+<r_c$. The associated electromagnetic potential sourced by such charged spacetime is
\begin{equation}
A=-\sqrt{\frac{d-2}{2(d-3)}}\frac{q_0}{r^{d-3}}dt.
\end{equation}
The propagation of a massive charged scalar field on a fixed $d-$dimensional RNdS background is governed by the Klein-Gordon equation 
\begin{equation}
(D^\nu D_\nu-\mu^2)\Psi=0, 
\end{equation}
where $D_\nu=\nabla_\nu-iqA_\nu$ is the covariant derivative and $\mu$, $q$ are the mass and charge of the field, respectively. By expanding $\Psi$ in terms of spherical harmonics with harmonic time dependence,
\begin{equation}
\Psi=\sum_{lm}\frac{\psi_{l m}(r)}{r^{\frac{d-2}{2}}}Y_{lm}(\chi)e^{-i\omega t},
\end{equation}
and dropping the subscripts on the radial functions, we obtain the master equation
\begin{equation}
\label{master_eq_RNdS1}
\frac{d^2 \psi}{d r_*^2}+\left[\omega^2-2\omega\Phi(r)-V(r)\right]\psi=0\,,
\end{equation}
where 
\begin{equation}
\Phi(r)=\frac{q_0 q}{\sqrt{\frac{2(d-3)}{d-2}}r^{d-3}},
\end{equation}
is the electrostatic potential, $dr_*=dr/f(r)$ the tortoise coordinate and 
\begin{align}
\label{RNdS_general potential1}
V(r)=f(r)\left(\mu^2+\frac{l(l+d-3)}{r^2}+\frac{(d-2)f^\prime(r)}{2r}+f(r)\frac{(d-4)(d-2)}{4r^2}\right)-\Phi(r)^2,
\end{align}
where $l$ is an angular number, corresponding to the eigenvalue of the spherical harmonics, and prime denotes the derivative with respect to the radial coordinate $r$. We are interested in the characteristic QNMs $\omega$ of such spacetime, obtained by imposing the boundary conditions~\cite{Berti:2009kk}
\begin{equation}
\label{bcs}
\psi \sim
\left\{
\begin{array}{lcl}
e^{-i (\omega-\Phi(r_+))r_* },\,\,\,\quad r \rightarrow r_+, \\
&
&
\\
 e^{+i(\omega-\Phi(r_c))r_*},\,\,\,\,\quad r \rightarrow r_c.
\end{array}
\right.
\end{equation}
The QN frequencies are characterized, for each $l$, by an integer $n\geq 0$ labeling the mode number. The fundamental mode $n=0$ corresponds, by definition, to the non-vanishing frequency with the smallest (by absolute value) imaginary part. 

The results shown in the following sections were obtained mostly with the Mathematica package of \cite{Jansen:2017oag} (based on methods developed in \cite{Dias:2010eu}), and checked in various cases with a WKB approximation \cite{Iyer:1986np} and a spectral code that was developed based on a non-grid based interpolation scheme \cite{KaiLin1}.
\begin{figure}[H]
\subfigure{\includegraphics[scale=0.166]{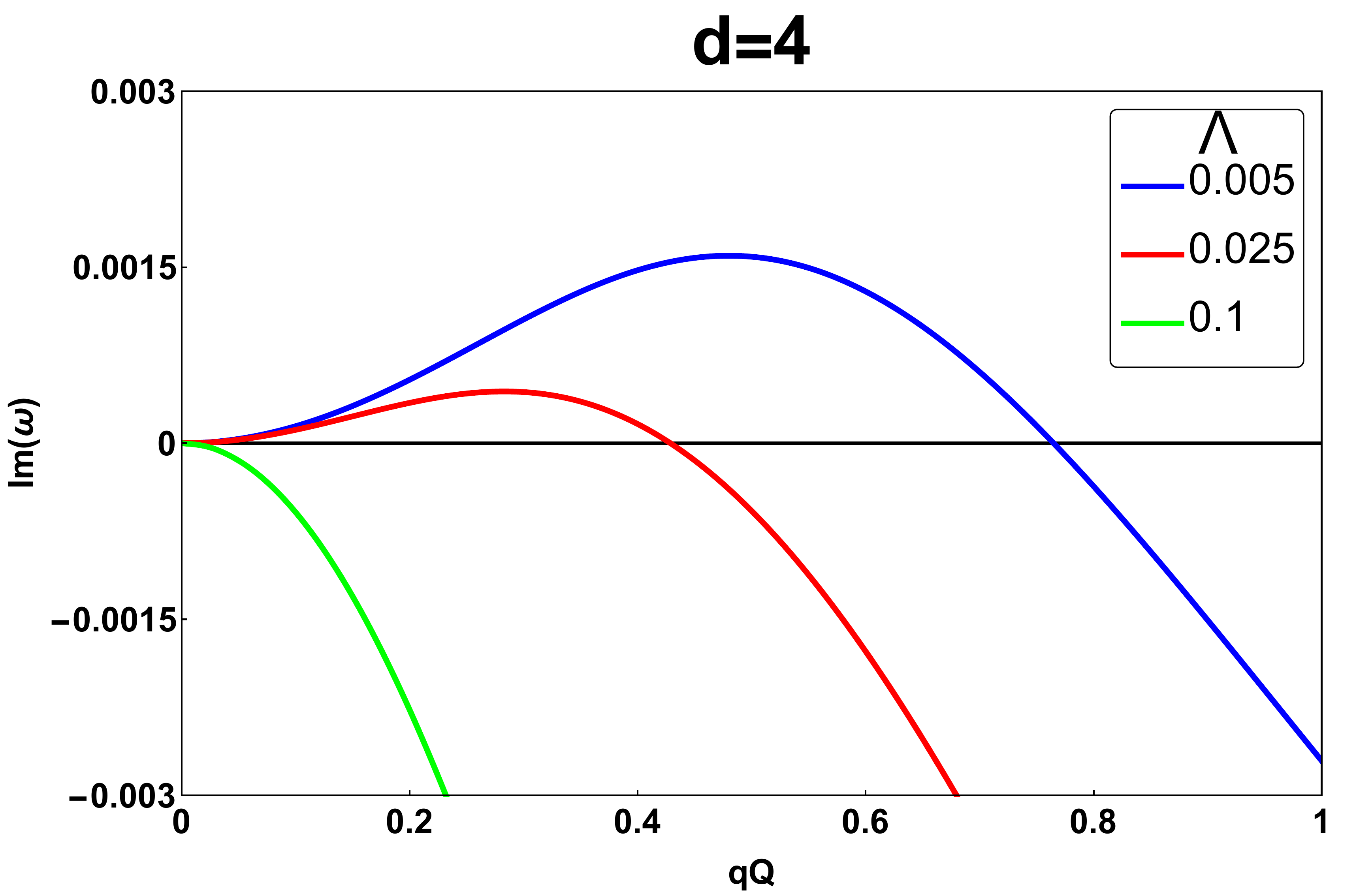}}
\hskip -2.5ex
\subfigure{\includegraphics[scale=0.15]{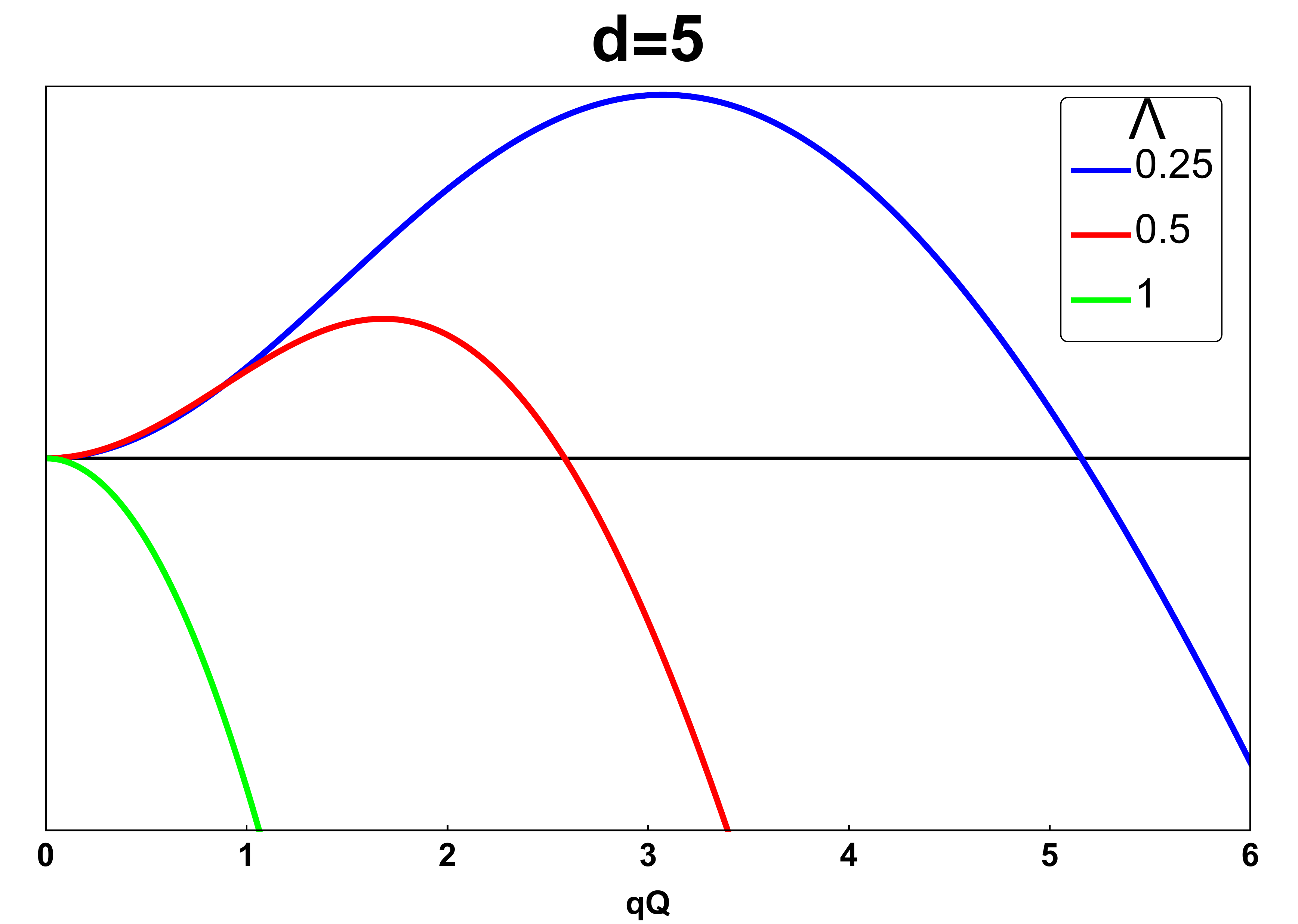}}
\hskip -2.5ex
\subfigure{\includegraphics[scale=0.15]{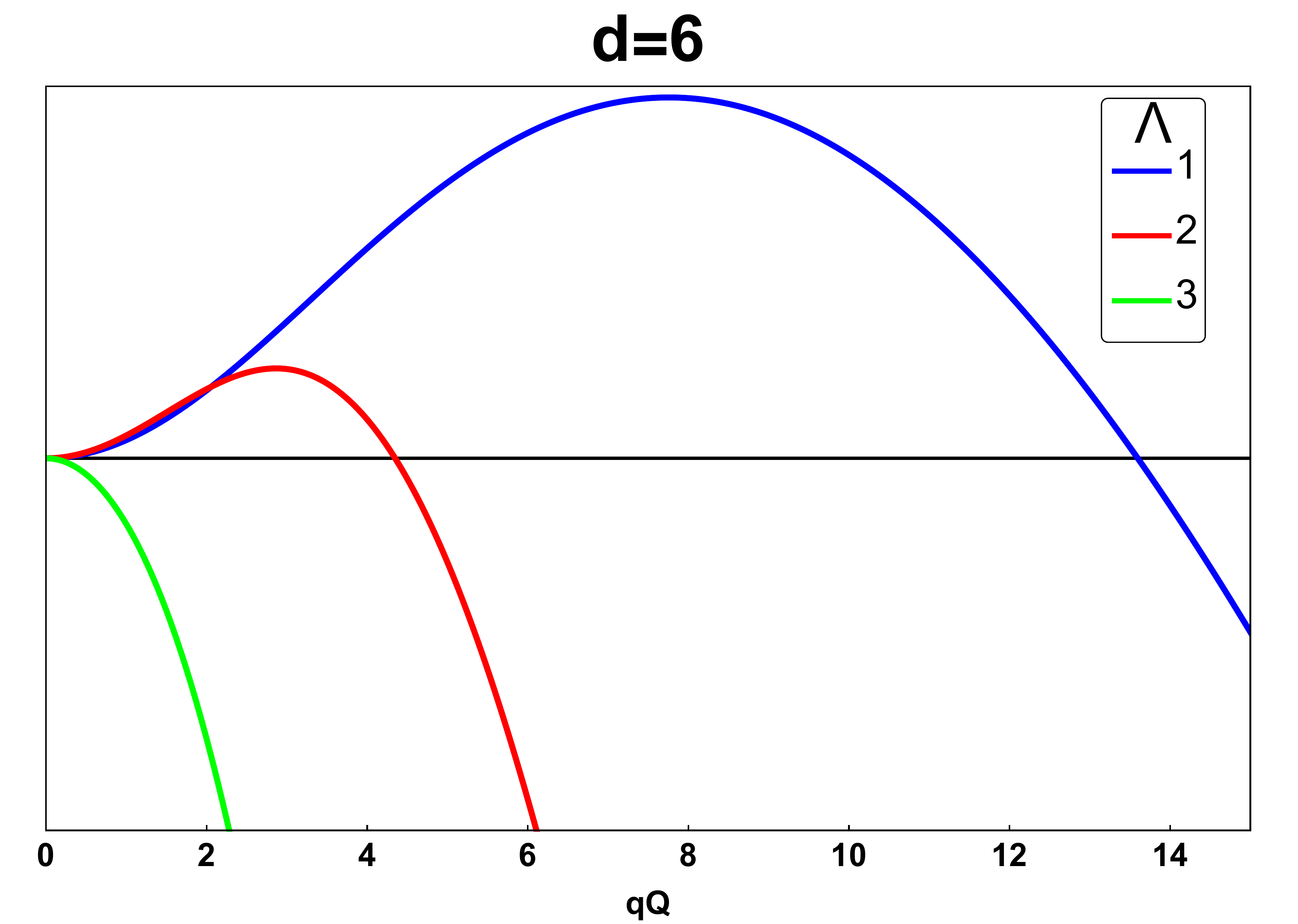}}
\vskip -3ex
\subfigure{\includegraphics[scale=0.166]{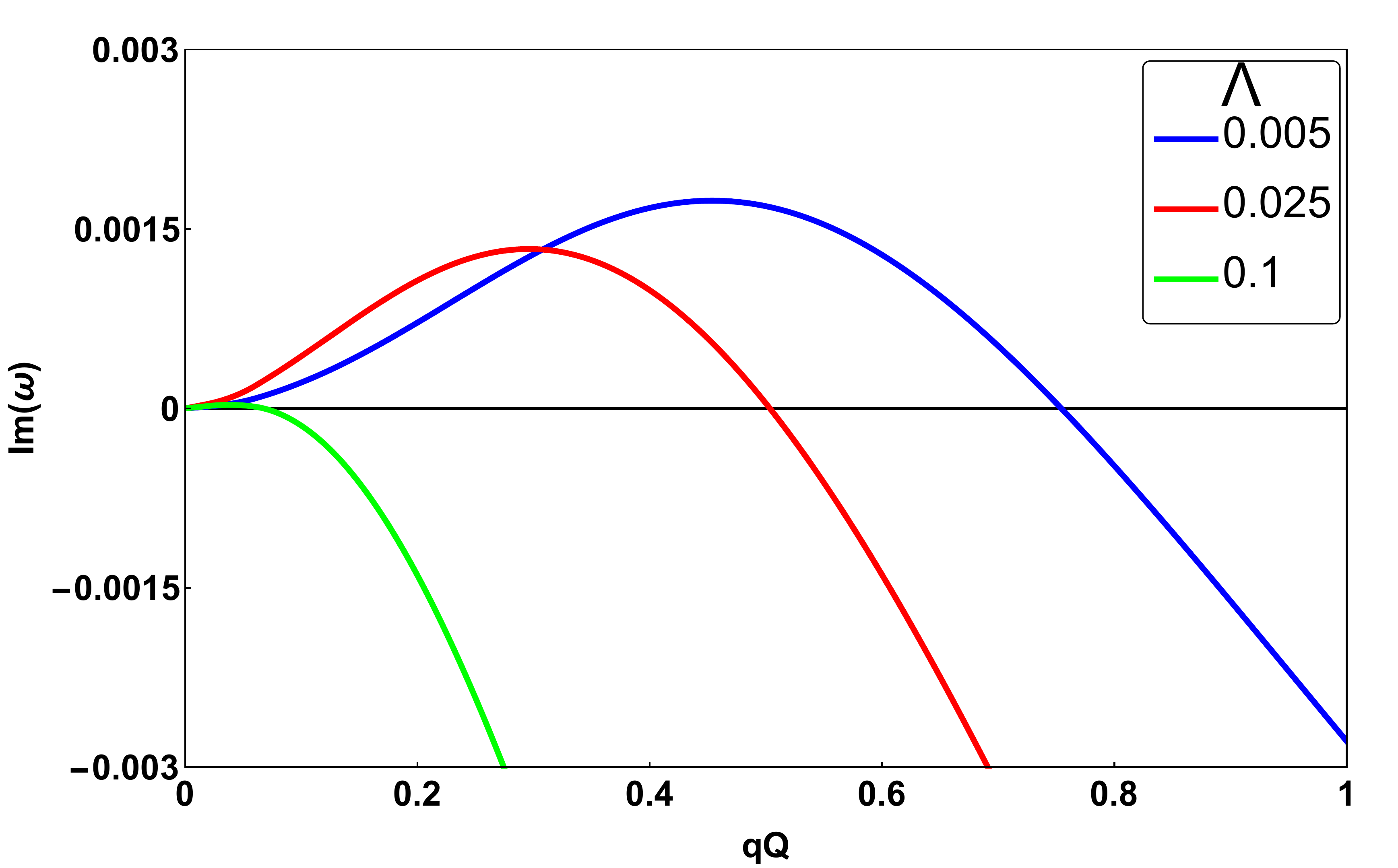}}
\hskip -2.5ex
\subfigure{\includegraphics[scale=0.15]{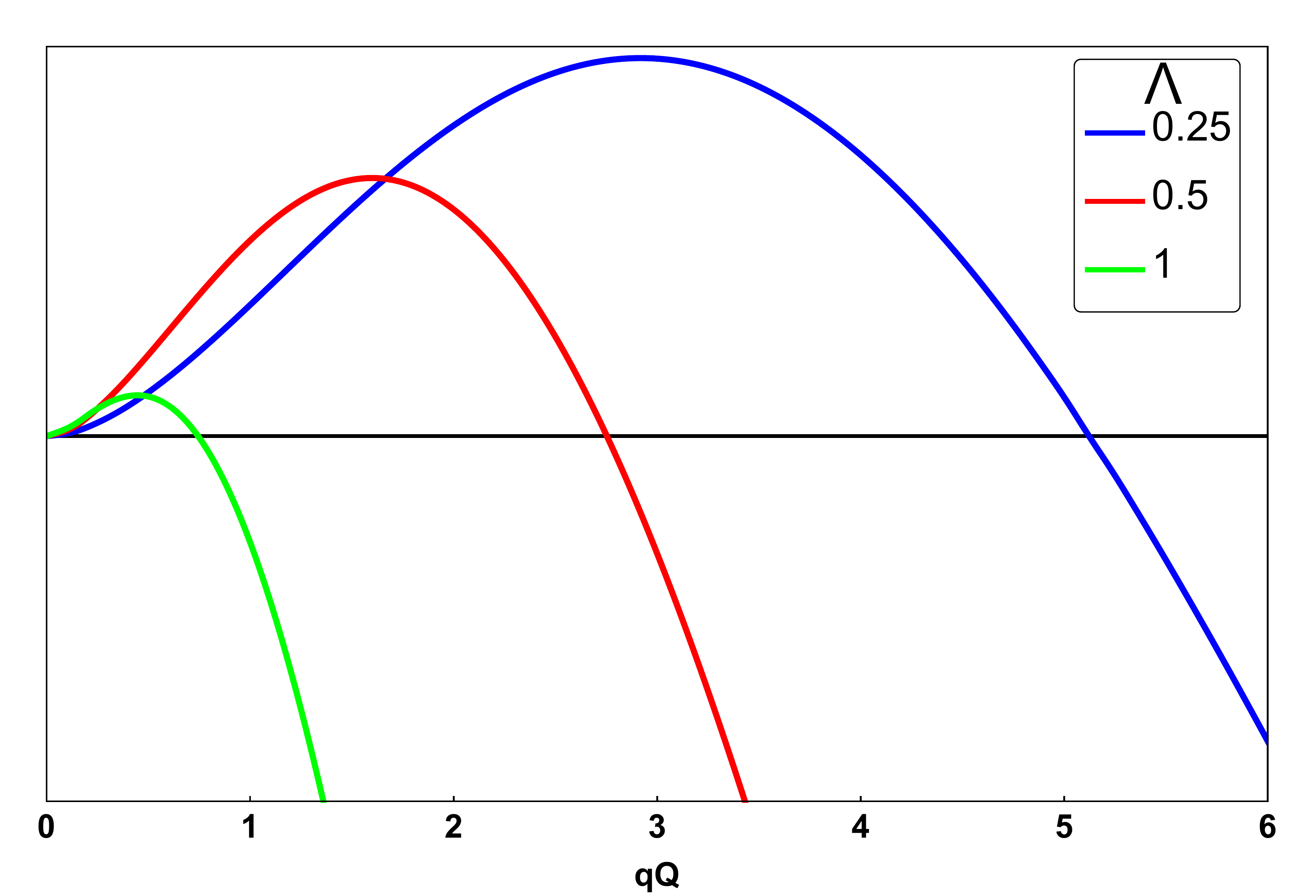}}
\hskip -2.5ex
\subfigure{\includegraphics[scale=0.15]{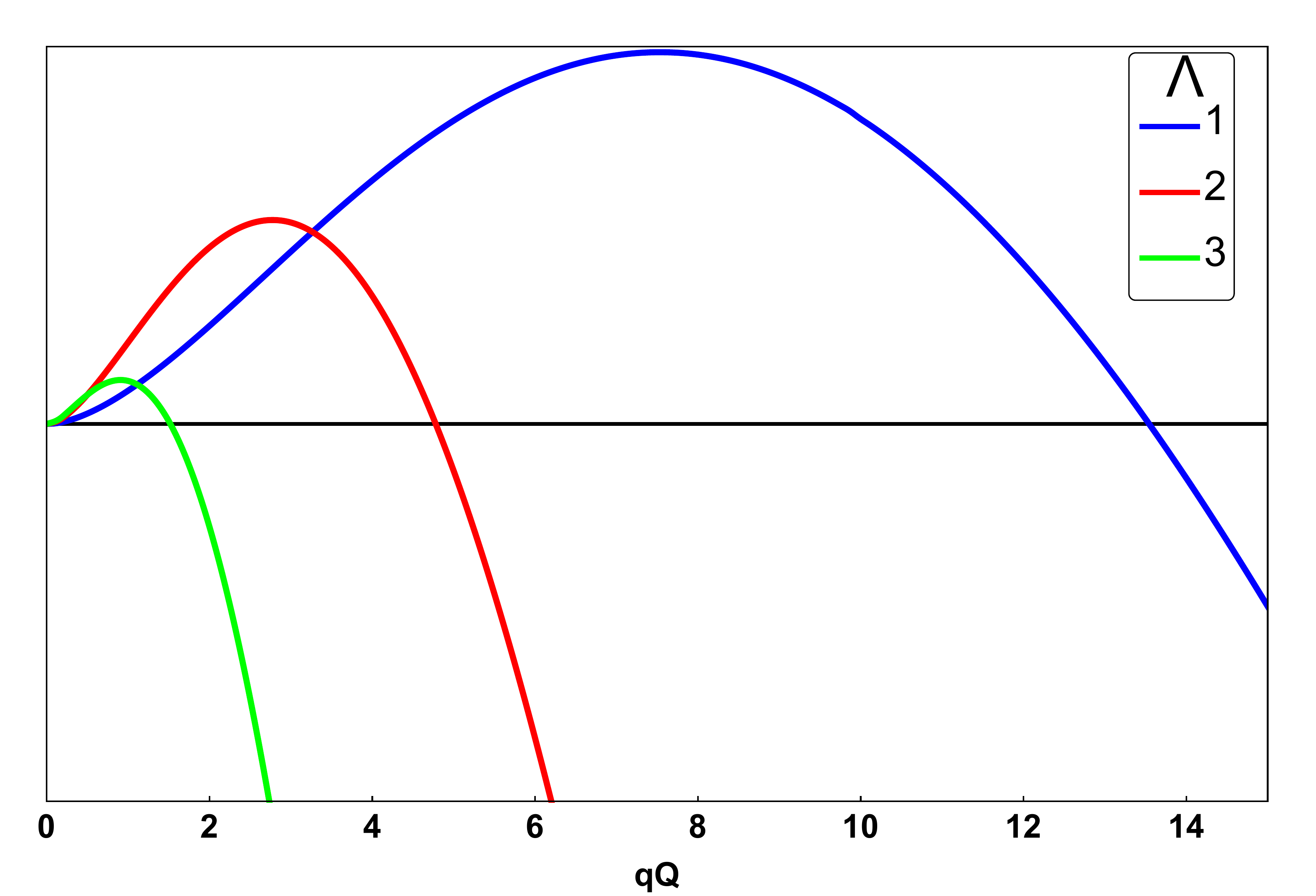}}
\caption{Imaginary part of a charged massless scalar perturbation with $l=0$ on a fixed $d-$dimensional RNdS BH with $M=1$, $Q=0.5$ (top panel) and $Q=0.999\, Q_\text{max}$ (bottom panel) versus the charge coupling $qQ$. Different colors designate distinct choices of cosmological constants $\Lambda$.}
\label{massless}
\end{figure}

\section{Superradiance in higher-dimensional RNdS}
The superradiant effect of charged massive scalar fields around $4-$dimensional RNdS spacetime has been analyzed in \cite{Zhu:2014sya,Konoplya:2014lha}. The study can be easily generalized in $d-$dimensions. For this analysis, we consider a wave-packet scattering off the BH effective potential. Such a scenario demands that we impose the following boundary conditions in (\ref{master_eq_RNdS1}):
\begin{equation}
\label{scat}
\psi \sim
\left\{
\begin{array}{lcl}
B e^{-i (\omega-\Phi(r_+))r_* },\,\,\,\quad\quad\quad\quad\quad\quad\,\, r \rightarrow r_+, \\
&
&
\\
 e^{-i(\omega-\Phi(r_c))r_*} + A e^{i(\omega-\Phi(r_c))r_*},\,\,\, r \rightarrow r_c.
\end{array}
\right.
\end{equation}
In fact, \eqref{master_eq_RNdS1} has a complex conjugate solution ${\psi^\dagger}$. It is easy to prove that the Wronskian of the independent solutions $\psi$, ${\psi^\dagger}$ is $r_*-$independent and therefore, conserved. Due to the conservation of the Wronskian, we derive the following relation associating the reflexion ($A$) and transmission ($B$) coefficients
\begin{equation}
|A|^2=1-\frac{\omega-\Phi(r_+)}{\omega-\Phi(r_c)}|B|^2.
\end{equation}
Hence, superradiance occurs when
\begin{equation}
\label{suprad}
\Phi(r_c)<\omega<\Phi(r_+),
\end{equation}
which designates that the amplitude of the reflected wave is larger than the amplitude of the incident wave. In \cite{Konoplya:2014lha} it was proven that the real part of $\omega$ satisfying (\ref{suprad}) is the necessary, but not sufficient, condition for the instability. Thus, the instability can only take place when (\ref{suprad}) is satisfied, but on the other hand, (\ref{suprad}) can also be satisfied by stable modes. As we will see below, the same holds for higher-dimensional RNdS spacetimes. This result is qualitatively different compared to the higher-dimensional asymptotically anti-de Sitter BHs, where the necessary condition is also the sufficient one \cite{Kodama:2009rq, Wang:2014eha}.

\section{Massless charged scalar fields}
In this section, we focus on the dominant unstable modes of charged massless scalar fields in $d-$dimensional RNdS. The dominant modes will be the ones that control the dynamical evolution of the perturbation at late times, thus defining the end-state of the perturbation. As shown in \cite{Zhu:2014sya,Konoplya:2014lha} the only unstable modes occur for $l=0$. Both \cite{Zhu:2014sya,Konoplya:2014lha} focus mostly on time evolutions of the perturbation. Although the onset of such an instability has been linked with a non-vanishing cosmological constant in $4-$dimensions, its origin was recently unraveled initially in Chapter \ref{PRD} and later discussed analytically (for $qQ\ll 1$) and numerically in \cite{Dias:2018ufh}. It was recently shown that a new distinct family of QNMs arise in the presence of a positive $\Lambda$ when considering neutral scalar perturbations around a $4-$dimensional (see Chapter \ref{PRL}) and higher-dimensional (see Chapter \ref{JHEP}) RNdS BH. This family of modes is purely imaginary, it directly relates to the existence and timescale of the cosmological horizon, and can be very well approximated by the scalar QNMs of pure $d-$dimensional de Sitter space \cite{LopezOrtega:2006my}:
\begin{eqnarray}
\label{zero mode}
\omega_{\rm pure \,\rm dS}/\kappa_c^{\rm dS} &=&-i (l+2n)\,,\\
\omega_{\rm pure \,\rm dS}/\kappa_c^{\rm dS} &=&-i(l+2n+d-1)\,.\label{pure_dS_scalar0}
\end{eqnarray}

Interestingly, these modes depend on the surface gravity of the cosmological horizon of pure $d-$dimensional de Sitter space $\kappa_c^\text{dS}=\sqrt{2\Lambda/(d-2)(d-1)}$ as opposed
to that of the cosmological horizon in the $d-$dimensional RNdS BH under consideration. This can, in principle, be explained by the fact that the accelerated expansion of RNdS spacetimes is also governed by $\kappa_c^{\rm dS}$~\cite{Brill:1993tw,Rendall:2003ks}. 
It is evident from (\ref{zero mode}) that the dominant $l=0$ mode is a stationary mode $\omega=0$ which does not depend on the dimensions $d$. This mode is responsible for the dynamical instability found in \cite{Zhu:2014sya,Konoplya:2014lha} when the scalar charge is turned on. 
\begin{figure}[H]
\subfigure{\includegraphics[scale=0.152]{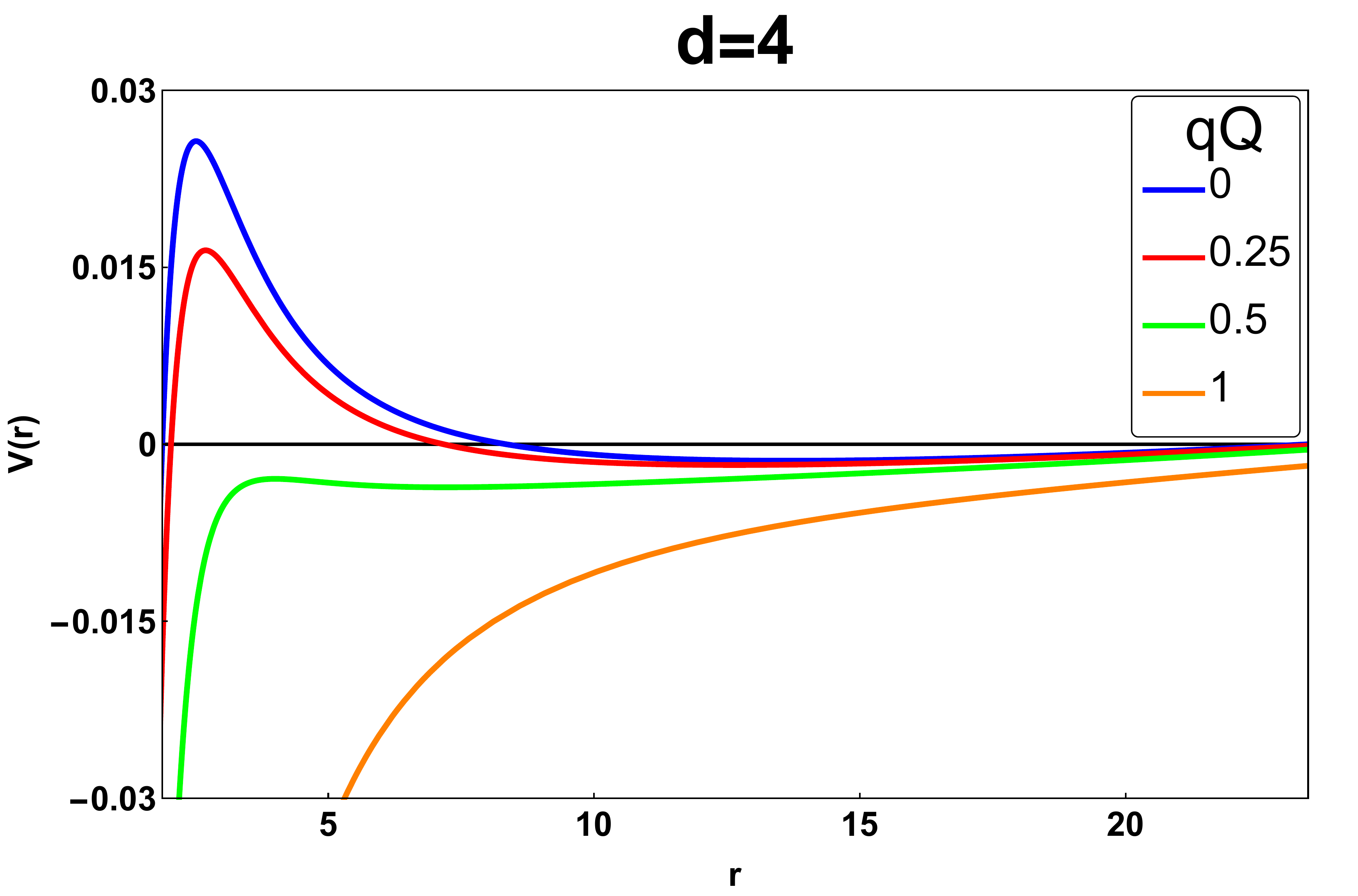}}\hskip -0.5ex
\subfigure{\includegraphics[scale=0.152]{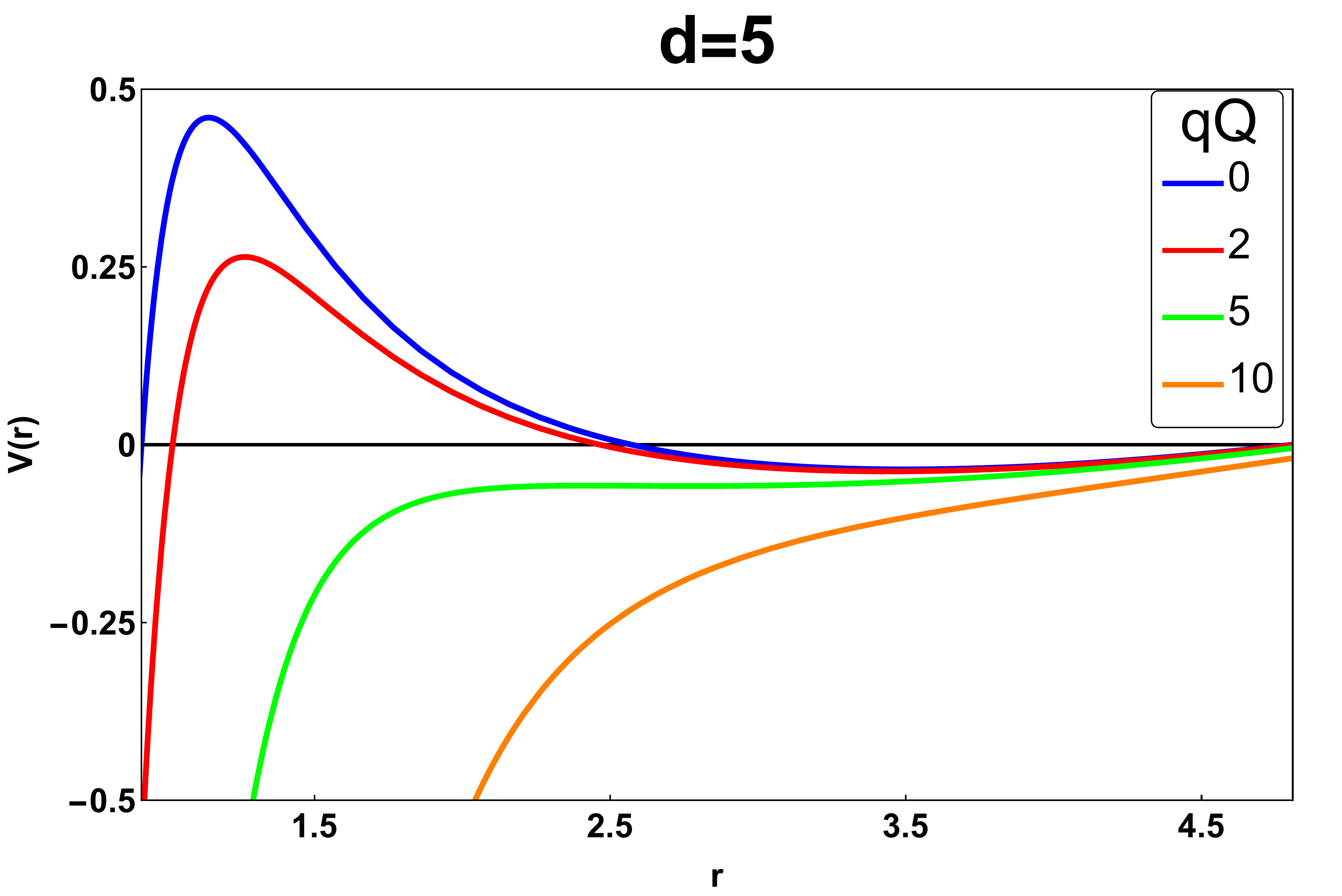}}\hskip -0.5ex
\subfigure{\includegraphics[scale=0.152]{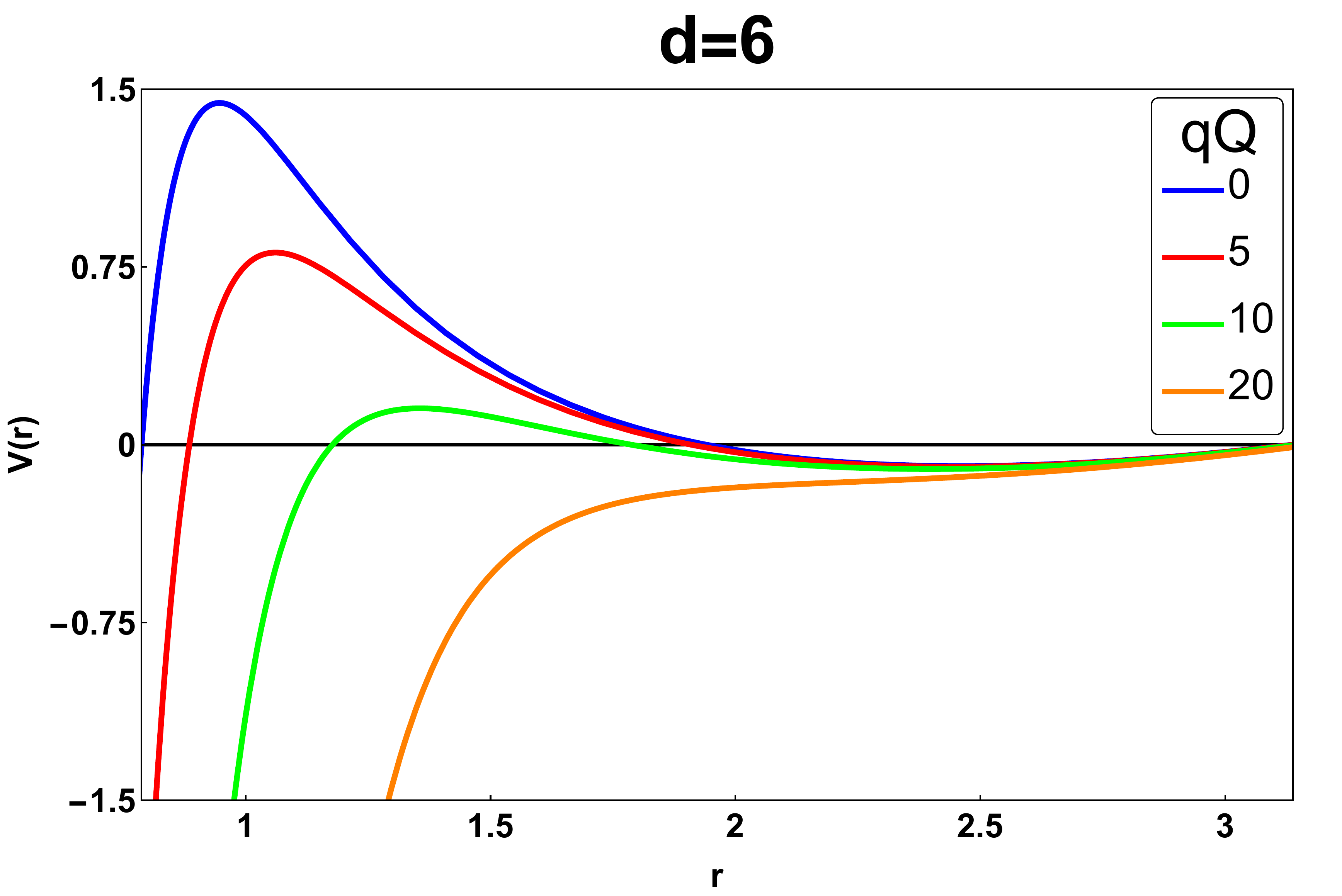}}
\caption{Effective potentials of the $l=0$ massless charged scalar perturbation on a $d-$dimensional RNdS BH with $M=1$ and $Q=0.5$ for various charge couplings $qQ$. The cosmological constants used, from left to right, are $\Lambda=0.005$, $\Lambda=0.25$ and $\Lambda=1$, respectively.}
\label{potentials}
\end{figure}
Here, we identify the existence of the zero (prone to instability) mode with the existence of a cosmological bound in the BH spacetime, which directly links to the QNMs of pure dS space. As shown in Fig. \ref{massless}, for $qQ\neq 0$ the zero modes evolves to QNMs with positive imaginary parts, thus unstable. As we increase the charge coupling, the instability is saturated and the family acquires a negative imaginary part, thus stability is restored. The stabilization due to the increment of the charge coupling can be explained through the form of the effective potential. 

In Fig. \ref{potentials}, we plot the effective potentials $V(r)$ for $d=4,5,6$ dimensions, for various charge couplings $qQ$. The $4-$dimensional picture of \cite{Zhu:2014sya} remains the same in higher dimensions. By increasing $qQ$, (\ref{RNdS_general potential1}) acquires a potential well close to the light ring which serves as a trapping region for waves to be captured and possibly be amplified. More increment of $qQ$ leads to a purely negative effective potential where the potential well vanishes. It is observed that with the increment of dimensions, though, the effective potential requires larger charge couplings to become purely negative. This can be explained by the fact that \eqref{RNdS_general potential1} becomes more positive by the increment of dimensions. This leads to the amplification of the instability as $d$ increases, as well as the enlargement of the parameter space region where the instability occurs. It is important to state that for $l\neq 0$ the potential well vanishes, therefore no instabilities are found.

The real part of the unstable modes increases monotonously with respect to $qQ$ (see Table \ref{table1}). The imaginary part decreases more rapidly for larger $\Lambda$ and smaller $Q$. A peak of instability seems to occur for every $\Lambda$. It is evident that as we approach extremality, the peak of instability is slightly increased.  

Interestingly enough, it has been shown \cite{Zhu:2014sya,Konoplya:2014lha} that all charged massless scalar modes in 4-dimensional RNdS fit the superradiant condition (\ref{suprad}). In Table \ref{table1} we show that $l=0$ perturbations in $d-$dimensional RNdS with arbitrarily small or large charge couplings, also fit the superradiant condition, even when the perturbations are decaying ($\text{Im}(\omega)<0$).

\section{Massive charged scalar fields}
\begin{figure}[H]
\subfigure{\includegraphics[scale=0.166]{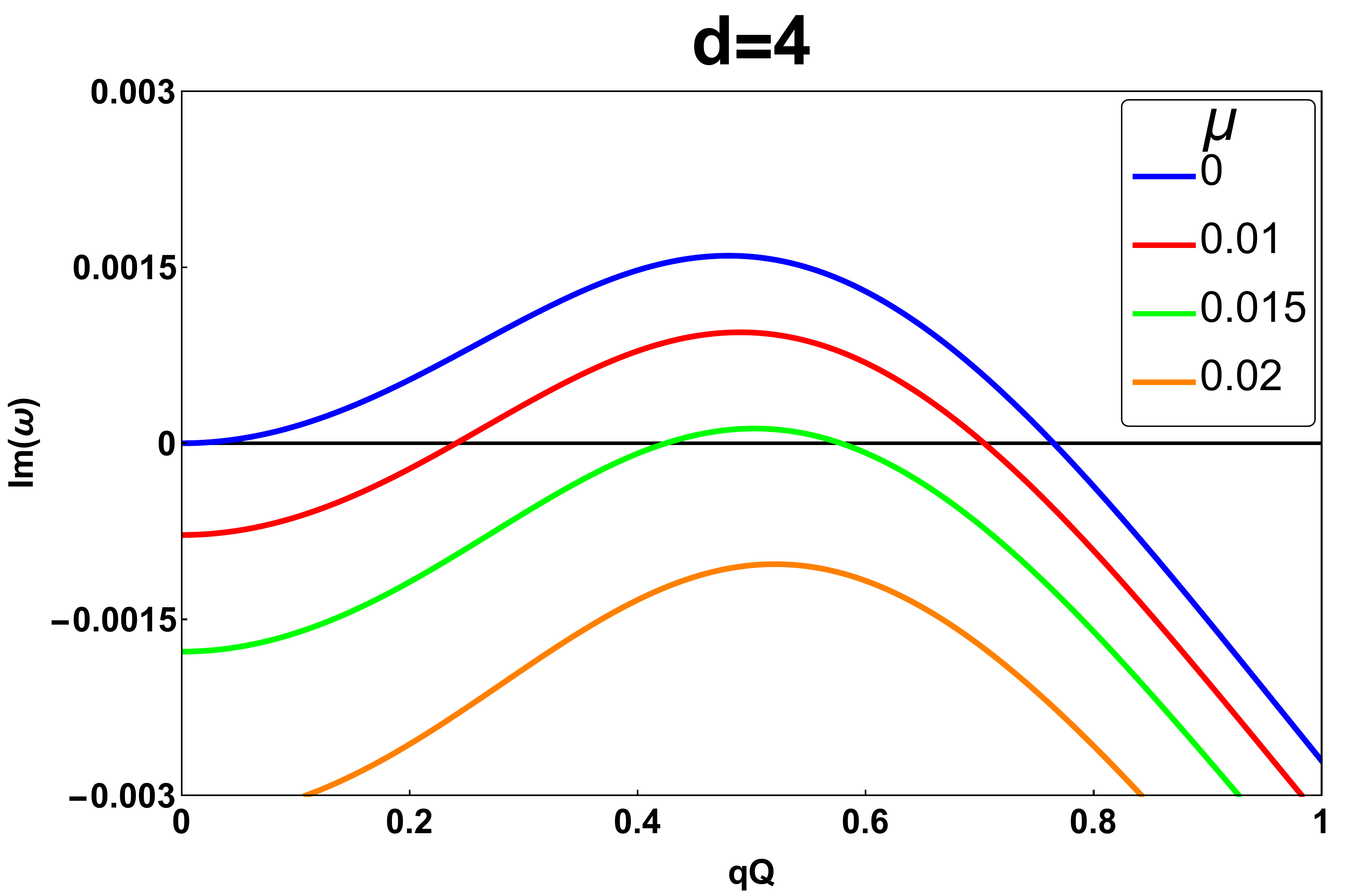}}
\hskip -2.5ex
\subfigure{\includegraphics[scale=0.15]{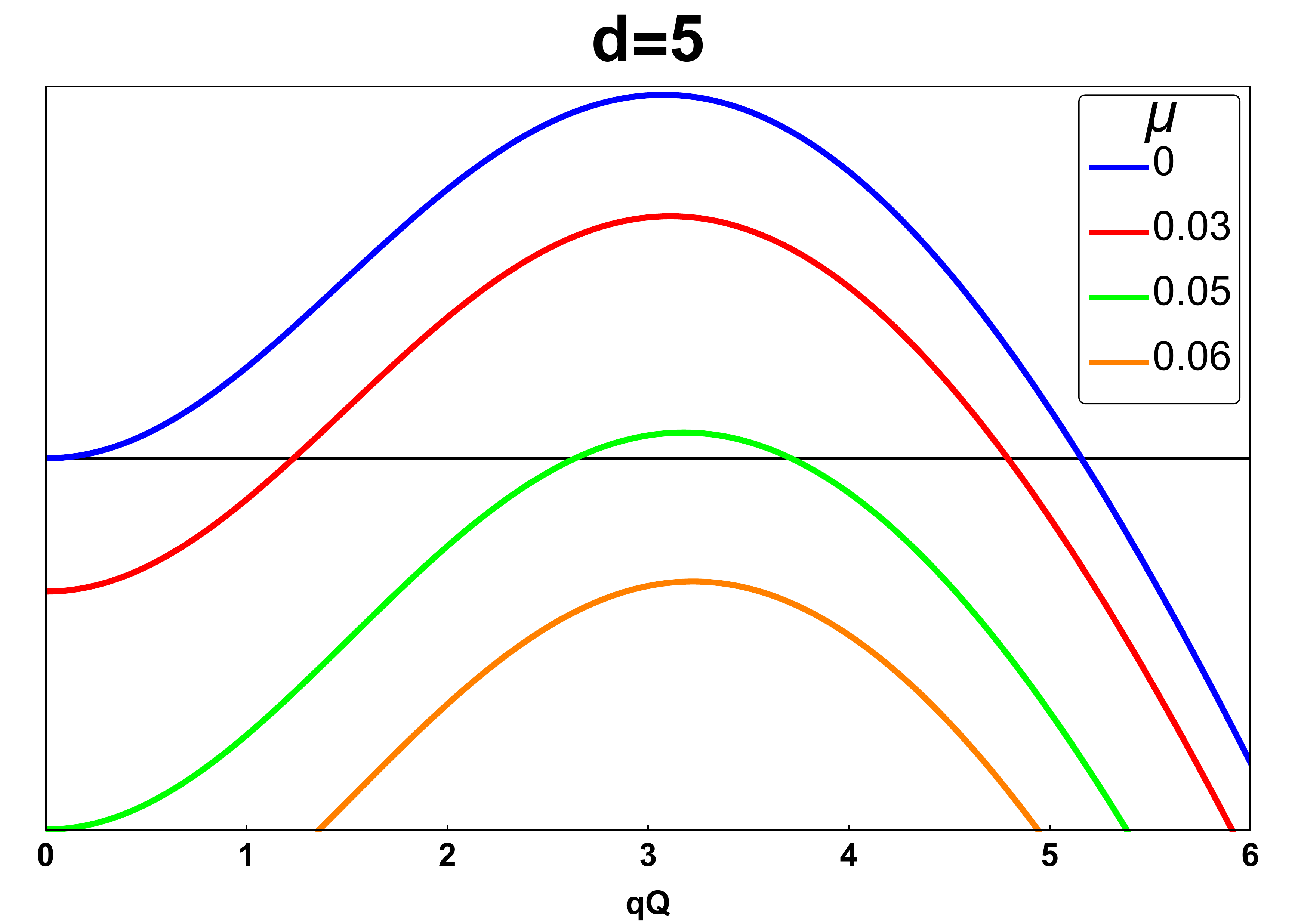}}
\hskip -2.5ex
\subfigure{\includegraphics[scale=0.15]{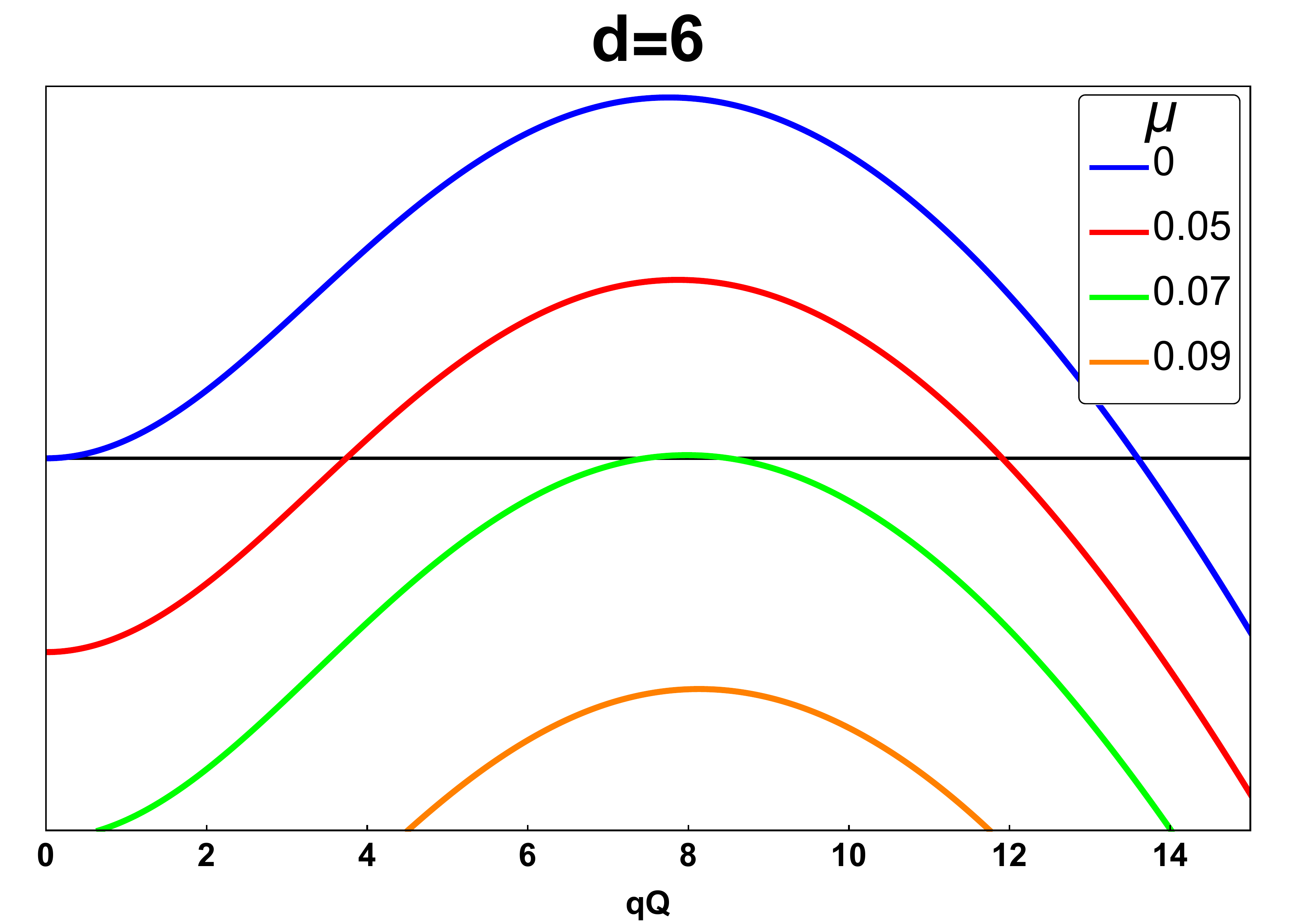}}
\caption{Imaginary part of a charged massive scalar perturbation with $l=0$ on a fixed $d-$dimensional RNdS BH with $M=1$, $Q=0.5$ versus the charge coupling $qQ$. Different colors designate distinct choices of scalar masses $\mu$. The cosmological constants used, from left to right, are $\Lambda=0.005$, $\Lambda=0.25$ and $\Lambda=1$, respectively.}
\label{massive}
\end{figure}

In this section, we perform a frequency domain analysis of the $l=0$ massive charged scalar perturbations in $d-$dimensional RNdS BHs and explore the superradiant instability as the scalar mass $\mu$ increases. The addition of a nonzero mass affects the unstable modes in a manner shown in Fig. \ref{massive}. The $\text{Im}(\omega)$ shift downwards as mass increases which leads to increasingly smaller regions of $qQ$ where unstable modes exist. In fact, the ``unstable" $l=0$ family will initially originate from stable purely imaginary modes for $qQ=0$. After a finite $\mu$, perturbations become stable for all $qQ$. This is due to the upwards shift of (\ref{RNdS_general potential1}) as $\mu$ increases which eliminates the potential well formed for $\mu=0$. 

From Fig. \ref{massive} we realize that two critical charge couplings exist, $qQ_c(\text{min})$ beyond which linear instabilities arise and $qQ_c(\text{max})$ beyond which stability is restored for $d=4,5,6$. As expected, $qQ_c(\text{min})=0$ and $qQ_c(\text{max})$ is maximized for $\mu=0$. With the increment of scalar mass $qQ_c(\text{min})$ increases and $qQ_c(\text{max})$ decreases until a finite mass where they coincide. Beyond that point they both vanish.

Curiously enough, unstable QNMs with nonzero mass exist in various regions of the parameter space and, even more strikingly, they still satisfy (\ref{suprad}). In Table \ref{table2} we show that the addition of a mass term can still lead to unstable and, more surprisingly, stable modes which are superradiant. More increment of $\mu$ leads to the cancellation of superradiance for $4$ and higher dimensions. This can be explained by the fact that the gravitational attraction between the massive field and the BH becomes dominant for large enough $\mu$ if one compares it with the electric repulsion between the BH and scalar field charges $qQ$.
\begin{table}[H]
\centering
\scalebox{0.45}{
\begin{tabular}{||c| c | c | c ||} 
\hline
  \multicolumn{4}{||c||}{$d=4$} \\
   \hline
\hline
  \multicolumn{4}{||c||}{$\Lambda=0.005$} \\
   \hline
    $qQ$ & $\omega$ & $\Phi(r_c)$ &$\Phi(r_+)$ \\ [0.5ex]
   \hline
    0.005 & 0.00023 +  3.7$\times$$10^{-7}$i  & 0.00021   &0.00266 \\
   \hline
    0.05 & 0.00229 + 0.00004 i  & 0.00213   &0.02663 \\
   \hline
   0.5 & 0.02577 + 0.00159 i  & 0.02134   &0.26625 \\
   \hline
   1 & 0.05322 - 0.00271 i & 0.04268&0.53251 \\ 
   \hline
   5 &0.21907 - 0.01736 i & 0.21338& 2.66253\\
   \hline
   10 &0.42966 - 0.01832 i &0.42677 &5.32507 \\
   \hline\hline
   \multicolumn{4}{||c||}{$\Lambda=0.05$} \\
    \hline
     $qQ$ & $\omega$ & $\Phi(r_c)$ &$\Phi(r_+)$ \\ [0.5ex]
   \hline
    0.005 & 0.00092 - 3.2$\times$$10^{-7}$i  & 0.00077   &0.00249 \\
    \hline
    0.05 & 0.00925 - 0.00003 i  & 0.00773   &0.02486 \\  
    \hline
    0.5 &  0.09222 - 0.00430 i & 0.07734   & 0.24859 \\
    \hline
    1 & 0.17777 - 0.01595 i &0.15468 &0.49719 \\ 
    \hline
    5 &0.78240 - 0.04099 i &0.77338 &2.48594 \\
    \hline
    10 &1.55129 - 0.04204 I &1.54676 &4.97188 \\
    \hline
\end{tabular}
}
\scalebox{0.45}{
\begin{tabular}{||c| c | c | c ||} 
\hline
  \multicolumn{4}{||c||}{$d=5$} \\
   \hline
\hline
  \multicolumn{4}{||c||}{$\Lambda=0.25$} \\
   \hline
    $qQ$ & $\omega$ & $\Phi(r_c)$ &$\Phi(r_+)$ \\ [0.5ex] 
  \hline
   0.05 &0.00081 + 2$\times 10^{-6}$ i &0.00069  &0.01900 \\
   \hline
   0.5  &0.00818 + 0.00020 i  &0.00688  &0.18998 \\
   \hline
   1   &0.01656 + 0.00073 i  & 0.01377 &0.37996 \\
   \hline
   5 & 0.09344 + 0.00040 i &0.06884 & 1.89978\\ 
   \hline
   10 &0.18206 - 0.01759 i &0.13768 &3.79956 \\
   \hline
   15 &0.25790 - 0.03625 i & 0.20651&5.69934 \\
   \hline\hline
   \multicolumn{4}{||c||}{$\Lambda=0.75$} \\
    \hline
     $qQ$ & $\omega$ & $\Phi(r_c)$ &$\Phi(r_+)$ \\ [0.5ex] 
    \hline
       0.05 & 0.00293 - $4\times 10^{-7}$ i &0.002260 &0.01735 \\
       \hline
       0.5 & 0.02945 - 0.00006 i  & 0.02260   &0.17353 \\
       \hline
       1 &  0.05922 - 0.00042 i & 0.04521   &0.34707 \\
       \hline
       5 & 0.28945 - 0.03062 i &0.22604 &1.73533 \\ 
       \hline
       10 &0.52736 - 0.07913 i & 0.45208&3.47066 \\
       \hline
       15 &0.74441 - 0.10774 i &0.67812 &5.20598\\
    \hline
\end{tabular}
}
\scalebox{0.45}{
\begin{tabular}{||c| c | c | c ||} 
\hline
  \multicolumn{4}{||c||}{$d=6$} \\
   \hline
\hline
  \multicolumn{4}{||c||}{$\Lambda=1$} \\
   \hline
    $qQ$ & $\omega$ & $\Phi(r_c)$ &$\Phi(r_+)$ \\ [0.5ex] 
  \hline
   0.05 &0.00032 + 4$\times 10^{-7}$ i &0.00026  &0.01630 \\
   \hline
   0.5  &0.00321 + 0.00004 i  &0.00258  &0.16297 \\
   \hline
   1   &0.00644 + 0.00015 i  & 0.00515 &0.32594 \\
   \hline
   5 & 0.03428 + 0.00222 i &0.02576 & 1.62972\\ 
   \hline
   10 &0.07291 + 0.00245 i &0.05152 &3.25944 \\
   \hline
   20 &0.14819 - 0.00775 i & 0.10304&6.51887 \\
   \hline\hline
   \multicolumn{4}{||c||}{$\Lambda=2$} \\
    \hline
     $qQ$ & $\omega$ & $\Phi(r_c)$ &$\Phi(r_+)$ \\ [0.5ex] 
    \hline
       0.05 & 0.00103 + $5\times 10^{-7}$ i &0.00076 &0.01505 \\
       \hline
       0.5 & 0.01028 + 0.00005 i  & 0.00764   &0.15054 \\
       \hline
       1 &  0.02062 - 0.00018 i & 0.01527   &0.30108 \\
       \hline
       5 & 0.10765 - 0.00084 i &0.07635 &1.50539 \\ 
       \hline
       10 &0.21518 - 0.01597 i & 0.15271&3.01079 \\
       \hline
       20 &0.40031 - 0.06012 i &0.30541 &6.02159\\
    \hline
\end{tabular}
}
\caption{Dominant $l=0$ massless charged scalar modes of $d-$dimensional RNdS spacetime with $M=1$, $Q=0.5$ for various parameters. All modes shown satisfy the superradiant condition (\ref{suprad}).}
\label{table1}
\end{table}

\section{Conclusions}
In the present chapter, we study a dynamical instability emerging from a spherically symmetric $l=0$ charged scalar perturbation propagating on a higher-dimensional RNdS background. Such an instability has a superradiant nature, therefore electromagnetic energy can be extracted from the BH.

By performing a thorough frequency domain analysis of higher-dimensional RNdS BHs under charged scalar perturbations we realize that the source of instability is directly linked with the existence of the cosmological horizon, as well as the QNMs of pure $d-$dimensional de Sitter space. As the potential barrier close to the event horizon vibrates when we probe it, leaking out energy in a particular manner, so the cosmological horizon fluctuates. These new ``oscillations" of course have a distinct nature, comparing to any other QN oscillatory mode. They originate from purely imaginary modes, very well approximated by the pure de Sitter space QNMs and even exhibit a stationary mode $\omega=0$ for any dimension with vanishing angular momentum. 
\begin{table}[H]
\centering
\scalebox{0.8}{
\begin{tabular}{||c| c | c | c ||} 
\hline
  \multicolumn{4}{||c||}{$d=4, \Lambda=0.005$, $qQ=0.45$} \\
   \hline
    $\mu $ & $\omega$ & $\Phi(r_c)$ &$\Phi(r_+)$ \\ [0.5ex] 
  \hline
   $10^{-3}$ &0.02305 + 0.00173 i &0.01919  &0.42925 \\
   \hline
   $10^{-2}$  & 0.02280 + 0.00107 i  &0.01919  &0.42925 \\
   \hline
   $3\times 10^{-2}$ &0.02054 - 0.00458 i  &0.01919  &0.42925 \\ 
   \hline
   $5\times 10^{-2}$ &0.01263 - 0.01796 i &0.01919  &0.42925 \\
   \hline
\end{tabular}
}
\scalebox{0.8}{
\begin{tabular}{||c| c | c | c ||} 
\hline
  \multicolumn{4}{||c||}{$d=5, \Lambda=0.25$, $qQ=3$} \\
   \hline
    $\mu $ & $\omega$ & $\Phi(r_c)$ &$\Phi(r_+)$ \\ [0.5ex] 
  \hline
   $10^{-3}$ &0.05381 + 0.00293 i &0.04130  &1.13987 \\
   \hline
   $10^{-2}$  & 0.05379 + 0.00282 i  &0.04130  &1.13987 \\
   \hline
   $ 10^{-1}$ &0.05093 - 0.00819 i  &0.04130  &1.13987 \\ 
   \hline
   $2\times 10^{-1}$ &0.04100 - 0.04427 i &0.04130  &1.13987 \\
   \hline
\end{tabular}
}
\scalebox{0.8}{
\begin{tabular}{||c| c | c | c ||} 
\hline
  \multicolumn{4}{||c||}{$d=6, \Lambda=1$, $qQ=7$} \\
   \hline
    $\mu$ & $\omega$ & $\Phi(r_c)$ &$\Phi(r_+)$ \\ [0.5ex] 
  \hline
   $10^{-3}$ &0.04949 + 0.00286 i &0.03606  &2.28161 \\
   \hline
   $10^{-2}$  & 0.04948 + 0.00280 i  &0.03606  &2.28161 \\
   \hline
   $2\times 10^{-1}$ &0.04496 - 0.02141 i  &0.03606  &2.28161 \\ 
   \hline
   $4\times 10^{-1}$ &0.03005 - 0.10116 i &0.03606  &2.28161 \\
   \hline
\end{tabular}
}
\caption{Dominant $l=0$ massive charged scalar modes of $d-$dimensional RNdS spacetime with $M=1$, $Q=0.5$ for various parameters.}
\label{table2}
\end{table}
When charge is introduced to the scalar field, the zero-mode evolves to a complex QNM with positive imaginary part and monotonously increasing real part. For a finite region of the charge coupling $qQ$ the family remains unstable which leads to a growing profile of the perturbation with respect to time. Such modes satisfy the superradiant condition even when stable configurations occur. We demonstrate that the source of superradiantly amplified modes occurs for spherically symmetric charged scalar perturbations due to the formation of a potential well between the photon sphere and the cosmological horizon of the BH, for a finite range of $qQ$. Increasing $qQ$ leads to the vanishing of the potential well and the dissipation of the superradiant instability.

With the increment of dimensions, the potential well deepens and therefore, the instability timescale is decreased leading to larger regions in the parameter space where unstable modes occur. Interestingly, even though the introduction of mass stabilized the system, there are still regions in the parameter space where the dominant modes remain superradiantly unstable. 

An open, and still interesting, problem is the nonlinear development of such a system to grasp the end-state of the evolving BH spacetime. A huge challenge in such nonlinear evolutions is the very large timescale of the instability which requires highly precise numerical developments. Since the increment of dimensions decreases the timescale of the instability, it would be more feasible for such an instability to be tested in higher-dimensional RNdS spacetime non-linearly and realize if it leads to a novel scalarized BH or to the evacuation of all matter.

\begin{appendices}
\appendixpage
\noappendicestocpagenum
\addtocontents{toc}{\protect\setcounter{tocdepth}{0}}
\chapter{Scalar perturbations in spherically symmetric spacetimes}\label{appA}
For the following calculation, the Christoffel symbols of spherically symmetric spacetimes of the form 
\begin{equation}
ds^2=-f(r)dt^2+\frac{1}{f(r)}dr^2 +r^2(d\theta^2+\sin^2\theta\, d\varphi^2)
\end{equation}
or, equivalently, in matrix form
\begin{equation}
\label{}
g_{\mu\nu}=\text{diag}\left(-f(r),\frac{1}{f(r)},r^2,r^2\sin^2\theta\right),
\end{equation}
should be known. Below, we include the nonzero components in Schwarzschild-like coordinates:
\begin{align}
\nonumber
\Gamma^0_{10}&=\Gamma^0_{01}=\frac{f^\prime(r)}{2f(r)},\,\,\,\,\,\,\,\,\,\,\,\,\,\,\Gamma^1_{00}=\frac{f(r)f^\prime(r)}{2},\,\,\,\,\,\,\,\,\,\,\,\,\,\,\Gamma^1_{11}=-\frac{f^\prime(r)}{2f(r)} \\
\label{symbols_FG}
\Gamma^1_{22}&=-rf(r),\,\,\,\,\,\, \Gamma^1_{33}=-r f(r)\sin^2\theta,\,\,\,\,\,\, \Gamma^2_{21}=\Gamma^2_{12}=\frac{1}{r},\\
\nonumber
\Gamma^2_{33}&=-\cos\theta\sin\theta\,\,\,\,\,\,\Gamma^3_{31}=\Gamma^3_{13}=\frac{1}{r},\,\,\,\,\,\,\,\,\,\,\,\,\Gamma^3_{32}=\Gamma^3_{23}=\frac{\cos\theta}{\sin\theta}.
\end{align}
The propagation of a charged scalar perturbation $\Psi$ with mass and charge $\mu$ and $q$, respectively, is described by the Klein-Gordon equation
\begin{equation}
\label{KG_charged}
(D^\nu D_\nu-\mu^2)\Psi=0 
\end{equation}
where $D_\nu=\nabla_\nu-iqA_\nu$ the covariant derivative associated with the gauge transformation $\nabla_\nu\rightarrow\nabla_\nu-iqA_\nu$. Here, $A_\nu=-\delta^0_\nu Q/r$, is the electrostatic potential originating from a source with charge $Q$. By expanding the covariant derivatives, (\ref{KG_charged}) reads
\begin{align}
\label{first_eq}
\nabla^\nu\left(\nabla_\nu\Psi\right)-iq\nabla^\nu \left(A_\nu\Psi\right)-iqA^\nu\nabla_\nu\Psi-q^2A^\nu A_\nu\Psi-\mu^2\Psi=0
\end{align}
By utilizing (\ref{symbols_FG}) and the following identities
\begin{align*}
\nabla_\nu\Psi&=\partial_\nu\Psi,\\
\nabla^\nu A_\nu&=g^{\mu\nu}\nabla_\mu g_{\kappa\nu}A^\kappa=\delta^\mu_\kappa\nabla_\mu A^\kappa=\nabla_\mu A^\mu=\partial_\mu A^\mu+\Gamma^\mu_{\mu\nu}A^\nu=0\\
A_\nu \partial^\nu&=g_{\mu\nu}A^\mu g^{\kappa\nu}\partial_\kappa=A^\mu\partial_\kappa\delta^\kappa_\mu=A^\mu\partial_\mu\\
\nabla^\nu\left(A_\nu\Psi\right)&=\Psi\nabla^\nu A_\nu+A_\nu\nabla^\nu\Psi=A_\nu\partial^\nu\Psi=A^\mu\partial_\mu\Psi\\
\nabla^\nu\left(\nabla_\nu\Psi\right)&=g^{\mu\nu}\nabla_\mu\left(\partial_\nu\Psi\right)=\frac{1}{\sqrt{-g}}\partial_\mu\left(g^{\mu\nu}\sqrt{-g}\,\partial_\nu\Psi\right)
\end{align*}
(\ref{first_eq}) becomes
\begin{align}
\nonumber
\frac{1}{\sqrt{-g}}\partial_\mu\left(g^{\mu\nu}\sqrt{-g}\,\partial_\nu\Psi\right)-2iq A^\nu\partial_\nu\Psi-q^2g^{\mu\nu}A_\mu A_\nu\Psi-\mu^2\Psi&=0\Leftrightarrow\\
\nonumber
-\frac{\partial_t^2}{f}\Psi+\frac{1}{r^2}\partial_r\left(fr^2\partial_r\Psi\right)+\frac{1}{r^2}\left(\frac{1}{\sin\theta}\partial_\theta\left(\sin\theta\partial_\theta\Psi\right)+\frac{1}{\sin^2\theta}\partial^2_\varphi\Psi\right)\\
\label{second_eq}
-\frac{2iqQ}{fr}\partial_t\Psi+\frac{q^2Q^2}{r^2f}\Psi-\mu^2\Psi&=0
\end{align}
By assuming that the scalar field can be expanded in temporal, radial and angular parts
\begin{equation}
\Psi\sim e^{-i\omega t}Y_{lm}(\theta,\varphi)\frac{\psi(r)}{r},
\end{equation}
and by utilizing the identity of the square of the orbital angular momentum operator
\begin{equation}
{L}^2Y_{lm}=-\left(\frac{1}{\sin\theta}\partial_\theta\left(\sin\theta\partial_\theta\right)+\frac{1}{\sin^2\theta}\partial^2_\varphi\right)Y_{lm}=l(l+1)Y_{lm},
\end{equation}
(\ref{second_eq}) becomes
\begin{align}
\label{third_eq}
\omega^2\psi+f^2\frac{d^2\psi}{dr^2}+ff^\prime\frac{d\psi}{dr}-\left(\frac{ff^\prime}{r}+f\frac{l(l+1)}{r^2}\right)\psi+\frac{q^2Q^2}{r^2}\psi-\frac{2qQ}{r}\omega\psi-f\mu^2\psi=0.
\end{align}
If we consider the tortoise coordinate $dr_*=dr/f$ then we can write (\ref{third_eq}) in a Schr\"{o}dinger-like form
\begin{equation}
\label{master scalars}
\frac{d^2\psi}{dr_*^2}+\left((\omega-\Phi(r))^2-V\right)\psi=0,
\end{equation}
where $\Phi(r)=qQ/r$ and 
\begin{equation}
V=f\left(\frac{l(l+1)}{r^2}+\frac{f^\prime}{r}+\mu^2\right),
\end{equation}
the effective potential. Eq. (\ref{master scalars}) holds for any spherically symmetric four-dimensional spacetime in Schwarzschild-like coordinates. 

\chapter{Fermionic perturbations in spherically symmetric spacetimes}\label{appB}
Fermion fields are described by spinors $\psi$, and in order to accommodate spinors in general relativity we need the tetrad formalism. A tetrad is a set of four linearly independent vectors that can be defined at each point in a (pseudo-)Riemannian spacetime. The tetrads by definition satisfy the relations
\begin{align}
e^{(a)}_\mu \,e^{\nu}_{(a)}&=\delta^\nu_\mu,\\
e^{(a)}_\mu \,e^{\mu}_{(b)}&=\delta^{(a)}_{(b)},
\end{align}
The choice of the tetrad field determines the metric through
\begin{align}
\label{tetradgmn}
g_{\mu\nu}&=e^{(a)}_\mu \,e^{(b)}_\nu\,\eta_{(a)(b)},\\
\label{minkowski}
\eta_{(a)(b)}&=e^\mu_{(a)}\,e^\nu_{(b)}\,g_{\mu\nu},
\end{align}
where $\eta_{(a)(b)}$ and $g_{\mu\nu}$ are the Minkowski and curved spacetime metric, respectively. We have the following rules for raising and lowering indices
\begin{align}
e_{(a)\mu}&=g_{\mu\nu}\,e^\nu_{(a)},\\
e^{(a)}_\mu&=\eta^{(a)(b)}e_{(b)\mu},
\end{align}
where latin letters in parentheses are raised and lowered by the flat metric while the Greek letters are raised and lowered by the curved metric. The components of tensors in the tetrad frame are given by the following relations
\begin{align}
V^{(a)}&=e^{(a)}_\mu V^\mu,\\T^{(a)}_{(b)}&=e^{(a)}_\mu e^\nu_{(b)}T^\mu_\nu,
\end{align}
and so on.

In order to write the Dirac equation in GR, we also need to introduce the spacetime dependent gamma matrices $G^\mu$. The $G^\mu$ matrices are related to the special relativity matrices, $\gamma^{(a)}$, by the relation
\begin{equation}
\label{gtet}
G^\mu=e^\mu_{(a)}\gamma^{(a)},
\end{equation}
and are chosen in a way so they satisfy the anti-commutation relations
\begin{align}
\label{anticom_flat}
\{\gamma^{(a)},\gamma^{(b)}\}&=\epsilon 2\eta^{(a)(b)},\\
\label{anticom_curved}
\{G^\mu, G^\nu\}&=\epsilon 2 g^{\mu\nu},
\end{align}
where $\epsilon=1$ for metric signature $(+,-,-,-)$ and $\epsilon=-1$ for $(-,+,+,+)$. If the metric is diagonal then $G^\mu$ anti-commute for $\mu\neq \nu$. 
\subsection{Dirac equation in Schwarschild-like coordinates}
For a spherically symmetric four-dimensional spacetime we have the line element
\begin{equation}
\label{metric3}
ds^2=-f(r)dt^2+\frac{1}{f(r)}dr^2 +r^2(d\theta^2+\sin^2\theta\, d\varphi^2)
\end{equation}
or in matrix form
\begin{equation}
\label{spherical metric}
g_{\mu\nu}=\text{diag}\left(-f(r),\frac{1}{f(r)},r^2,r^2\sin^2\theta\right),
\end{equation}
while the flat Minkowski metric is 
\begin{equation}
\label{mink}
\eta_{(a)(b)}=\text{diag}(-1,1,1,1).
\end{equation}
We can satisfy  the anti-commutation relations (\ref{anticom_flat}) by choosing the standard Dirac representation of $\gamma-$matrices
\begin{equation}
\label{dirac_gamma}
\gamma^{(0)}=\begin{pmatrix}
\mathbf{1}&0\\
0&-\mathbf{1}
\end{pmatrix},\,\,\,\,\,\,\,\,\,
\gamma^{(k)}=\begin{pmatrix}
0&\sigma^{k}\\
-\sigma^{k}&0
\end{pmatrix}
\end{equation}
where $\sigma^k$ the Pauli matrices
\begin{equation}
\sigma^1=\begin{pmatrix}
0&1\\1&0
\end{pmatrix},\,\,\,\,\,\sigma^2=\begin{pmatrix}
0&-i\\i&0
\end{pmatrix},\,\,\,\,\,\sigma^3=\begin{pmatrix}
1&0\\0&-1
\end{pmatrix}.
\end{equation}
It is easy to verify that 
\begin{equation}
\label{gamma}
\left(\gamma^{(0)}\right)^2=\mathbf{1},\,\,\,\,\,\,\,\,\,\left(\gamma^{(k)}\right)^2=-\mathbf{1}.
\end{equation}
For the anti-commutations relations (\ref{anticom_curved}) to be satisfied we can define the curved $\gamma$-matrices $G^\mu$ with respect to a fixed (Cartesian) tetrad
\begin{align}
\begin{split}
\label{tetrad}
e^t_{(a)}&=\left(\frac{1}{\sqrt{f(r)}},0,0,0\right)\\
e^r_{(a)}&=\left(0,\sqrt{f(r)}\sin\theta\cos\phi,\sqrt{f(r)}\sin\theta\sin\phi,\sqrt{f(r)}\cos\theta\right)\\
e^\theta_{(a)}&=\left(0,\frac{1}{r}\cos\theta \cos\phi,\frac{1}{r}\cos\theta \sin\phi,-\frac{\sin\theta}{r}\right)\\
e^\phi_{(a)}&=\left(0,-\frac{\sin\phi}{r\,\sin\theta},\frac{\cos\phi}{r\,\sin\theta},0\right)
\end{split}
\end{align}
to be 
\begin{align}
G^t&=e^t_{(a)}\gamma^{(a)}=\frac{\gamma^t}{\sqrt{f(r)}},\,\,\,\,\,\,G^r=e^r_{(a)}\gamma^{(a)}=\sqrt{f(r)}\gamma^r,\\
G^\theta&=e^\theta_{(a)}\gamma^{(a)}=\gamma^\theta,\,\,\,\,\,\,\,\,\,\,\,\,\,\,\,\,G^\varphi=e^\varphi_{(a)}\gamma^{(a)}=\gamma^\varphi,
\end{align}
where $\gamma^t,\,\gamma^r,\,\gamma^\theta$ and $\gamma^\varphi$ are the $\gamma-$matrices in ``polar coordinates"\footnote{Since the original $\gamma-$matrices are expressed in Cartesian coordinates with respect to the flat Minkowski spacetime, we must express them into spherical coordinates by the relation $\gamma^\mu=\vec{\gamma}\hat{\mu}$, where $\hat{\mu}=\hat{r},\hat{\theta},\hat{\varphi}$, $\vec{\gamma}=\gamma^{(1)}\hat{x}+\gamma^{(2)}\hat{y}+\gamma^{(3)}\hat{z}$, and also take into account the normalization factors that come up from the anti-commutation relation $\{\gamma^\mu, \gamma^\nu\}=\epsilon 2 g^{\mu\nu}$ where $g^{\mu\nu}$ the inverse metric tensor of the flat Minkowski space in spherical coordinates $g_{\mu\nu}=\text{diag}\left(-1,1,r^2,r^2\sin^2\theta\right)$. }
\begin{align}
\gamma^t&=\gamma^{(0)},\\
\label{gr}
\gamma^r&=\sin\theta\cos\varphi\,\gamma^{(1)}+\sin\theta\sin\varphi\,\gamma^{(2)}+\cos\theta\,\gamma^{(3)}\\
\label{gth}
\gamma^\theta&=\frac{1}{r}\left(\cos\theta\cos\varphi\,\gamma^{(1)}+\cos\theta\sin\varphi\,\gamma^{(2)}-\sin\theta\,\gamma^{(3)}\right),\\
\label{gf}
\gamma^\varphi&=\frac{1}{r\,\sin\theta}\left(-\sin\varphi\,\gamma^{(1)}+\cos\varphi\,\gamma^{(2)}\right).
\end{align}
It is easy to verify that
\begin{align}
\left(\gamma^t\right)^2=\mathbf{1},\,\,\,\,\, \left(\gamma^r\right)^2=-\mathbf{1},\,\,\,\,\,\left(\gamma^\theta\right)^2=-\frac{\mathbf{1}}{r^2},\,\,\,\,\,\left(\gamma^\varphi\right)^2=-\frac{\mathbf{1}}{r^2\sin^2\theta}.
\end{align}
To consider fluctuations of a spin $1/2$ particle of mass $m_f$ we utilize the Dirac equation in curved spacetime \cite{Fock:1929vt}
\begin{equation}
\label{dirac equation}
(iG^\mu \mathcal{D}_\mu-m_f)\Psi=(i\gamma^{(a)}e^\mu_{(a)}\mathcal{D}_\mu-m_f)\Psi=0,
\end{equation}
with the covariant derivative 
\begin{equation}
\mathcal{D}_\mu=\partial_\mu-iqA_\mu+\Gamma_\mu,
\end{equation}
where $q$ the charge of the Dirac particle, $A=-(Q/r) dt$ the electromagnetic potential sourced by a charge $Q$, and $\Gamma_\mu$ the spin connection coefficients defined as \cite{Carroll:2004st,Collas:2018jfx}
\begin{equation}
\label{spin}
\Gamma_\mu=\frac{\epsilon}{4}\omega_{(a)(b)\mu}\gamma^{(a)}\gamma^{(b)}=\frac{\epsilon}{8}\omega_{(a)(b)\mu}\left[\gamma^{(a)},\gamma^{(b)}\right].
\end{equation}
The spin connection $\omega_{(a)(b)\mu}$ is defined as \cite{Carroll:2004st,Collas:2018jfx}
\begin{align}
\label{omega}
\omega_{(a)(b)\mu}&=\eta_{(a)(c)}\omega^{(c)}_{\,\,\,\,(b)\mu}\\
&=\eta_{(a)(c)}\left(e^{(c)}_\nu e^\lambda_{(b)}\Gamma^\nu_{\mu\lambda}-e^\lambda_{(b)}\partial_\mu e^{(c)}_\lambda\right)
\end{align}
with $e^\lambda_{(b)}$, $e^{(c)}_\nu$ defined by (\ref{tetrad}) and the inverse of (\ref{tetrad}), respectively, and $\Gamma^\nu_{\mu\lambda}$ the Christoffel symbols of (\ref{metric3}). Using $\epsilon=-1$ and (\ref{mink}), (\ref{spin}), (\ref{omega}) we obtain the following spin connection coefficients 
\begin{align*}
\Gamma_0&=\frac{f^\prime(r)}{4}\gamma^{(0)}\gamma^{r},\,\,\,\,\,\,\,\,\,\,\,\,\,\,\,\,\,\,\,\,\,\,\,\,\,\,\,\,\,\,\,\,\,\,\,\,\,\,\,\,\Gamma_1=0,\\ \Gamma_2&=\frac{1}{2}\left(\sqrt{f(r)}-1\right)r\gamma^{r}\gamma^{\theta},\,\,\,\,\,\,\,\,\,\,\,\,\,\,\Gamma_3=\frac{1}{2}\left(\sqrt{f(r)}-1\right)r\sin^2\theta\gamma^r\gamma^\varphi,
\end{align*}
where 
\begin{align}
\gamma^r\gamma^\theta&=\frac{1}{r}\left(\cos\varphi\,\gamma^{(3)}\gamma^{(1)}+\sin\varphi\,\gamma^{(3)}\gamma^{(2)}\right),\\
\gamma^r\gamma^\varphi&=\frac{1}{r\sin\theta}\left(\cos\theta\sin\varphi\,\gamma^{(1)}\gamma^{(3)}+\cos\theta\cos\varphi\,\gamma^{(3)}\gamma^{(2)}+\sin\theta\,\gamma^{(1)}\gamma^{(2)}\right).
\end{align}
Plugging in all the above into the Dirac equation (\ref{dirac equation}) we get
\begin{align}
\nonumber
i\frac{\gamma^t}{\sqrt{f}}\frac{\partial\Psi}{\partial t}+i\gamma^{\theta}\frac{\partial\Psi}{\partial\theta}+i\gamma^{\varphi}\frac{\partial\Psi}{\partial\varphi}+i\gamma^{r}\left(\sqrt{f(r)}\frac{\partial}{\partial r}+\frac{\sqrt{f(r)}-1}{r}+\frac{\left(\sqrt{f(r)}\right)^\prime}{2}\right)\Psi\\
\label{dirac_1}
+\frac{\gamma^{t}}{\sqrt{f}}q A_t\Psi-m\Psi=0.
\end{align}
By choosing the ansatz $\Psi\rightarrow f^{-1/4}r^{-1}\psi$, and utilizing $A_\mu$, (\ref{dirac_1}) can be written in a simpler form
\begin{equation}
\label{dirac_2}
\frac{i}{\sqrt{f}}\gamma^t\frac{\partial}{\partial t}\psi+i\sqrt{f}\gamma^r\frac{\partial}{\partial r}\psi-\frac{i}{r}\gamma^r\psi+i\left(\gamma^\theta\frac{\partial}{\partial\theta}+\gamma^\varphi\frac{\partial}{\partial\varphi}\right)\psi-\gamma^{t}\frac{q Q}{r\sqrt{f}}\psi-m_f\psi=0.
\end{equation}
\subsection{Separation of the angular and time dependence}
Since the external fields are spherically symmetric and time-independent, we can separate out the angular and time dependence of the wave functions via spherical harmonics and plane waves, respectively. We start by defining the angular momentum operator together with a compilation of some formulas:
\begin{align}
\vec{L}&=-i\left(\vec{r}\times\vec{\nabla}\right),\\
L_\pm&=L_x\pm i L_y,\\
L^2&=-\Delta_{S^2}=L_+ L_- + L^2_z -L_z=L_- L_+ L^2_z +L_z,
\end{align}
where $\Delta_{S^2}$ the Laplacian in spherical coordinates. The spherical harmonics $Y^k_l$, $l=0,1,\dots$, $k=-l,\dots,l$ form a basis for square integrable functions over $S^2$, that is every square integrable function over $S^2$ can be expressed as a linear combination of spherical harmonics. They are simultaneous eigenfunctions of $L^2$ and $L_z$, namely
\begin{equation}
L^2 Y^k_l=l(l+1)Y^k_l,\,\,\,\,\,\,\,\,\,\,\,\,\,L_z Y^k_l=k Y^k_l.
\end{equation}
They are orthonormal,
\begin{equation}
\label{orthoY}
\int_{S^2}^{} (Y^k_l)^\dagger Y^{k^\prime}_{l^\prime}d\Omega=\delta_{ll^\prime}\delta^{kk^\prime}
\end{equation}
and the operators $L_\pm$ serve as ladder operators,
\begin{equation}
L_\pm Y^k_l=\sqrt{l(l+1)-k(k\pm 1)}Y^{k\pm 1}_l.
\end{equation}
In analogy to (\ref{gr})-(\ref{gf}), we define the Pauli matrices in "spherical coordinates" as follows
\begin{align}
\label{sr}
\sigma^r&=\sin\theta\cos\varphi\,\sigma^{1}+\sin\theta\sin\varphi\,\sigma^{2}+\cos\theta\,\sigma^{3}\\
\label{sth}
\sigma^\theta&=\frac{1}{r}\left(\cos\theta\cos\varphi\,\sigma^{1}+\cos\theta\sin\varphi\,\sigma^{2}-\sin\theta\,\sigma^{3}\right),\\
\label{sf}
\sigma^\varphi&=\frac{1}{r\,\sin\theta}\left(-\sin\varphi\,\sigma^{1}+\cos\varphi\,\sigma^{2}\right),
\end{align}
with
\begin{equation*}
\left(\sigma^r\right)^2=\mathbf{1},\,\,\,\,\,\,\,\,\,\,\,\,\,\left(\sigma^\theta\right)^2=\frac{\mathbf{1}}{r^2},\,\,\,\,\,\,\,\,\,\,\,\,\,\left(\sigma^\varphi\right)^2=\frac{\mathbf{1}}{r^2\sin^2\theta}.
\end{equation*}
From (\ref{dirac_2}) we immediately see that the angular part of the Dirac equation is located in the terms $i(\gamma^{\theta}\partial_\theta+\gamma^{\varphi}\partial_\phi)$ which will translate into
\begin{align}
\nonumber
\sigma^\theta\partial_\theta+\sigma^{\varphi}\partial_\phi&=\vec{\sigma}
\vec{\nabla}-\sigma^r\partial_r=\frac{\sigma^r}{r}(\vec{\sigma} \vec{r})(\vec{\sigma} \vec{\nabla}-\sigma^r\partial_r)\\
\label{sL}
&=\frac{\sigma^r}{r}(r\partial_r+i\vec{\sigma}(\vec{r}\times\vec{\nabla})-r\partial_r)=-\frac{\sigma^r}{r}\vec{\sigma}\vec{L},
\end{align}
where the position vector, Pauli vector and differential operator in spherical coordinates are
\begin{align*}
\vec{r}&=r\hat{r},\\
\vec{\sigma}&=\sigma^1\hat{x}+\sigma^2\hat{y}+\sigma^3\hat{z}=\sigma^r \hat{r}+r\sigma^\theta \hat{\theta}+r\sin\theta\sigma^\varphi \hat{\varphi},\\
\vec{\nabla}&=\partial_r\hat{r}+\frac{\hat{\theta}}{r}\partial_\theta+\frac{\hat{\varphi}}{r\sin\theta}\partial_\varphi,
\end{align*}
respectively, and the identity $(\vec{\sigma}\mathbf{A})(\vec{\sigma}\mathbf{B})=\mathbf{A}\mathbf{B}+i\vec{\sigma}(\mathbf{A}\times\mathbf{B})$ was used. Thus, from (\ref{sL}) we get
\begin{equation}
\label{sl}
\vec{\sigma}\vec{L}=-r\sigma^r(\sigma^\theta\partial_\theta+\sigma^{\varphi}\partial_\phi).
\end{equation}
For $j=\frac{1}{2},\frac{3}{2},\dots$ and $k=-j,-j+1,\dots,j$ we introduce the spinor spherical harmonics \cite{Finster:1998ak}
\begin{align}
\label{phi_j-}
\phi^k_{j-1/2}&=\begin{pmatrix}
\sqrt{\frac{j+k}{2j}}Y^{k-1/2}_{j-1/2}\\
\sqrt{\frac{j-k}{2j}}Y^{k+1/2}_{j-1/2}
\end{pmatrix}\,\,\,\,\,\,\,\,\,\,\,\,\,\,\,\,\,\,\,\,\,\text{for}\,\,\,\,j=l+\frac{1}{2}\\
\label{phi_j+}
\phi^k_{j+1/2}&=\begin{pmatrix}
\sqrt{\frac{j+1-k}{2j+2}}Y^{k-1/2}_{j+1/2}\\
-\sqrt{\frac{j+1+k}{2j+2}}Y^{k+1/2}_{j+1/2}
\end{pmatrix}\,\,\,\,\,\,\,\,\,\,\,\text{for}\,\,\,\,j=l-\frac{1}{2}
\end{align}
If $l=0$, only $j=1/2$ is possible for (\ref{phi_j-}) and (\ref{phi_j+}) is omitted. These spinors are orthonormal in accordance with (\ref{orthoY}),
\begin{align}
\label{orthoF}
\int_{S^2}^{} (\phi^k_{j\pm 1/2})^\dagger \phi^{k^\prime}_{j^\prime\pm 1/2}d\Omega&=\delta_{jj^\prime}\delta^{kk^\prime},\\
\int_{S^2}^{} (\phi^k_{j\pm 1/2})^\dagger \phi^{k^\prime}_{j^\prime\mp 1/2}d\Omega&=0,
\end{align}
they form a basis for square integrable functions over $S^2$ for each of the two components of the spinor and are eigenvectors of the operator $K=\vec{\sigma}\vec{L}+\mathbf{1}$. More precisely,
\begin{align}
\nonumber
K \phi^k_{j-1/2}&=\left(\vec{\sigma}\vec{L}+\mathbf{1}\right)\phi^k_{j-1/2}=(\sigma^1L_x+\sigma^2L_y+\sigma^3L_z+\mathbf{1})\phi^k_{j-1/2}
\\
\label{K1}
&=\begin{pmatrix}
L_z+1&L_-\\
L_+&-L_z+1
\end{pmatrix}\phi^k_{j-1/2}=(j+\frac{1}{2})\phi^k_{j-1/2}\\
\label{K2}
K \phi^k_{j+1/2}&=-(j+\frac{1}{2})\phi^k_{j+1/2}
\end{align}
Furthermore, multiplication with $\sigma^r$ gives an eigenvector of K, namely
\begin{align}
\nonumber
K\sigma^r\phi^k_{j-1/2}&=\left(-r\sigma^r\left(\sigma^\theta\partial_\theta+\sigma^\varphi\partial_\varphi\right)+1\right)\sigma^r\phi^k_{j-1/2}\\
\nonumber
&=-\sigma^r\phi^k_{j-1/2}-r\sigma^r\left(\sigma^\theta\sigma^r\partial_\theta+\sigma^\varphi\sigma^r\partial_\varphi\right)\phi^k_{j-1/2}\\
\nonumber
&=-\sigma^r\phi^k_{j-1/2}-\sigma^r\left(\vec{\sigma}\vec{L}\right)\phi^k_{j-1/2}\\
\label{Ks}
&=-\sigma^r K\phi^k_{j-1/2}=-(j+\frac{1}{2})\sigma^r\phi^k_{j-1/2},
\end{align}
where we used
\begin{align*}
\sigma^\theta\sigma^r&=-i\sin\theta\sigma^\varphi,\\
\sigma^\varphi\sigma^r&=\frac{i}{\sin\theta}\sigma^\theta,\\
\vec{L}&=-i\left(\vec{r}\times\vec{\nabla}\right)=i\left(\frac{\hat{\theta}}{\sin\theta}\partial_\varphi-\partial_\theta\hat{\varphi}\right),\\
\vec{\sigma}\vec{L}&=ir\left(\frac{\sigma^\theta}{\sin\theta}\partial_\varphi-\sin\theta\sigma^\varphi\partial_\theta\right).
\end{align*}
Taking into account the normalization factors (\ref{orthoF}), we immediately see from (\ref{K1})-(\ref{Ks}) that
\begin{align}
\label{ident1}
K(\sigma^r\phi^k_{j-1/2})&=-(j+\frac{1}{2})(\sigma^r\phi^k_{j-1/2})\Longleftrightarrow
\sigma^r\phi^k_{j-1/2}=\phi^k_{j+1/2},\\
\label{ident2}
K(\sigma^r\phi^k_{j+1/2})&=(j+\frac{1}{2})(\sigma^r\phi^k_{j+1/2})\Longleftrightarrow\sigma^r\phi^k_{j+1/2}=\phi^k_{j-1/2}.
\end{align}
\subsection{Radial Dirac equation}
For the Dirac wavefunctions, we choose the ansatz
\begin{align}
\label{spinor1}
\psi^+_{jk\omega}&=e^{-i\omega t}\begin{pmatrix}
\phi^k_{j-1/2}\Phi^+_1(r)\\
i\phi^k_{j+1/2}\Phi^+_2(r)
\end{pmatrix},\\
\label{spinor2}
\psi^-_{jk\omega}&=e^{-i\omega t}\begin{pmatrix}
\phi^k_{j+1/2}\Phi^-_1(r)\\
i\phi^k_{j-1/2}\Phi^-_2(r)
\end{pmatrix}.
\end{align}
A general solution of the Dirac equation can be written as a linear combination of these wave functions, because one can obtain every combination of spherical harmonics in the four spinor components. Substituting (\ref{spinor1}) into (\ref{dirac_2}) gives
\begin{align}
\nonumber
\frac{\omega-qQ/r}{\sqrt{f}}\begin{pmatrix}
\mathbf{1}&0\\
0&-\mathbf{1}
\end{pmatrix}\begin{pmatrix}
\phi^k_{j-1/2}\Phi^+_1\\
i\phi^k_{j+1/2}\Phi^+_2
\end{pmatrix}+i\sqrt{f}\begin{pmatrix}
0 &\sigma^r\\
-\sigma^r&0
\end{pmatrix}\begin{pmatrix}
\phi^k_{j-1/2}\frac{d\Phi^+_1}{dr}\\
i\phi^k_{j+1/2}\frac{d\Phi^+_2}{dr}
\end{pmatrix}\\
-\frac{i}{r}\begin{pmatrix}
0 &\sigma^r\\\nonumber
-\sigma^r&0
\end{pmatrix}\begin{pmatrix}
\phi^k_{j-1/2}\Phi^+_1\\
i\phi^k_{j+1/2}\Phi^+_2
\end{pmatrix}+i\begin{pmatrix}
0&\sigma^\theta\partial_\theta+\sigma^\varphi\partial_\varphi\\
-(\sigma^\theta\partial_\theta+\sigma^\varphi\partial_\varphi)&0
\end{pmatrix}\begin{pmatrix}
\phi^k_{j-1/2}\Phi^+_1\\
i\phi^k_{j+1/2}\Phi^+_2
\end{pmatrix}\\\label{psi+}
-m_f\begin{pmatrix}
\mathbf{1}&0\\
0&\mathbf{1}
\end{pmatrix}\begin{pmatrix}
\phi^k_{j-1/2}\Phi^+_1\\
i\phi^k_{j+1/2}\Phi^+_2
\end{pmatrix}=\begin{pmatrix}
0\\0
\end{pmatrix}
\end{align}
By utilizing (\ref{sl}) and (\ref{K1})-(\ref{ident2}), (\ref{psi+}) becomes
\begin{align*}
\frac{\omega-qQ/r}{\sqrt{f}}\begin{pmatrix}
\phi^k_{j-1/2}\Phi^+_1\\
-i\phi^k_{j+1/2}\Phi^+_2
\end{pmatrix}-\sqrt{f}\begin{pmatrix}
\phi^k_{j-1/2}\frac{d\Phi^+_2}{dr}\\
i\phi^k_{j+1/2}\frac{d\Phi^+_1}{dr}
\end{pmatrix}+\frac{1}{r}\begin{pmatrix}
-(j+\frac{1}{2})\phi^k_{j-1/2}\Phi^+_2\\
i(j+\frac{1}{2})\phi^k_{j+1/2}\Phi^+_1
\end{pmatrix}\\
-m_f\begin{pmatrix}
\phi^k_{j-1/2}\Phi^+_1\\
i\phi^k_{j+1/2}\Phi^+_2
\end{pmatrix}=\begin{pmatrix}
0\\0
\end{pmatrix}
\end{align*}
which gives the two following coupled differential equations
\begin{align}
\label{1}
\sqrt{f}\frac{d\Phi^+_1}{dr}+\frac{\omega-qQ/r}{\sqrt{f}}\Phi^+_2-\frac{(j+1/2)}{r}\Phi^+_1+m_f\Phi^+_2&=0,\\
\label{2}
\sqrt{f}\frac{d\Phi^+_2}{dr}-\frac{\omega-qQ/r}{\sqrt{f}}\Phi^+_1+\frac{(j+1/2)}{r}\Phi^+_2+m_f\Phi^+_1&=0.
\end{align}
Following exactly the same procedure for (\ref{spinor2}) we get
\begin{align}
\label{3}
\sqrt{f}\frac{d\Phi^-_1}{dr}+\frac{\omega-qQ/r}{\sqrt{f}}\Phi^-_2+\frac{(j+1/2)}{r}\Phi^-_1+m_f\Phi^-_2&=0,\\
\label{4}
\sqrt{f}\frac{d\Phi^-_2}{dr}-\frac{\omega-qQ/r}{\sqrt{f}}\Phi^-_1-\frac{(j+1/2)}{r}\Phi^-_2+m_f\Phi^-_1&=0.
\end{align}
Since (\ref{1}), (\ref{3}) and (\ref{2}), (\ref{4}) differ only on the sign of $(j+1/2)$ we can compactify them by considering
\begin{equation}
\xi=\left\{\begin{matrix}
\xi_+=j+\frac{1}{2}\,\,\,\,\,\,\,\,\,\,\\
\xi_-=-(j+\frac{1}{2})
\end{matrix}\right.
\end{equation}
so when $\xi$ is positive we treat the system (\ref{1})-(\ref{2}) and when it is negative we treat (\ref{3})-(\ref{4}). Since the goal is to study wavefunctions with $j=\frac{1}{2},\,\frac{3}{2},\dots$, the integer $\xi$ will take values $\xi=\pm 1,\,\pm 2,\dots$.\footnote{$\xi_+$ appear when $K$ acts on (\ref{phi_j-}) with $l=j-1/2$ and $\xi_-$ appear when $K$ acts on (\ref{phi_j+}) with $l=j+1/2$. So, considering the spinors (\ref{spinor1}), (\ref{spinor2}), if $j=\frac{1}{2}$, then $\xi_+=1$ for $l=0$, since $l=j-1/2$, and $\xi_-=-1$ for $l=1$, since $l=j+1/2$. If $j=\frac{3}{2}$, then $\xi_+=2$ for $l=1$ and $\xi_-=-2$ for $l=2$ and so on.} This leads to the final coupled system of ordinary differential equations
\begin{align}
\label{final1}
f\frac{\partial F}{\partial r}-\frac{\xi\sqrt{f}}{r}F+\left(\omega -\frac{qQ}{r}\right)G+m_f\sqrt{f}G&=0,\\
\label{final2}
f\frac{\partial G}{\partial r}+\frac{\xi\sqrt{f}}{r}G-\left(\omega -\frac{qQ}{r}\right)F+m_f\sqrt{f}F&=0,
\end{align}
where we set $\Phi_1^+(r)=\Phi_1^-(r)=F(r)$ and $\Phi_2^+(r)=\Phi_2^-(r)=G(r)$ for simplicity.

A complementary way to write the system (\ref{final1}), (\ref{final2}) in a different form comes by performing the following transformation
\begin{align}
\label{trans1}
R_+=F-iG&\rightarrow iR_+=iF+G\\
\label{trans2}
R_-=-F-iG&\rightarrow iR_-=-iF+G.
\end{align}
Then, by performing (\ref{final1})$-i$(\ref{final2}) we obtain
\begin{align}
\frac{dR_+}{dr_*}+i\left(\omega-\frac{qQ}{r}\right)R_+ +\frac{\xi\sqrt{f}}{r}R_-+im_f\sqrt{f}R_-=0,
\end{align}
and by performing -(\ref{final1})$-i$(\ref{final2}) we obtain
\begin{equation}
\frac{dR_-}{dr_*}-i\left(\omega-\frac{qQ}{r}\right)R_- +\frac{\xi\sqrt{f}}{r}R_+-im_f\sqrt{f}R_+=0,
\end{equation}
where $r_*$ the tortoise coordinate defined as
\begin{equation}
\label{tort}
dr_*=\frac{dr}{f(r)}.
\end{equation}
\subsection{The Dirac equation in Schr\"{o}dinger-like form}
To study the propagation of massless charged fermions in a spherically symmetric background we set $m_f=0$ in (\ref{final1}) and (\ref{final2}) to obtain
\begin{align}
\label{finalm01}
f\frac{\partial F}{\partial r}-\frac{\xi\sqrt{f}}{r}F+\left(\omega -\frac{qQ}{r}\right)G&=0,\\
\label{finalm02}
f\frac{\partial G}{\partial r}+\frac{\xi\sqrt{f}}{r}G-\left(\omega -\frac{qQ}{r}\right)F&=0.
\end{align}
By introducing a new coordinate
\begin{equation}
\label{new_tort}
d\bar{r}_*=\frac{\left(1-\frac{qQ}{r\omega}\right)}{f}dr,
\end{equation}
(\ref{finalm01}), (\ref{finalm02}) become
\begin{align}
\label{almdec1}
\frac{dF}{d\bar{r}_*}-WF+\omega G&=0,\\
\label{almdec2}
\frac{dG}{d\bar{r}_*}+WG-\omega F&=0,
\end{align}
where 
\begin{equation}
W=\frac{\xi\sqrt{f}}{r\left(1-\frac{qQ}{r\omega}\right)}.
\end{equation}
(\ref{almdec1}) and (\ref{almdec2}) are coupled first order ordinary differential equations which can be decoupled by solving (\ref{almdec1}) with respect to $G$ and plug it into (\ref{almdec2}) and vice versa. By doing so we get the decoupled equations
\begin{align}
\label{schroedinger1}
\frac{d^2F}{d\bar{r}_*^2}+\left(\omega^2-V_+\right)F&=0,\\
\label{schroedinger2}
\frac{d^2G}{d\bar{r}_*^2}+\left(\omega^2-V_-\right)G&=0,
\end{align}
with 
\begin{equation}
V_\pm=\pm\frac{dW}{d\bar{r}_*}+W^2.
\end{equation}

It has been proven \cite{Anderson:1991kx} that potentials related in this manner are supersymmetric partners and are equispectral.

\subsection{The Dirac equation in Eddington-Finkelstein coordinates}
The ingoing Eddington-Finkelstein coordinates are obtained by replacing the $t$ coordinate with the new coordinate $\upsilon=t+r_*$, where $r_*$ the tortoise coordinate defined in the previous section. The metric in these coordinates can be written as
\begin{equation}
\label{ingoing3}
ds^2=-f(r)d\upsilon^2+2d\upsilon dr +r^2(d\theta^2+\sin^2\theta\, d\varphi^2),
\end{equation}
with the associated electromagnetic potential\footnote{Here, if we want to transform from the coordinate system $(t,r,\theta,\varphi)$ to $(\upsilon,r,\theta,\varphi)$ then the components of the potential will be $A_\mu=\text{diag}(-Q/r,-Q/rf,0,0)$ while if we are already working on the coordinate system $(\upsilon,r,\theta,\varphi)$ then the components of the potential will be ${A}_\mu=\text{diag}(-Q/r,0,0,0)$.}
\begin{equation}
\label{inpot3}
{A}=-\frac{Q}{r}d\upsilon=-\frac{Q}{r}(dt+dr_*)=-\frac{Q}{r}\left(dt+\frac{dr}{f(r)}\right).
\end{equation}
Similarly, the outgoing Eddington-Finkelstein coordinates are obtained by replacing $t$ with $u=t-r_*$ to get the metric
\begin{equation}
\label{outgoing3}
ds^2=-f(r)du^2-2du dr +r^2(d\theta^2+\sin^2\theta\, d\varphi^2),
\end{equation}
with the associated electromagnetic potential\footnote{Here, if we want to transform from the coordinate system $(t,r,\theta,\varphi)$ to $(u,r,\theta,\varphi)$ then the components of the potential will be $A_\mu=\text{diag}(-Q/r,Q/rf,0,0)$ while if we are already working on the coordinate system $(u,r,\theta,\varphi)$ then the components of the potential will be ${A}_\mu=\text{diag}(-Q/r,0,0,0)$.}
\begin{equation}
\label{outpot3}
{A}=-\frac{Q}{r}du=-\frac{Q}{r}(dt-dr_*)=-\frac{Q}{r}\left(dt-\frac{dr}{f(r)}\right).
\end{equation}
A straightforward way to write down the Dirac equation (\ref{dirac equation}) in the new coordinates is to choose a new tetrad that reproduces either (\ref{ingoing3}) or (\ref{outgoing3}) with (\ref{tetradgmn}) and follow the same process as before. An alternative way is to transform (\ref{dirac_2}) in the new coordinates $(\upsilon,r)$ or $(u,r)$ with the associated transformed electromagnetic potential (\ref{inpot3}) or (\ref{outpot3}), respectively. 

Following the same procedure as before we write (\ref{dirac_2}) now with non-zero $A_r$ component
\begin{equation}
\label{dirac_33}
\frac{i}{\sqrt{f}}\gamma^t\frac{\partial}{\partial t}\psi+i\sqrt{f}\gamma^r\frac{\partial}{\partial r}\psi-\frac{i}{r}\gamma^r\psi+i\left(\gamma^\theta\frac{\partial}{\partial\theta}+\gamma^\varphi\frac{\partial}{\partial\varphi}\right)\psi+\frac{q}{\sqrt{f}}\left(\gamma^{t} A_t+f\gamma^r A_r\right)\psi-m_f\psi=0.
\end{equation}
\subsubsection{Ingoing Eddington-Finkelstein coordinates}
Performing the coordinate transformation $\upsilon=t+r_*$ we have the following chain rule expressions
\begin{align}
\frac{\partial}{\partial t}&=\frac{\partial}{\partial\upsilon}\frac{\partial \upsilon}{\partial t}+\frac{\partial}{\partial r}\frac{\partial r}{\partial t}=\partial_\upsilon,\\
\frac{\partial}{\partial r}&=\frac{\partial}{\partial\upsilon}\frac{\partial \upsilon}{\partial r}+\frac{\partial}{\partial r}\frac{\partial r}{\partial r}=\frac{\partial r_*}{\partial r}\partial_\upsilon+\partial_r=\frac{\partial_\upsilon}{f}+\partial_r,
\end{align}
and $A_\mu=\text{diag}(-Q/r,-Q/rf,0,0)$, which will replace the corresponding derivatives in (\ref{dirac_33}), leading to
\begin{align}
\nonumber
\frac{i\gamma^t}{\sqrt{f}}\partial_\upsilon\psi+i\sqrt{f}\gamma^r\left(\frac{\partial_\upsilon}{f}+\partial_r\right)\psi-\frac{i}{r}\gamma^r\psi+i\left(\gamma^\theta\frac{\partial}{\partial\theta}+\gamma^\varphi\frac{\partial}{\partial\varphi}\right)\psi-\gamma^{t}\frac{qQ}{r\sqrt{f}}\psi-\gamma^r \frac{qQ }{r \sqrt{f}}\psi\\\label{dirac_43}-m_f\psi=0.
\end{align}
By utilizing the ansatz
\begin{align}
\label{spinorin33}
\psi^+&=e^{-i\omega\upsilon}\begin{pmatrix}
\phi^k_{j-1/2}\Phi^+_1(r)\\
i\phi^k_{j+1/2}\Phi^+_2(r)
\end{pmatrix},\\
\label{spinorin43}
\psi^-&=e^{-i\omega\upsilon}\begin{pmatrix}
\phi^k_{j+1/2}\Phi^-_1(r)\\
i\phi^k_{j-1/2}\Phi^-_2(r)
\end{pmatrix}.
\end{align}
we get
\begin{align}
\label{ed13}
\frac{\omega}{\sqrt{f}}\Phi_1^++i\frac{\omega}{\sqrt{f}}\Phi_2^+-\sqrt{f}\partial_r\Phi_2^+-\frac{\xi}{r}\Phi_2^+-\frac{q Q}{r\sqrt{f}}\Phi_1^+-i\frac{qQ}{r\sqrt{f}}\Phi_2^+-m_f\Phi^+_1&=0,\\
\label{ed23}
\frac{\omega}{\sqrt{f}}\Phi_2^+-i\frac{\omega}{\sqrt{f}}\Phi_1^++\sqrt{f}\partial_r\Phi_1^+-\frac{\xi}{r}\Phi_1^+-\frac{q Q}{r\sqrt{f}}\Phi_2^++i\frac{qQ}{r\sqrt{f}}\Phi_1^++m_f\Phi_2^+&=0,
\end{align}
and
\begin{align}
\label{ed33}
\frac{\omega}{\sqrt{f}}\Phi_1^-+i\frac{\omega}{\sqrt{f}}\Phi_2^--\sqrt{f}\partial_r\Phi_2^-+\frac{\xi}{r}\Phi_2^--\frac{q Q}{r\sqrt{f}}\Phi_1^--i\frac{qQ}{r\sqrt{f}}\Phi_2^--m_f\Phi^-_1&=0,\\
\label{ed43}
\frac{\omega}{\sqrt{f}}\Phi_2^--i\frac{\omega}{\sqrt{f}}\Phi_1^-+\sqrt{f}\partial_r\Phi_1^-+\frac{\xi}{r}\Phi_1^--\frac{q Q}{r\sqrt{f}}\Phi_2^-+i\frac{qQ}{r\sqrt{f}}\Phi_1^-+m_f\Phi_2^-&=0,
\end{align}
respectively, where $\xi=j+1/2$. Since (\ref{ed23}), (\ref{ed43}) and (\ref{ed13}), (\ref{ed33}) differ only in the sign of $\xi$ we can compactify them, as follows, to obtain the Dirac equation for spin $1/2$ particles in ingoing Eddington-Finkelstein coordinates
\begin{align}
\label{edf13}
f\partial_rF-\frac{\xi\sqrt{f}}{r}F+\left(\omega-\frac{qQ}{r}\right)G-i\left(\omega-\frac{qQ}{r}\right)F+m_f\sqrt{f}G&=0,\\
\label{edf23}
f\partial_rG+\frac{\xi\sqrt{f}}{r}G-\left(\omega-\frac{q Q}{r}\right) F-i\left(\omega-\frac{qQ}{r}\right)G+m_f\sqrt{f}F&=0,
\end{align}
where we set $\Phi_1^+=\Phi_1^-=F$, $\Phi_2^+=\Phi_2^-=G$ and $\xi=\pm 1,\,\pm 2,\dots$. Utilizing (\ref{trans1}), (\ref{trans2}) and performing (\ref{edf13})$-i$(\ref{edf23}) and -(\ref{edf13})$-i$(\ref{edf23}) we can rewrite the previous set of equation in a simpler form
\begin{align}
f\partial_r R_++\frac{\xi\sqrt{f}}{r}R_-+im_f\sqrt{f}R_-&=0,\\
f\partial_r R_-+\frac{\xi\sqrt{f}}{r}R_+-2i\left(\omega-\frac{qQ}{r}\right)R_--im_f\sqrt{f}R_+&=0.
\end{align}
\subsubsection{Outgoing Eddington-Finkelstein coordinates}
Performing the coordinate transformation $u=t-r_*$ we have the following chain rule expressions
\begin{align}
\frac{\partial}{\partial t}&=\frac{\partial}{\partial u}\frac{\partial  u}{\partial t}+\frac{\partial}{\partial r}\frac{\partial r}{\partial t}=\partial_u,\\
\frac{\partial}{\partial r}&=\frac{\partial}{\partial u}\frac{\partial  u}{\partial r}+\frac{\partial}{\partial r}\frac{\partial r}{\partial r}=-\frac{\partial r_*}{\partial r}\partial_u+\partial_r=-\frac{\partial_u}{f}+\partial_r,
\end{align}
and $A_\mu=\text{diag}(-Q/r,Q/rf,0,0)$, which will replace the corresponding derivatives in (\ref{dirac_33}), leading to
\begin{align}
\nonumber
\frac{i\gamma^t}{\sqrt{f}}\partial_u\psi+i\sqrt{f}\gamma^r\left(-\frac{\partial_u}{f}+\partial_r\right)\psi-\frac{i}{r}\gamma^r\psi+i\left(\gamma^\theta\frac{\partial}{\partial\theta}+\gamma^\varphi\frac{\partial}{\partial\varphi}\right)\psi-\gamma^{t}\frac{qQ}{r\sqrt{f}}\psi+\gamma^r \frac{qQ }{r \sqrt{f}}\psi\\\label{dirac_54}-m_f\psi=0.
\end{align}
By utilizing the ansatz
\begin{align}
\label{spinorout34}
\psi^+&=e^{-i\omega\upsilon}\begin{pmatrix}
\phi^k_{j-1/2}\Phi^+_1(r)\\
i\phi^k_{j+1/2}\Phi^+_2(r)
\end{pmatrix},\\
\label{spinorout44}
\psi^-&=e^{-i\omega\upsilon}\begin{pmatrix}
\phi^k_{j+1/2}\Phi^-_1(r)\\
i\phi^k_{j-1/2}\Phi^-_2(r)
\end{pmatrix}.
\end{align}
we get
\begin{align}
\label{edout14}
\frac{\omega}{\sqrt{f}}\Phi_1^+-i\frac{\omega}{\sqrt{f}}\Phi_2^+-\sqrt{f}\partial_r\Phi_2^+-\frac{\xi}{r}\Phi_2^+-\frac{q Q}{r\sqrt{f}}\Phi_1^++i\frac{qQ}{r\sqrt{f}}\Phi_2^+-m_f\Phi_1^+&=0,\\
\label{edout24}
\frac{\omega}{\sqrt{f}}\Phi_2^++i\frac{\omega}{\sqrt{f}}\Phi_1^++\sqrt{f}\partial_r\Phi_1^+-\frac{\xi}{r}\Phi_1^+-\frac{q Q}{r\sqrt{f}}\Phi_2^+-i\frac{qQ}{r\sqrt{f}}\Phi_1^++m_f\Phi_2^+&=0,
\end{align}
and
\begin{align}
\label{edout34}
\frac{\omega}{\sqrt{f}}\Phi_1^--i\frac{\omega}{\sqrt{f}}\Phi_2^--\sqrt{f}\partial_r\Phi_2^-+\frac{\xi}{r}\Phi_2^--\frac{q Q}{r\sqrt{f}}\Phi_1^-+i\frac{qQ}{r\sqrt{f}}\Phi_2^--m_f\Phi_1^-&=0,\\
\label{edout44}
\frac{\omega}{\sqrt{f}}\Phi_2^-+i\frac{\omega}{\sqrt{f}}\Phi_1^-+\sqrt{f}\partial_r\Phi_1^-+\frac{\xi}{r}\Phi_1^--\frac{q Q}{r\sqrt{f}}\Phi_2^--i\frac{qQ}{r\sqrt{f}}\Phi_1^-+m_f\Phi_2^-&=0,
\end{align}
respectively, where $\xi=j+1/2$. Since (\ref{edout24}), (\ref{edout44}) and (\ref{edout14}), (\ref{edout34}) differ only in the sign of $\xi$ we can compactify them, as follows, to obtain the Dirac equation for spin $1/2$ particles in outgoing Eddington-Finkelstein coordinates
\begin{align}
\label{edoutf14}
f\partial_rF-\frac{\xi\sqrt{f}}{r}F+\left(\omega-\frac{qQ}{r}\right)G+i\left(\omega-\frac{qQ}{r}\right)F+m_f\sqrt{f}G&=0,\\
\label{edoutf24}
f\partial_rG+\frac{\xi\sqrt{f}}{r}G-\left(\omega-\frac{q Q}{r}\right) F+i\left(\omega-\frac{qQ}{r}\right)G+m_f\sqrt{f}F&=0,
\end{align}
where we set $\Phi_1^+=\Phi_1^-=F$, $\Phi_2^+=\Phi_2^-=G$ and $\xi=\pm 1,\,\pm 2,\dots$. Utilizing (\ref{trans1}), (\ref{trans2}) and performing (\ref{edoutf14})$-i$(\ref{edoutf24}) and -(\ref{edoutf14})$-i$(\ref{edoutf24}) we can rewrite the previous set of equation in a simpler form
\begin{align}
f\partial_r R_++\frac{\xi\sqrt{f}}{r}R_-+2i\left(\omega-\frac{qQ}{r}\right)R_++im_f\sqrt{f}R_-&=0,\\
f\partial_r R_-+\frac{\xi\sqrt{f}}{r}R_+-im_f\sqrt{f}R_+&=0.
\end{align}
\chapter{The classical definition of $\beta$ for complex scalar fields in Reissner-Nordstr\"{o}m-de Sitter spacetime}\label{appC}

The RNdS spacetime is described by the metric 
\begin{equation}
\label{metricC}
ds^2=-f(r)dt^2+\frac{1}{f(r)}dr^2+r^2\left(d\theta^2+\sin^2\theta\,d\phi^2\right).
\end{equation}
The metric function reads
\begin{equation}
 f(r)=1-\frac{2M}{r}+\frac{Q^2}{r^2}-\frac{\Lambda r^2}{3}=\frac{\Lambda}{3r^2}\left(r-r_0\right)(r-r_-)(r-r_+)(r_c-r),
\end{equation}
where $M,\, Q$ the mass and charge of the black hole, $\Lambda$ the positive cosmological constant and $r_0<r_-<r_+<r_c$ with $r_0=-(r_-+r_++r_c)$ where $r_-,\,r_+,\,r_c$ the Cauchy, event and cosmological horizon radius, respectively. The associated electromagnetic potential of such a charged spacetime is
\begin{equation}
A=-\frac{Q}{r}dt,
\end{equation}
so
\begin{align}
A_\nu&=-\delta^0_\nu \frac{Q}{r},\\
A^\nu&=g^{\mu\nu}A_\mu=\delta^\nu_0\frac{Q}{rf(r)}.
\end{align}
To determine the regularity of the metric up to the CH we study the regularity of mode solutions of the wave equation $\Box\Psi=P\Psi=0$, there, where $P$ a differential operator to be determined. To do so, we change to outgoing Eddington-Finkelstein coordinates 
\begin{equation}
u=t-r_*,\,\,\,\,\,\,\,\,\,\,\,dr_*=dr/f(r),
\end{equation}
which are regular across the CH. By computing $dt=du+dr/f(r)$ the metric transforms to
\begin{align}
ds^2=-f(r)du^2-2du dr+r^2\left(d\theta^2+\sin^2\theta\, d\phi^2\right),
\end{align}
where the level sets of $u$ are null hypersurfaces transversal to the CH (parallel to the event horizon). The inverse metric reads
\begin{equation}
\partial s^{2}=f(r)\partial_r^2-2\partial_u\partial_r+r^{-2}\left(\partial_\theta^{2}+\sin^{-2}\theta\, \partial_\phi^2\right).
\end{equation}
The electromagnetic potential in the new coordinates will be
\begin{equation}
A=-\frac{Q}{r}dt=-\frac{Q}{r}du-\frac{Q}{rf(r)}dr,
\end{equation}
which immediately give rise to a singularity at $f(r_-)=0$. If we add a pure gauge term to the potential then at $r=r_-$ we have
\begin{equation}
\tilde{A}_\nu=A_\nu+\nabla_\nu\chi,
\end{equation}
with $\chi=Qr_*/r_-$, $\partial_\nu\chi=\delta^1_\nu Q/r_-f(r)$ then
\begin{equation}
\tilde{A}_\nu=-\delta^0_\nu\frac{Q}{r}-\delta^1_\nu\frac{Q}{rf(r)}+\partial_\nu \chi=-\delta^0_\nu\frac{Q}{r},
\end{equation}
or simply
\begin{equation}
\tilde{A}=-\frac{Q}{r}du,
\end{equation}
and we get rid of the singular terms. Consequently,
\begin{align}
P\Psi=\nabla^\nu\nabla_\nu\Psi-iq\nabla^\nu\left(\tilde{A}_\nu\Psi\right)-iq\tilde{A}^\nu\nabla_\nu\Psi-q^2\tilde{A}^\nu\tilde{A}_\nu\Psi -\mu^2\Psi.
\end{align}
The Christofel symbols in outgoing Eddington-Finkelstein coordinates are:
\begin{align}
\nonumber
\Gamma^0_{00}&=-\frac{f^\prime(r)}{2},\,\,\,\,\,\,\,\,\,\,\,\,\,\,\Gamma^0_{22}=r,\,\,\,\,\,\,\,\,\,\,\,\,\,\,\Gamma^0_{33}=r\sin^2\theta \\
\nonumber
\Gamma^1_{00}&=\frac{f(r)f^\prime(r)}{2},\,\,\,\,\,\, \Gamma^1_{10}=\Gamma^1_{01}=\frac{f^\prime(r)}{2},\,\,\,\,\,\, \Gamma^1_{22}=-r f(r),\,\,\,\,\,\,\Gamma^1_{33}=-r f(r)\sin^2\theta\\
\label{symbols}
\Gamma^2_{21}&=\Gamma^2_{12}=\frac{1}{r},\,\,\,\,\,\,\,\,\,\,\,\,\Gamma^2_{33}=-\cos\theta\sin\theta\\
\nonumber
\Gamma^3_{13}&=\Gamma^3_{31}=\frac{1}{r},\,\,\,\,\,\,\,\,\,\,\,\,\Gamma^3_{23}=\Gamma^3_{32}=\frac{\cos\theta}{\sin\theta}
\end{align}
Utilizing
\begin{align*}
\nabla_\nu\Psi&=\partial_\upsilon\Psi,\\
\nabla^\nu\tilde{A}_\nu&=g^{\mu\nu}\nabla_\mu g_{\kappa\nu}\tilde{A}^\kappa=\nabla_\mu\tilde{A^\mu}=\partial_\mu\tilde{A}^\nu+\Gamma^\mu_{\mu\nu}\tilde{A}^\nu=\partial_\mu g^{\kappa\nu}\tilde{A}_\kappa+\Gamma^\mu_{\mu\nu}g^{\lambda\nu}\tilde{A}_\lambda\\&=\partial_r g^{01}\tilde{A}_0+\Gamma^2_{21}g^{01}\tilde{A}_0+\Gamma^3_{31}g^{01}\tilde{A}_0=\frac{Q}{r^2},\\
\nabla^\nu\left(\tilde{A}_\nu\Psi\right)&=\Psi \nabla^\nu\tilde{A}_\nu+\tilde{A}_\nu\nabla^\nu\Psi=\frac{Q}{r^2}\Psi+\tilde{A}_\nu\partial^\nu\Psi=\frac{Q}{r^2}\Psi+g_{\mu\nu}\tilde{A}^\mu g^{\kappa\nu}\partial_\kappa\Psi\\&=\frac{Q}{r^2}\Psi+\delta^\kappa_\mu\tilde{A}^\mu\partial_\kappa\Psi=\frac{Q}{r^2}\Psi+\tilde{A}^\mu\partial_\mu\Psi,\\
\tilde{A}^\nu\tilde{A}_\nu&=g^{\mu\nu}\tilde{A}_\mu\tilde{A}_\nu=0,
\end{align*}
we get
\begin{align}
\nonumber
P\Psi&=g^{\mu\nu}\nabla_\mu\left(\partial_\nu\Psi\right)-\frac{iqQ}{r^2}\Psi-2iq\tilde{A}^\nu\partial_\nu\Psi-\mu^2\Psi\\
&=\frac{1}{\sqrt{-g}}\partial_\mu \left(g^{\mu\nu}\sqrt{-g}\partial_\nu\Psi\right)-\frac{iqQ}{r^2}\Psi-2iq\,g^{\mu\nu}\tilde{A}_\mu\partial_\nu\Psi-\mu^2\Psi,
\end{align}
where $g=\det(g_{\mu\nu})$. Proceeding with the calculation we get
\begin{align}
\nonumber
P\Psi&=-\frac{1}{r^2}\partial_u\left(r^2\partial_r\Psi\right)-\frac{1}{r^2}\partial_r\left(r^2\partial_u\Psi\right)+\frac{1}{r^2}\partial_r\left( fr^2\partial_r\Psi\right)+\frac{1}{r^2}\left(\frac{1}{\sin\theta}\partial_\theta(\sin\theta\partial_\theta)+\frac{1}{\sin^2\theta}\partial_\varphi^2\right)\Psi\\\nonumber
&-\frac{iqQ}{r^2}\Psi-2iq\frac{Q}{r}\partial_r\Psi-\mu^2\Psi\\
&=-\partial_u\partial_r\Psi-\frac{1}{r^2}\left(2r\partial_u+r^2\partial_r\partial_u\right)\Psi+\frac{1}{r^2}\partial_r\left( fr^2\partial_r\right)\Psi+\frac{\boldsymbol{L}^2}{r^2}\Psi-\frac{iqQ}{r^2}\Psi-2iq\frac{Q}{r}\partial_r\Psi-\mu^2\Psi,
\end{align}
where ${L}^2$ the square angular momentum operator in spherical coordinates and $\partial_t=\partial_u$. Finally, we find
\begin{equation}
fP=\frac{f}{r^2}\partial_r(fr^2\partial_r)-\frac{2f}{r}\partial_r(r\partial_u)+\frac{f{L}^2}{r^2}-f\frac{iqQ}{r^2}-2f\frac{iqQ}{r}\partial_r-f\mu^2.
\end{equation}
Acting on modes of the form $\Psi\sim e^{-i\omega u}\psi(r,\theta,\varphi)$ the operator $fP$ becomes
\begin{equation}
fP\Psi=\frac{f}{r^2}\partial_r(fr^2\partial_r)\psi+f\frac{2i\omega}{r}\partial_rra+\frac{f{L}^2}{r^2}\psi-f\frac{iqQ}{r^2}\psi-2\frac{iqQ}{r}(f\partial_r)\psi-f\mu^2\psi.
\end{equation}
It can be shown that mode solution of $P$ are conormal at $r=r_-$, meaning that they grow at the same rate $|r-r_-|^\lambda$ (see Section \ref{section weak solutions} for the asymptotic behavior of mode solutions at the CH). Thus, if $\psi\sim|r-r_-|^\lambda$ then the following terms have regularity
\begin{align*}
\frac{f{L}^2}{r^2}\psi\sim |r-r_-|^\lambda |r-r_-|\sim |r-r_-|^{\lambda+1},\\
f\frac{iqQ}{r^2}\psi\sim |r-r_-|^\lambda |r-r_-|\sim |r-r_-|^{\lambda+1},\\
f\mu^2\psi\sim \sim |r-r_-|^\lambda |r-r_-|\sim |r-r_-|^{\lambda+1},
\end{align*}
where $f \sim |r-r_-|$ near the CH. This means that these terms have one order higher regularity than the rest so they can be neglected.\footnote{For example, if $\lambda=-1/2$ then $f^\lambda=1/\sqrt{f}$ which diverges rapidly at $r=r_-$ while $f^{\lambda+1}=\sqrt{f}$ which doesn't diverge at $r=r_-$.} Neglecting these terms we get regular-singular ordinary differential equation near $r=r_-$ of the form
\begin{equation}
\label{operatorC}
\tilde{P}=(f\partial_r)^2+2i\omega(f\partial_r)-\frac{2iqQ}{r}(f\partial_r).
\end{equation}
It is convenient to use $f$ as a radial coordinate instead of $r$, so $\partial_r=f^\prime\partial_f=f^\prime(r_-)\partial_f$ near the CH plus irrelevant terms. Moreover, the surface gravity at the CH is $\kappa_-=-f^\prime(r_-)/2$ so $f\partial_r=-2\kappa_-(f\partial_f)$. Thus, (\ref{operatorC}) becomes
\begin{equation}
\label{indicialC}
\frac{\tilde{P}}{4\kappa_-^2}=(f\partial_f)^2-\frac{i\omega}{\kappa_-}(f\partial_f)+\frac{iqQ}{\kappa_-r_-}(f\partial_f)=f\partial_f\left(f\partial_f-\left(\frac{i\omega}{\kappa_-}-\frac{iqQ}{\kappa_-r_-}\right)\right).
\end{equation}
It remains to calculate the allowed growth rates $\lambda$, i.e. indicial roots of the differential operator (\ref{indicialC}). Acting with $|f|^\lambda$ we get
\begin{equation}
\label{polynomialC}
\frac{\tilde{P}}{4\kappa_-^2}|f|^\lambda=\lambda\left(\lambda-\left(\frac{i\omega}{\kappa_-}-\frac{iqQ}{\kappa_-r_-}\right)\right)|f|^\lambda,
\end{equation}
where $f\partial_f|f|^\lambda=\lambda|f|^\lambda$ and $(f\partial_f)^2|f|^\lambda=(f\partial_f)(f\partial_f)|f|^\lambda=(f\partial_f)\lambda|f|^\lambda=\lambda^2|f|^\lambda$. The indicial roots are the roots of the quadratic polynomial (\ref{polynomialC}), namely
\begin{equation}
\lambda_1=0,\,\,\,\,\,\,\,\,\,\,\,\,\,\lambda_2=\frac{i\omega}{\kappa_-}-\frac{iqQ}{\kappa_-r_-}.
\end{equation}
The root $\lambda_1=0$ corresponds to mode solutions which are approximately constant, i.e. remain smooth at the CH and are irrelevant for SCC, while $\lambda_2$ corresponds to asymptotics
\begin{equation}
|f|^{\lambda_2}\sim|r-r_-|^{\frac{i\omega}{\kappa_-}}|r-r_-|^{-\frac{iqQ}{\kappa_-r_-}}.
\end{equation}
If we consider QNMs of the form $\omega=\omega_R+i\omega_I$, with $\omega_I<0$, then
\begin{equation}
|f|^{\lambda_2}\sim|r-r_-|^{-\frac{\omega_I}{\kappa_-}}|r-r_-|^{i\left(\frac{\omega_R}{\kappa_-}-\frac{qQ}{\kappa_-r_-}\right)}.
\end{equation}
The second factor is purely oscillatory, so the only relevant factor for SCC is $|r-r_-|^\frac{\alpha}{\kappa_-}$ with $\alpha:=-\text{Im}(\omega)$ the spectral gap. This function lies in the Sobolev space $H^m$ for all $m<\frac{1}{2}+\frac{\alpha}{\kappa_-}$. For the function to belong in $H^1_\text{loc}$ the following should hold
\begin{equation}
\beta\equiv\frac{\alpha}{\kappa_-}>1-\frac{1}{2}=\frac{1}{2}.
\end{equation}

\chapter{The classical definition of $\beta$ for scalar fields in higher-dimensional Reissner-Nordstr\"{o}m-de Sitter spacetime}\label{appE}
To determine the regularity of the metric up to the CH we study the regularity of QNMs at the CH. To do so, we change to outgoing Eddington-Finkelstein coordinates which are regular there.  The outgoing Eddington-Finkelstein coordinates are obtained by replacing $t$ with $u=t-r_*$ in (\ref{dspace}) to get
\begin{equation}
\label{outgoingD}
ds^2=-f(r)du^2-2du dr +r^2d\Omega_{d-2}^2.
\end{equation}
By expanding the Klein-Gordon equation
\begin{equation}
\Box\Psi=0,
\end{equation}
we get $P\Psi=0$ where the operator $P$ reads
\begin{equation}
\label{int1}
P\Psi=-2\partial_u\partial_r\Psi-\frac{d-2}{r}\partial_u\Psi+\frac{1}{r^{d-2}}\partial_r\left(f r^{d-2}\partial_r\Psi\right)+\frac{\Delta_{\Omega_{d-2}}}{r^{d-2}}\Psi,
\end{equation}
where $\Delta_{\Omega_{d-2}}$ the Laplace-Beltrami operator \cite{Berti:2009kk}. By acting on mode solutions of the form $\Psi\sim e^{-i\omega u} \psi$ we obtain
\begin{equation}
\label{intD}
f P\Psi=2i\omega f\partial_r\psi +\frac{i\omega(d-2)}{r^{d-2}}f\psi+\frac{1}{r^{d-2}}f\partial_r\left(f r^{d-2}\partial_r\psi\right)+\frac{\Delta_{\Omega_{d-2}}}{r^{d-2}}f\psi.
\end{equation}
It can be shown that the mode solutions of (\ref{intD}) are conormal at $r=r_-$, meaning that they grow at the same rate $|r-r_-|^\lambda$. Thus, if $\psi\sim|r-r_-|^\lambda$ then the second and last term have higher regularity than the rest, since $f \sim |r-r_-|$ near the CH. This means that these terms can be neglected, which leads to a regular-singular ordinary differential equation near $r=r_-$ of the form $\tilde{P}\psi=fP\psi=0$ with the operator
\begin{equation}
\label{operatorD}
\tilde{P}=2i\omega f\partial_r+\left(f\partial_r\right)^2.
\end{equation}
It is convenient to use $f$ as a radial coordinate instead of $r$, so $\partial_r=f^\prime\partial_f=f^\prime(r_-)\partial_f$ near the CH modulo irrelevant terms. Moreover, the surface gravity at the CH is $\kappa_-=-f^\prime(r_-)/2$ so $f\partial_r=-2\kappa_-(f\partial_f)$. Thus, (\ref{operatorD}) becomes
\begin{equation}
\label{indicialD}
\frac{\tilde{P}}{4\kappa_-^2}=(f\partial_f)^2-\frac{i\omega}{\kappa_-}\left(f\partial_f\right)=f\partial_f\left(f\partial_f-\frac{i\omega}{\kappa_-}\right).
\end{equation}
It remains to calculate the allowed growth rates $\lambda$, i.e. indicial roots of the differential operator (\ref{indicialD}). Acting with $|f|^\lambda$ we get
\begin{equation}
\label{polynomialD}
\frac{{\tilde{P}}}{4\kappa_-^2}|f|^\lambda=\lambda\left(\lambda-\frac{i\omega}{\kappa_-}\right)|f|^\lambda.
\end{equation}
The indicial roots are the roots of the quadratic polynomial (\ref{polynomialD}), namely
\begin{equation}
\lambda_1=0,\,\,\,\,\,\,\,\,\,\,\,\,\,\lambda_2=\frac{i\omega}{\kappa_-}.
\end{equation}
The root $\lambda_1=0$ corresponds to mode solutions which are approximately constant, i.e. remain smooth at the CH and are irrelevant for SCC, while $\lambda_2$ corresponds to asymptotics
\begin{equation}
|f|^{\lambda_2}\sim|r-r_-|^{\frac{i\omega}{\kappa_-}}.
\end{equation}
If we consider QNMs of the form $\omega=\omega_R+i\omega_I$, with $\omega_I<0$ then
\begin{equation}
|f|^{\lambda_2}\sim|r-r_-|^{-\frac{\omega_I}{\kappa_-}}|r-r_-|^\frac{i\omega_R}{\kappa_-}.
\end{equation}
The second factor is purely oscillatory, so the only relevant factor for SCC is $|r-r_-|^\frac{\alpha}{\kappa_-}$ with $\alpha:=-\text{Im}({\omega})$ the spectral gap. This function lies in the Sobolev space $H^s$ for all $s<\frac{1}{2}+\frac{\alpha}{\kappa_-}$.

Since we are considering scalar fields, we require locally
square integrable gradient of the scalar field at the CH\footnote{The energy-momentum tensor for scalar fields is $T_{\mu\nu}\sim (\partial\psi)^2$.}, i.e., the mode solutions should belong to the Sobolev space $H^1_\text{loc}$ for the Einstein's field equations to be satisfied weakly at the CH. This justifies our search for $\beta=-\text{Im}(\omega)/\kappa_->1/2$.
\chapter{The classical definition of $\beta$ for charged fermions in Reissner-Nordstr\"{o}m-de Sitter spacetime}\label{appD}

To determine the regularity of the metric up to the CH we study the regularity of mode solutions there. To do so, we change to outgoing Eddington-Finkelstein coordinates which are regular at the CH.  The outgoing Eddington-Finkelstein coordinates are obtained by replacing $t$ with $u=t-r_*$ in (\ref{metric3}) to get
\begin{equation}
\label{outgoing}
ds^2=-f(r)du^2-2du dr +r^2(d\theta^2+\sin^2\theta\, d\varphi^2),
\end{equation}
with the associated electromagnetic potential\footnote{Here, if we want to transform from the coordinate system $(t,r,\theta,\varphi)$ to $(u,r,\theta,\varphi)$ then the components of the potential will be $A_\mu=\text{diag}(-Q/r,+Q/rf,0,0)$ while if we are already working on the coordinate system $(u,r,\theta,\varphi)$ then the components of the potential will be ${A}_\mu=\text{diag}(-Q/r,0,0,0)$.}
\begin{equation}
\label{outpot}
{A}=-\frac{Q}{r}du=-\frac{Q}{r}(dt-dr_*)=-\frac{Q}{r}\left(dt-\frac{dr}{f(r)}\right).
\end{equation}
A straightforward way to write down the Dirac equation (\ref{dirac equation}) in the new coordinates is to choose a new tetrad that reproduces (\ref{outgoing}) by satisfying (\ref{tetradgmn}) and follow the same process as before. An alternative way is to transform (\ref{dirac_2}) in the new coordinates $(u,r)$ with the associated transformed electromagnetic potential (\ref{outpot}).

Following the same procedure as above we write (\ref{dirac_2}) now with non-zero $A_r$ component
\begin{equation}
\label{dirac_3}
\frac{i}{\sqrt{f}}\gamma^t\frac{\partial}{\partial t}\psi+i\sqrt{f}\gamma^r\frac{\partial}{\partial r}\psi-\frac{i}{r}\gamma^r\psi+i\left(\gamma^\theta\frac{\partial}{\partial\theta}+\gamma^\varphi\frac{\partial}{\partial\varphi}\right)\psi+\frac{q}{\sqrt{f}}\left(\gamma^{t} A_t+f\gamma^r A_r\right)\psi-m_f\psi=0.
\end{equation}
Performing the coordinate transformation $u=t-r_*$ we have the following chain rule expressions
\begin{align}
\frac{\partial}{\partial t}&=\frac{\partial}{\partial u}\frac{\partial  u}{\partial t}+\frac{\partial}{\partial r}\frac{\partial r}{\partial t}=\partial_u,\\
\frac{\partial}{\partial r}&=\frac{\partial}{\partial u}\frac{\partial  u}{\partial r}+\frac{\partial}{\partial r}\frac{\partial r}{\partial r}=-\frac{\partial r_*}{\partial r}\partial_u+\partial_r=-\frac{\partial_u}{f}+\partial_r,
\end{align}
which will replace the corresponding derivatives in (\ref{dirac_3}), leading to
\begin{align}
\nonumber
\frac{i\gamma^t}{\sqrt{f}}\partial_u\psi+i\sqrt{f}\gamma^r\left(-\frac{\partial_u}{f}+\partial_r\right)\psi-\frac{i}{r}\gamma^r\psi+i\left(\gamma^\theta\frac{\partial}{\partial\theta}+\gamma^\varphi\frac{\partial}{\partial\varphi}\right)\psi\\\label{dirac_5}+\frac{q}{\sqrt{f}}\left(\gamma^{t} A_t+f\gamma^r A_r\right)\psi-m_f\psi=0.
\end{align}
By utilizing the ansatz
\begin{align}
\label{spinorout3}
\psi^+&=e^{-i\omega\upsilon}\begin{pmatrix}
\phi^k_{j-1/2}F(r)\\
i\phi^k_{j+1/2}G(r)
\end{pmatrix},\\
\label{spinorout4}
\psi^-&=e^{-i\omega\upsilon}\begin{pmatrix}
\phi^k_{j+1/2}F(r)\\
i\phi^k_{j-1/2}G(r)
\end{pmatrix},
\end{align}
we get
\begin{align}
\label{edoutf1}
f\partial_rF-\frac{\xi\sqrt{f}}{r}F+\left(\omega-\frac{qQ}{r}\right)G+i\left(\omega-\frac{qQ}{r}\right)F+m_f\sqrt{f}G&=0,\\
\label{edoutf2}
f\partial_rG+\frac{\xi\sqrt{f}}{r}G-\left(\omega-\frac{q Q}{r}\right) F+i\left(\omega-\frac{qQ}{r}\right)G+m_f\sqrt{f}F&=0,
\end{align}
where $\xi=\pm 1,\pm 2,\dots$. Since the mass-to-charge ratio of the electron is very small ($\sim 10^{-11\,\text{kg}/\text{C}}$) we will consider massless fermions propagating on the fixed background of RNdS. By setting $m_f=0$ in (\ref{edoutf1}) and (\ref{edoutf2}) and solving (\ref{edoutf1}) with respect to $G$, we can plug in the resulting equation into (\ref{edoutf2}) to get
\begin{align}
\nonumber
\left[-\frac{2\xi r \omega f^{3/2}}{qQ-r\omega}-\xi r \sqrt{f}f^\prime-2\xi^2 f\right]F+\left[\frac{2qQrf^2}{qQ-r\omega}+2rf\left(-2iqQ+2ir\omega +rf^\prime\right)\right]F^\prime(r)\\+2r^2f^2 F^{\prime\prime}(r)=0.
\end{align}
Using
\begin{equation}
2r^2f\left(f F^\prime(r)\right)^\prime=2r^2ff^\prime F^\prime(r)+2r^2f^2F^{\prime\prime}(r)
\end{equation}
we can rewrite the previous equation as \footnote{Here, we have already acted on modes of the spinor form (\ref{spinorout3}) and (\ref{spinorout4}) to decouple the angular and radial part of the Dirac equation so we end up with a radial equation for the field $F(r)$ or $G(r)$, respectively.}
\begin{align}
\nonumber
-\frac{2\xi \omega f^{3/2}}{r(qQ-r\omega)} F-\frac{\xi\sqrt{f}}{r}f^\prime F-2\frac{\xi^2 f}{r^2}F+\frac{2qQf}{r(qQ-r\omega)}f\partial_r F-\frac{4iqQ}{r}f\partial_r F+4i\omega f\partial_rF\\\label{int}+2(f\partial_r)^2F=0.
\end{align}
It can be shown that mode solution of (\ref{int}) are conormal at $r=r_-$, meaning that they grow at the same rate $|r-r_-|^\lambda$. Thus, if $F\sim|r-r_-|^\lambda$ then the first four terms have regularity
\begin{align*}
\frac{2\xi \omega f^{3/2}}{r_-(qQ-r_-\omega)} F &\sim |r-r_-|^\lambda |r-r_-|^{3/2}\sim |r-r_-|^{\lambda+3/2},\\
\frac{\xi\sqrt{f}}{r_-}f^\prime(r_-) F&\sim |r-r_-|^\lambda|r-r_-|^{\lambda+1/2}\sim |r-r_-|^{\lambda+1/2},\\
\frac{2\xi^2 f}{r_-^2}F&\sim |r-r_-|^\lambda |r-r_-|\sim |r-r_-|^{\lambda+1},\\
\frac{2qQf}{r_-(qQ-r_-\omega)}f\partial_r F&\sim \lambda |r-r_-|^{\lambda-1}|r-r_-|^2\sim |r-r_-|^{\lambda+1},
\end{align*}
where $f \sim |r-r_-|$ near the CH. This means that these terms have regularity of higher order comparing to the rest so they can be neglected. Neglecting these terms we get a regular-singular ordinary differential equation near $r=r_-$ of the form $PF=0$ with the operator\footnote{The same operator arises for the field $G$ with $PG=0$ by following exactly the same steps.}
\begin{equation}
\label{operator}
P=(f\partial_r)^2+2i\omega(f\partial_r)-\frac{2iq\omega}{r}(f\partial_r).
\end{equation}
It is convenient to use $f$ as a radial coordinate instead of $r$, so $\partial_r=f^\prime\partial_f=f^\prime(r_-)\partial_f$ near the CH plus irrelevant terms. Moreover, the surface gravity at the CH is $\kappa_-=-f^\prime(r_-)/2$ so $f\partial_r=-2\kappa_-(f\partial_f)$. Thus, (\ref{operator}) becomes
\begin{equation}
\label{indicial}
\frac{{P}}{4\kappa_-^2}=(f\partial_f)^2-\frac{i\omega}{\kappa_-}(f\partial_f)+\frac{iqQ}{\kappa_-r_-}(f\partial_f)=f\partial_f\left(f\partial_f-\left(\frac{i\omega}{\kappa_-}-\frac{iqQ}{\kappa_-r_-}\right)\right).
\end{equation}
It remains to calculate the allowed growth rates $\lambda$, i.e. indicial roots of the differential operator (\ref{indicial}). Acting with $|f|^\lambda$ we get
\begin{equation}
\label{polynomial}
\frac{{P}}{4\kappa_-^2}|f|^\lambda=\lambda\left(\lambda-\left(\frac{i\omega}{\kappa_-}-\frac{iqQ}{\kappa_-r_-}\right)\right)|f|^\lambda,
\end{equation}
where $f\partial_f|f|^\lambda=\lambda|f|^\lambda$ and $(f\partial_f)^2|f|^\lambda=(f\partial_f)(f\partial_f)|f|^\lambda=(f\partial_f)\lambda|f|^\lambda=\lambda^2|f|^\lambda$. The indicial roots are the roots of the quadratic polynomial (\ref{polynomial}), namely
\begin{equation}
\lambda_1=0,\,\,\,\,\,\,\,\,\,\,\,\,\,\lambda_2=\frac{i\omega}{\kappa_-}-\frac{iqQ}{\kappa_-r_-}.
\end{equation}
The root $\lambda_1=0$ corresponds to mode solutions which are approximately constant, i.e. remain smooth at the CH and are irrelevant for SCC, while $\lambda_2$ corresponds to asymptotics
\begin{equation}
|f|^{\lambda_2}\sim|r-r_-|^{\frac{i\omega}{\kappa_-}}|r-r_-|^{-\frac{iqQ}{\kappa_-r_-}}.
\end{equation}
If we consider quasinormal modes of the form $\omega=\omega_R+i\omega_I$, with $\omega_I<0$, then
\begin{equation}
|f|^{\lambda_2}\sim|r-r_-|^{-\frac{\omega_I}{\kappa_-}}|r-r_-|^{i\left(\frac{\omega_R}{\kappa_-}-\frac{qQ}{\kappa_-r_-}\right)}.
\end{equation}
The second factor is purely oscillatory, so the only relevant factor for SCC is $|r-r_-|^\frac{\alpha}{\kappa_-}$ with $\alpha:=-\text{Im}{\omega}$ the spectral gap. This function lies in the Sobolev space $H^m$ for all $m<\frac{1}{2}+\frac{\alpha}{\kappa_-}$. In this case, the Einstein-Hilbert stress-energy tensor of the massless fermionic field $\Psi$ lying on the right-hand-side of Einstein's equations has the form \cite{Toth:2015cda}
\begin{equation}
\label{tmn}
T_{\mu\nu}=\frac{i}{4}\left(\Psi^\dagger G^\mu (\mathcal{D}_\nu\Psi)+\Psi^\dagger G^\nu (\mathcal{D}_\mu\Psi)-(\mathcal{D}_\mu\Psi)^\dagger G^\nu\Psi-(\mathcal{D}_\nu\Psi)^\dagger G^\mu\Psi\right).
\end{equation}
and again this leads to the requirement of square integrability of the gradient of the fermionic field $\Psi$.\footnote{$\int dr\, \Psi^\dagger G^r D_r\Psi\leq \frac{1}{2}\int (\Psi^\dagger G^r)^2+\frac{1}{2}\int (D_r\Psi)^2$ where $(G^r)^2=0$ at the CH.} Thus, the mode solutions should belong to the Sobolev space $H^1_\text{loc}$ for our metric to make sense as a weak solution of Einstein's field equations at the CH. This provides the justification for our search for BH parameters for which $\beta\equiv\alpha/\kappa_->1/2$.
\chapter{Superradiant instability of charged scalar fields in Reissner-Nordstr\"{o}m-de Sitter spacetime}\label{superradiance}
In this appendix, we take a thorough look at the instability that arises in spherically-symmetric charged scalar perturbations of RNdS spacetime. The setup is the $4-$dimensional RNdS spacetime with line element
\begin{equation}
ds^2=-f(r)dt^2+\frac{1}{f(r)}dr^2+r^2\left(d\theta^2+\sin^2\theta\,d\phi^2\right).
\end{equation}
The metric function is
\begin{align}
 f(r)&=1-\frac{2M}{r}+\frac{Q^2}{r^2}-\frac{\Lambda r^2}{3},
\end{align}
where $M,\, Q$ the mass and charge of the black hole, $\Lambda>0$ the cosmological constant. The associated electromagnetic potential sourced by such charged spacetime is
\begin{equation}
A=-\frac{Q}{r}dt.
\end{equation}
The propagation of a massive charged scalar field on a fixed RNdS background is governed by the Klein-Gordon equation
\begin{equation}
(D^\nu D_\nu-\mu^2)\Psi=0, 
\end{equation}
where $\mu$, $q$ the mass and charge of the field, respectively, and $D_\nu=\nabla_\nu-iqA_\nu$ the covariant derivative. By expanding $\Psi$ in terms of spherical harmonics with harmonic time dependence we obtain the radial master equation
\begin{equation}
\label{master_eq_RNdS}
\frac{d^2 \psi}{d r_*^2}+\left[\left(\omega-\Phi(r)\right)^2-V(r)\right]\psi=0\,,
\end{equation}
where $\Phi(r)=q Q/r$ is the electrostatic potential and $dr_*=dr/f(r)$ the tortoise coordinate.
\begin{figure}[H]
\subfigure{\includegraphics[scale=0.217]{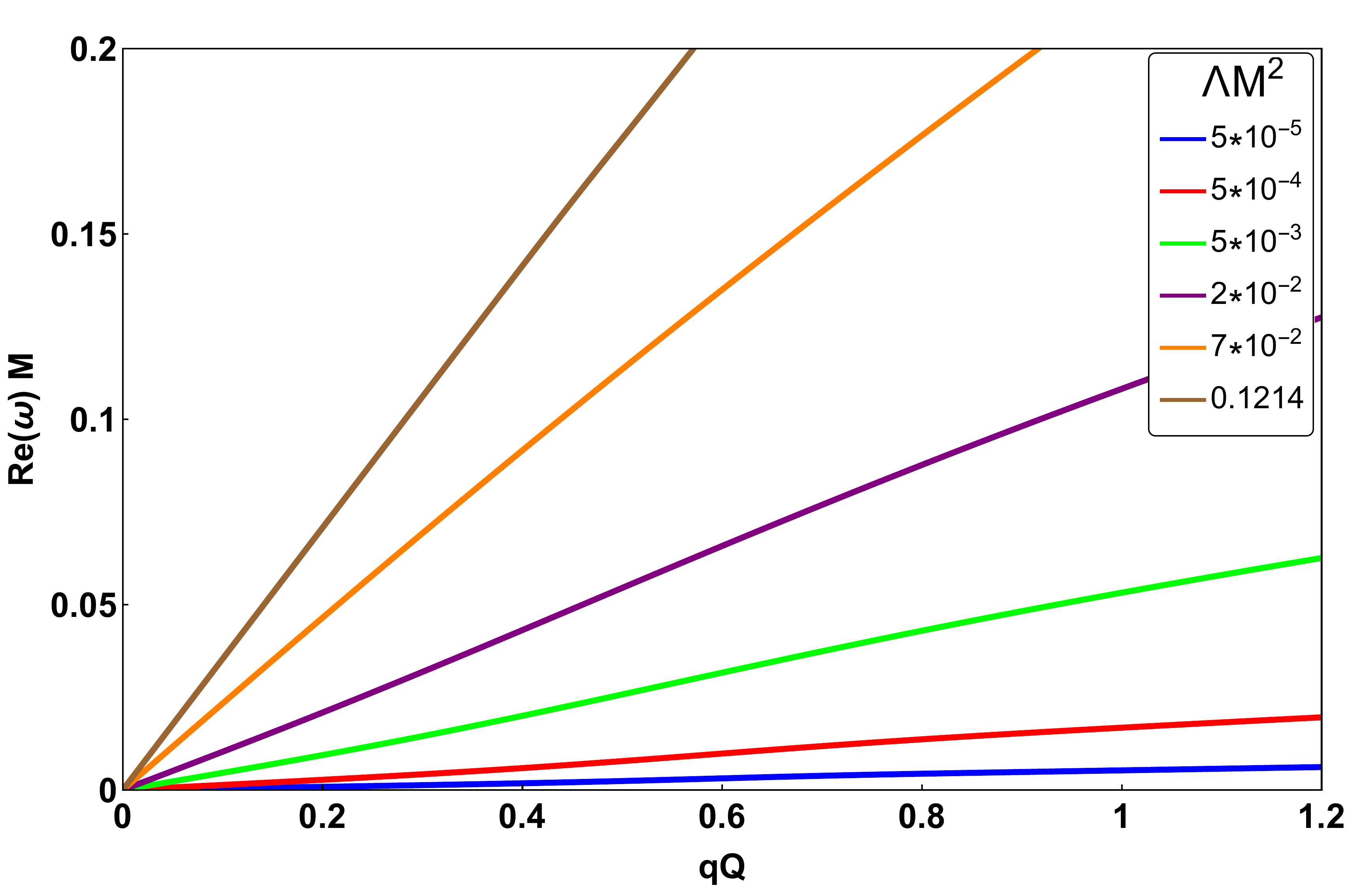}}\qquad
\subfigure{\includegraphics[scale=0.22]{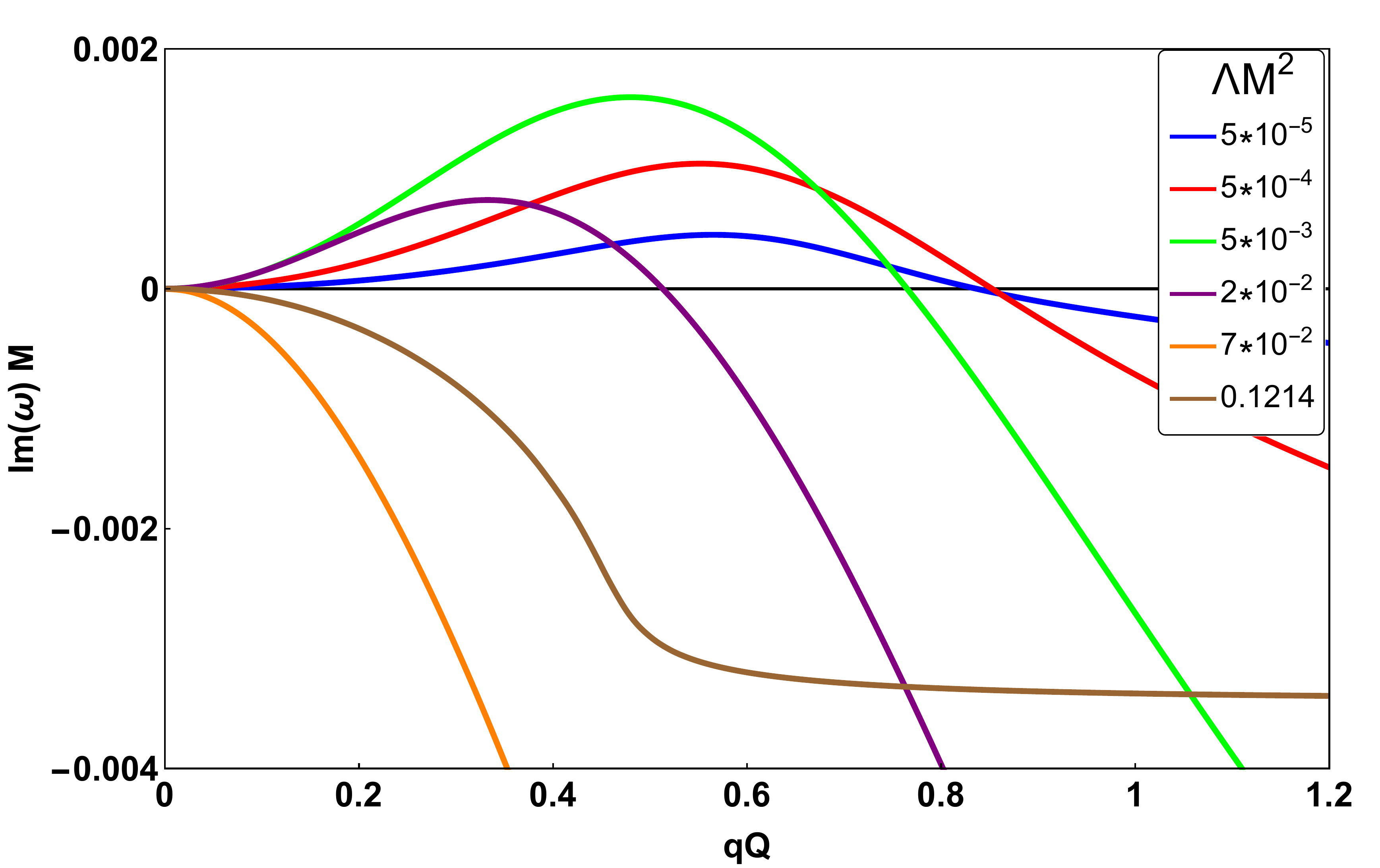}}
\subfigure{\includegraphics[scale=0.217]{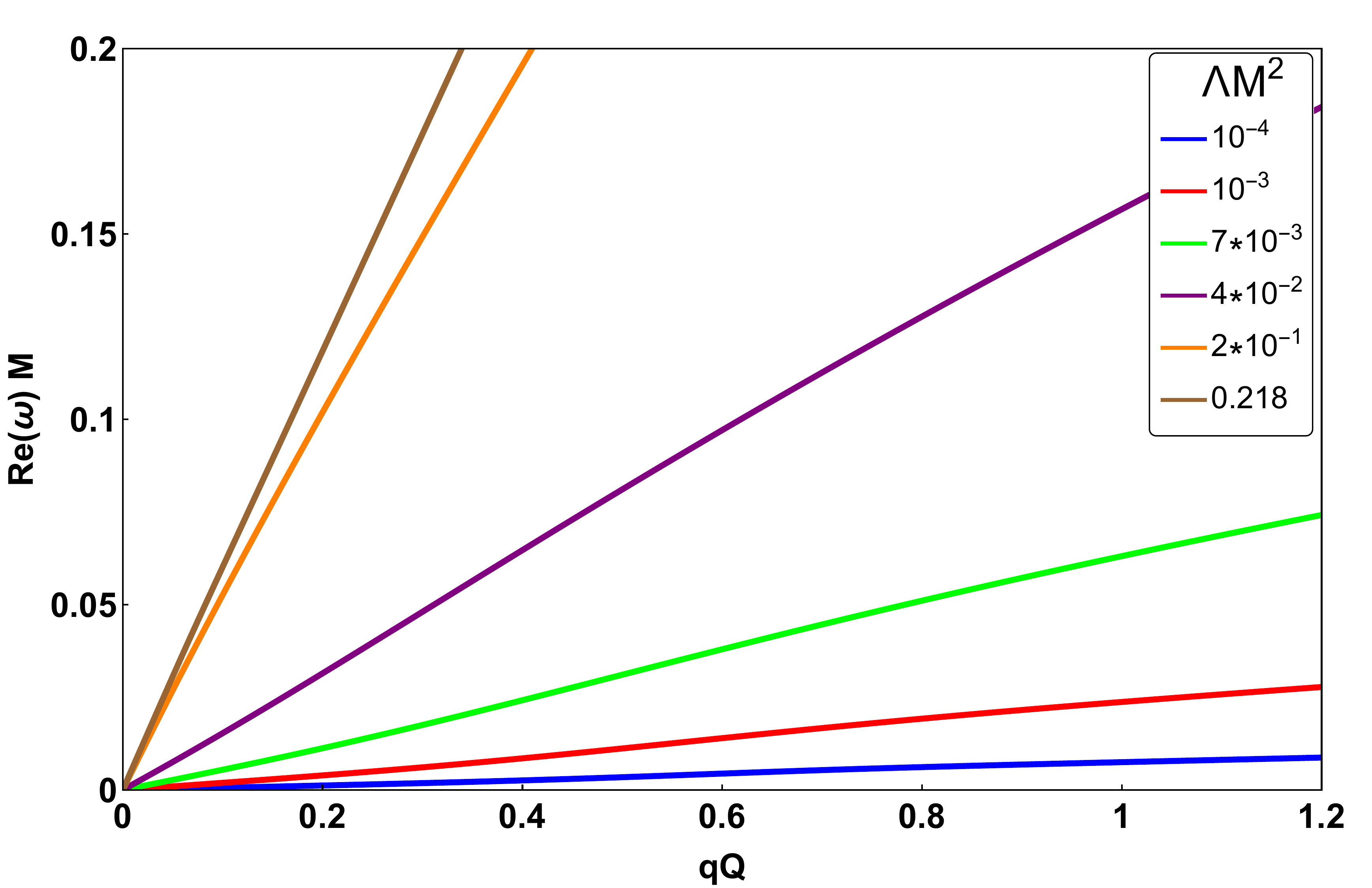}}\qquad
\subfigure{\includegraphics[scale=0.22]{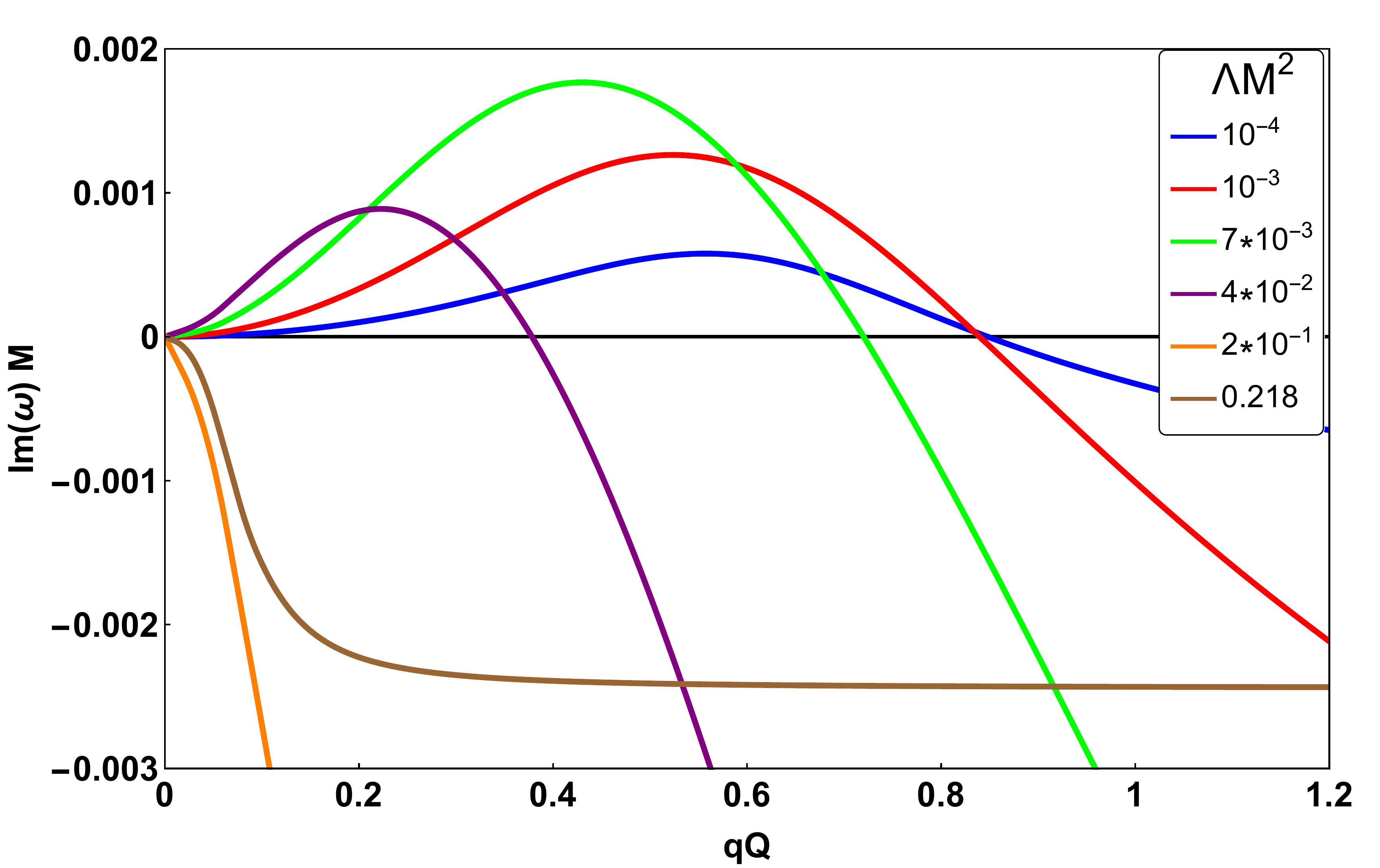}}
\caption{Real (left) and imaginary (right) part of a charged massless scalar perturbation with $l=0$ on a fixed RNdS with $Q/M=0.5$ (top panel) and $Q=0.999\, Q_\text{max}$ (bottom panel) versus the charge coupling $qQ$. Different colors designate distinct choices of cosmological constants $\Lambda M^2$.}
\label{Q05}
\end{figure}
The effective potential for massive scalar perturbations is
\begin{equation}
\label{RNdS_general potential}
V(r)=f(r)\left(\mu^2+\frac{l(l+1)}{r^2}+\frac{f^\prime(r)}{r}\right).
\end{equation}
\begin{figure}[H]
\subfigure{\includegraphics[scale=0.215]{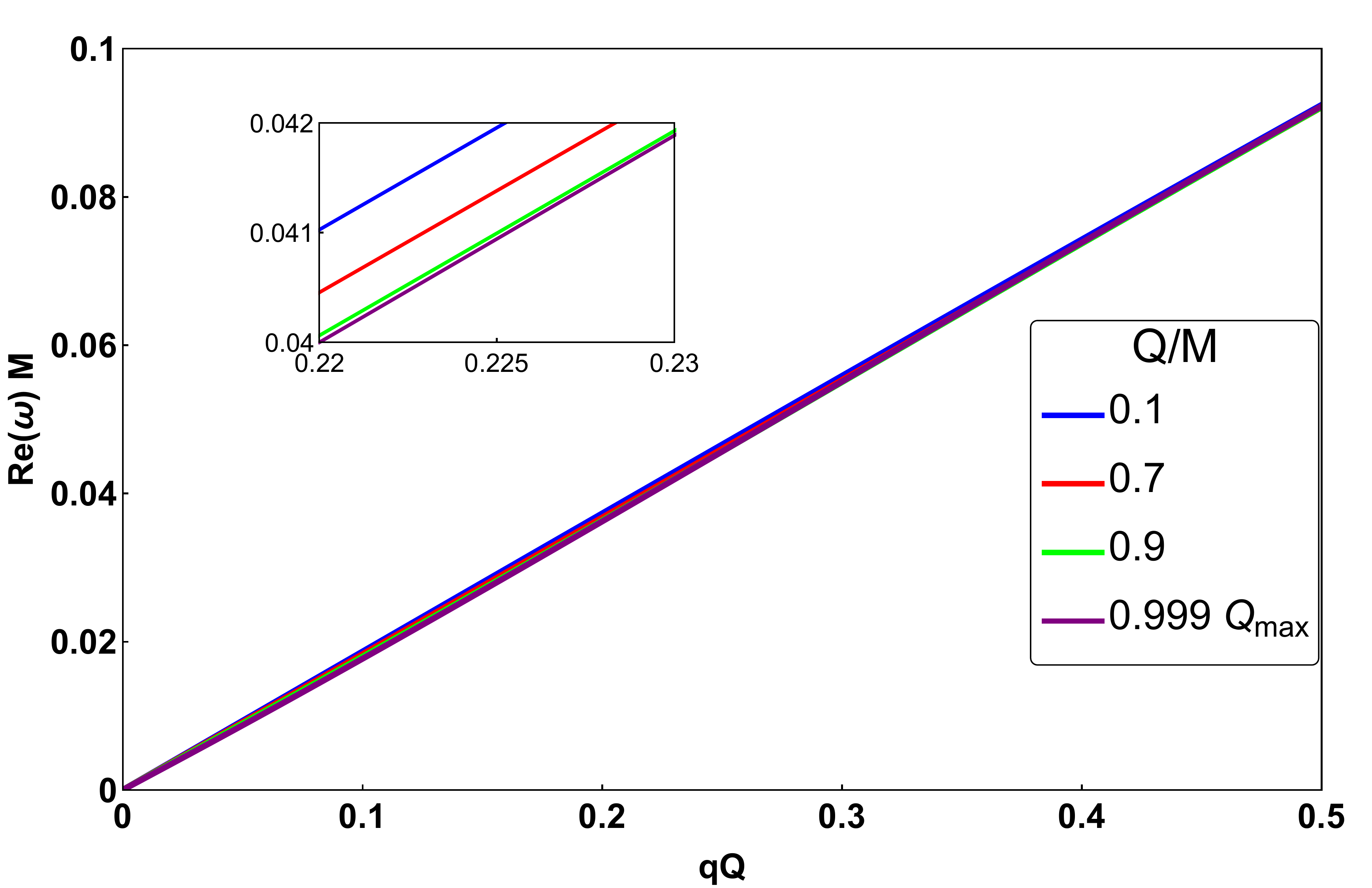}}\quad
\subfigure{\includegraphics[scale=0.22]{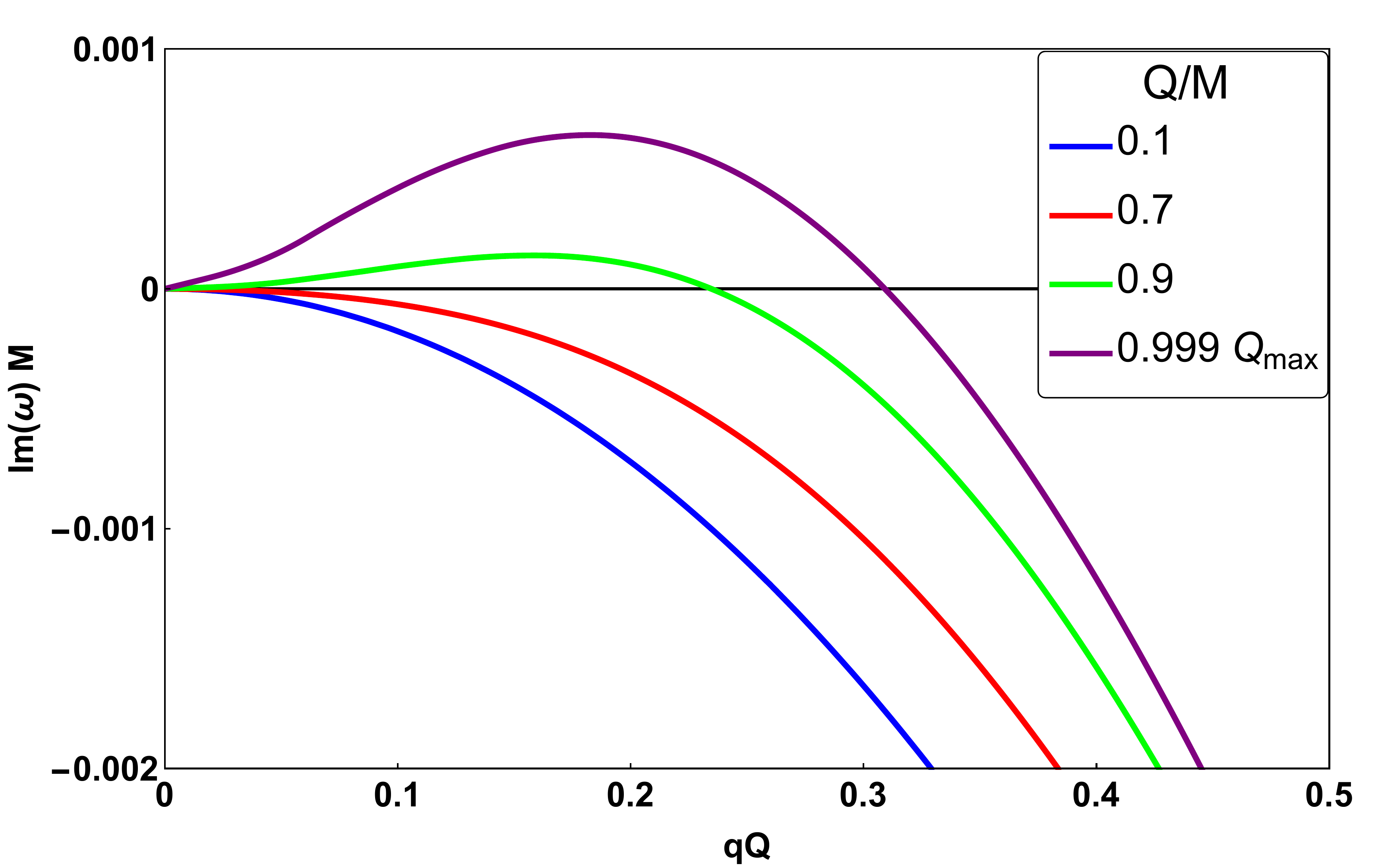}}
\caption{Real (left) and imaginary (right) part of a charged massless scalar perturbation with $l=0$ on a fixed RNdS with $\Lambda M^2=0.05$ versus the charge coupling $qQ$. Different colors designate distinct choices of black hole charges $Q/M$.}
\label{L05}
\end{figure}
The boundary conditions for QNMs are
\begin{equation}
\label{bcs}
\psi \sim
\left\{
\begin{array}{lcl}
e^{-i (\omega-\Phi(r_+))r_* },\,\,\,\quad r \rightarrow r_+, \\
&
&
\\
 e^{+i(\omega-\Phi(r_c))r_*},\,\,\,\,\quad r \rightarrow r_c.
\end{array}
\right.
\end{equation}
Due to the underlying symmetry of (\ref{master_eq_RNdS}) $\text{Re}(\omega)\rightarrow-\text{Re}(\omega)$ and $\Phi(r)\rightarrow-\Phi(r)$, we will consider only cases where $qQ>0$.

It has been shown in Chapter \ref{higher instability} that for superradiance to occur in $d-$dimensional RNdS the following inequality must hold:
\begin{equation}
\label{suprad1}
\Phi(r_c)<\omega<\Phi(r_+),
\end{equation}
which designates that the amplitude of the reflected wave is larger than the amplitude of the incident wave.

In the following figures, we present the superradiant instability found for the $l=0$ charged scalar perturbation in \cite{PhysRevD.98.104007} and we analyze it for various BH parameters.

In figure \ref{Q05} we show the unstable $l=0$ perturbation versus the charge coupling $qQ$ for various cosmological constants. The modes originate from $\omega=0$, the zero QNM of pure dS space. The real part originates from $\omega_R=0$, increases monotonously with $qQ$ and grown faster for larger $\Lambda M^2$. The imaginary part originates from $\omega_I=0$, grows to a maximum and then decreases till $\omega_I$ becomes negative again. The region where $\omega_I>0$ is the region of instability where superradiance can occur. It evident that as $\Lambda M^2$ increases, the unstable mode reaches a maximal imaginary part and beyond that point decreases while for sufficiently large cosmological constants no instabilities are found.

The real parts of the unstable modes are very weakly dependent on the BH charge $Q/M$. For a large cosmological constant, and small $Q/M$ we can see that no instabilities arise. With the increment of the BH charge, though, instabilities still arise. The former is demonstrated in Fig. \ref{L05}.

\begin{figure}[H]
\subfigure{\includegraphics[scale=0.22]{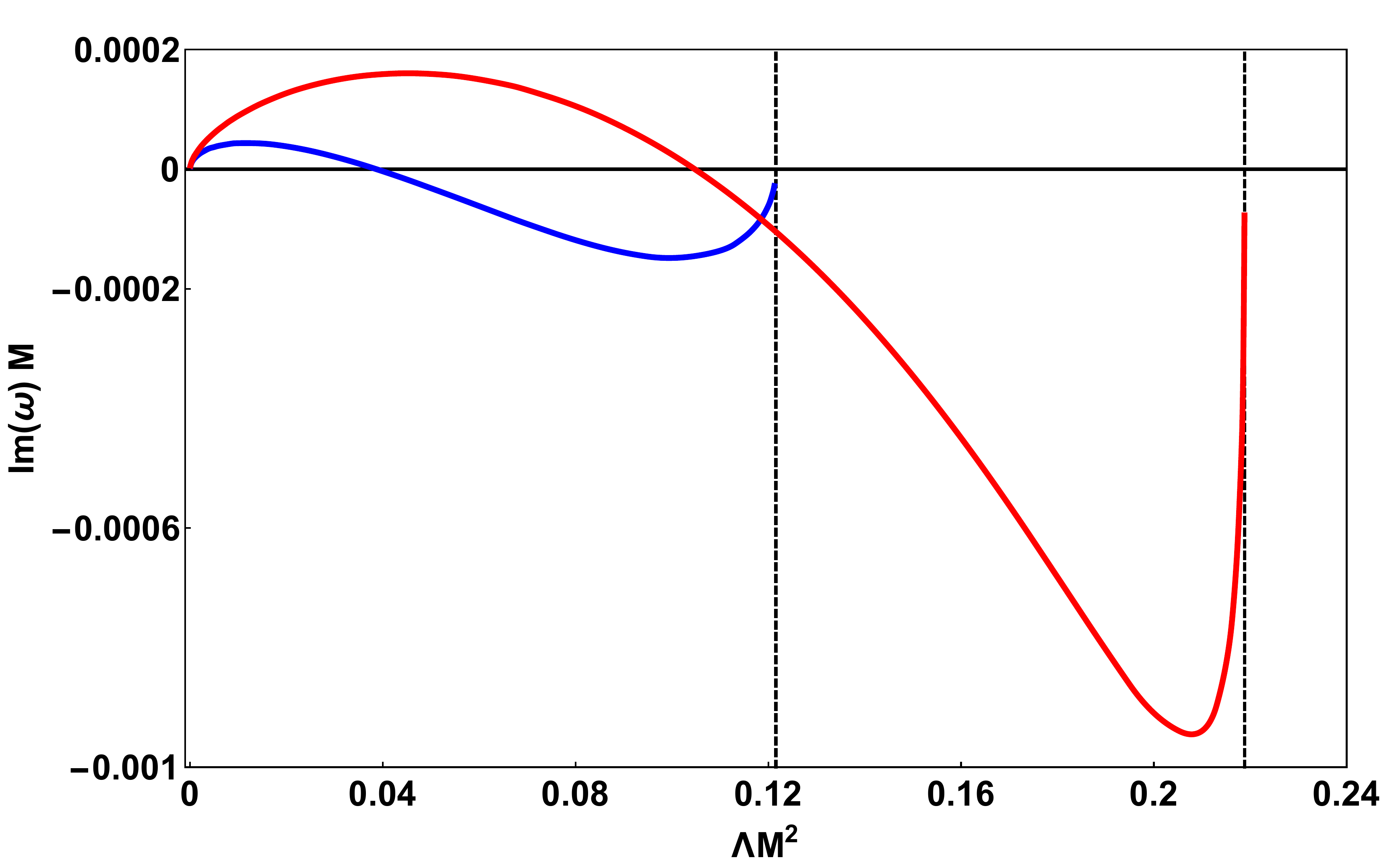}}\quad
\subfigure{\includegraphics[scale=0.22]{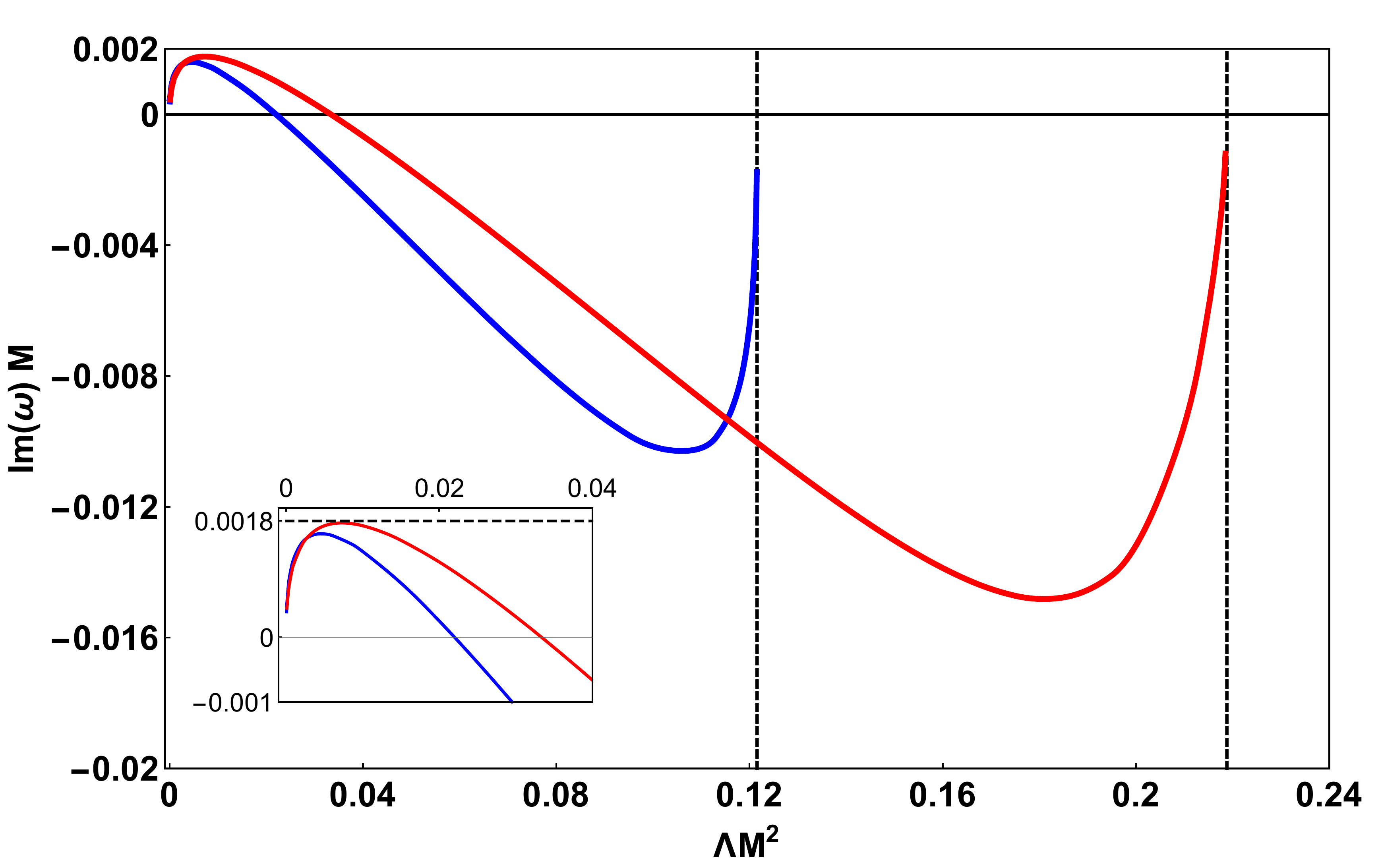}}
\caption{{\bf Left:} Imaginary part of the $l=0$ charged massless scalar perturbation on a fixed RNdS with $Q=0.5$ (blue) and $Q=0.999\, Q_\text{max}$ (red) versus the cosmological constant $\Lambda M^2$. The charge coupling is $qQ=0.05$. {\bf Right:} Imaginary part of the $l=0$ charged massless scalar perturbation on a fixed RNdS with $Q=0.5$, $qQ=0.48$ (blue) and $Q=0.999\, Q_\text{max}$, $qQ=0.43$ (red) versus the cosmological constant $\Lambda M^2$. The charge couplings for each $Q$ are chosen to maximize the imaginary part of $\omega$. The maximal $\text{Im}(\omega)\sim 0.0018$ is achieved at extremality for $\Lambda M^2=0.007$ and $qQ=0.43$.  The vertical dashed lines designate the maximal cosmological constant for each choice of $Q$ ($\Lambda_\text{max}M^2=0.121555$ for $Q/M=0.5$ and $\Lambda_\text{max}M^2=0.218815$ for $Q=0.999Q_\text{max}$).}
\label{peak}
\end{figure}

The peak of instability seems to occur close to extremality, and for that reason, we depict a set of specific parameters that minimizes the instability timescale and track the unstable mode for various $\Lambda M^2$. In Fig. \ref{peak} (right panel) we demonstrate that the maximal imaginary part is obtained at extremality, with $\omega_I\sim 0.0018$. As it turns out, the parameters chosen in \cite{PhysRevD.98.104007} are a very good choice to test SCC and at the same time demonstrate the contribution of the superradiant instability.

\begin{figure}[H]
\subfigure{\includegraphics[scale=0.223]{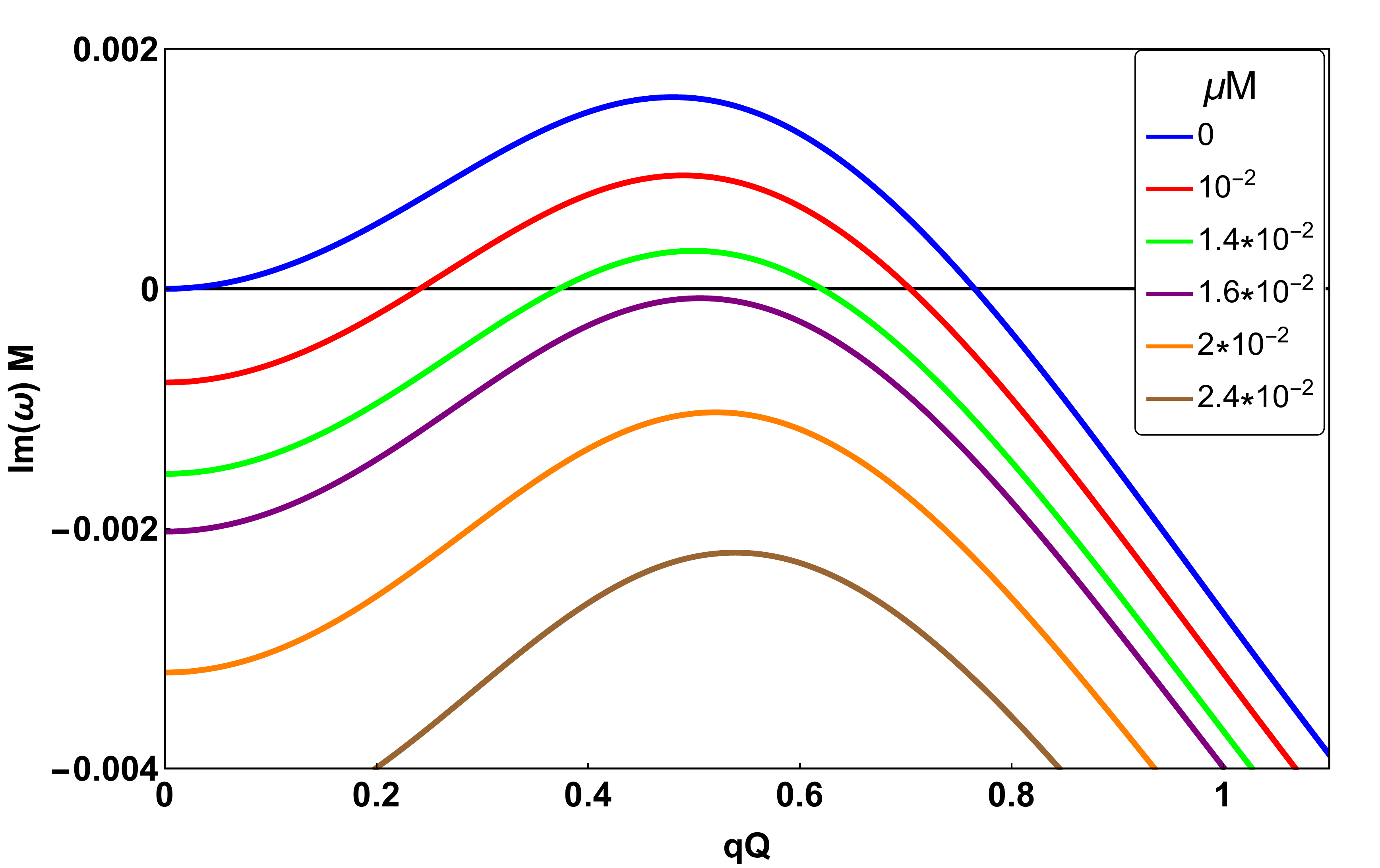}}
\subfigure{\includegraphics[scale=0.22]{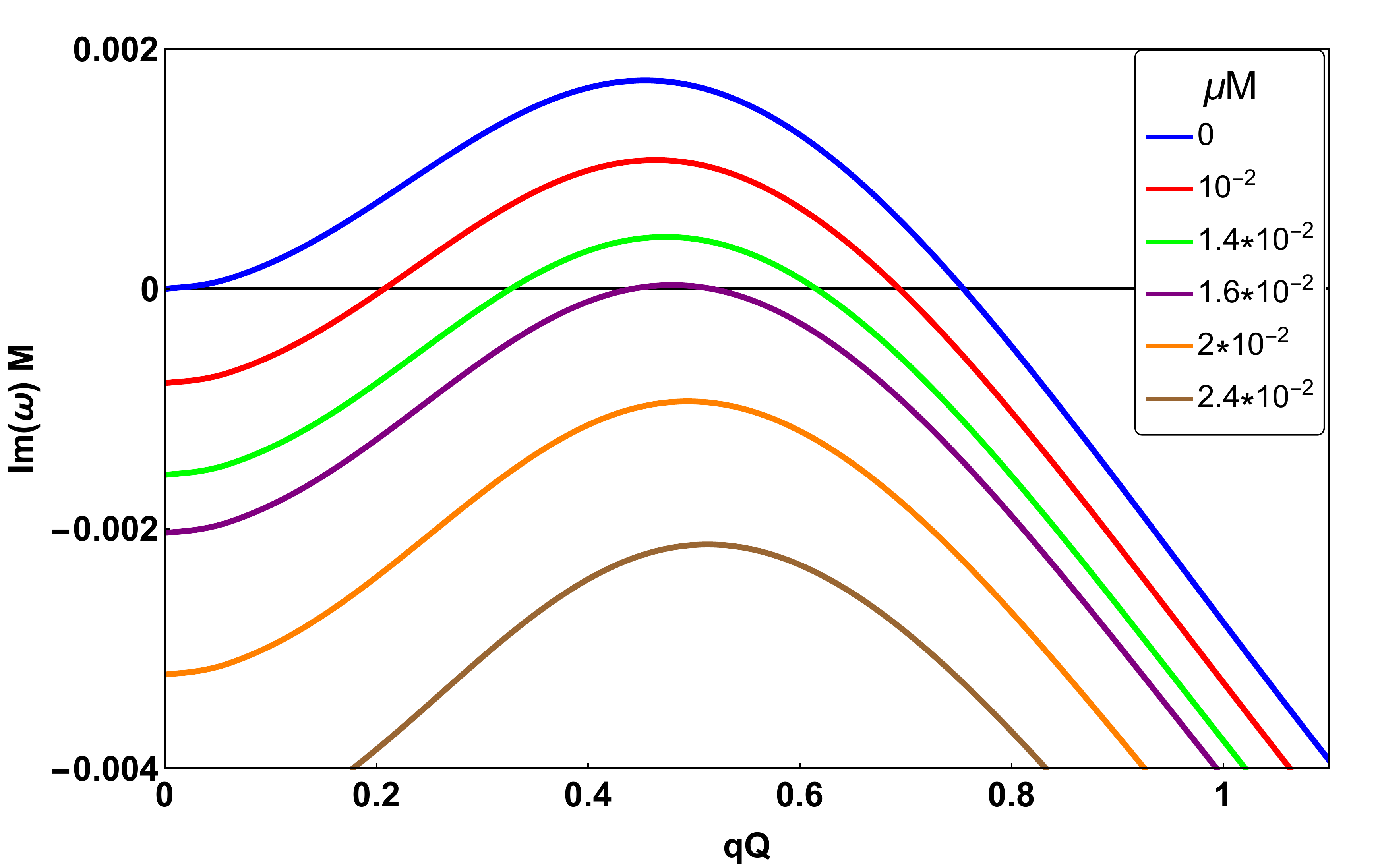}}
\caption{Imaginary part of the $l=0$ charged scalar perturbation on a fixed RNdS with $Q/M=0.5$ (left) and $Q=0.999\, Q_\text{max}$ (right) versus the charge coupling $qQ$. The cosmological constant is $\Lambda M^2=0.005$. Different colors designate distinct choices of scalar mass $\mu M$.}
\label{massive_L005_Im}
\end{figure}

\begin{figure}[H]
\subfigure{\includegraphics[scale=0.22]{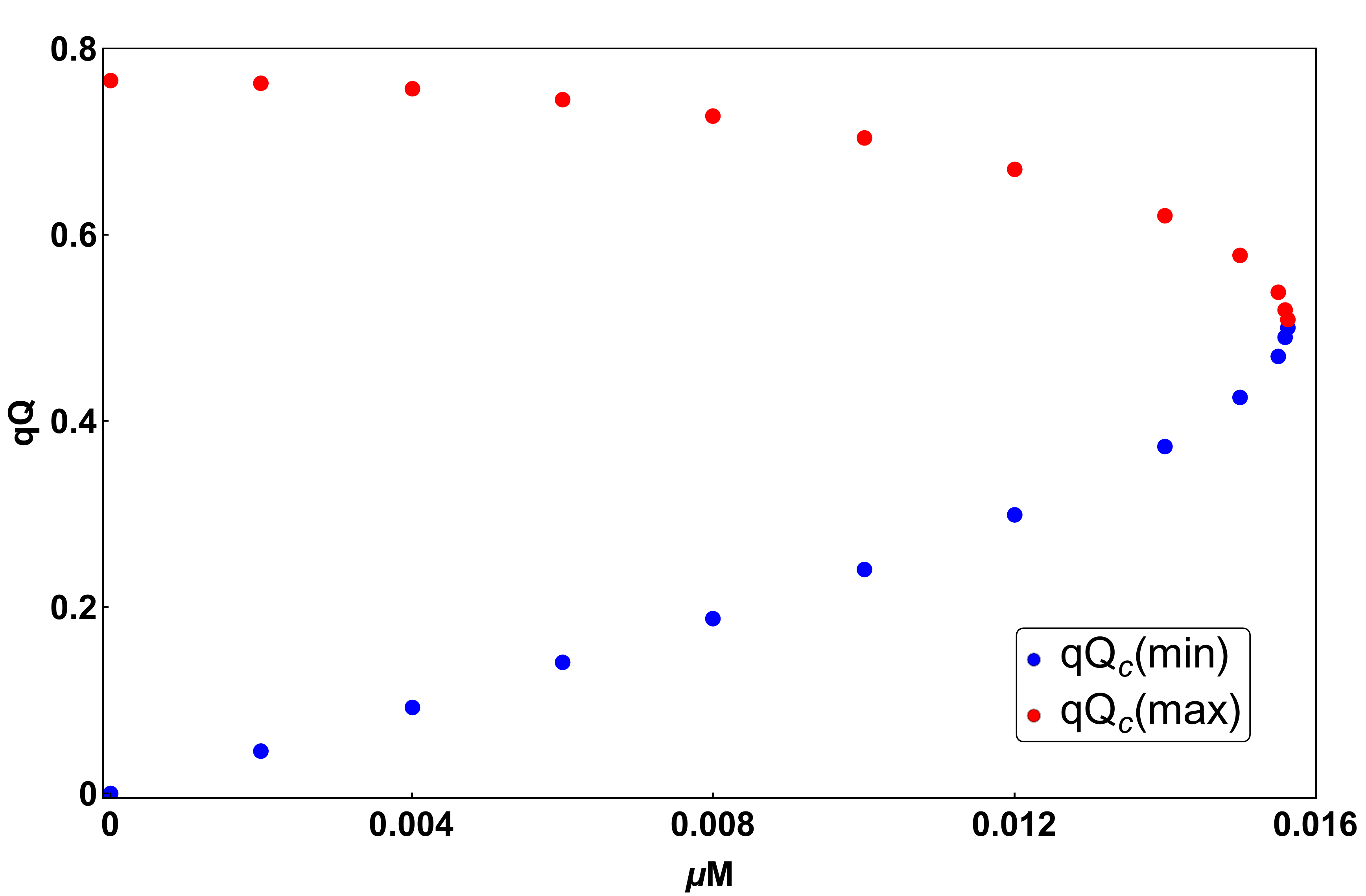}}\qquad
\subfigure{\includegraphics[scale=0.22]{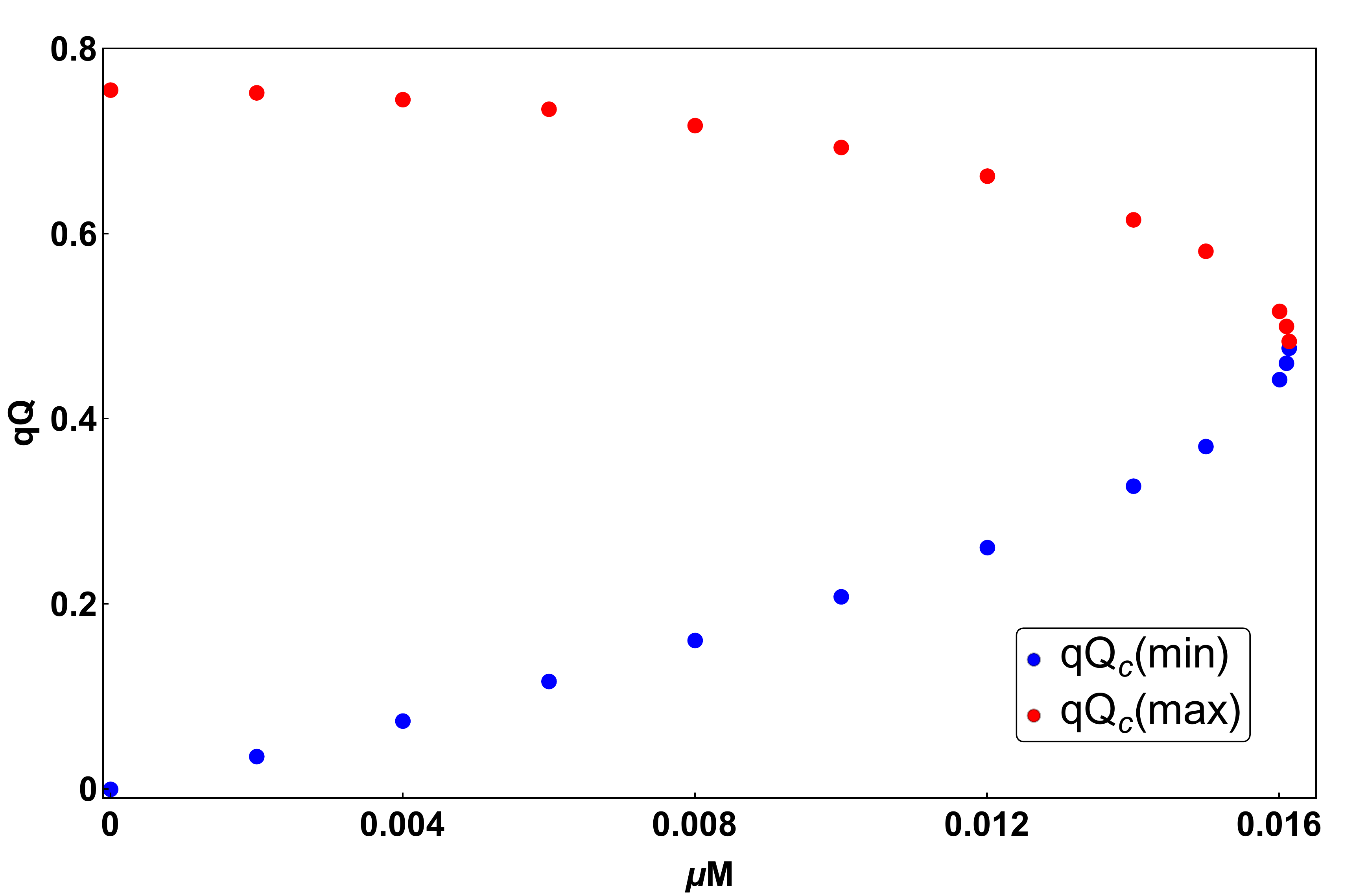}}
\caption{Critical charges $qQ_c(\text{min})$ and $qQ_c(\text{max})$ versus the scalar mass $\mu M$. The perturbation has angular momentum $l=0$ and propagates on a fixed RNdS with $\Lambda M^2=0.005$, $Q=0.5$ (left) and $Q=0.999\, Q_\text{max}$ (right).}
\label{massive_L005_qc}
\end{figure}

In Fig. \ref{massive_L005_Im} we demonstrate that even massive charged scalar perturbations can give rise to instabilities, but a small scalar mass $\mu M$ is proven to be enough to stabilize the system. As the scalar mass increases, two critical charge couplings arise, $qQ_\text{min}$ beyond which instabilities arise and $qQ_\text{max}$, beyond which stability is restored (see Fig. \ref{massive_L005_qc}). For a proper scalar mass and BH parameters, $qQ_\text{min}=qQ_\text{max}$, which designates that at this point there is only a purely oscillatory mode that does not decay. This might be a proper set of parameters where spontaneous scalarization of the BH can happen.

\begin{figure}[H]
\subfigure{\includegraphics[scale=0.22]{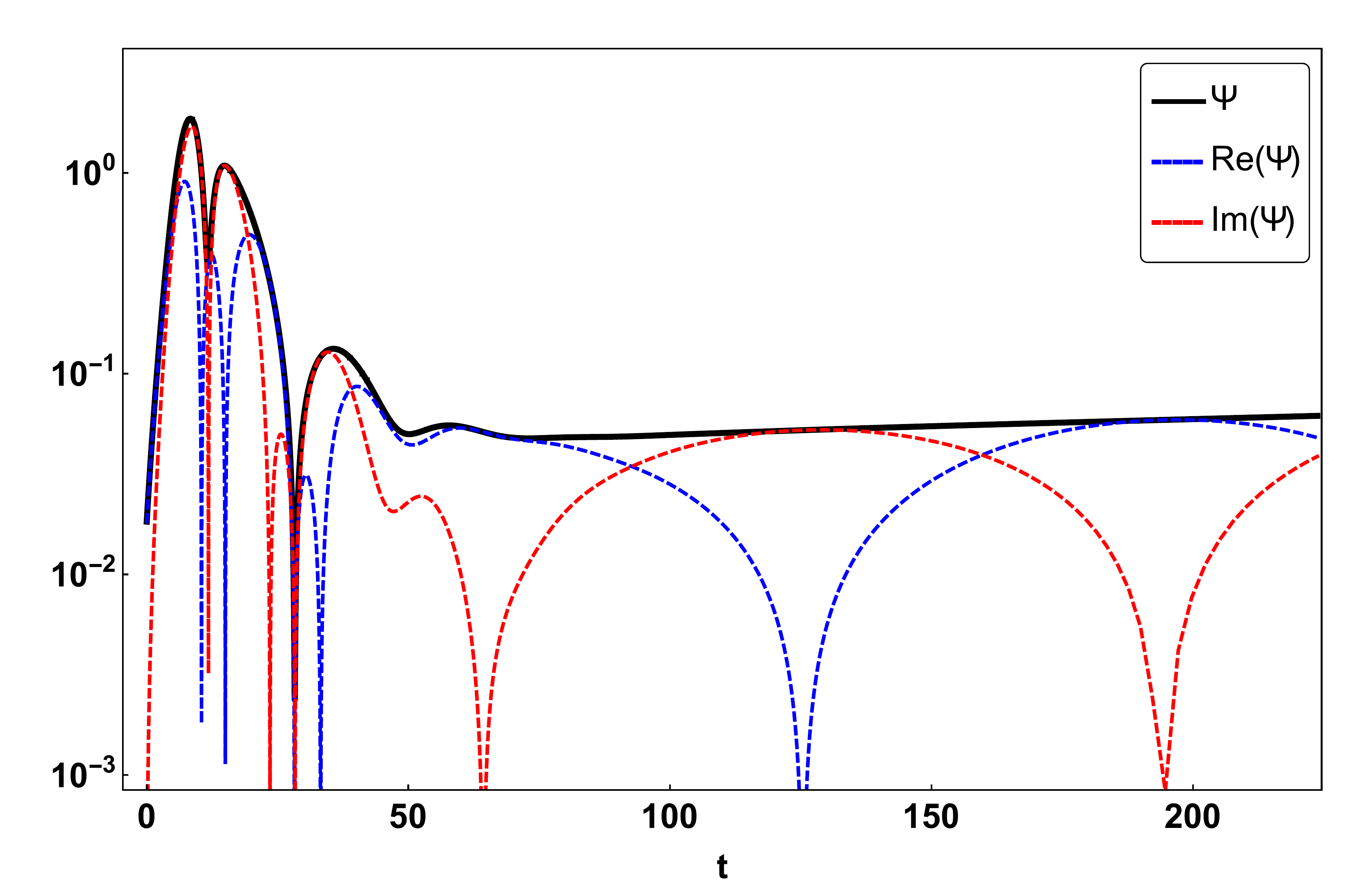}}\qquad
\subfigure{\includegraphics[scale=0.22]{plots/massless_evolution}}
\caption{Time evolution of the $l=0$ charged scalar perturbation on a fixed RNdS background, where $Q/M=0.5$, $qQ=0.45$, $\Lambda M^2=0.005$ and $\mu M=0$ (left), $\mu M=10^{-2}$ (right). The spectral prediction yields $\omega_\text{spc}=0.0229+0.0016i$ while the time evolution yields $\omega_\text{evl}=0.0228+0.0016i$ for the massless case and $\omega_\text{spc}=0.0226+0.0009i$, $\omega_\text{evl}=0.0226+0.0009i$ for the massive case. Both modes satisfy the superradiant condition.}
\label{evolutions}
\end{figure}

Although the calculations arise from highly precise spectral methods which has been tested through time, we performed another test to justify our results by comparing the unstable QNMs with the ones extracted by time evolutions based on the method in \cite{Luna:2018jfk}. The extracted modes from the time evolution data show great agreement with our spectral results. They also justify that even for $l=0$ massive, charged scalar perturbations in RNdS, we still have growing evolution of the perturbations with respect to time (see Fig. \ref{evolutions}).
In table \ref{table12} we show that all massless perturbations satisfy the superradiant relation \eqref{suprad1}. The same hold for some massive perturbations as well. The increment of the scalar beyond a specific value leads to stable modes that are not superradiant.

\begin{table}[H]
\centering
\scalebox{0.6}{
\begin{tabular}{||c| c | c | c ||} 
\hline
  \multicolumn{4}{||c||}{$Q/M=0.5$} \\
   \hline
\hline
  \multicolumn{4}{||c||}{$\Lambda M^2=0.005$} \\
   \hline
    $qQ$ & $\omega$ & $qQ/r_c$ &$qQ/r_+$ \\ [0.5ex]
   \hline
    0.005 & 0.00023 +  3.7$\times$$10^{-7}$i  & 0.00021   &0.00266 \\
   \hline
    0.05 & 0.00229 + 0.00004 i  & 0.00213   &0.02663 \\
   \hline
   0.5 & 0.02577 + 0.00159 i  & 0.02134   &0.26625 \\
   \hline
   1 & 0.05322 - 0.00271 i & 0.04268&0.53251 \\ 
   \hline
   5 &0.21907 - 0.01736 i & 0.21338& 2.66253\\
   \hline
   10 &0.42966 - 0.01832 i &0.42677 &5.32507 \\
   \hline\hline
   \multicolumn{4}{||c||}{$\Lambda M^2=0.05$} \\
    \hline
     $qQ$ & $\omega$ & $qQ/r_c$ &$qQ/r_+$ \\ [0.5ex]
   \hline
    0.005 & 0.00092 - 3.2$\times$$10^{-7}$i  & 0.00077   &0.00249 \\
    \hline
    0.05 & 0.00925 - 0.00003 i  & 0.00773   &0.02486 \\  
    \hline
    0.5 &  0.09222 - 0.00430 i & 0.07734   & 0.24859 \\
    \hline
    1 & 0.17777 - 0.01595 i &0.15468 &0.49719 \\ 
    \hline
    5 &0.78240 - 0.04099 i &0.77338 &2.48594 \\
    \hline
    10 &1.55129 - 0.04204 I &1.54676 &4.97188 \\
    \hline
\end{tabular}
}
\scalebox{0.6}{
\begin{tabular}{||c| c | c | c ||} 
\hline
  \multicolumn{4}{||c||}{$Q/Q_\text{max}=0.999$} \\
   \hline
\hline
  \multicolumn{4}{||c||}{$\Lambda M^2=0.005$} \\
   \hline
    $qQ$ & $\omega$ & $qQ/r_c$ &$qQ/r_+$ \\ [0.5ex] 
  \hline
   0.005 &0.00022 + 7$\times 10^{-7}$ i &0.00021  &0.00477 \\
   \hline
   0.05  &0.00224 + 0.00006 i  &0.00213  &0.04769 \\
   \hline
   0.5   &0.02599 + 0.00169 i  & 0.02132 &0.47694 \\
   \hline
   1 & 0.05320 - 0.00278 i &0.04264 & 0.95388\\ 
   \hline
   5 &0.21892 - 0.01739 i &0.21322 &4.76941 \\
   \hline
   10 &0.42934 - 0.01836 i & 0.42643&9.53881 \\
   \hline\hline
   \multicolumn{4}{||c||}{$\Lambda M^2=0.05$} \\
    \hline
     $qQ$ & $\omega$ & $qQ/r_c$ &$qQ/r_+$ \\ [0.5ex] 
    \hline
       0.005 & 0.00086 + $2.2\times 10^{-6}$ i &0.00076    &0.00461 \\
       \hline
       0.05 & 0.00872 + 0.00016 i  & 0.00761   &0.04608 \\
       \hline
       0.5 &  0.09225 - 0.00311 i & 0.07612   &0.46082 \\
       \hline
       1 & 0.17715 - 0.01602 i &0.15225 &0.92163 \\ 
       \hline
       5 &0.77125 - 0.04331 i & 0.76123&4.60816 \\
       \hline
       10 &1.52749 - 0.04455 i &1.52245 &9.21631 \\
    \hline
\end{tabular}
}
\scalebox{0.65}{
\begin{tabular}{||c| c | c | c ||} 
\hline
  \multicolumn{4}{||c||}{$\Lambda M^2=0.005$, $Q/Q_\text{max}=0.999$, $qQ=0.45$} \\
   \hline
    $\mu M$ & $\omega$ & $qQ/r_c$ &$qQ/r_+$ \\ [0.5ex] 
  \hline
   $10^{-3}$ &0.02305 + 0.00173 i &0.01919  &0.42925 \\
   \hline
   $10^{-2}$  & 0.02280 + 0.00107 i  &0.01919  &0.42925 \\
   \hline
   $2\times 10^{-2}$   & 0.02202 - 0.00098 i &0.01919  &0.42925 \\
   \hline
   $3\times 10^{-2}$ &0.02054 - 0.00458 i  &0.01919  &0.42925 \\ 
   \hline
   $4\times 10^{-2}$ &0.01789 - 0.01007 i &0.01919  &0.42925 \\
   \hline
   $5\times 10^{-2}$ &0.01263 - 0.01796 i &0.01919  &0.42925 \\
   \hline
   \multicolumn{4}{||c||}{$\Lambda M^2=0.05$, $Q/Q_\text{max}=0.999$, $qQ=0.2$} \\
    \hline
     $\mu M$ & $\omega$ & $qQ/r_c$ &$qQ/r_+$ \\ [0.5ex] 
    \hline
       $10^{-3}$ &0.03624 + 0.00063 i  &0.03045  &0.18433 \\
       \hline
       $10^{-2}$ & 0.03619 + 0.00044 i  &0.03045    &0.18433 \\
       \hline
       $5\times 10^{-2}$ & 0.03497 - 0.00418 i  & 0.03045   &0.18433 \\
       \hline
       $10^{-1}$ & 0.03008 - 0.02012 i  &0.03045 &0.18433 \\ 
       \hline
       $1.2\times 10^{-1}$ &0.02604 - 0.03086 i &0.03045 &0.18433 \\
       \hline
       $1.4\times 10^{-1}$ &0.01911 - 0.04555 i &0.03045 &0.18433 \\
    \hline
\end{tabular}
}
\caption{Dominant $l=0$ massless/massive charged scalar modes in RNdS spacetime for various parameters.}
\label{table12}
\end{table}
\end{appendices}
\bibliographystyle{unsrt}
\bibliography{all_references}
\end{document}